\documentclass[default]{sn-jnl}% Default with double column layout

\jyear{2022}%
\theoremstyle{thmstyleone}%
\newtheorem{theorem}{Theorem}%  meant for continuous numbers
\newtheorem{proposition}[theorem]{Proposition}%

\theoremstyle{thmstyletwo}%

\theoremstyle{thmstylethree}%

\usepackage{amsmath}
\usepackage{graphicx,psfrag,epsf}
\usepackage{enumerate}
\usepackage{thumbpdf,lmodern}
\usepackage{amsmath,amssymb}
\usepackage{natbib}
\setcitestyle{authoryear,open={(},close={)}}

\usepackage{mathtools}

\usepackage{mathtools,amssymb,lipsum}

\usepackage{graphicx}
\usepackage{amsthm}
\usepackage{float}
\usepackage{dsfont}
\usepackage[thinlines]{easytable}
\usepackage{caption}
\usepackage{subcaption}
\usepackage{subcaption}
\usepackage{multirow}
\usepackage{makecell}
\usepackage{booktabs}
\usepackage{rotating}

\newtheorem*{assumption*}{\assumptionnumber}
\providecommand{\assumptionnumber}{}
\makeatletter
\newenvironment{assumption}[2]
 {%
  \renewcommand{\assumptionnumber}{Assumption #1#2}%
  \begin{assumption*}%
  \protected@edef\@currentlabel{#1#2}%
 }
 {%
  \end{assumption*}
 }
\makeatother

\foreach \x in {a,...,z}{
    \expandafter\xdef\csname bf\x
        \endcsname{
            \noexpand\ensuremath{\noexpand\mathbf{\x}}
        }
}

\foreach \x in {A,...,Z}{
    \expandafter\xdef\csname bf\x
        \endcsname{
            \noexpand\ensuremath{\noexpand\mathbf{\x}}
            }
    \expandafter\xdef\csname cal\x
        \endcsname{
            \noexpand\ensuremath{\noexpand\mathcal{\x}}
            }
    \expandafter\xdef\csname bb\x
        \endcsname{
            \noexpand\ensuremath{\noexpand\mathbb{\x}}
            }
    \expandafter\xdef\csname frak\x
        \endcsname{
            \noexpand\ensuremath{\noexpand\mathfrak{\x}}
            }
}
\makeatletter
\newcommand*{\rom}[1]{\expandafter\@slowromancap\romannumeral #1@}
\makeatother
\renewcommand{\d}{\mathrm{d}}
\def\toolbox{\href{structure\_factor}{https://github.com/For-a-few-DPPs-more/structure-factor}}
\renewcommand\i{\mathrm{i}}

\newcommand{\Leb}{\calL^{d}}

\newcommand{\Var}{\bbV\text{ar}}
\newcommand{\leb}[1]{ \lvert #1 \rvert }
\begin{document}

\title[Article Title]{On estimating the structure factor of a point process, with applications to hyperuniformity}

\author*[1,2]{\fnm{Diala} \sur{Hawat} }\email{diala.hawat@univ-lille.fr}

\author[1]{\fnm{Guillaume} \sur{Gautier}}
\author[1]{\fnm{R\'emi} \sur{Bardenet}}
\author[2]{\fnm{Rapha\"el} \sur{Lachi\`eze-Rey}}

\affil[1]{\orgname{Universit\'e de Lille, CNRS, Centrale Lille, UMR 9189 -- CRIStAL} \orgaddress{\city{Lille}, \country{France}}}
%\\ \orgaddress{\postcode{F-59000 Lille}, \city{Lille}, \postcode{100190}, \country{France}}
\affil[2]{ \orgname{Universit\'e Paris Cit\'e, MAP5}, \orgaddress{\city{Paris}, \country{France}}}

\abstract{
  Hyperuniformity is the study of stationary point processes with a sub-Poisson variance in a large window.
  In other words, counting the points of a hyperuniform point process that fall in a given large region yields a small-variance Monte Carlo estimation of the volume.
  Hyperuniform point processes have received a lot of attention in statistical physics, both for the investigation of natural organized structures and the synthesis of materials.
  Unfortunately, rigorously proving that a point process is hyperuniform is usually difficult.
  A common practice in statistical physics and chemistry is to use a few samples to estimate a spectral measure called the structure factor.
  Its decay around zero provides a diagnostic of hyperuniformity.
  Different applied fields use however different estimators, and important algorithmic choices proceed from each field's lore.

  This paper provides a systematic survey and derivation of known or otherwise natural estimators of the structure factor.
  We also leverage the consistency of these estimators to contribute the first asymptotically valid statistical test of hyperuniformity.
  We benchmark all estimators and hyperuniformity diagnostics on a set of examples.
  In an effort to make investigations of the structure factor and hyperuniformity systematic and reproducible, we further provide the \texttt{Python} toolbox \toolbox{}, containing all the estimators and tools that we discuss.}



\maketitle

\newpage
\tableofcontents
\newpage

\section{Introduction}\label{sec1}
Condensed matter physicists have observed that, for some random particle systems, the variance of the number of points in a large window scales slower than the volume of that window, a phenomenon called \emph{hyperuniformity} \citep{Torquato:2018}.
This statistical property, in turn, implies desirable physical properties for materials \citep[Section 14]{Torquato:2018}.
Outside physics, hyperuniform point processes have generated broad interest in statistics, machine learning, and probability.
By definition, a hyperuniform point process leads to a fast-decaying Monte Carlo error when estimating volumes.
This variance reduction property makes hyperuniformity a natural concept for many statistical applications.
Projection determinantal point processes (DPPs), a particular class of hyperuniform point processes, have thus been proposed for Monte Carlo integration \citep*{BaHa20,CoMaAm21,BeBaCh19,BeBaCh20}, subsampling stochastic gradients \citep*{BaGhLi21}, or feature selection \citep*{BeBaCh20b}.

More generally, spectral properties of point configurations have been linked to variance reduction in Monte Carlo integration \citep*{Bhavya:2018,Pilleboue+al:2015}.
In probability, hyperuniform point processes also appear across all fields.
The Ginibre ensemble, arguably one of the most famous DPPs arising from random matrix theory \citep*{Anderson+Guionnet+Zeitouni:2010}, is a typical example of hyperuniform point processes.
In stochastic geometry, and beyond DPPs, hyperuniform point processes appear, e.g., in the zeros of Gaussian analytic functions \citep*{Hough+al:2013}, or matching constructions \citep*{Klatt+Last+Yogeshwaran:2020}.
Intriguingly, there is also empirical evidence that repeatedly applying an algorithm involving local repulsion of points to an initial point process leads to a hyperuniform point process \citep{Klatt+al:2019}.
Across all these scientific fields, there is a need for statistical diagnostics of hyperuniformity.

In theory, under mild assumptions, a point process in $\mathbb{R}^d$ is hyperuniform if and only if its structure factor $S(\bfk)$ decays to zero as $\|\bfk \|_2$ goes to zero \citep{Coste:2021}.
The structure factor is the Fourier transform of a measure that encodes the pairwise correlations of a point process, and the behavior in zero of the structure factor reflects long-range correlations in the original point process.
Given a realization of a point process, the standard empirical diagnostic of hyperuniformity thus involves estimating and plotting the structure factor of the underlying point process \citep{Torquato:2018,Klatt+al:2019}.
This graphical assessment is however not standardized and often not described in full reproducible detail in the literature, with implementation choices and statistical properties often part of each field's folklore.
There is also no standard software to estimate structure factors.
In this paper, we contribute $(i)$ a survey of existing estimators of the structure factor and their main properties, $(ii)$ a new asymptotically valid statistical test of hyperuniformity, and $(iii)$ a modular, open-source Python package\footnote{\url{https://github.com/For-a-few-DPPs-more/structure-factor}} that implements all estimators, diagnostics, and the test discussed in the paper.

In Section~\ref{sec:point_processes_hyperuniformity_and_the_structure_factor}, we introduce hyperuniformity and the relevant background.
In Section~\ref{sec:Estimators of the structure factor}, we rederive two families of estimators of the structure factor.
The first family assumes that the underlying point process is stationary, while the second further requires isotropy.
We also review existing hyperuniformity diagnostics and contribute our test of hyperuniformity in Section~\ref{sec:The coupled sum estimator and a test of hyperuniformity}, using the seminal debiasing techniques of \cite{Rhee+Glynn:2015}.
We demonstrate estimators and diagnostics using our companion Python toolbox \toolbox{} in Section~\ref{sec:Illustrating the toolbox}, on four point processes exhibiting behaviors such as repulsion, spatial independence, and clustering.
In Section~\ref{sec:Comparison of the estimators}, we numerically compare the accuracy of the estimators.
We conclude with a few research leads in Section~\ref{Summary and discussion}.
% section introduction (end)
\paragraph{Related work}
The closest work to the survey part of our paper is the preprint of \citet*{Rajala+Olhede+John:2020}.
They introduce important new estimators based on the idea of tapering in time series analysis \citep{Percival+Walden:2020}, and investigate central limit theorems for their estimators.
On our end, we limit ourselves to a survey --~including  the estimators of \cite{Rajala+Olhede+John:2020}~-- and simpler properties like asymptotic unbiasedness and its relation to implicit implementation choices in statistical physics papers.
One reason for this is our motivation for the study of hyperuniformity: hyperuniform point processes are unlikely to satisfy the assumptions\footnote{E.g., Hypothesis (H4) of \citet{Biscio+Waagepetersen:2019}, when the linear statistic is the number of points, contradicts hyperuniformity.}
behind the central limit theorems referenced by \citet{Rajala+Olhede+John:2020}.
Moreover, our survey includes a broader choice of estimators, including numerical quadratures of Hankel transforms, and a companion Python package.
Overall, we believe that the survey part of our paper and the paper of \citet{Rajala+Olhede+John:2020} are complementary.\footnote{
    We note that during the reviewing process of our paper, a second version of the preprint \citep{Rajala+Olhede+John:2020} has been arXived \citep{Rajala+Olhede+John_V2:2020}.
    The changes in the new version do not seem to impact our work.
  }

\paragraph{Notation}
Throughout this paper, bold lowercase letters like $\bfx, \bfr$ indicate vectors in $\bbR^d$, and the corresponding non-bold characters are scalars.
In particular, $\bfx = (x_1, \ldots, x_d)$.
Whenever not confusing, we use the same letter in different fonts for a vector and its Euclidean norm, i.e., $r=\Vert \bfr \Vert_2$ and $k=\Vert \bfk \Vert_2$.

\section{Point processes and their structure factor} % (fold)
\label{sec:point_processes_hyperuniformity_and_the_structure_factor}

Investigating the hyperuniformity of a point process commonly goes through the visual inspection of its \emph{structure factor}, also known as the \emph{scaled spectral density function}.
In this section, we introduce the key definitions to understand that procedure.
For a general reference on point processes, we refer to \citet*{Chiu+Stoyan+al:2013}.
For hyperuniformity and structure factors, we primarily refer to \citet{Torquato:2018} for readers with a physics background, and to \citet{Coste:2021} for readers with a mathematics background.

\subsection{Point processes and their correlation measures} % (fold)
\label{sub:point_processes}

Loosely speaking, a point process is a random set of points, such that we can count the number of points falling in any observation window.
More formally, let a \emph{configuration} of $\bbR^{d}$ be a locally finite set $\calX\subset \bbR^{d}$, i.e., for any compact $K$ of $\bbR^d$, the cardinality $\calX(K)$ of $\calX\cap K$ is finite.
The family $\calN$ of configurations of $\bbR^{d}$ is endowed by the $\sigma$-algebra generated by the mappings $\calX\mapsto \calX (K)$, for $K$ compact.
A \emph{point process} is a random element of $\calN$.

In this paper, we restrict ourselves to \emph{simple} and \emph{stationary (a.k.a. homogeneous)} point processes.
By \emph{simple}, we mean that the considered point process almost surely contains only distinct points.
By \emph{stationary}, we mean that the law of the point process $\calX$ is identical to that of
$\calX + \bfy
  \triangleq \{\bfx + \bfy ; \; \bfx \in \calX\}$,
for all $\bfy \in \bbR^{d}$. In that case, the \emph{intensity measure} $\rho^{(1)}$ of $\calX$, defined by
\begin{align*}
  \rho^{(1)}(A) = \bbE\left[\calX(A)\right],
\end{align*}
is proportional to the Lebesgue measure.
We then write $\rho^{(1)}(\d  \bfx) = \rho\, \d  \bfx$, and $\rho\geq 0$ is called the \emph{intensity} of $\calX$.

Assessing pairwise correlations in a point process, and --~as we shall see~-- hyperuniformity, usually goes through estimating the two-point correlation measure $\rho^{(2)}$ of $\calX$.
It is formally defined by
\begin{equation*}
  \bbE
  \left[
    \sum_{\bfx, \bfy \in \calX }^{\neq}
    f(\bfx, \bfy)
    \right]
  = \int_{\bbR^d \times \bbR^d}
  f(\bfx, \bfy)
  \rho^{(2)}(\d \bfx, \d \bfy),
\end{equation*}
for any nonnegative measurable bounded function $f$ with compact support.
If $\calX$ is stationary with intensity $\rho\geq 0$, the previous expression can be factorized in
\begin{equation}
  \label{eq:pcf_by_expectation}
  \int_{\bbR^d \times \bbR^d} f(\bfx + \bfy, \bfy)\rho^{2} g_{2}(\d\bfx) \d\bfy,
\end{equation}
where $g_2$ is called the \emph{pair correlation measure}.

Intuitively, $g_2(\d \bfx)$ is the probability that $\calX$ has a point at location $\d \bfx$, given that $\calX$ already contains the origin.
If in addition $g_2$ has a density $g$ w.r.t.\ the Lebesgue measure, i.e., $g_{2}(\d\bfx) = g(\bfx) \d\bfx$, then $g$ is called the \emph{pair correlation function} of $\calX$.
Informally, $g(\bfr)$ counts the pairs $(\bfx, \bfy) \in \calX \times \calX$ such that $\bfx-\bfy=\bfr\in \bbR^d$.
Finally, when $\cal X$ is assumed both stationary and \emph{isotropic}, i.e., the distribution of $\calX$ is both, translation- and rotation-invariant, then the pair correlation function $g$ is radial.
In that case, we abusively write $g(\bfr) = g(r)$, where $r=\Vert \bfr\Vert_2$.
Finally, note that throughout the paper, the assumptions of stationarity and isotropicy can be straightforwardly weakened to assuming, e.g., that the intensity measure is invariant to translations or that the pair correlation function only depends on the inter-point distance.
Yet, for simplicity, we prefer to stick with assuming that the point process under scrutiny is itself stationary/isotropic.

\subsection{The Fourier transform} % (fold)
\label{sub:fourier_transform}

The Fourier transform $\calF$ of an integrable function $f : \bbR^d \to \bbC$ is the square integrable function
\begin{equation*}
  \calF(f)(\bfk)
  =
  \int_{\bbR^d}
  f(\bfx)
  e^{-\i \left\langle \bfk, \bfx \right\rangle}
  \d \bfx,
\end{equation*}
where $\left\langle \bfk, \bfx \right\rangle$ is the dot product of the \emph{wavevector} $\bfk$ by $\bfx$.
If $f$ is furthermore a radial function, denoted abusively by $f(\bfr)=f(r)$, where $r = \|\bfr\|_2$, then the Fourier transform of $f$ is the corresponding \emph{symmetric Fourier transform} $\calF_{s}$, namely
\begin{align}
  \label{eq:radial_fourier}
  \calF_{s}(f)(k) &= \calF(f)(\bfk)= \nonumber\\
  &(2 \pi)^{d/2}
  \int_0^{\infty}
  r^{d/2}f(r)
  \frac{J_{d/2 -1}(kr)}{k^{d/2 -1}}
  \d r,
\end{align}
where $k=\|\bfk\|_2$ is called a \emph{wavenumber}, and $J_\nu$ is the Bessel function of the first kind of order $\nu$ \citep{Osgood:2014}.
If we further define the \emph{Hankel transform} of order $\nu\geq -1/2$ as
\begin{equation}
  \label{eq:hankel_tranfrom}
  \calH_\nu(f)(k)
  = \int_0^\infty f(r) J_\nu(kr) r \d r,
  \quad \text{for } k\geq 0,
\end{equation}
then \eqref{eq:radial_fourier} rewrites as
\begin{equation}
  \label{eq:hankel_and_Fourier}
  \calF_{s}(f)(k)
  =
  \frac
  {(2\pi)^{d/2}}
  {k^{d/2 -1}}
  \calH_{d/2 -1}(\tilde f)(k),
\end{equation}
where $\tilde{f}: x\mapsto f(x) x^{d/2 -1}$.
Finally, note that the Fourier transform can be generalized to tempered distributions through duality \cite[Appendix B]{Coste:2021}.

We now introduce a function, the Fourier transform of which will be central in studying the properties of spectral estimators based on point processes.
Let $W \subset \bbR^d$ be a convex set, later called \emph{observation window}, and denote by $\leb{W}$ its Lebesgue measure, i.e., its volume.
The \emph{scaled intersection volume} is the function
\begin{equation}
  \label{eq:alpha_2}
  \alpha_0(\bfr, W)
  = \frac{1}{\leb{W}}
  \int_{\bbR^d} \mathds{1}_W(\bfr + \bfy)\mathds{1}_W(\bfy)d \mathrm{\bfy},
\end{equation}
and its Fourier transform is given by
\begin{equation}
  \label{eq:fourier_of_alpha2}
  \calF(\alpha_0)(\bfk, W)
  = \frac{1}{\leb{W}} \left( \calF(\mathds{1}_W)(\bfk)\right)^2.
\end{equation}
%see also \citet[Section 3.1]{Torquato:2018}.
In particular, if $W=B^d(\mathbf{0}, R)$ is the Euclidean ball of radius $R$, the scaled intersection volume function \eqref{eq:alpha_2} is radial \citep[Section 3.1.1]{Torquato:2018}.
In this case, we abusively write $\alpha_0(\mathbf{r}, W) = \alpha_0(r, W)$ and
\begin{equation}
  \label{eq:ft_alpha0_radial}
  \calF_s(\alpha_0)(k, W) = 2^d \pi^{d/2}\frac{\Gamma(1+d/2)}{k^d}J^2_{d/2}(kR),
\end{equation}
where $\Gamma$ is Euler's Gamma function.
On the other hand, if $W=\prod_{j=1}^d[-L_j/2, L_j/2]$, Equation~\eqref{eq:fourier_of_alpha2} simplifies to
\begin{equation}
  \label{eq:ft_alpha_0_box}
  \calF(\alpha_0)(\bfk, W)
  = \left(
  \prod_{j=1}^d \frac{\sin(k_j L_j/2)}{k_j \sqrt{L_j}/2}
  \right)^2.
\end{equation}

\subsection{The structure factor} % (fold)
\label{sub:structure_factor}

The \emph{structure factor measure} $\calS$ of a stationary point process $\calX$ in $\bbR^d$ with intensity $\rho > 0$ is the measure on $\bbR^{d}$, when it exists, defined by
\begin{equation}
  \label{eq:structure_factor_measure}
  \calS
  = \Leb + \rho \calF[g_2 -\Leb]
  = \calF(\delta_{0} + \rho (g_{2} - \Leb)),
\end{equation}
where $\Leb$ is the $d$-dimensional Lebesgue measure and $\delta _{0}$ is the Dirac mass in $0$.
We note that in the spectral analysis of stochastic processes $\calS$, up to a scale factor, is also called the \textit{Bartlett spectral measure \citep*[Section 5.2]{Bremaud:2014}}.
%The behavior of $\calS$ depends on how $g_2$ fluctuates around $1$.

When
%$g_2 - \Leb$ has a bounded density $g - 1$ w.r.t.\ the Lebesgue measure, then
the measure $\calS$ is absolutely continuous w.r.t.\ the Lebesgue measure, i.e.,
$\calS(\d \bfk)
  = S(\bfk) \d \bfk$,
we call $S$ the \emph{structure factor}.\footnote{
  The literature is inconsistent as to whether the structure factor is the measure $\calS$ or its density $S$. We choose the density, which is also sometimes known as the \emph{scaled spectral density function}.
}
When $g_2$ is absolutely continuous w.r.t.\ the Lebesgue measure and $g-1$ is integrable, $S$ can be explicitly written as
\begin{equation}
  \label{eq:stucture_factor_function}
  S(\bfk) = 1 + \rho \calF(g-1)(\bfk).
\end{equation}
If $\calX$ is further assumed to be isotropic with intensity $\rho > 0$, then both the pair correlation function $g$ and the structure factor $S$ are radial functions.
Abusively denoting $S(k) = S(\bfk)$, Equation~\eqref{eq:stucture_factor_function} reads
\begin{equation}
  \label{eq:sf_as_hankel_tranform}
  S(k)= 1 + \rho \calF_s(g-1)(k),
\end{equation}
see \citet[Section 2]{Torquato:2018}, which can be expressed analytically using \eqref{eq:hankel_and_Fourier} as
\begin{equation}
  \label{eq:simplified_radial_structure_factor}
  S(k) = 1 + \rho \frac{ (2\pi)^{d/2}}{k^{d/2-1}}
  \int_{0}^{\infty} (g(r) -1) r^{d/2} J_{d/2 - 1}(kr) ~\d r.
\end{equation}
Finally, if the pair correlation function $g$ exists and is smooth, then one can expect that
$$
  S(\bfk) \xrightarrow[\|\bfk\|_2 \to \infty]{} 1.
$$
On the other hand, the behavior of $S$ in zero measures the fluctuations of $g$ around $1$ at large scales $\Vert \mathbf{r}\Vert_2\gg 1$, which can in turn help quantify properties like hyperuniformity.

%end section

\subsection{Hyperuniformity} % (fold)
\label{sub:hyperuniformity}

A point process $\calX$ of $\bbR^d$ is often said to be \emph{hyperuniform} (or \emph{super-homogeneous}) if the variance of the number of points that fall in a Euclidean ball scales slower than the volume of that ball, i.e.
\begin{equation}
  \label{eq:variance_hyp}
  \lim_{R\to \infty}
  \frac
  {\mathrm{Var} \left[ \calX( B(\mathbf{0}, R) ) \right]} {\leb{B(\mathbf{0}, R)}}
  = 0.
\end{equation}
Some comments are in order.
Although hyperuniformity \emph{a priori} depends on the shape of the window, e.g., a ball in \eqref{eq:variance_hyp}, mild technical assumptions allow to show that the definition is robust to the choice of the growing window \citep[Section 2]{Coste:2021}.
Second, being hyperuniform is not a standard feature of point processes; a homogeneous Poisson process, for instance, is not hyperuniform, as the ratio in \eqref{eq:variance_hyp} is a positive constant.
Third, the most general definition of hyperuniformity goes through the structure factor of a point process \eqref{eq:structure_factor_measure};
under the mild assumption that $g_2-1$ is a signed measure, \eqref{eq:variance_hyp} is equivalent to $S(\mathbf{0})=0$, see \citet[Proposition 2.2]{Coste:2021} and \citet[Section 5.3.1]{Torquato:2018}.
Moreover, \citet[Section 5.3.2]{Torquato:2018} states that if the structure factor undergoes a power decay $\vert S(\bfk)\vert\sim c \Vert \bfk \Vert_2^\alpha$ in the neighborhood of zero, the process can be classified into three categories depending on how $\alpha$ compares to 1, as summarized in Table~\ref{tab:class_of_hyperuniformity}.

\begin{table}[!ht]
  \centering
  \caption{Classes of hyperuniformity} \label{tab:class_of_hyperuniformity}
  \begin{tabular}{|l|c|l|}
    \hline
    Class
     & $\alpha$
     & $\mathrm{Var}\left[\calX(B(0,R))\right]$
    \\
    \hline
    \rom{1}
     & $> 1$
     & $\mathcal{O}(R^{d-1})$
    \\
    \rom{2}
     & $= 1$
     & $\mathcal{O}(R^{d-1}\log(R))$
    \\
    \rom{3}
     & $\in ]0,1[$
     & $\mathcal{O}(R^{d-\alpha})$              \\
    \hline
  \end{tabular}
  %\footnotetext{Classes of hyperuniformity for point processes satisfying $\vert S(\bfk)\vert \underset{\bfk \to 0}{\sim} c ~ \Vert \bfk \Vert_2^\alpha$.}
\end{table}

Since the variance cannot grow more slowly than $R^{d-1}$ for a spherical window \citep{Beck:1987}, Class \rom{1} represents the \emph{strongest} form of hyperuniformity. It includes, for instance, the Ginibre point process, as we shall see in Section~\ref{sub:point_processes}.
Class \rom{2} hyperuniform processes include, for instance, the sine process, a central object in random matrix theory \citep{Anderson+Guionnet+Zeitouni:2010}.
By contrast, systems that fall in Class \rom{3} present the \emph{weakest} form of hyperuniformity.
Many stationary point processes are believed to be Class \rom{3}, although few rigorous proofs exist; one can mention the asymptotic result of \citep{Boursier:2021} for the one-dimensional Riesz gaz.

A final remark about  \eqref{eq:variance_hyp} is that it is in contradiction with superlinear variance, which is sometimes defined as \textit{long memory} or \textit{long-range dependence} \citep{Daley+al:1997,Samo}. We find the latter definition to be confusing given that a hyperuniform point process can also have long-range interactions through a non-summable reduced covariance function, such as the sine process; see \citep[Section 4.3.5]{Hough+al:2013}.

\subsection{Empirical diagnostics of hyperuniformity} % (fold)
\label{sub:Effective hyperuniformity}

As suggested in Section~\ref{sub:hyperuniformity}, the behavior in zero of the structure factor quantifies hyperuniformity.
Investigating whether one or several samples come from a hyperuniform point process is thus often carried out by estimating the structure factor, and then either visually inspecting the resulting plots around zero, or regressing the estimated values.
Say for a stationary and isotropic point process, one option is to regress $\log S$ onto $\log k$ around zero, to assert a potential value for the decay rate $\alpha$ in Table~\ref{tab:class_of_hyperuniformity}.

Another criterion of \emph{effective} hyperuniformity has been proposed; see \citet[Supplementary material]{Klatt+al:2019} and \citet[Section 11.1.6]{Torquato:2018}.
For a stationary and isotropic hyperuniform point process, given a set of estimated values
$$
  \{(k_1,\widehat{S}(k_1)),\dots,(k_n, \widehat{S}(k_n))\}
$$
with $0<k_1< \dots < k_n$,
the $H$-index is defined by
\begin{equation}
  \label{eq:H_index}
  H = \frac{\widehat{S}(0)}{\widehat{S}(k_{\mathrm{peak}})},
\end{equation}
where $\widehat{S}(0)$ is a linear extrapolation of the structure factor in $k=0$ based on the estimated values of $S$, and $k_{\mathrm{peak}}$ is the location of the first dominant peak value of the estimated structure factor, defined here as
\begin{align*}
  k_\mathrm{peak} = \inf_i \big \{ k_i; \widehat{S}(k_i)>1, \widehat{S}(k_{i-1})<\widehat{S}(k_i),
   \text{ and }\widehat{S}(k_{i+1})<\widehat{S}(k_i)\big\}.
  \end{align*}
If the set is empty, we set $\widehat{S}(k_{\mathrm{peak}})=1$ in \eqref{eq:H_index}.
% \begin{align*}
% S(k_\mathrm{peak}) = 1\wedge S\Big(\inf_i \big \{& k_i; \widehat{S}(k_i)>1, \widehat{S}(k_{i-1})<\widehat{S}(k_i),\\
%  &\text{ and }\widehat{S}(k_{i+1})<\widehat{S}(k_i)\big\}\Big).
% \end{align*}
When $H<10^{-3}$, the process is called \emph{effectively hyperuniform} by \citet[Section 11.1.6]{Torquato:2018}.
Note that the linear extrapolation is chosen for simplicity and not based on model selection.
Like the threshold of $10^{-3}$, the definition of a dominant peak is also somewhat arbitrary.
In Section~\ref{sec:The coupled sum estimator and a test of hyperuniformity}, we will introduce a novel statistical test that bypasses some of these arbitrary choices.

Finally, both an estimate of the decay rate from Table~\ref{tab:class_of_hyperuniformity} and the $H$-index \eqref{eq:H_index} require estimators of the structure factor.
Before we survey these estimators, we introduce a few benchmark point processes and discuss their structure factors.

\subsection{Benchmark point processes}
\label{sub:Some point processes and their structure factor}

\begin{figure*}[!ht]
  \centering
  \begin{subfigure}{0.24\textwidth}
    \centering
    \includegraphics[width=0.8\linewidth]{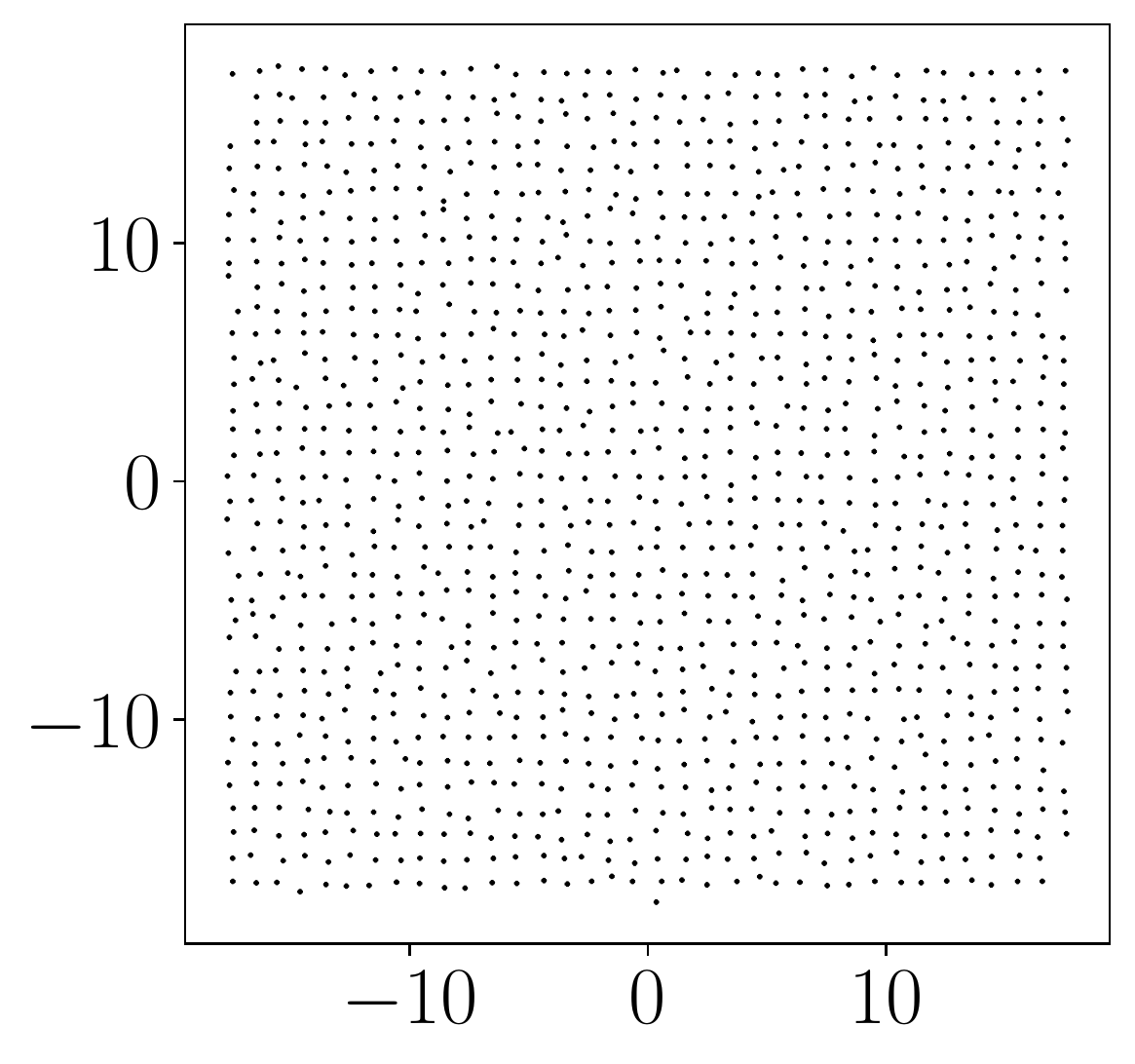}
    \caption{KLY }
    \label{fig:kly_pp}
  \end{subfigure}
  \begin{subfigure}{0.24\textwidth}
    \centering
    \includegraphics[width=0.8\linewidth]{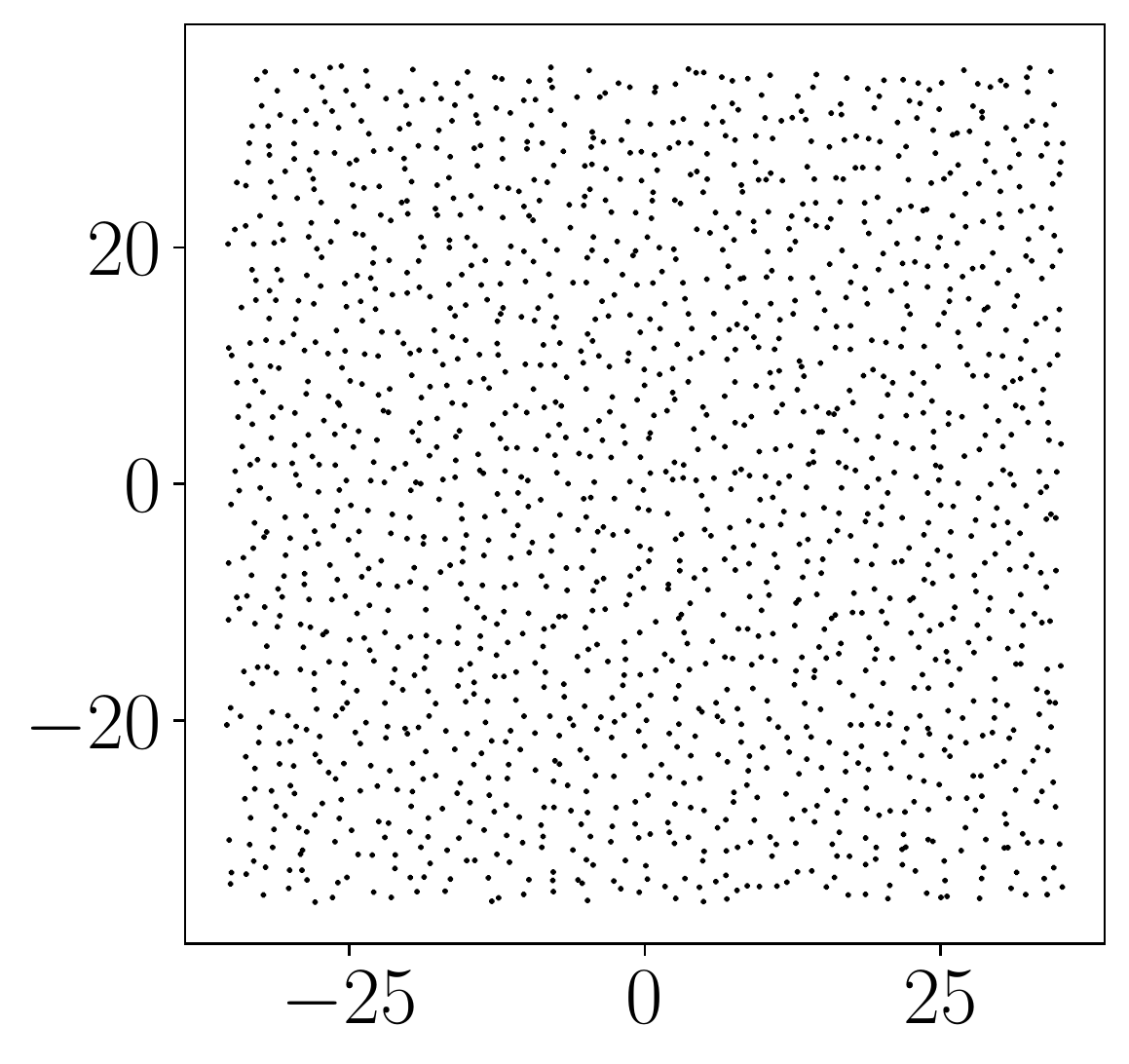}
    \caption{Ginibre }
    \label{fig:ginibre_pp}
  \end{subfigure}
  \begin{subfigure}{0.24\textwidth}
    \centering
    \includegraphics[width=0.8\linewidth]{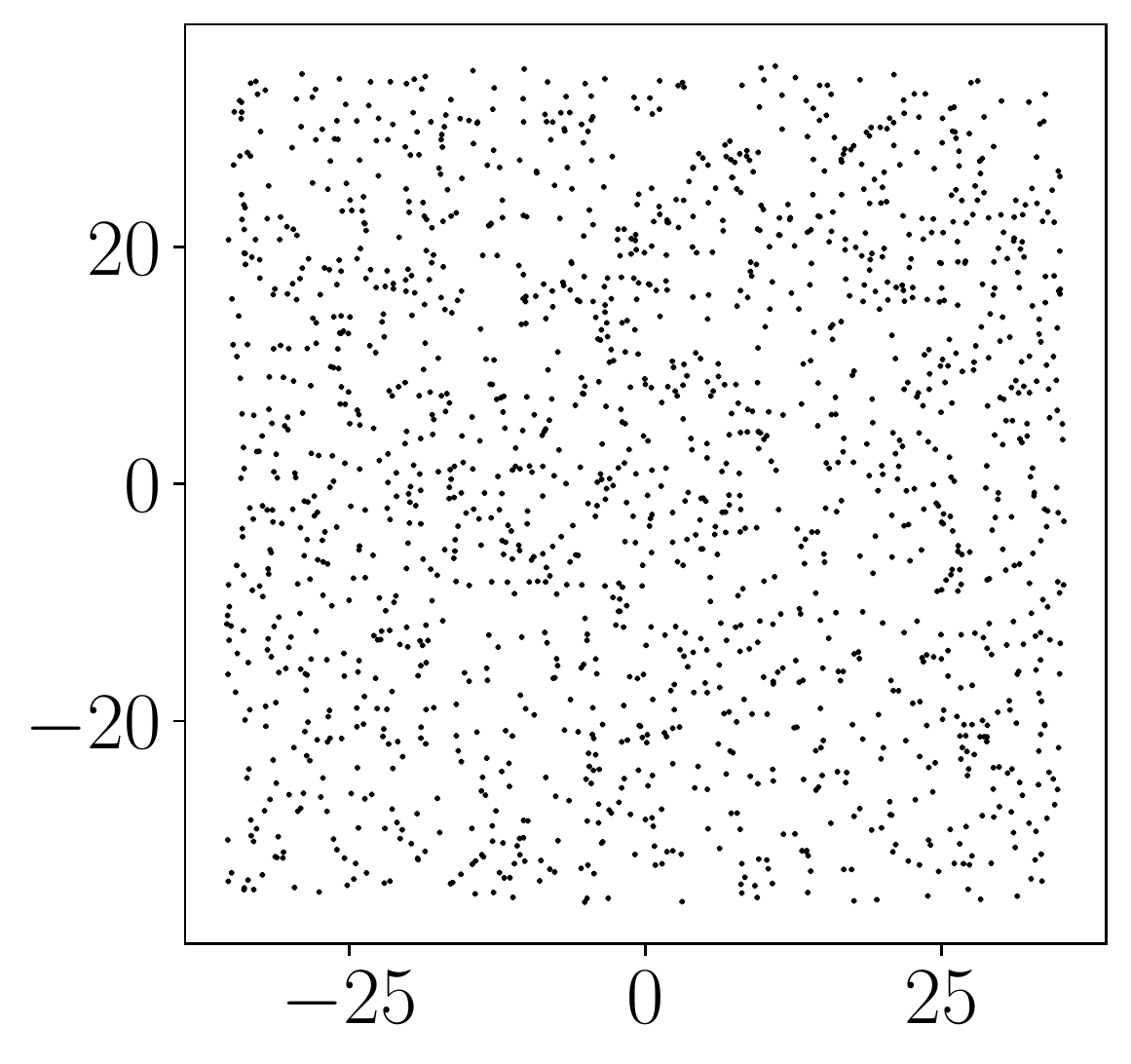}
    \caption{Poisson}
    \label{fig:poisson_pp}
  \end{subfigure}
  \begin{subfigure}{0.24\textwidth}
    \centering
    \includegraphics[width=0.8\linewidth]{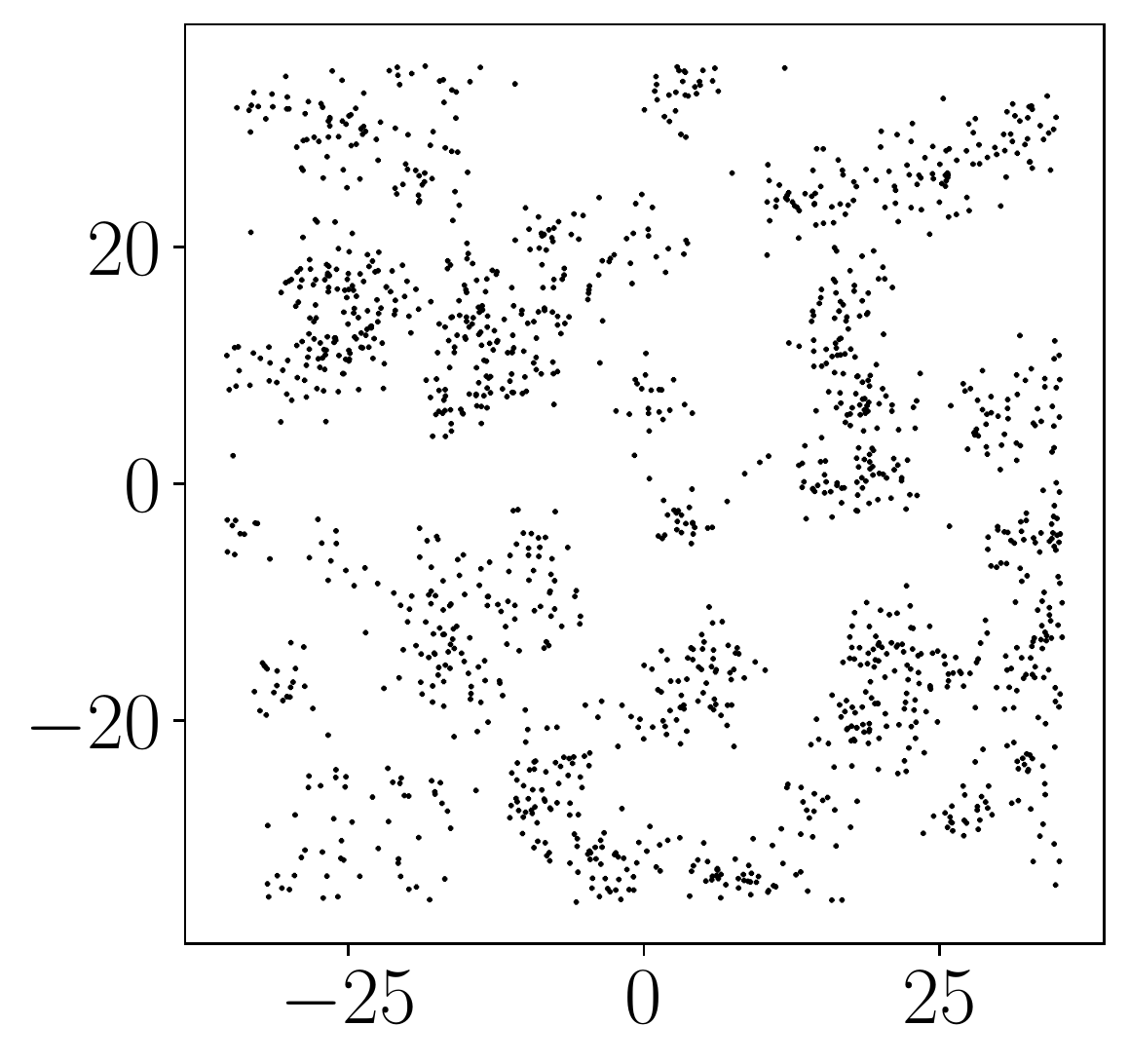}
    \caption{Thomas }
    \label{fig:thomas_pp}
  \end{subfigure}
  \caption{A sample of each of our four benchmark point processes}
\end{figure*}

\begin{figure*}[!ht]
  \centering
  \begin{subfigure}{0.24\textwidth}
    \centering
    \includegraphics[width=0.8\linewidth]{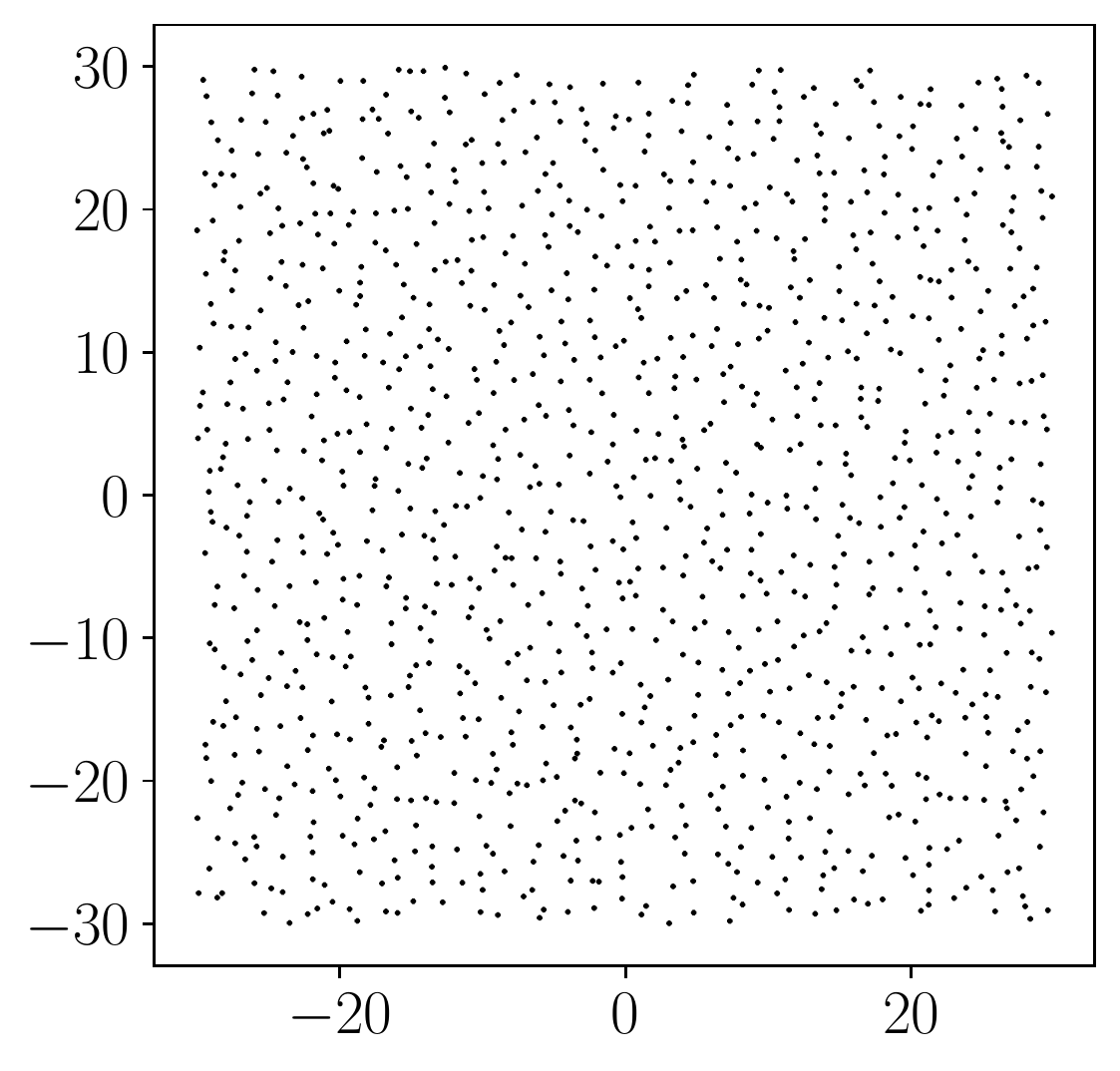}
    \caption{Ginibre}
    %\label{fig:ginibre}
  \end{subfigure}
  \begin{subfigure}{0.24\textwidth}
    \centering
    \includegraphics[width=0.8\linewidth]{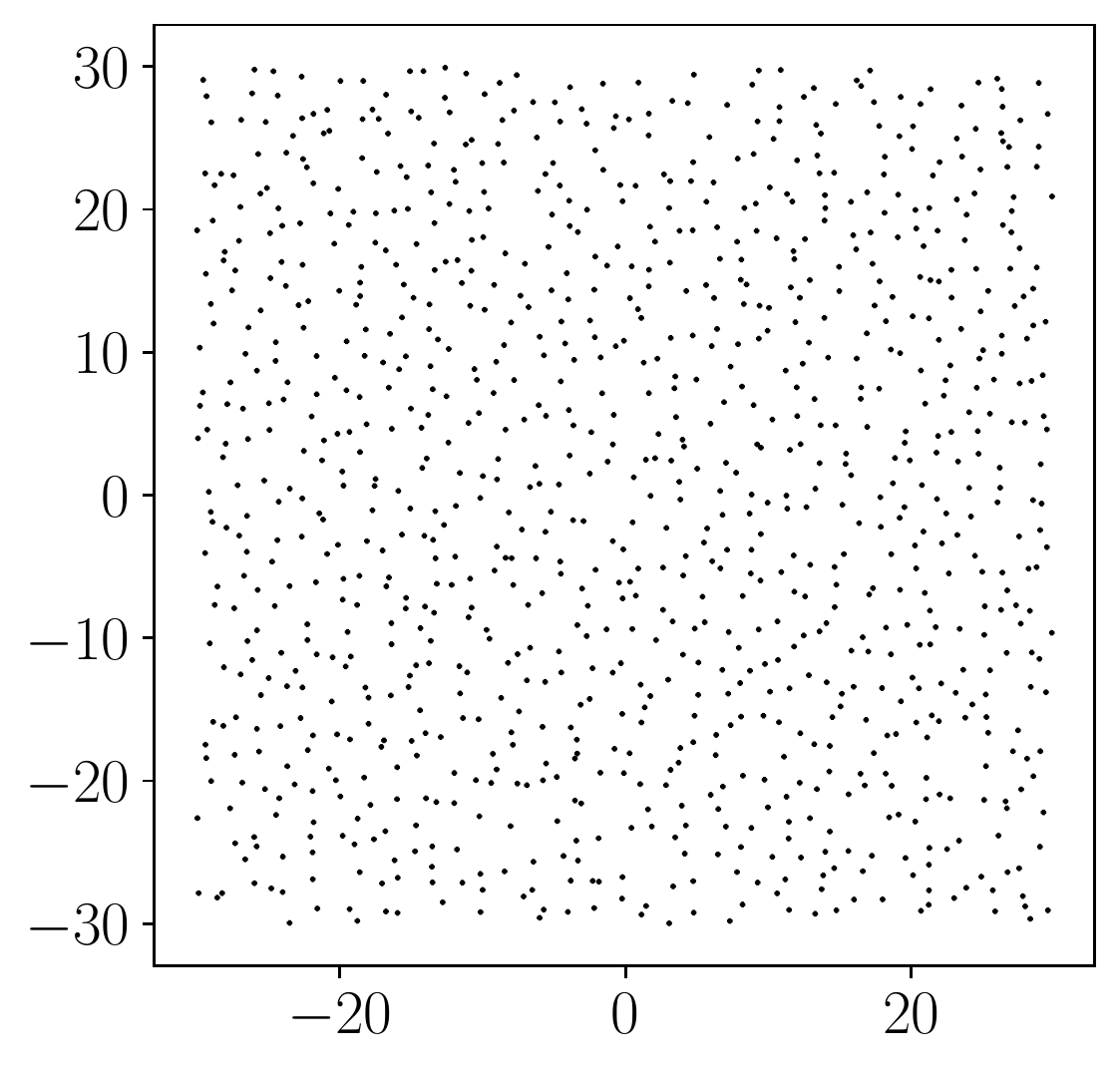}
    \caption{$p = 0.9$}
    %\label{fig:ginibre_pp}
  \end{subfigure}
  \begin{subfigure}{0.24\textwidth}
    \centering
    \includegraphics[width=0.8\linewidth]{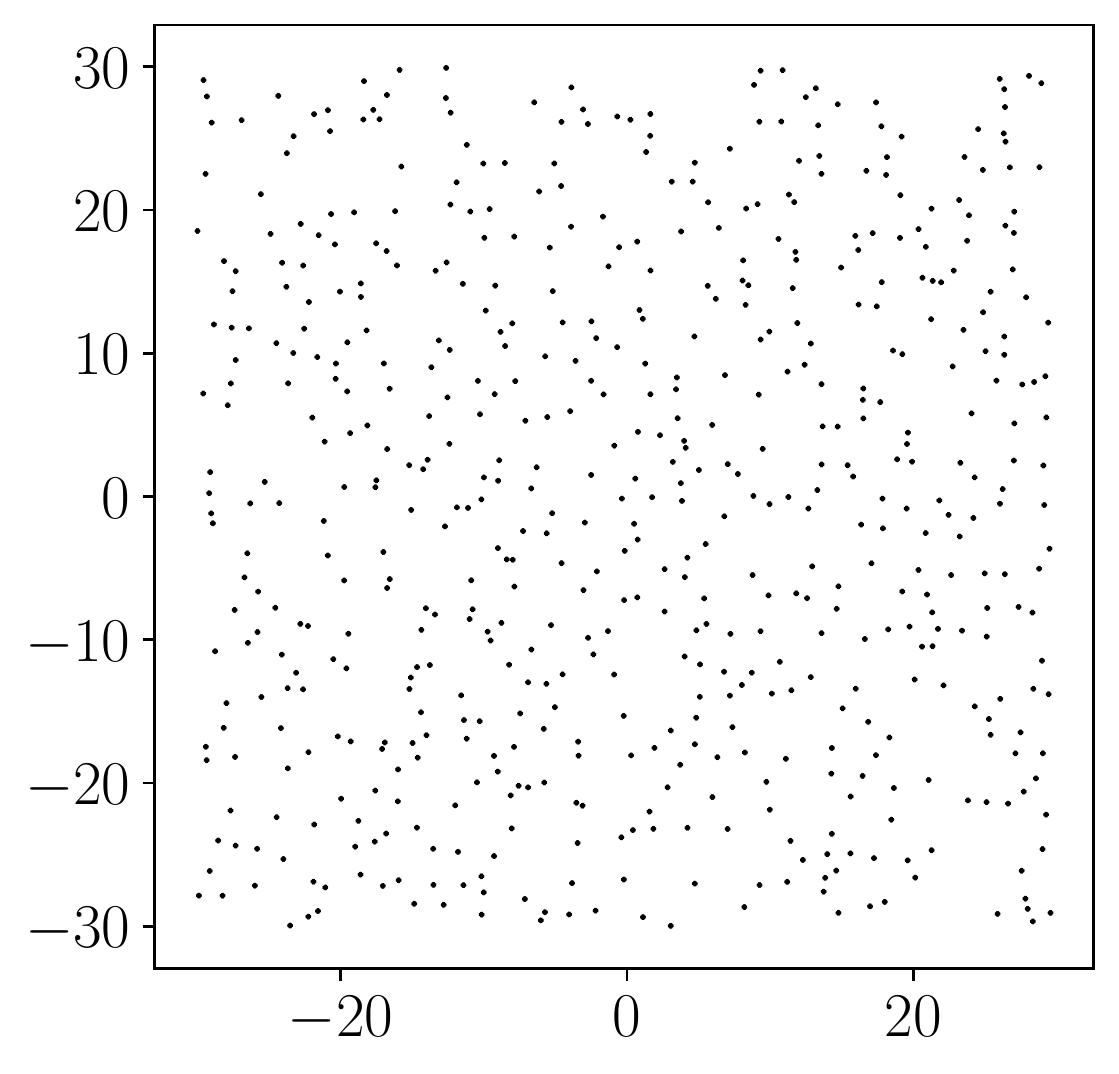}
    \caption{$p = 0.5$}
    %\label{fig:poisson_pp}
  \end{subfigure}
  \begin{subfigure}{0.24\textwidth}
    \centering
    \includegraphics[width=0.8\linewidth]{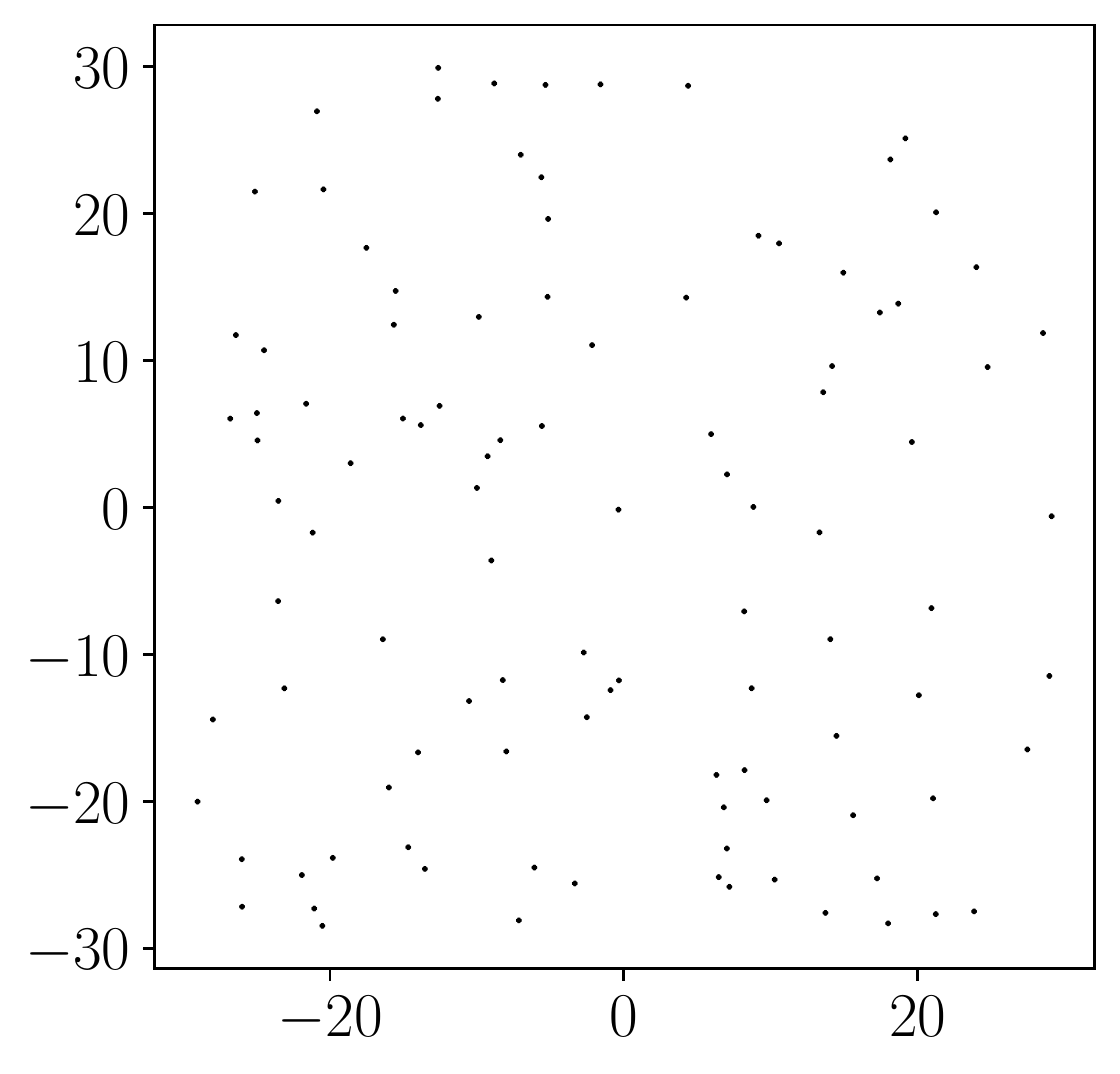}
    \caption{$p = 0.1$}
    %\label{fig:thomas_pp}
  \end{subfigure}
  \caption{A sample from the Ginibre ensemble, and the samples obtained after independent thinning with different retaining probabilities $p$}
  \label{fig:thinning_ginibre}
\end{figure*}

\subsubsection{The KLY process} % (fold)
\label{ssub:The KLY process}

Our first benchmark point process is the result of a matching algorithm proposed by \citet*{Klatt+Last+Yogeshwaran:2020}.
Loosely speaking, each point of $\bbZ^d$ is matched with a close-by point of a user-defined point process in $\mathbb{R}^d$, like a homogeneous Poisson point process.
The KLY process is an example of a point process that is known to be stationary, ergodic, and hyperuniform (but not   isotropic).
However the corresponding pair correlation function, structure factor, and hyperuniformity class are unknown.

\citet{Klatt+Last+Yogeshwaran:2020} use hyperuniformity diagnostics (Section~\ref{sub:Effective hyperuniformity}) to assess the degree of hyperuniformity, and we shall reproduce their experiment with our software and all available estimators of the structure factor.
Figure~\ref{fig:kly_pp} shows a sample from the KLY process generated by matching a subset of $\bbZ^2$ with a realization of a Poisson point process with intensity $\rho=11$.
The intensity of the resulting point process is equal to $1$.
The same parameters are used to generate the samples from the KLY process used in Section~\ref{sec:Illustrating the toolbox}.

% subsubsectionThe KLY process (end)

\subsubsection{The Ginibre ensemble} % (fold)
\label{ssub:The Ginibre ensemble (GPP)}

Our second hyperuniform point process is the Ginibre ensemble in $\bbC\approx \bbR^2$, both stationary and isotropic.
The Ginibre ensemble can be defined (and approximately sampled) as the limit of the set of eigenvalues of matrices filled with i.i.d.\ standard complex Gaussian entries, as the size of the matrix goes to infinity \citep[Theorem 4.3.10]{Hough+al:2013}.
A sample is shown in Figure~\ref{fig:ginibre_pp} as well.

Its intensity is equal to $\rho_{\text{Ginibre}} = 1/\pi$, and
its pair correlation function (Figure~\ref{fig:summery_pcf}) is
\begin{equation}
  \label{eq:pcf_ginibre}
  g_{\text{Ginibre}}(r) = 1 - \exp(-r^2).
\end{equation}
The structure factor (Figure~\ref{fig:summery_sf}) can be computed exactly,
\begin{equation}
  \label{eq:sf_ginibre}
  S_{\text{Ginibre}}(k) = 1 - \exp (-k^2/4).
\end{equation}
In particular, we have,
\begin{equation}
  \label{eq:hyperuniformity_class_ginibre}
  S_{\text{Ginibre}}(k) \sim k^2, \quad \text{for } k\rightarrow 0.
\end{equation}
Thus, according to Table~\ref{tab:class_of_hyperuniformity}, the Ginibre ensemble is a hyperuniform point process of Class~\rom{1}, with $\alpha=2$.
Much is known about the Ginibre ensemble, and we use it to benchmark our toolbox and its different estimators, rather than to infer new results about it.

% subsubsectionThinning point process (end)
\subsubsection{The homogeneous Poisson point process}
\label{ssub:The Poisson point process (PPP)}
The homogeneous Poisson point process of intensity $\rho$ of $\bbR^d$ is stationary and isotropic.
It is often thought of as having as little structure as possible and can be defined as the limit of $N$ uniform i.i.d.\ points in a window of volume $N/\rho$, as $N$ goes to infinity.
In particular, for any collection of mutually disjoint measurable subsets of $\bbR^d$, the number of points that fall in these subsets are independent; Figure~\ref{fig:poisson_pp} displays a realization from a Poisson process with the same intensity $\rho_{\text{Poisson}}=1/\pi$ as the Ginibre ensemble.
In the same vein, the pair correlation function and the structure factor of the Poisson point process are both equal to one; see Figures~\ref{fig:summery_pcf} and \ref{fig:summery_sf}.
The Poisson process is not hyperuniform.
% subsubsection The Poisson point process (PPP) (end)

\subsubsection{The Thomas point process} % (fold)
\label{ssub:The Thomas point process (TPP)}

\begin{figure}[!ht]
  \centering
  \begin{subfigure}{0.42\textwidth}
    \centering
    \includegraphics[width=0.7\linewidth]{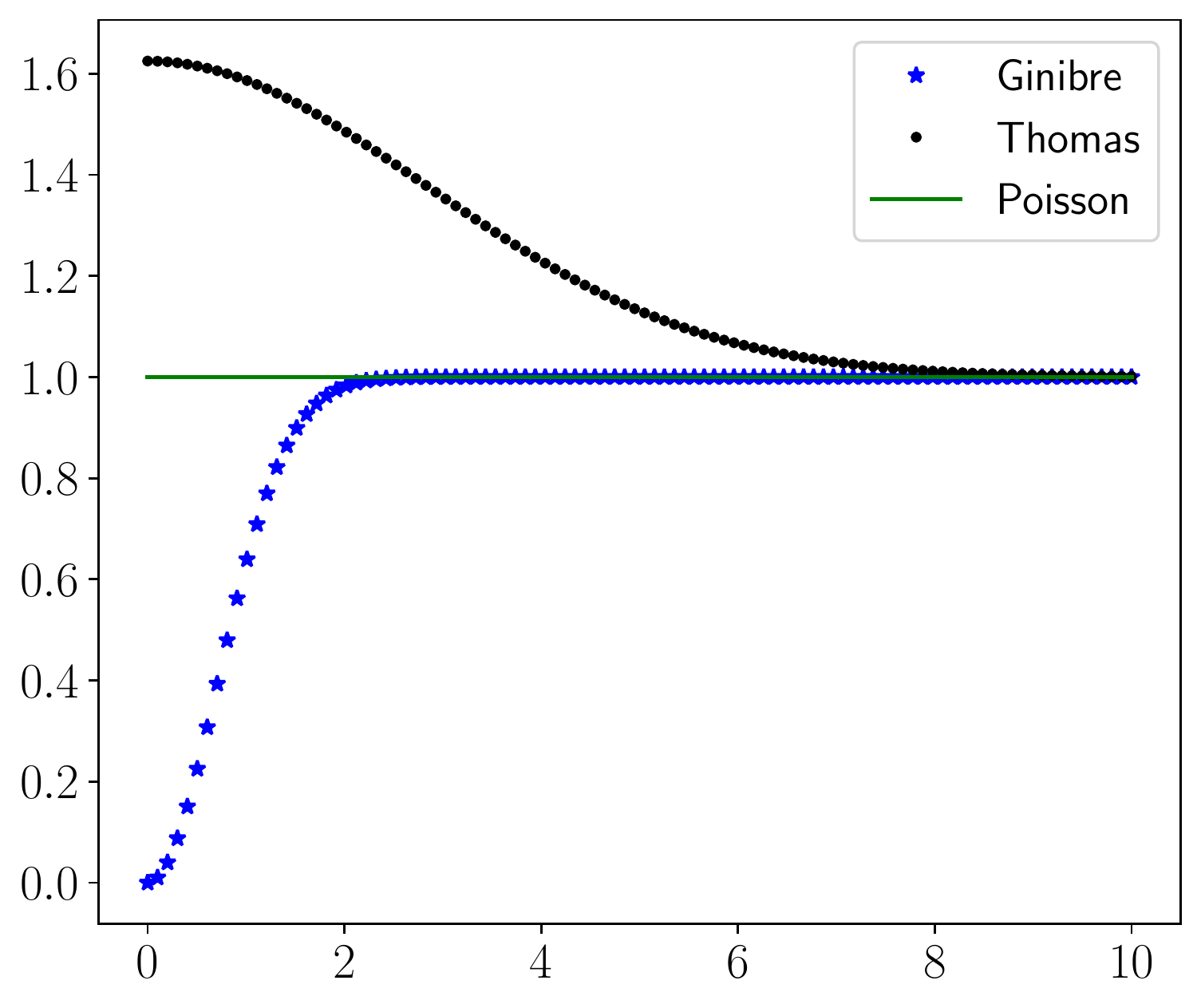}
    \caption{$g(r)$}
    \label{fig:summery_pcf}
  \end{subfigure}
  \begin{subfigure}{0.42\textwidth}
    \centering
    \includegraphics[width=0.7\linewidth]{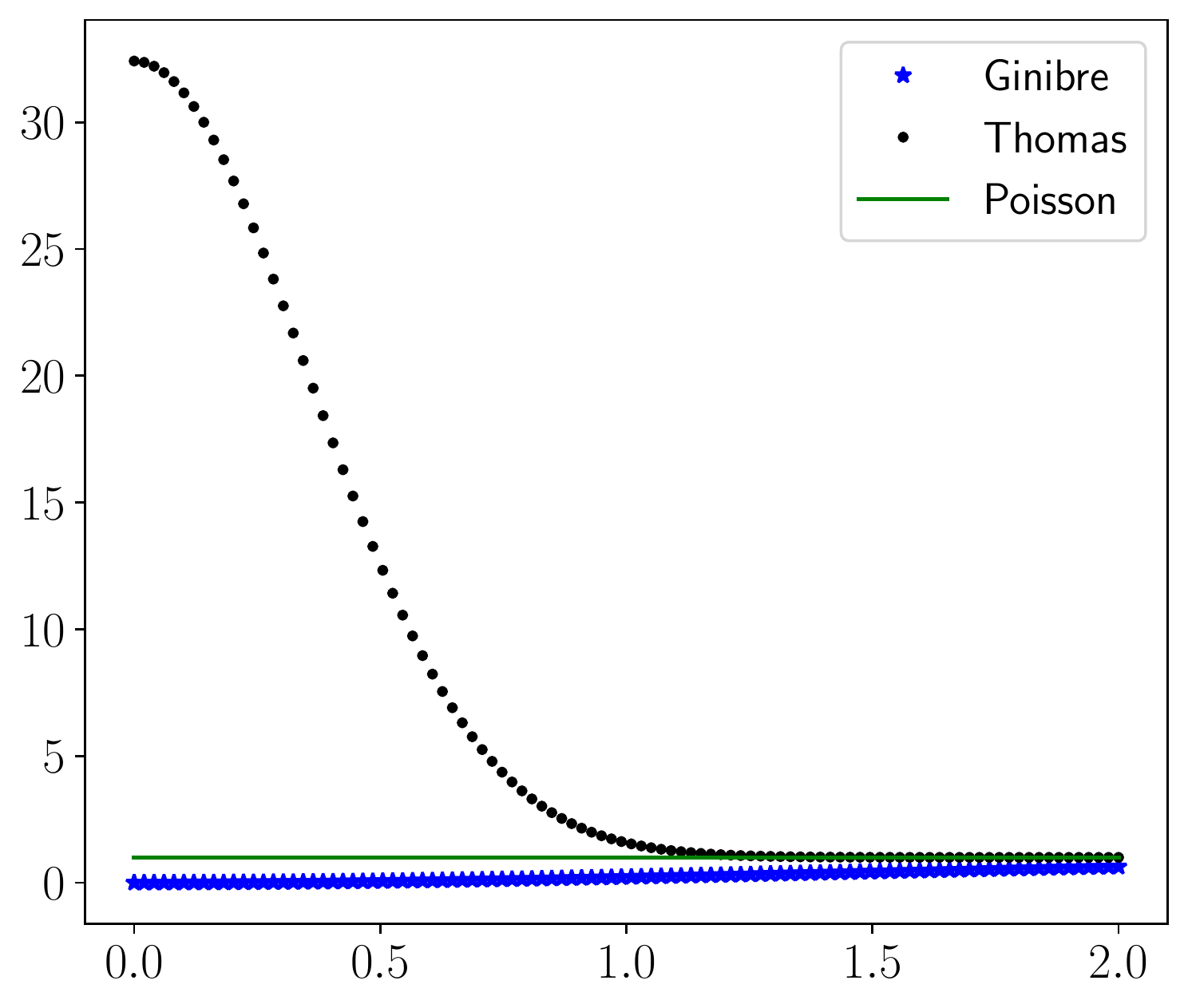}
    \caption{$S(k)$}
    \label{fig:summery_sf}
  \end{subfigure}
  \caption{Pair correlation functions $g(r)$ and structure factors $S(k)$ of some point processes}
\end{figure}

The Thomas point process is a point process in $\bbR^d$ that is reminiscent of a mixture of Gaussians and typically exhibits clusters \citep[Section 5.3]{Chiu+Stoyan+al:2013}.
Formally, consider a homogeneous Poisson point process $\calX_\text{parent}$ of intensity $\rho_{\text{parent}}$, and let $\lambda,\sigma $ be positive.
Conditionally to $\calX_\text{parent}$, let $(N_{\bfx})_{\bfx \in \calX_\text{parent}}$ be i.i.d.\ Poisson variables with mean $\lambda $.
For any $\bfx \in \calX_\text{parent}$, conditionally to $N_{\bfx}$, let $(Y_{\bfx,i})_{i=1}^{N_{\bfx}}$ be i.i.d.\ centered two-dimensional isotropic Gaussian vectors with variance $\sigma $.
The resulting Thomas process is given by
\begin{align*}
  \calX_\text{Thomas}=\cup _{\bfx\in \calX_\text{parent}}\{\bfx+Y_{\bfx,i},i=1,\dots ,N_{\bfx}\},
\end{align*}
and its intensity is given by $\rho_{\text{Thomas}} = \rho_{\text{parent}} \times \lambda$.
Figure~\ref{fig:thomas_pp} shows a realization of a Thomas point process of intensity $1/\pi$, generated from a parent Poisson point process of intensity $\rho_{\text{parent}}=1/(20 \pi)$, and a standard deviation $\sigma = 2$.
Since $\sigma$ is small enough compared to $\lambda =20$, clusters naturally appear.
The pair correlation function and the structure factor of the Thomas process, see Figures~\ref{fig:summery_pcf} and \ref{fig:summery_sf}, are both radial functions \citep[Equation 5.52]{Chiu+Stoyan+al:2013}, given by
\begin{equation}
  \label{eq:TPP_pcf}
  g_{\text{Thomas}}(r)
  =
  1
  + \frac
  {1}
  {\rho_{\text{parent}} (\sqrt{4\pi\sigma^2})^{d}}
  \exp(-\frac{r^2}{4 \sigma^2}),
\end{equation}
and
\begin{equation}
  \label{eq:s_thomas}
  S_{\text{Thomas}}(k) = 1 + \lambda\exp\left(-k^2 \sigma^2\right).
\end{equation}
The intuition that the Thomas point process is not hyperuniform is confirmed by the expression of the structure factor \eqref{eq:s_thomas} which is even larger than 1.
% subsubsectionThe Thomas point process (TPP) (end)

\subsubsection{Thinning the Ginibre ensemble} % (fold)
\begin{figure}[!h]
  \centering
    \includegraphics[width=0.5\linewidth]{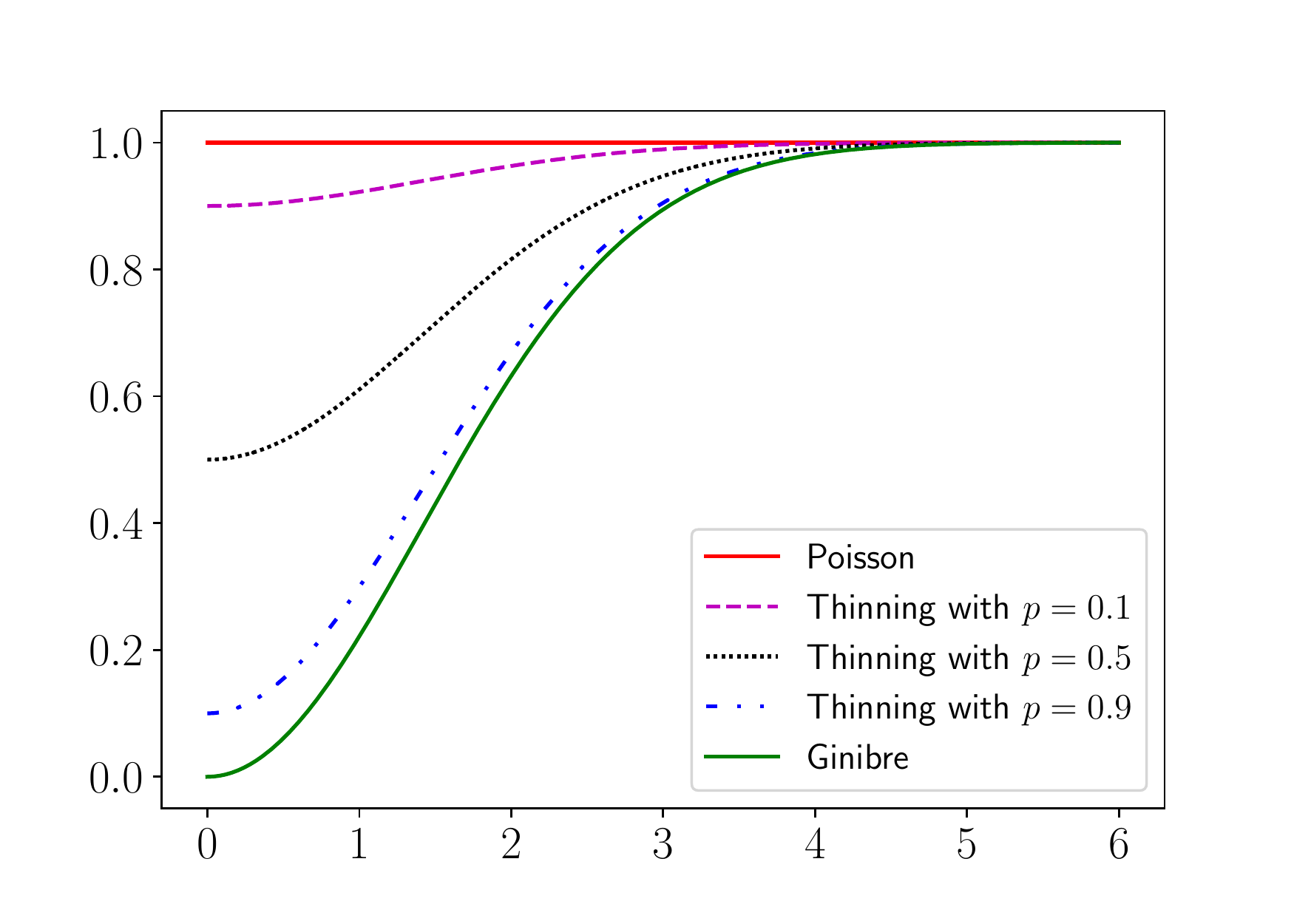}
    \caption{Structure factor of the Ginibre before and after applying independent thinning with different retaining probabilities $p$}
    \label{fig:sf_ginibre_and _thinning}
\end{figure}

\label{ssub:Thinning the Ginibre point process}
  To benchmark diagnostics of hyperuniformity, it is useful to have a parametrized non-hyperuniform point process with $S(0)$ arbitrarily close to zero.
  To create such a point process, we sample a Ginibre point process and keep each point independently with probability $p \in (0,1)$.
  The resulting point process is simple, stationary, and isotropic, with intensity $\rho_{p} = p\times \rho_{\text{Ginibre}}$.
  Figure \ref{fig:thinning_ginibre} shows such thinned Ginibre samples, with retaining probabilities $p \in \{0.1, 0.5, 0.9\}$.
  The pair correlation function of the thinned point process is independent of $p$, and actually equal to the Ginibre counterpart $g_{p}(r)= g_{\text{Ginibre}}(r)$.
  However, the structure factor becomes
  \begin{equation}
  \label{eq:sf_thinning_ginibre}
    S_{p}(k) =  pS_{\text{Ginibre}}(k) + 1 - p.
  \end{equation}
  Clearly, $0<S_{p}<1$ so the obtained point process is not hyperuniform, and as $p$ increases, $S_p(0)$ decreases, making it harder to distinguish hyperuniformity based on an approximate value of the structure factor.
  The structure factor $S_p$ is shown in Figure \ref{fig:sf_ginibre_and _thinning} for a few values of $p$.
% Decreasing the value of $p$ gives a family of point processes with a structure factor lying between $S_{\text{Ginibre}}$ and 1 (which is the structure factor of the Poisson point process described in Section \ref{ssub:The Poisson point process (PPP)}); See
% subsubsectionThinning the Ginibre point process (end)
% end Section 2

% Section 3
\section{Estimators of the structure factor} % (fold)
\label{sec:Estimators of the structure factor}

Starting from the theoretical definition \eqref{eq:stucture_factor_function} of the structure factor, we derive all known estimators and add a few other natural candidates based on numerical quadratures of the symmetric Fourier transform.
We pay particular attention to the sources of asymptotic (in the size of the window) bias for each estimator.
We split the estimators according to whether they only assume stationarity or stationarity \emph{and} isotropy.

\subsection{Assuming stationarity} % (fold)
\label{sub:Estimators assuming stationarity}
The most common estimator of the structure factor is the so-called \emph{scattering intensity} \citep{Torquato:2018, Klatt+al:2019, Klatt+Last+Yogeshwaran:2020, Coste:2021}.
Its name stems from its origins in the physics of diffraction, and it corresponds to a scaled version of Bartlett's periodogram in time series analysis.
After introducing the scattering intensity, we follow \citet{Rajala+Olhede+John:2020}, who generalize it to so-called \emph{tapered} estimators.
We group our assumptions and notation in Assumption~\ref{ass:1}.

\begin{assumption}{A}{1} \label{ass:1}
  $\calX$ is a stationary simple point process of $\bbR^d$ with intensity $\rho > 0$.
  Its pair correlation function $g$ exists, and $ \bfr \mapsto g(\bfr) -1 $ is integrable on $\mathbb{R}^{d}$.
  Moreover, we only observe a realization of
  $$ \calX\cap  W = \{\bfx_1,\dots,\bfx_N\} $$
  in a centered, rectangular window
  $$
  W=\prod_{j=1}^d \limits[-L_j/2,L_j/2].
  $$
  We write $\bfL = \left(L_{1}, \dots, L_{d}\right)$.
\end{assumption}

\subsubsection{The scattering intensity} % (fold)
\label{ssub:The scattering intensity}
In the physics literature, the following derivation is often assumed to be known to reader , and it seemed worthwhile to us to make it explicit.
This allows, in particular, to understand the role played by the so-called \emph{allowed wavevectors}.
Note that \citet{Rajala+Olhede+John:2020} provide similar derivations.

The basic idea is to introduce the scaled intersection volume $\alpha_0$ \eqref{eq:alpha_2} in the definition \eqref{eq:stucture_factor_function} of the structure factor.
We obtain
\begin{align}
  S&(\bfk)=
  1 + \rho \int_{\bbR^d} (g(\bfr) -1) e^{-\i \langle\bfk, \bfr\rangle} ~\d \bfr
 \nonumber \\
   &= 1 + \rho \int_{\bbR^d} \lim_{W \uparrow \bbR^d} (g(\bfr) -1) \alpha_0(\bfr, W) e^{-\i \langle\bfk, \bfr\rangle} ~\d \bfr \nonumber
  \\
   &= 1 + \lim_{W \uparrow \bbR^d} \rho \int_{\bbR^d} (g(\bfr) -1) \alpha_0(\bfr, W) e^{-\i \langle\bfk, \bfr\rangle} ~\d \bfr \label{eq:1} ,
\end{align}
where we used dominated convergence and the limit $ \lim_{W \uparrow \bbR^d} \alpha_0(\bfr, W)  = 1$.
In the notation $W \uparrow \bbR^d$, the limit is taken so that the window progressively covers the whole space at roughly equal speed in all directions, i.e., $\min_j L_j\rightarrow \infty$ in Assumption~\ref{ass:1}.
Splitting the integral from \eqref{eq:1} into two, we shall recognize an expectation under our censored point process $\calX\cap W$ and a bias term,
\begin{align*}
  &S(\bfk) -1
   = \\
  & \lim_{W \uparrow \bbR^d} \bigg[
    \frac{\rho}{\leb{W}}
    \int_{\bbR^d} \int_{\bbR^d}
    \underbrace{
      e^{-\i \langle\bfk, \bfr\rangle} \mathds{1}_{W \times W}  (\bfr + \bfy, \bfy)
    }_{
      f(\bfr + \bfy, \bfy)
    }
    g(\bfr)
    \d \bfy
    \d \bfr
    - \underbrace{\rho \calF(\alpha_0)(\bfk, W)}_{\epsilon_0(\bfk, \bfL)} \bigg].
\end{align*}
Now, by definition \eqref{eq:pcf_by_expectation} of the pair correlation measure, and still for any $\bfk\in\mathbb{R}^d$, $S(\bfk)-1$ is the limit as $W \uparrow \bbR^d$ of
$$
\frac{\rho}{\rho^2 \leb{W}} \mathbb{E}  	\left[ \sum_{\bfx,\bfy \in \calX}^{\bfx \neq \bfy} \mathds{1}_W(\bfx) \mathds{1}_{W}(\bfy)e^{-\i 	\langle\bfk, \bfx - \bfy \rangle} \right]  - \epsilon_0(\bfk, \bfL),
$$
so that
\begin{align}
 S&(\bfk) =  \lim_{W \uparrow \bbR^d} \frac{1}{ \rho \leb{W}} \mathbb{E} \left[ \sum_{\bfx,\bfy \in \calX \cap W} e^{-\i \langle\bfk, \bfx - \bfy \rangle}\right]  - \epsilon_0(\bfk, \bfL)       \nonumber                                               \\
&= \lim_{W \uparrow \bbR^d}\frac{1}{ \rho \leb{W}} \mathbb{E}  \left[\left \vert \sum_{\bfx \in \calX \cap W} e^{-\i \langle\bfk, \bfx \rangle}\right\vert^2 \right]  - \epsilon_0(\bfk, \bfL).
  \label{e:unbiasedness_of_SI}
\end{align}
Note that, by \eqref{eq:ft_alpha_0_box}, the bias term satisfies
\begin{align*}
  \lvert \epsilon_0(\bfk, \bfL) \rvert
   & = \rho  \frac{\calF^2(\mathds{1}_W)(\bfk)}{\leb{W}}
  = \rho  \bigg(\prod_{j=1}^d\frac{\sin(k_j L_j/2)}{ \sqrt{L_j}k_j/2}\bigg)^2
  \\
   & \leq
  \begin{cases}
    0
     & \text{if } \bfk \in \bbA_{\bfL},
    \\
    \rho \prod_{j=1}^d \limits L_j
     & \text{as } \Vert \bfk\Vert_2 \rightarrow 0,
    \\
    2^{2d}
    \prod_{j=1}^d \limits
    \frac{1}{L_j k_j^2}
     & \text{as } \Vert \bfk\Vert_2 \rightarrow \infty,
  \end{cases}
\end{align*}
where we defined
\begin{align}
\mathbb{A}_\bfL= \Big \{ & \bfk \in (\bbR^\ast)^d \text{ such that }
 k_j = \frac{2\pi n}{L_j} \text{ for some }
  j\in\{1,\dots,d\} \text{ and some } n \in \bbZ^\ast\Big \} \label{eq:allowed_wave}.
\end{align}
We have thus proved the following proposition.
\begin{proposition} \label{prop:scattering_intensity_conv}
  Under Assumption~\ref{ass:1}, for $\mathbf{k}\in\bbA_\bfL$, the scattering intensity estimator
  \begin{equation}
    \label{eq:s_si}
    \widehat{S}_{\mathrm{SI}} \left(\bfk \right)
    \triangleq
    \frac{1}{\rho \leb{W}}
    \left \lvert
    \sum_{j=1}^N
    e^{-\i \langle\bfk, \bfx_j\rangle}
    \right\rvert^2
  \end{equation}
  is asymptotically unbiased, i.e.,
  $$
    \sup_{\bfk \in \bbA_\bfL}
    \left \lvert
    \mathbb{E} [\widehat{S}_{\mathrm{SI}} (\bfk)]
    - S(\mathbf{k})
    \right \rvert
    \xrightarrow[W \uparrow \bbR^d]{} 0.
  $$
\end{proposition}
This motivates restricting to $\bfk \in \bbA_\bfL$ if one is interested in estimating the behavior of $S$ in zero.
In the literature, the scattering intensity is actually often evaluated on a subset of $\bbA_\bfL$, namely
\begin{align}
 \bbA_{\bfL}^{\text{res}} =
  \left\{ \left( \frac{2\pi n_1}{L_1}, \ldots, \frac{2\pi n_d}{L_d}\right );  \mathbf{n}
  \in (\mathbb{Z}^\ast)^d
  \right\} .
\label{eq:res_allowed_wave}
\end{align}
The set $\bbA_{\bfL}^{\text{res}}$ is called the set of \emph{allowed wavevectors} in physics \citep[Section 10]{Klatt+Last+Yogeshwaran:2020}, or part of the \emph{dual lattice} of fundamental cell $W$ in sampling theory \citep[Chapter 5]{Osgood:2014}, or \emph{Fourier grid} in time series analysis \citep{Rajala+Olhede+John:2020}.
We add the superscript \emph{res} to underline that it is actually a restriction of the set $\bbA_\bfL$ of wavevectors justified by the cancellation of the asymptotic bias.

It is unclear to us why one should consider $\bbA_{{\bfL}}^{\text{res}}$ instead of $\bbA_{\bfL}$.
In particular, the minimal \emph{wavenumber} (the norm of a wavevector) in $\bbA_{{\bfL}}^{\text{res}}$  is
$$
k_{\min}^{\text{res}}= 2 \pi \sqrt{\sum_{j=1}^d L_j^{-2}},
$$
while working with $\bbA_\bfL$ in \eqref{eq:allowed_wave} relaxes this threshold to a $k_{\min}$ satisfying
$$
  \frac{2 \pi}{\text{max}_j L_j} < k_{\min} < k^{\text{res}}_{\min}.
$$
To see how far we can hope going down in $k>0$, we give evidence below as to why the scattering intensity should never be evaluated for $\|\bfk\|_2 \leq \frac{ \pi }{\sqrt{d}\max_{j}L_{j}}$.

Finally, when $\rho$ is unknown, and $\calX$ is further assumed to be ergodic, it is common to replace the denominator $\rho \leb{W}$ by $N$ in \eqref{eq:s_si}, leading to the self-normalized scattering intensity estimator
\begin{equation}
  \label{eq:s_si_s}
  \widehat{S}_{\mathrm{SI, s}} (\bfk)
  \triangleq
  \frac{1}{N}
  \left\vert
  \sum_{j=1}^N e^{-\i\langle\bfk, \bfx_j\rangle}
  \right\vert^2.
\end{equation}
Indeed, by ergodicity,
\begin{equation*}
  \frac
  {\widehat{S}_{\mathrm{SI, s}}}
  {\widehat{S}_{\mathrm{SI}}}
  \xrightarrow[W \uparrow \bbR^d]{\text{ a.s.}}
  1.
\end{equation*}
Many authors define the scattering intensity as its self-normalized version \citep{Torquato:2018, Klatt+al:2019, Coste:2021}, but we argue that when $\rho$ is known, it is not clear whether this estimator has a smaller mean squared error than \eqref{eq:s_si}.

For the reader interested in second order properties of this estimator, its second moment on Poisson input is derived in Appendix~\ref{appB: condition_of_prop3}.

\paragraph{A lower bound for $k_{\min}$.}
It is clear that when $W$ is fixed, the scattering intensity \eqref{eq:s_si_s} is not relevant when $\bfk$ is too close to $0$, since for a fixed sample $\calX_N=\{\bfx_{1},\dots ,\bfx_{N}\}$,
\begin{align*}
  \lim_{\bfk\to 0}
  \widehat{S}_{\mathrm{SI, s}} (\bfk)
  = N.
\end{align*}
The theoretical convergence to $0$ for hyperuniform processes is dictated by the compensation occurring between exponential terms $e^{-\i \langle \bfk,\bfx_{j}  \rangle }$ on large portions of the space.
One can safely infer that if a large portion of the terms gives a positive contribution, then the compensation does not occur and the global result will not be accurate.
Still with the notation in Assumption~\ref{ass:1},
if $\|\bfk\|_2< \frac{ \pi }{\sqrt{d}\max_{j}  L_{j}}$, then for $\bfx\in W$,
$\lvert \langle \bfx ,\bfk  \rangle  \rvert \leq \pi/2$,
so that there exists $\epsilon>0$ such that
\begin{align*}
  \frac{1}{N}
  \left \lvert
  \sum_{j = 1}^N
  e^{-\i \langle \bfk, \bfx_j \rangle}
  \right \rvert ^{2}
  \geq \frac{(N \epsilon)^2 }{N}.
\end{align*}
Thus for any $ \|\bfk\|_2< \frac{ \pi }{\sqrt{d}\max_{j}  L_{j}}$, we have a lower bound of $\widehat{S}_{\mathrm{SI, s}}(\bfk)$ that is independent of the point process, and which diverges as the number of points goes to infinity.
Consequently, we argue that $\pi/\sqrt{d}\max_{j}L_{j}$ is a lower bound for accessible wavenumbers, which one might improve with a finer study of the estimator bias.
\subsubsection{Tapered variants of the scattering intensity} % (fold)
\label{ssub:The tapered estimator}

The derivation of the scattering intensity can be generalized to the \emph{tapered estimator}
\begin{equation}
  \label{eq:s_t}
  \widehat{S}_{\mathrm{T}}(t, \bfk)
  \triangleq
  \frac{1}{\rho}
  \left\vert
  \sum_{j=1}^N
  t(\bfx_j, W) e^{-\i \langle\bfk, \bfx_j\rangle}
  \right\vert^2,
  \quad \mathbf{k}\in\bbR^d,
\end{equation}
where $t(\cdot, W)$ is a uniformly (in $W$) square-integrable function supported on the observation window $W$, called a \emph{taper}.
The tapered estimator \eqref{eq:s_t} corresponds to a scaled version of what is called a tapered periodogram in the signal processing literature \citep[Section 3]{Rajala+Olhede+John:2020}.
The vocabulary is adapted from the spectral analysis of time series, where tapers are now well established \citep{Percival+Walden:2020}.

In particular, one recovers the scattering intensity \eqref{eq:s_si} from the tapered formulation \eqref{eq:s_t} by plugging the taper
\begin{equation}
  \label{eq:t_0}
  t_0(\bfx, W)
  \triangleq
  \frac{1}{\sqrt{\leb{W}}} \mathds{1}_W(\bfx).
\end{equation}
To follow the derivation of \eqref{e:unbiasedness_of_SI}, we further let
\begin{equation}
  \label{eq:alpha_t}
  \alpha_t(\bfr, W)\triangleq  \int_{\bbR^d} t(\bfr + \bfy, W)t(\bfy, W ) \d  \bfy,
\end{equation}
and require that
\begin{equation}
  \label{eq:limite_alpha}
  \lim_{W \uparrow \bbR^d} \alpha_t(\bfr, W) = 1 \quad \text{and} \quad \alpha_t(\mathbf{0}, W) = 1,
\end{equation}
where the limit is again taken as $\min_j L_j\rightarrow \infty$ under Assumption~\ref{ass:1}.
Note first that our requirement that \eqref{eq:limite_alpha} holds differs from the treatment of \cite{Rajala+Olhede+John:2020}.
We find \eqref{eq:limite_alpha} to be a more natural generalization of the scattering intensity arguments of \cite{Torquato:2018}.
Note also that, for simplicity, we denoted $\alpha_{t_0}$ by $\alpha_{0}$ in Section \ref{sub:fourier_transform}, and we shall stick to this simplified notation.

Now, Cauchy-Schwarz inequality and the uniform integrability of $t$ guarantee that  $\alpha_t(\cdot,W)$ is uniformly (in $W$) bounded, so that by dominated convergence,
$$
  S(\bfk)
  = 1
  + \rho
  \lim_{W \uparrow \bbR^d}
  \int_{\bbR^d} (g(\bfr) -1)
  \alpha_t(\bfr, W) e^{-\i \langle\bfk, \bfr\rangle}
  ~\d \bfr.
$$
Following the lines of \eqref{e:unbiasedness_of_SI}, we obtain
\begin{align}
  S(\bfk)
  = \lim_{W \uparrow \bbR^d}
  \left(
  \mathbb{E}\left[
  \widehat{S}_{\mathrm{T}}(t, \bfk)
  \right]
  - \underbrace{\rho\left \lvert \calF(t)(\bfk, W)\right \rvert^2}_{\epsilon_t(\bfk, \bfL)}
  \right) \label{eq:4}.
\end{align}
where the tapered estimator $\widehat{S}_{\mathrm{T}}$ is defined in \eqref{eq:s_t}.
To eliminate the asymptotic bias $\epsilon_t(\bfk, \bfL)$, one can restrict oneself again to a set of allowed wavevectors as we did in \eqref{eq:allowed_wave}, i.e., the zeros of $\epsilon_t(\cdot, \bfL)$.
For general tapers, however, finding these zeros is not straightforward, and an alternative way to escape the bias is to correct it, as in
\begin{align}
\widehat{S}_{\mathrm{UDT}} (t, \bfk) \triangleq \frac{1}{\rho}
  \left\vert
  \sum_{j=1}^N
  t(\bfx_j, W) e^{-\i \langle\bfk, \bfx_j\rangle}
  \right\vert^2
  - \rho\left \lvert \calF(t)(\bfk, W)\right \rvert ^2.  \label{eq:s_UDT}
\end{align}
We refer to $\widehat{S}_{\mathrm{UDT}}$ as the \emph{undirectly debiased tapered estimator}, which is a scaled version of what \citet{Rajala+Olhede+John:2020} define.
The major issue of Estimator \eqref{eq:s_UDT} is that it may give negative values.
To remedy this, \citet{Rajala+Olhede+John:2020} propose to remove the bias inside the summation before taking the squared modulus, namely, to define the \emph{directly debiased tapered estimator}
\begin{align}
 \widehat{S}_{\mathrm{DDT}} (t, \bfk) \triangleq \frac{1}{\rho}
  \left\vert
  \sum_{j=1}^N t(\bfx_j, W) e^{-\i \langle\bfk, \bfx_j\rangle} - \rho\calF(t)(\bfk, W)
  \right\vert^2 . \label{eq:s_DT}
\end{align}
Finally, note that these debiasing techniques naturally apply to the special case of the scattering intensity, and thus offer an alternative to using allowed values \eqref{eq:allowed_wave}.

\subsubsection{The multitapered estimator} % (fold)
\label{ssub:The multitapered estimator}

In the spectral analysis of time series, \emph{multitapering} (MT) was first introduced by \citet{Thomson:1982}; see also \citet{Percival+Walden:2020} for a modern reference.
The idea is to average a periodogram over many tapers, in the hope to reduce the variance of the resulting estimator.
\citet{Rajala+Olhede+John:2020} propose to adapt the method to point processes, and we follow their lines.

For any $\bfk \in \bbR^d$ and $P \in \bbN^\ast$, and under Assumption \ref{ass:1}, \cite{Rajala+Olhede+John:2020} define the \emph{multitapered estimator} $\widehat{S}_{\mathrm{MT}}$ by
\begin{equation}
  \label{eq:s_mt}
  \widehat{S}_{\mathrm{MT}}((t_{q})_{q=1}^P, \bfk)
  = \frac{1}{P} \sum_{q=1}^{P} \widehat{S}(t_{q}, \bfk),
\end{equation}
where the $P$ tapers $(t_{_q})_{q=1}^{P}$ are typically taken to be pairwise orthogonal square integrable functions, and $\widehat{S}(t_q,\cdot)$ is any of the tapered estimators, whether undebiased \eqref{eq:s_t}, undirectly debiased \eqref{eq:s_UDT}, or directly debiased \eqref{eq:s_DT}.
The directly and undirectly debiased versions of $\widehat{S}_{ \mathrm{MT}}$ will be respectively denoted by $\widehat{S}_{ \mathrm{DDMT}}$ and $\widehat{S}_{ \mathrm{UDMT}}$.

\subsubsection{On the choice of tapers}
\label{ssub:The sinusoidal tapers}

Common taper choices in time series analysis are Slepian tapers, sinusoidal tapers, and minimum bias tapers \citep{Riedel+Sidorenko:1995}.
For instance, still assuming a centered rectangular window $W=\prod_{j=1}^d[-L_j/2, L_j/2]$, the family of sinusoidal tapers $(t_q)_{q\geq 1}$ supported on $W$ is defined by
\begin{align}
  t_q& (\bfx, W) = t(\bfx, \bfp^q, W) \triangleq \frac{\mathds{1}_W(\bfx)}{\sqrt{\leb{W}}}  \prod_{j=1}^d \sqrt{2}\sin \left(\frac{\pi p^q_j}{L_j}(x_j + \frac{L_j}{2})\right), \label{eq:sine_taper}
\end{align}
where $\bfp^q = (p^q_1, \ldots , p^q_d) \in (\bbN^d)^\ast$ and $\bfx = (x_1, \ldots, x_d) \in \bbR^d$.
The sinusoidal tapers are pairwise orthogonal, and an easy direct computation shows that they also satisfy \eqref{eq:limite_alpha}.
% For $\bfr \in (\bbR^d)^{+}$ and $q \geq 1$, we have
% \begin{align}
%  & \alpha_{t_q}(\bfr, W)  = \nonumber \\
%   & \prod_{j=1}^{d}\frac{2}{L_j}\int_{-L_j/2}^{ L_j/2-r_j}   \sin \left(\frac{\pi p^q_j}{L_j}(x_j + \frac{L_j}{2})\right) \nonumber \\
%    &\text{\hspace{2.5cm}}  \sin \left(\frac{\pi p^q_j}{L_j}(x_j + r_j + \frac{L_j}{2})\right) \d x_j  =\nonumber  \\
%   &\prod_{j=1}^{d} \bigg[ \frac{L_j-r_j}{L_j} \cos\left(\frac{\pi p_j^qr_j}{L_j}\right) \; \; - \nonumber\\
%   &  \text{\hspace{0.5cm}}\frac{1}{2 \pi p_j^q} \left[\sin\left(\frac{\pi p_j^q}{L_j} (2L_j -r_j)\right) - \sin\left(\frac{\pi p_j^q}{L_j}r_j\right)\right] \bigg] \label{eq:40}.
% \end{align}
% Trivially, $\alpha_{t_q}(\mathbf{0} , W)=1$.
% When $\min_{j} \limits L_j \rightarrow \infty$, the left-hand side of the product in \eqref{eq:40} converges to 1, and the right-hand side converges to zero.
% Hence, $\alpha_{t_q}(\bfr, W) \xrightarrow[]{W \uparrow \bbR^d} 1$, and the sinusoidal tapers satisfy \eqref{eq:limite_alpha}.
Moreover, for $\bfk = (k_1, \ldots, k_d) \in \bbR^d$ the Fourier transform $\calF(t_q)(\bfk, W)$
of $t_q$ for any $q$ is
% \begin{strip}
% \noindent\rule{\textwidth}{0.1pt}
\begin{align*}
  \frac{1}{\sqrt{\leb{W}}} & \prod_{j=1}^d \sqrt{2} i^{(p_j^q +1)} \bigg[\frac{\sin\left(\left(k_j - \frac{\pi p^q_j}{L_j}\right)\frac{L_j}{2}\right)}{k_j - \frac{\pi p^q_j}{L_j}}
  - \left(-1\right)^{p_j^q} \frac{\sin\left(\left(k_j + \frac{\pi p^q_j}{L_j}\right)\frac{L_j}{2}\right)}{k_j + \frac{\pi p^q_j}{L_j}} \bigg].
\end{align*}
This closed-form expression can thus be used in any debiasing scheme.
Note that this analytical tractability, along with the absence of a sensitive parameter like a lengthscale, lead us to choose the sine taper over, say, a multidimensional generalization of Slepian tapers \citep{Percival+Walden:2020,Rajala+Olhede+John:2020}.
\subsection{Assuming stationarity and isotropy} % (fold)
\label{sub:Estimators assuming stationarity and isotropy}

For isotropic point processes, a common approach is simply to numerically rotation-average the structure factor estimators presented in Section~\ref{sub:Estimators assuming stationarity}.
Alternatively, one could start from the analytical expression \eqref{eq:sf_as_hankel_tranform} of the structure factor as symmetric Fourier transform --~a univariate integral~-- involving the pair correlation function.
Then again, two approaches have been identified.
First, identifying an expectation under the point process as in Section~\ref{sub:Estimators assuming stationarity} leads to a natural estimator originally derived by \citet{Bartlett:1964}.
Second, estimation of the pair correlation followed by numerical quadrature leads to at least two natural estimators, depending on whether the quadrature is that of \citet{Ogata:2005}, or that of \citet{Baddour+Chouinard:2015}.
We review all these estimators in turn.
All point processes in Section~\ref{sub:Estimators assuming stationarity and isotropy} satisfy Assumption~\ref{ass:2}.
\begin{assumption}{A}{2}\label{ass:2}
  $\calX$ is a simplex stationary isotropic point process of $\bbR^d$, of intensity $\rho$. Its pair correlation function $g$ exists, and $ r \mapsto g(r) -1 $ is integrable.
  Moreover, we only observe a realization of
  $
  \calX\cap W = \{\bfx_1,\dots,\bfx_N\}
  $ in the centered ball $W=B^d(\mathbf{0}, R)$.
\end{assumption}
% subsectionIsotropic estimators (end)

\subsubsection{Bartlett's isotropic estimator} % (fold)
\label{ssub:bartletts_isotropic_estimator}

From the observation that the scaled intersection volume $\alpha_0$ \eqref{eq:alpha_2} is a radial function, $\alpha_0(\bfr, W) = \alpha_0(r, W)$, and following the lines of Section~\ref{sub:Estimators assuming stationarity}, dominated convergence yields
\begin{align}
 & S(k) -1 = \nonumber \\
   & \lim_{R \rightarrow \infty} \bigg[
    \rho
    \frac{ (2\pi)^{\frac{d}{2}}}{k^{\frac{d}{2}-1}}
    \int_{0}^{\infty}
    \alpha_0(r, W)
    ~
    g(r)
    r^{\frac{d}{2}}
    J_{d/2 - 1}(kr)
    \d r - \underbrace{
      \rho
      \calF(\alpha_0)(k, W)
    }_{\triangleq \epsilon_1(k, R)}
    \bigg].
  \label{e:bartlett_intermediate}
\end{align}
Now, precisely because $\alpha_0$ is radial, we have
\begin{equation}
  \alpha_0(r, W)
  =
  \frac{1}{\omega_{d-1}}
  \int_{S^{d-1}}
  \alpha_0(r\bfu, W) \d\bfu,
  \label{e:radial}
\end{equation}
where $\d\bfu$ is the $(d-1)$-dimensional Hausdorff measure and $S^{d-1}$ is the unit sphere of $\bbR^d$, with surface area $\omega_{d-1}$.
Plugging \eqref{e:radial} into \eqref{e:bartlett_intermediate} yields
\begin{align*}
   S (k) - 1
   & = \lim_{R \rightarrow \infty}
  \frac{\rho  (2\pi)^{\frac{d}{2}}k^{1 -\frac{d}{2}}}{\leb{W} \omega_{d-1}}
  \int_{0}^{\infty}
  \int_{S^{d-1}}
  \int_{\bbR^d}
  r^{\frac{d}{2}}
  J_{d/2 - 1}(kr)\\
 & \text{\hspace{1cm}}g(r)
  \mathds{1}_{W\times W}(r \bfu + \bfy,\bfy)
  ~\d \bfy
  \d \bfu
  \d r
  - \epsilon_1(k, R)
  \\
   & = \lim_{R \rightarrow \infty}
  \frac{\rho (2\pi)^{\frac{d}{2}}}{\leb{W} \omega_{d-1}}
  \int_{0}^{\infty}
  \int_{S^{d-1}}
  \int_{\bbR^d} \frac {J_{d/2 - 1}(kr)}{(kr)^{d/2 - 1}} g(r)
\\
  &\text{\hspace{1cm}} \mathds{1}_{W\times W}(r \bfu + \bfy,\bfy)
   ~\d \bfy
  \d \bfu
  ~ r^{d-1}
  \d r
  - \epsilon_1(k, R)
  \\
   & = \lim_{R \rightarrow \infty} \frac{\rho (2\pi)^{\frac{d}{2}}}{\leb{W} \omega_{d-1}}
  \int_{\bbR^d}
  \int_{\bbR^d}
  \mathds{1}_{W\times W}(\bfr + \bfy, \bfy)\\
 &\text{\hspace{1cm}} \frac
  {J_{d/2 - 1}(k \|\bfr\|_2)}
  {(k\|\bfr\|_2)^{\frac{d}{2} - 1}}
  g(\|\bfr\|_2)
  ~\d \bfy
  \d \bfr
  - \epsilon_1(k, R).
\end{align*}
We now recognize an expectation using \eqref{eq:pcf_by_expectation}, so that $S(k) -1$ rewrites as
\begin{equation}
  \label{eq:isotropic_estimator}
  \lim_{R \rightarrow \infty}
  \frac{ (2\pi)^{\frac{d}{2}}}{\rho \leb{W} \omega_{d-1}}
  \mathbb{E}
  \left[
    \sum_{\bfx, \bfy \in \calX \cap W}^{\bfx \neq \bfy}
    \frac{J_{d/2 - 1}(k \|\bfx - \bfy\|_2)}{(k \|\bfx - \bfy\|_2)^{d/2 - 1}}
    \right]
  - \epsilon_1(k, R).
\end{equation}
We thus define a new estimator
\begin{equation}
  \label{eq:s_BI}
  \widehat{S}_{\mathrm{BI}}(k)
  =
  1
  + \frac{(2\pi)^{\frac{d}{2}}}{\rho \leb{W} \omega_{d-1}}
  \sum_{ \substack{i, j=1 \\ i \neq j } }^{N}
  \frac{J_{d/2 - 1}(k \|\bfx_i - \bfx_j\|_2)}{(k \|\bfx_i - \bfx_j\|_2)^{d/2 - 1}}
  ,
\end{equation}
along with its self-normalized version
\begin{equation}
  \label{eq:s_BI_self_normalized}
  \widehat{S}_{\mathrm{BI,s}}(k)
  =
  1
  +
  \frac{ (2\pi)^{\frac{d}{2}} }{N \omega_{d-1}}
  \sum_{ \substack{i, j=1 \\ i \neq j } }^{N}
  \frac{J_{d/2 - 1}(k \|\bfx_i - \bfx_j\|_2)}{(k \|\bfx_i - \bfx_j\|_2)^{d/2 - 1}}
  ,
\end{equation}
as in the case of the self-normalized scattering intensity \eqref{eq:s_si_s}.
When $d=2$, $\widehat{S}_{\mathrm{BI,s}}$ corresponds to Bartlett's isotropic estimator \citep{Bartlett:1964}.

Here also, there are two sources of bias in the estimator \eqref{eq:s_BI}.
The first one is due to the restriction of the point process to a bounded observation window, which shall disappear as the window grows. The second source of bias $\epsilon_1(k,R)$  is again related to the Fourier transform of the scaled intersection volume $\alpha_0(\cdot, W)$.
\citet{Diggle:1987} observed that $\vert \epsilon_1(k,R)\vert$ is larger when $k>0$ is small, and proposed to artificially set the value of the estimator to some constant when $k$ is smaller than a certain threshold \citep[Equation~3.4]{Diggle:1987}.
Obviously, this correction is inadequate to study hyperuniformity, i.e., the behavior of $S$ near zero.

An alternative to Diggle's clipping procedure is to proceed as done for the scattering intensity \eqref{eq:s_si} and estimate the structure factor only at a set of allowed wavenumbers, defined as the zeros of $\epsilon_1(\cdot,R )$.
Using \eqref{eq:ft_alpha0_radial}, it comes, for fixed $d$,
\begin{align*}
  \epsilon_1(k, R)
  &= \calF(\alpha_0)(k)\\
  &= 2^d \pi^{d/2}\frac{\Gamma(d/2 +1)}{k^d}J^2_{\frac{d}{2}}(kR) \\
   &=\begin{cases}
    0                       & \text{if } k \in \{ \frac{x}{R} ; J_{\frac{d}{2}}(x)=0 \}, \\
    \calO (R^d)             & \text{as } k \rightarrow 0,                        \\
    \calO\left(\frac{1}{k^d(Rk)^{2/3}}\right) & \text{as } k \rightarrow \infty.
  \end{cases}
\end{align*}
The two bounds respectively come from the fact that $J_\nu(x) \sim \frac{1}{\Gamma(\nu +1)}(\frac{x}{2})^\nu$ in the neighborhood of zero, and that for all $\nu>0$ and $x \in \bbR$, $\lvert J_\nu(x) \rvert  \leq c \lvert x \rvert ^{-1/3}$ (with $c\approx 0.8$) \citep{Landau:2000}.
Thus, for the estimator \eqref{eq:s_BI}, we let the set of allowed wavenumbers associated to the window $W=B^d(\mathbf{0}, R)$ be
\begin{equation}
  \label{eq:allowed_k_isotropic}
  \bbA_R = \left\{ \frac{x}{R} \in \bbR, \text{ s.t. }  J_{d/2}(x)=0 \right\}.
\end{equation}

\begin{proposition}
  Under Assumption~\ref{ass:2}, for $k \in\bbA_R$, the estimator $\widehat{S}_{\mathrm{BI}}$ is asymptotically unbiased, i.e.,
  $$
    \sup_{k \in \bbA_R}
    \left \lvert
    \mathbb{E}
    \left[
    \widehat{S}_{\mathrm{BI}} (k)
    \right]
    - S(k)
    \right \rvert
    \xrightarrow[R \rightarrow \infty]{} 0.
  $$
\end{proposition}
Note that we can also define debiased tapered and multitapered versions of Bartlett's estimator, as done in Section~\ref{sub:Estimators assuming stationarity}, but the choice of the taper(s) requires more attention, as they must be separable and radial.
% end subsubsection

\subsubsection{Using Ogata's quadrature} % (fold)
\label{ssub:Estimating the structure factor using Ogata quadrature}

Still working under Assumption~\ref{ass:2}, we can define alternative estimators of the structure factor \eqref{eq:sf_as_hankel_tranform} by first approximating the pair correlation function from a realization of $\calX$, and then approximating the Hankel transform \eqref{eq:hankel_and_Fourier}.

Estimators of the pair correlation function have been thoroughly investigated; see \citep{Baddeley+Rubak+Turner:2013}.
In a nutshell, they divide in, on one side, numerical derivatives of estimates of Ripley's $K$ function, and on the other side, direct kernel density estimators based on the collection of pairwise distances in the sample.
Both families come with sophisticated edge correction techniques, and, at least for small sample sizes, it seems reasonable to build on this previous work.
Henceforth, we assume that an estimator of the pair correlation function is available, and defer the discussion of which estimator to use
to Section~\ref{sec:Illustrating the toolbox}.

It remains to perform a numerical quadrature on a Hankel transform.
\citet{Ogata:2005} approximates integrals of the form
\begin{align}
  &\calI_\nu(f)
  = \int_0^\infty f(x) J_\nu(x) \d  x \nonumber
\end{align}
as
\begin{align}
  \pi
  \sum_{j=1}^{\infty}
  w_{\nu, j}
  f( \frac{\pi}{h}
  \psi(h \xi_{\nu, j}))
  J_{\nu}(\frac{\pi}{h}
  \psi(h \xi_{\nu, j}))
  \psi^{\prime}(h \xi_{\nu, j}),
  \label{eq:Ogata_quadrature}
\end{align}
with $w_{\nu, j}
= \frac{Y_{\nu}(\pi \xi_{\nu, j})}{J_{\nu +1}(\pi \xi_{\nu, j})}$ and
$
\psi(t)
  = t \times \tanh
  (\frac{\pi}{2} \sinh(t))
$.
$Y_\nu$ is the Bessel function of the second kind of order $\nu$, $h$ is a positive constant called the \emph{stepsize}, and $(\xi_{\nu, j})_{j\geq 1}$ are the positive zeros of the Bessel function $J_{\nu}(\pi x)$ of the first kind of order $\nu$, arranged in increasing order.
In practice, the infinite sum on the right-hand side of \eqref{eq:Ogata_quadrature} can be truncated at a small number of function evaluations since the quadrature nodes approach the zeros of $J_\nu(x)$, that is $\frac{\pi}{h} \psi(h \xi_{\nu, j}) \sim \pi \xi_{\nu, j}$, very fast as $j \rightarrow \infty$.

Ogata's quadrature applies to the Hankel transform \eqref{eq:hankel_tranfrom} of an integrable function $f$, since
\begin{equation}
  \label{eq:hankel_and_ogata}
  \calH_\nu (f)(k)
  = \calI_\nu \left( x \mapsto \frac{x}{k^2} f\left( x / k \right) \right).
\end{equation}
In particular, it applies to the computation of structure factors since, by \eqref{eq:sf_as_hankel_tranform} and \eqref{eq:hankel_and_Fourier},
\begin{align*}
  S& (k)
   = 1 + \rho \calF_s(g-1)(k)                  \\
   & = 1 + \rho \frac{(2 \pi)^{d/2}}{k^{d/2 -1}}
  \calI_{d/2 -1}\left( x\mapsto \frac{x^{d/2}}{k^{d/2 +1}} g(x/k)  -1  \right).
\end{align*}
We thus define the Hankel-Ogata estimator of the structure factor as
\begin{equation}
  \label{eq:s_ho}
  \widehat{S}_\mathrm{H O}(k) = 1 + \rho \frac{(2 \pi)^{d/2}\pi}{k^{\nu}}
  \sum_{j=1}^{N}
  w_{\nu, j}
  \tilde{h}_k\left(\frac{\pi}{h}
  \psi(h \xi_{\nu, j}) \right)
  J_{\nu} \left(\frac{\pi}{h}
  \psi(h \xi_{\nu, j}) \right)
  \psi^{\prime}(h \xi_{\nu, j}),
\end{equation}
with $\nu=d/2-1$, $N\in\mathbb{N}$, $\tilde{h}_k(x) = \frac{x^{d/2}}{k^{d/2 +1}} (\hat{g}(x/k)-1)$, and $\hat{g}$ an estimator of the pair correlation function.
Finally, note that Ogata's quadrature is also implemented in the \texttt{Python} package \href{hankel}{https://pypi.org/project/hankel/} of \citet{Murray+Poulin:2019}.
\paragraph{Relation between $k_{\min}$ and $r_{\max}$}
There exists a hidden inverse proportionality relation in Equation \eqref{eq:s_ho}, between the minimal wavenumber $k_{\min}$ for which we can hope the estimator to be accurate and the maximal radius $r_{\max}$ at which the pair correlation function has been estimated.
Truncating the infinite sum after $N$ terms in Equation~\eqref{eq:Ogata_quadrature} has been informally justified by
\begin{equation}
  \label{eq:condition on N}
  \psi(h \xi_{d/2-1, N}) \approx h \xi_{\frac{d}{2}-1, N}
\end{equation}
The maximum radius $r_{\max}$ at which $\hat{g}$ is available should in turn satisfy
\begin{equation}
  \label{eq:condition on r_max}
  r_{\max}
  = \max_j
  \left\{
  \frac{\pi}{hk} \psi(h \xi_{\frac{d}{2}-1, j});
   k \in \bbR^\ast
    \right\}
\end{equation}
Together, \eqref{eq:condition on N} and \eqref{eq:condition on r_max} entail that
\begin{equation}
  \label{eq:condition on k_min ogata}
  k_{\min} \approx \frac{\pi \xi_{d/2-1, N} }{r_{\max}}.
\end{equation}
Thus $k_{\min}$ is not only proportional to $1/r_{\max}$ but also to the largest considered zero of the Bessel function $J_{d/2-1}(x)$.

\subsubsection{Using the quadrature of Baddour and Chouinard} % (fold)
\label{ssub:Estimating the structure factor using the discrete Hankel transform (DHT)}

Instead of using the quadrature of \citet{Ogata:2005}, one can estimate Hankel transforms more directly, similarly to how the discrete Fourier transform is used to approximate Fourier transforms.
Intuitively, assuming that either $f$ or its Hankel transform $\mathcal{H}_\nu(f)$ \eqref{eq:hankel_tranfrom} has bounded support allows rewriting it as a Fourier-Bessel series, with coefficients involving evaluations of either $\mathcal H_\nu(f)$ or $f$, respectively.
Truncating the resulting Fourier-Bessel series yields approximate direct and inverse Hankel transforms.
This discrete Hankel transform (DHT) was derived by \citet{Baddour+Chouinard:2015}, and originally implemented in \href{Matlab}{https://sourceresearchsoftware.metajnl.com/articles/10.5334/jors.82/}.
Moreover, \citet{GuGu04} developed a \texttt{Python} package, \href{pyhank}{https://pypi.org/project/pyhank/}, based on the same idea.

In detail, let $N>0$ and $f:\mathbb{R}^+ \rightarrow \mathbb{R}$ be a continuous function, \citet{Baddour+Chouinard:2015} approximate
\begin{equation}
  \label{eq:DHT}
  \calH_\nu(f)(k_m) \approx
  \frac{r_{\max}^2}{\eta_{\nu, N}} \sum_{j=1}^{N-1} \frac{2 J_\nu \left( \frac{\eta_{\nu, m} \eta_{\nu, j}}{\eta_{\nu, N}}\right)f(r_j)}{\eta_{\nu, N} J^2_{\nu+1}(\eta_{\nu, j})},
  \end{equation}
where $\eta_{\nu, m} = \pi \xi_{\nu, m}$ is the $m^{th}$ positive zero of the Bessel function $J_\nu(x)$ of the first kind, $1\leq j,m\leq N-1$ and
\begin{equation}
  \label{e:BC_evaluations}
  r_j = \frac{\eta_{\nu, j}}{\eta_{\nu, N}} r_{\max},
  \quad
  k_m = \frac{\eta_{\nu, m}}{\eta_{\nu, N}} k_{\max},
  \quad
  k_{\max} = \frac{\eta_{\nu, N}}{r_{\max}},
  \quad\text{with}\quad
  r_{\max} > 0.
\end{equation}
The user thus needs to specify both $N$ and $r_{\max}$. Intuitively, the choice of $r_{\max}$ is governed by how far on the positive axis one has been able to evaluate $f$.
Once $r_{\max}$ is fixed, $N$ decides how large $k_{\max}$ is, that is, how high in frequency one wishes to estimate the Hankel transform.

To conclude, given an estimator $\hat{g}$ of the pair correlation function, we define yet another estimator of the structure factor, called the Hankel-Baddour-Chouinard estimator,
\begin{equation}
  \label{eq:s_hbc}
  \widehat{S}_\mathrm{H BC}(k_m) =
   1 + \rho (2\pi)^{\frac{d}{2}} \frac{r_{\max}^2}{\eta_{\nu, N}} \sum_{j=1}^{N-1} \frac{2 J_\nu\left( \frac{\eta_{\nu, m} \eta_{\nu, j}}{\eta_{\nu, N}}\right)}{\eta_{\nu, N} J^2_{d/2}(\eta_{\nu, j})k_m^{\nu}}\tilde{h}(r_j),
\end{equation}
where $\nu=d/2-1$, $\tilde{h}(x) = x^{\nu} (\hat g(x) -1)$, and the set of wavenumbers $\{k_m\}_m$ is fixed by \eqref{e:BC_evaluations}.
Finally, we can deduce from \eqref{e:BC_evaluations} that the minimal wavenumber of  $\widehat{S}_{\mathrm{HBC}}$ \eqref{eq:s_hbc} is $k_1 = k_1^{d/2-1} = \frac{\eta_{d/2-1, 1}}{r_{\max}}$.
Comparing $k_1$ with the minimal wavenumber $k_{\min}$ \eqref{eq:condition on k_min ogata} of $\widehat{S}_{\mathrm{HO}}$ \eqref{eq:s_ho}, for the same number of points $N$ and the same $r_{\max}$, we observe that $k_1 < k_{\min}$, as $k_1$ is proportional to the first zero of the Bessel function $J_{d/2 - 1}(x)$ while $k_{\min}$ is proportional to the $N^{th}$ zero of $J_{d/2 - 1}(x)$.
For the study of hyperuniformity, this suggests an advantage to using $\widehat{S}_{\mathrm{HBC}}$.
%End Section

\section{A statistical test of hyperuniformity} % (fold)
\label{sec:The coupled sum estimator and a test of hyperuniformity}
In Section~\ref{sec:Estimators of the structure factor}, we have surveyed estimators $\widehat{S}$ of the structure factor $S$ of a stationary point process $\calX$.
In the best case, we had an asymptotically unbiased estimator of $S$.
Furthermore, hyperuniformity diagnostics like the effective hyperuniformity surveyed in Section~\ref{sub:Effective hyperuniformity} require subjective algorithmic choices and do not come with a statistical guarantee.

In this section, we use the debiasing device of \cite{Rhee+Glynn:2015} to turn realizations of any of the nonnegative, asymptotically unbiased estimators of Section~\ref{sec:Estimators of the structure factor}, on windows of increasing size, into an unbiased estimator of a truncated equivalent to $S(\mathbf{0})$.
We then propose the first asymptotically valid test of hyperuniformity.
We group our assumptions and notation for this section in Assumption~\ref{ass:3}.
\begin{assumption}{A}{3} \label{ass:3}
  $\calX$ is a stationary point process of $\bbR^d$ with intensity $\rho > 0$.
  Its pair correlation function $g$ exists, and $ \bfr \mapsto g(\bfr) -1 $ is integrable on $\mathbb{R}^{d}$.
 \end{assumption}

 \subsection{The coupled sum estimator} % (fold)
 \label{sub:The coupled sum estimator}

Consider a stationary point process $\calX$ of $\bbR^d$, of which we observe the intersection of a single realization with multiple increasing windows.
Formally, we consider an increasing sequence of sets $(\calX \cap W_m)_{m \geq 1}$, with $W_s \subset W_r $ for all $0< s<r$, and $W_\infty = \bbR^d$.
For simplicity, we assume that the windows are centered and either rectangular as in Assumption~\ref{ass:1}, or ball-shaped as in
Assumption~\ref{ass:2} if $\calX$ is isotropic.

We consider any of the positive, asymptotically unbiased estimators $\widehat{S}_m$ of $S$ listed in Section~\ref{sec:Estimators of the structure factor}, applied to $\calX\cap W_m$.
All such estimators are asymptotically unbiased on a set of values $\mathbb{B}_m$, possibly in the sense of Proposition~\ref{prop:scattering_intensity_conv}, and there exists $\bfk_m^{\text{min}}\in\mathbb{B}_m$ with $ \bfk_m^{\text{min}} \rightarrow \bf0$.
The reader can keep in mind the scattering intensity of Section~\ref{ssub:The scattering intensity}, with, say, the minimum restricted allowed value $\mathbf{k}_m^{\text{min}} =(2\pi/L_{m_1} ,\dots,2\pi/L_{m_d})\in \mathbb{Z}^d$, but the proofs hold more generally.

  % where $\widehat{S}_m$ is one of the positive, asymptotically unbiased estimators of $S$ listed in Section~\ref{sec:Estimators of the structure factor}, applied on the observation $\calX\cap W_m$, $\mathbf{k}_m^{\text{min}}$ is the minimum allowed wavevector associated with $W_m$ and the chosen estimator, so that $\mathbf{k}_m^{\text{min}} \rightarrow \mathbf{0}$ as $m$ goes to infinity.
  % The reader can keep in mind the scattering intensity of Section~\ref{ssub:The scattering intensity}, with, say, the minimum restricted allowed value $\mathbf{k}_m^{\text{min}} =(2\pi/L_1 ,\dots,2\pi/L_d)\in \mathbb{Z}^d$, but the proofs hold more generally.

We define the sequence of random variables
\begin{equation}
  \label{e:capped_estimates}
  Y_m = 1\wedge \widehat{S}_m(\mathbf{k}_m^{\text{min}}), \quad m\geq 1,
\end{equation}
 We cap the estimators at $1$ arbitrarily to make them uniformly bounded.
 The idea is to use the sequence $(Y_m)$ to test whether $S(\mathbf{0}) = 0$.

Following \cite[Section 2]{Rhee+Glynn:2015}, we first define a new sequence
\begin{equation}
\label{eq:Z_m}
  Z_m= \sum_{j=1}^{m\wedge M} \frac{Y_j - Y_{j-1}}{\mathbb{P}(M\geq j)},\quad m\geq 1,
\end{equation}
where $M$ is an $\mathbb{N}$-valued random variable such that $\mathbb{P}(M \geq j)>0$ for all $j$, and $Y_{0}=0$ by convention.
\citet{Rhee+Glynn:2015} observed that
\begin{equation*}
\label{eq:}
  \bbE[Z_m]= \bbE[Y_m] \quad \text{and} \quad Z_m \xrightarrow[m \rightarrow \infty]{\text{a.s.} } Z,
\end{equation*}
where $Z$ is the \emph{coupled sum estimator}
\begin{equation}
\label{eq:coupeled_sum_estimator}
  Z = \sum_{j=1}^{M} \frac{Y_j - Y_{j-1}}{\mathbb{P}(M\geq j)}.
\end{equation}
Under some conditions, \cite{Rhee+Glynn:2015} proved that if $Y_m \rightarrow Y$ in $L^2$, then $Z \in L^2$ is an unbiased estimator of $\bbE [Y]$ with a square root convergence rate.
For our choice of $(Y_m)_m$, the assumptions of \cite[Theorem 1]{Rhee+Glynn:2015}, especially the $L^2$ convergence of $Y_m$, may not hold for non-hyperuniform point processes.
We thus use weaker assumptions that are still enough to build a hyperuniformity test.

\begin{proposition} \label{prop:hu_test}
Under Assumption~\ref{ass:3}, with $Z$ defined in \eqref{eq:coupeled_sum_estimator}, assume that $M \in L^p$ for some $p\geq 1$.
Then $Z\in L^p$ and $Z_m \rightarrow Z$ in $L^p$.
Moreover,
 \begin{enumerate}
   \item If $\calX$ is hyperuniform, then $\bbE[Z] =0$.
   \item If $\calX$ is not hyperuniform and
   \begin{align}
    \label{e:boundedness_condition}
   \sup_{m }\limits  \bbE  [\widehat{S}^2_m(\bfk_m^{\mathrm{min}})] < \infty,
   \end{align}
  then $\bbE [Z] \neq 0$.
 \end{enumerate}
\end{proposition}
The proof is deferred to Appendix~\ref{proof:prop3}.
Assumption~\eqref{e:boundedness_condition} bears on the estimator that we use for the structure factor and the point process.
We believe it not to be too strong and we prove it in Appendix~\ref{appB: condition_of_prop3} for the scattering intensity for a Poisson point process.
Proposition~\ref{prop:hu_test} naturally leads to a test of hyperuniformity.

\subsection{A multiscale test} % (fold)
\label{sub:Multiscale hyperuniformity test}
We apply Proposition~\ref{prop:hu_test} with $p=2$, say $M$ is a Poisson random variable with mean $\lambda>0$.
Then $\Var [Z]< \infty$, and we can apply the central limit theorem to build a standard test comparing $\mathbb E [Z]$ to zero.

Consider $A$ i.i.d. pairs $(\calX_a, M_a)_{a=1}^A$ of realizations of $(\calX, M)$, and let $Z_1, \cdots Z_A$ be the $A$ corresponding i.i.d. copies of $Z$.
Now, denote the sample mean and sample standard deviation of $Z$ by $\bar{Z}_A$
and $\bar{\sigma}_A$.
Since the variance of $Z$ is finite, Slutsky's lemma yields the usual asymptotic confidence interval $CI[\bbE [Z]]$ of level $\zeta$ for $\bbE [Z]$,
\begin{equation}
  \left[\bar{Z}_A - z \bar{\sigma}_A A^{-1/2}, \bar{Z}_A + z \bar{\sigma}_A A^{-1/2} \right],
\label{e:confidence_interval}
\end{equation}
where $z$ is chosen such that $\bbP (-z< \calN (0,1) <z) =  \zeta$, and $\calN (0,1)$ denotes the standard normal distribution.
By Proposition~\ref{prop:hu_test}, for an estimator of the structure factor satisfying \eqref{e:boundedness_condition}, a test of hyperuniformity of asymptotic level $\zeta$ consists in assessing whether $0$ lies in the interval \eqref{e:confidence_interval}.
Since the estimators $Z$ correspond to windows of different sizes, we call the test \emph{multiscale}.

\section{Demonstrating all estimators} % (fold)
\label{sec:Illustrating the toolbox}

We have implemented all estimators of Section~\ref{sec:Estimators of the structure factor}, as well as the regression diagnostics of Section~\ref{sub:Effective hyperuniformity}, in an open-source \texttt{Python} toolbox called \toolbox\footnote{\url{https://github.com/For-a-few-DPPs-more/structure-factor}}.

In this section, we quickly demonstrate the toolbox on the four point processes described in Section~\ref{sub:Some point processes and their structure factor}, that is,
the KLY process of intensity $\rho_{\text{KLY}}=1$, the Ginibre ensemble of intensity $\rho_{\text{Ginibre}} = 1/\pi$, the Poisson process of intensity $\rho_{\mathrm{Poisson}}=1/\pi$, and the Thomas process with intensity $\rho_{\text{Thomas}} = 1/\pi$, $\rho_{\text{parent}}=1/(20\pi)$ and $\sigma=2$.
Our choice of observation window depends on the intensity and makes sure that we get samples of around $5800$ points.
The dimension is always $d=2$.
All figures in this section can be reproduced by following our demonstration \href{notebook}{https://github.com/For-a-few-DPPs-more/structure-factor/tree/main/notebooks} \footnote{\url{https://github.com/For-a-few-DPPs-more/structure-factor/tree/main/notebooks}}.
% subsectionInstallation (end)

\subsection{Basic software objects} % (fold)
\label{sub:Initialization}
While we refer to our online \href{documentation}{https://for-a-few-dpps-more.github.io/structure-factor/}\footnote{\url{https://for-a-few-dpps-more.github.io/structure-factor/}} for details, we believe that a quick overview of the main objects of our package is useful.
All estimators from Section~\ref{sec:Estimators of the structure factor} are methods of the class \texttt{StructureFactor}.
The class constructor takes as input an object of type \texttt{PointPattern}.
In a nutshell, for a stationary point process $\calX$ of intensity $\rho$, a \texttt{PointPattern} encapsulates a sample of $\calX \cap W = \{x_1, \ldots, x_N\}$, the observation window $W$, and the intensity $\rho$ (optional).
If the intensity of $\calX$ is not provided by the user, it is automatically approximated by the asymptotically unbiased estimator $ \widehat{\rho} = \frac{N}{\leb{W}}$.
Finally, to comply with the requirements of estimators on specific windows, we provide a \texttt{restrict\_to\_window()} method for the class \texttt{PointPattern}.

\subsection{Demonstrating estimators that only assume stationarity} % (fold)
\label{ssub:test  The scattering intensity}

\paragraph{The scattering intensity}
\begin{figure*}[!ht]
  %\vspace{-0.9cm}
  \begin{tabular}{p{\dimexpr 0.03\textwidth-\tabcolsep}p{\dimexpr 0.23\textwidth-\tabcolsep}p{\dimexpr 0.23\textwidth-\tabcolsep}p{\dimexpr 0.23\textwidth-\tabcolsep}p{\dimexpr 0.23\textwidth-\tabcolsep}}
    \multirow{9}{*}{\rotatebox[origin=l]{90}{Point process}}                                  &
    \raisebox{-\height}{\includegraphics[width=0.9\linewidth]{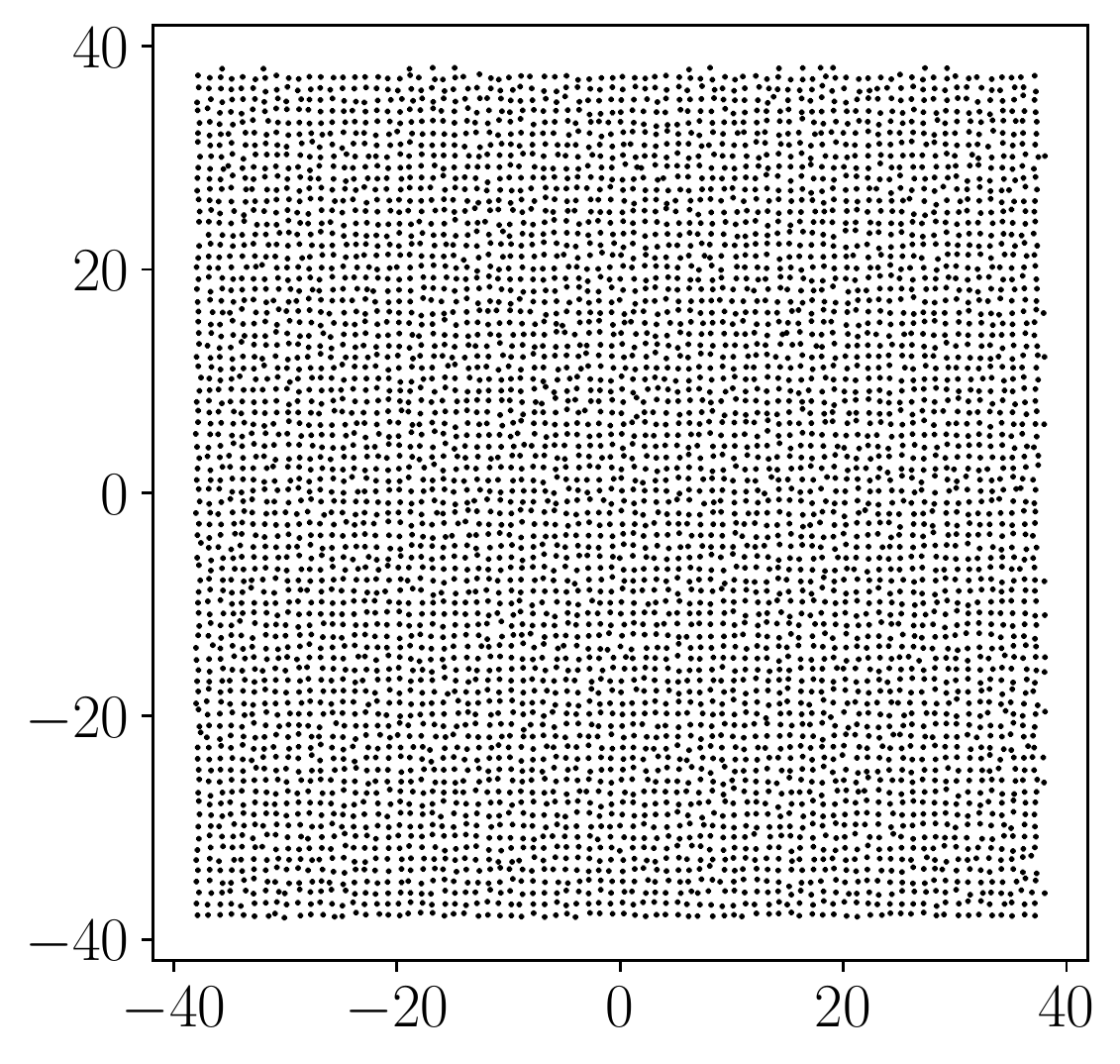}}     &
    \raisebox{-\height}{\includegraphics[width=0.9\linewidth]{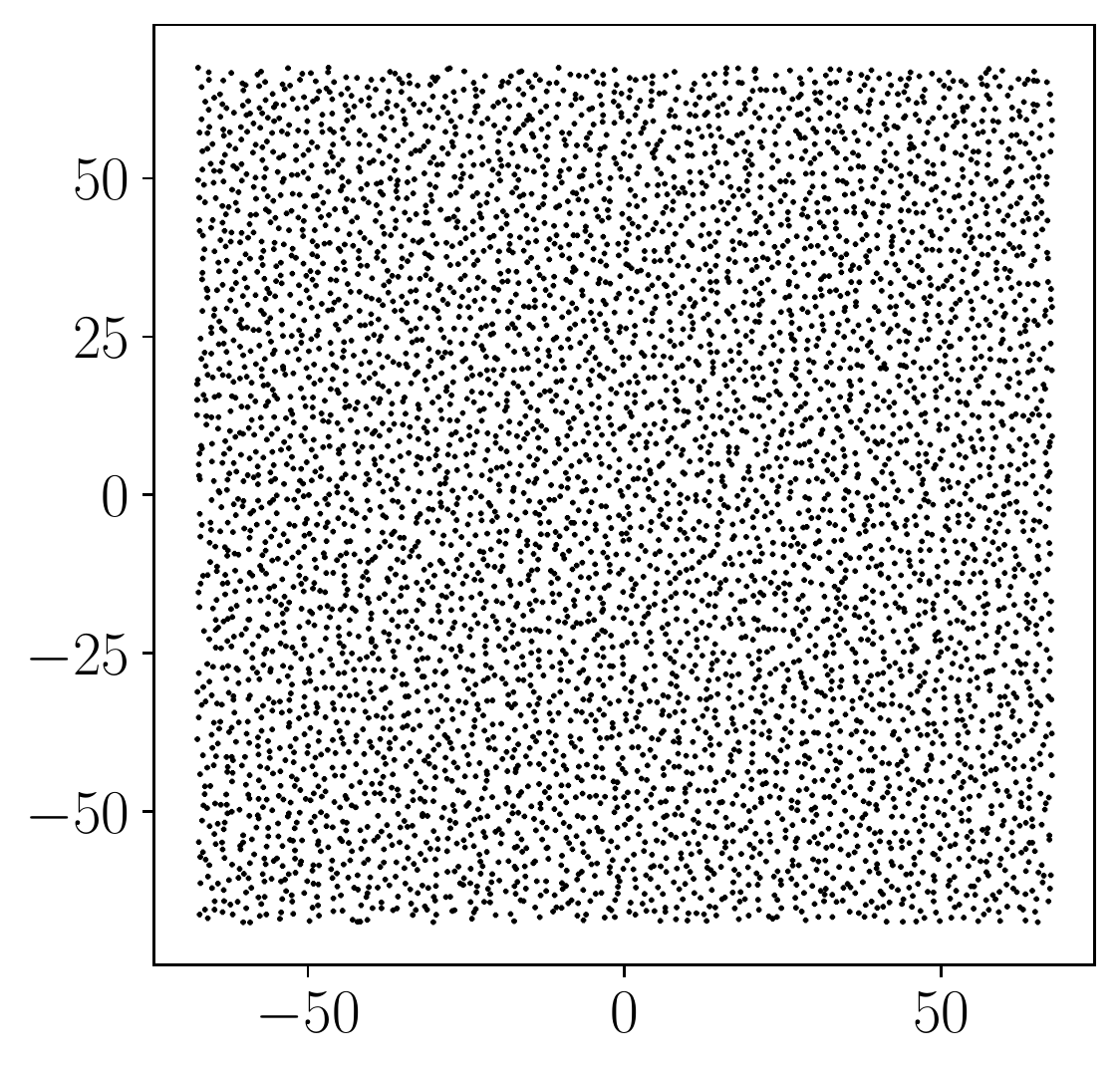}} &
    \raisebox{-\height}{\includegraphics[width=0.9\linewidth]{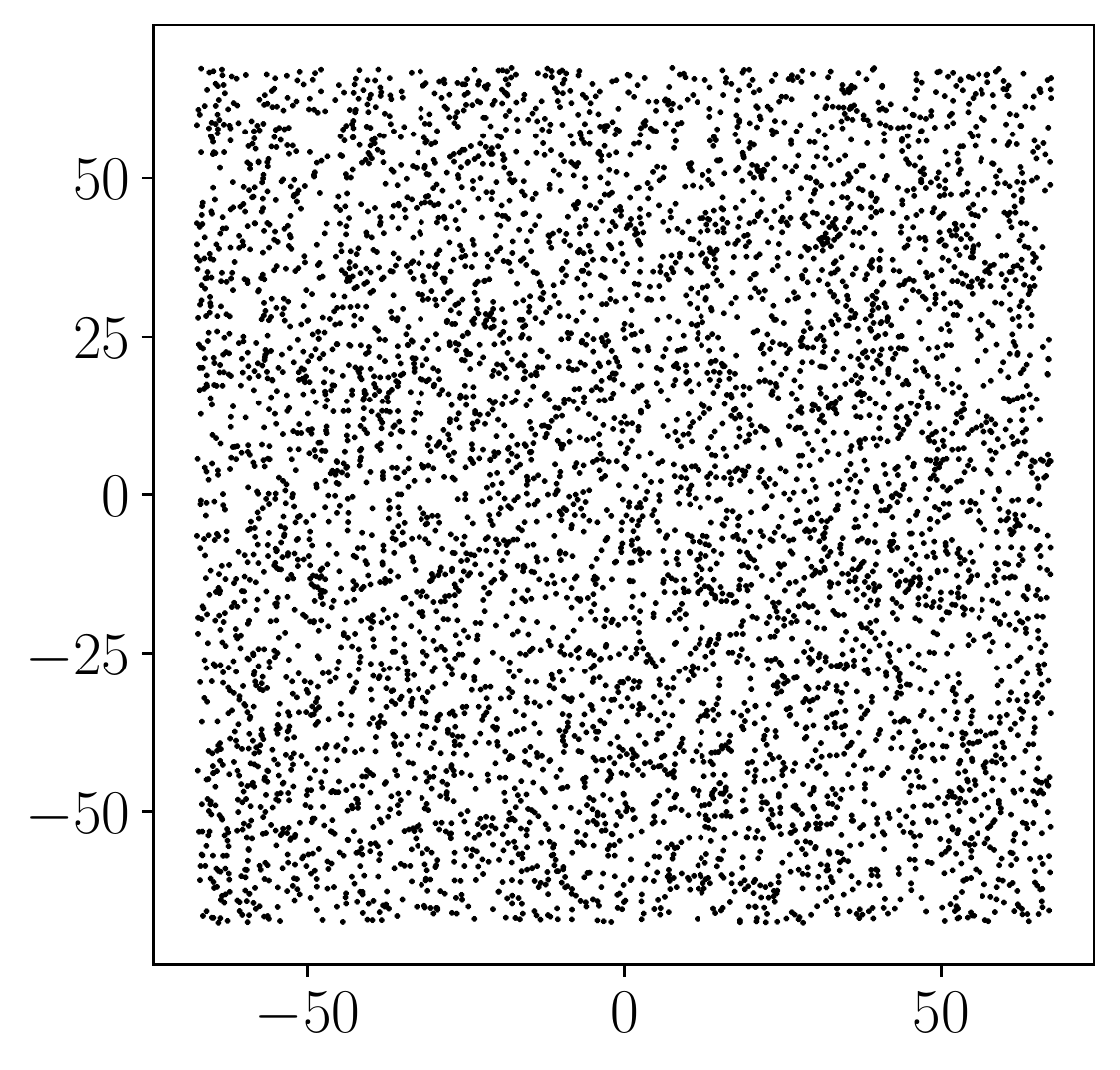}} &
    \raisebox{-\height}{\includegraphics[width=0.9\linewidth]{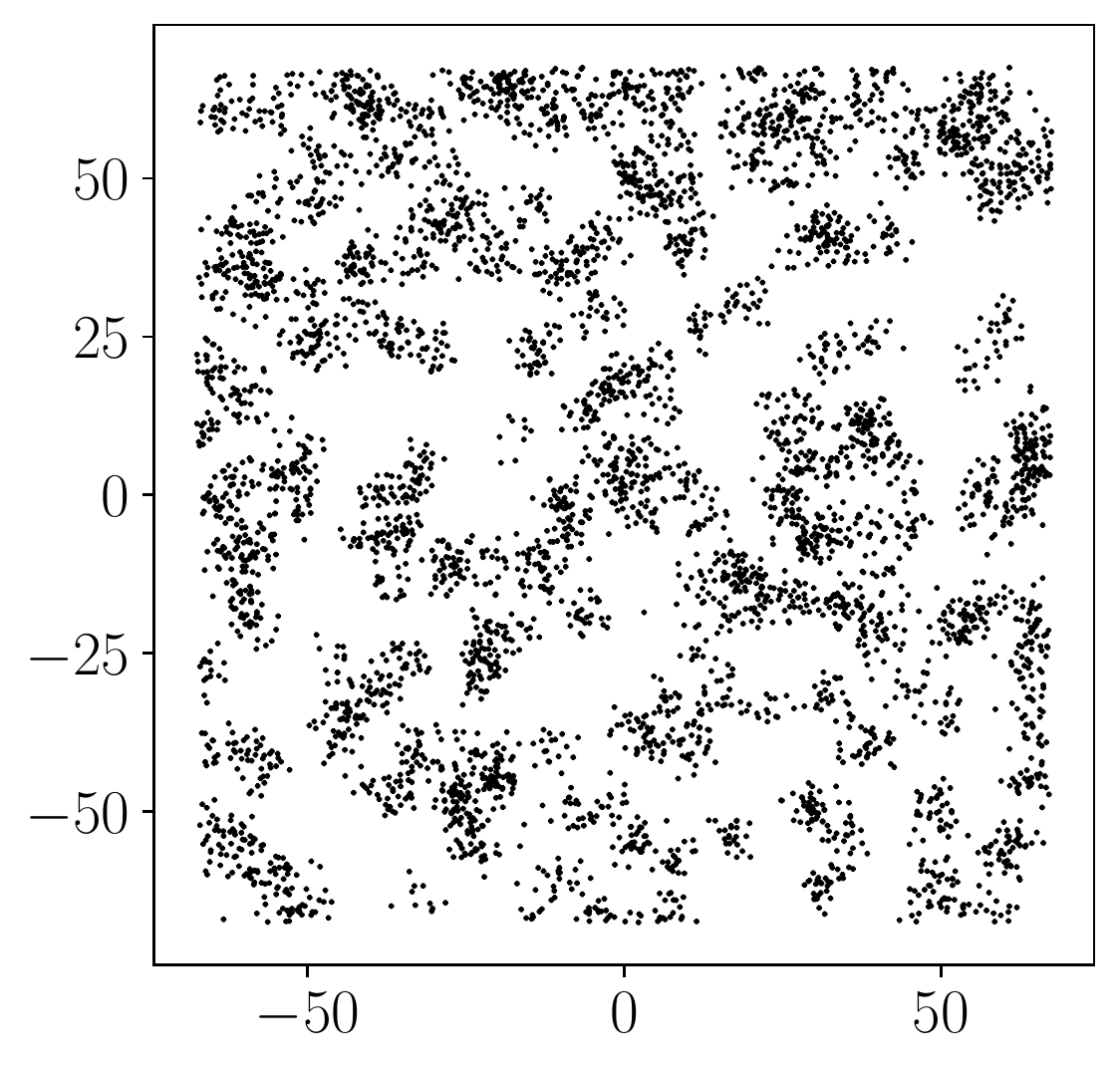}}
  \end{tabular}
  \vspace{-0.2cm}
  \begin{tabular}{p{\dimexpr 0.03\textwidth-\tabcolsep}p{\dimexpr 0.23\textwidth-\tabcolsep}p{\dimexpr 0.23\textwidth-\tabcolsep}p{\dimexpr 0.23\textwidth-\tabcolsep}p{\dimexpr 0.23\textwidth-\tabcolsep}}
    \multirow{9}{*}{\rotatebox[origin=l]{90}{$\widehat{S}_{\mathrm{SI}}(\bfk )$}}         &
    \raisebox{-\height}{\includegraphics[width=1\linewidth]{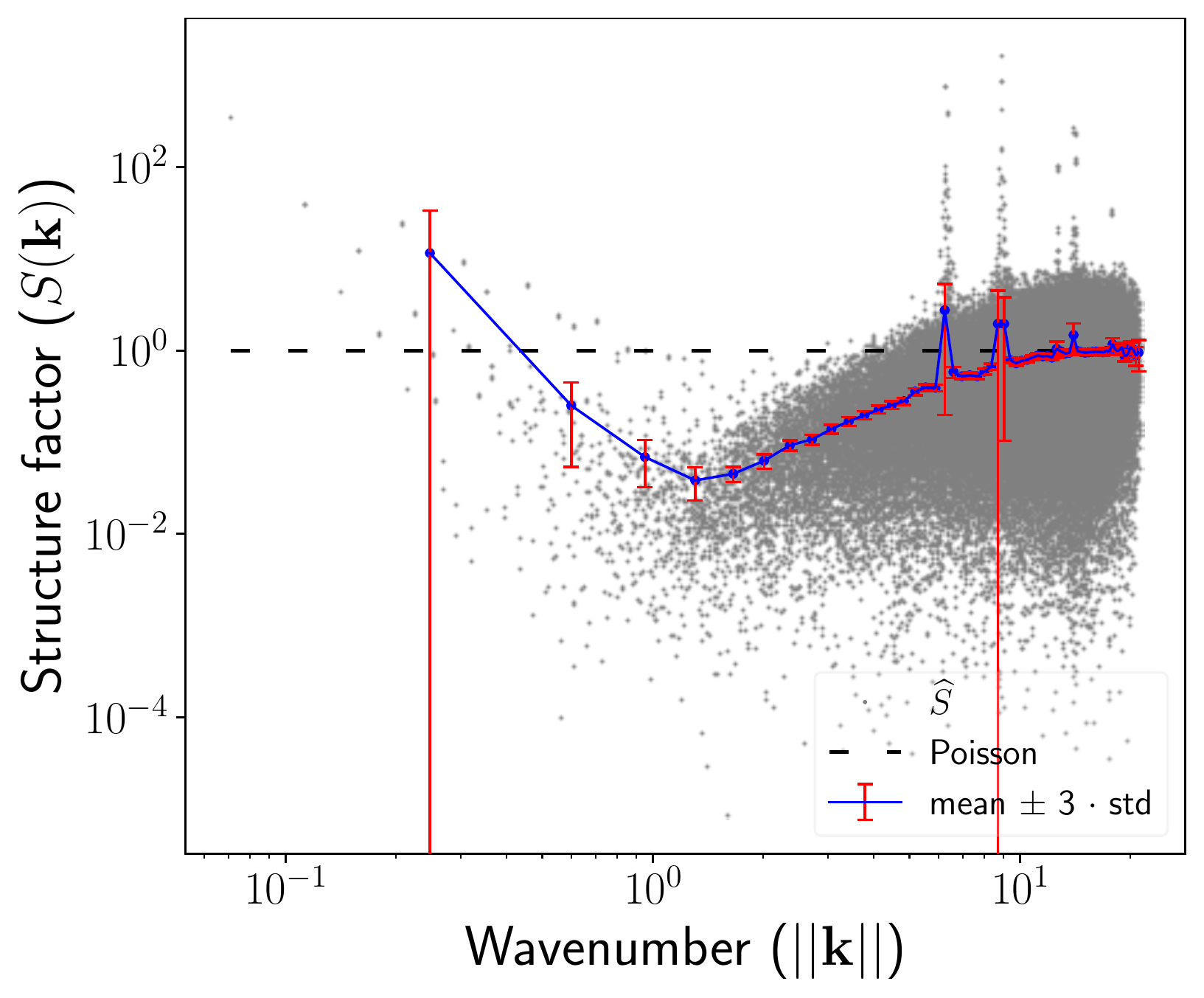}}     &
    \raisebox{-\height}{\includegraphics[width=1\linewidth]{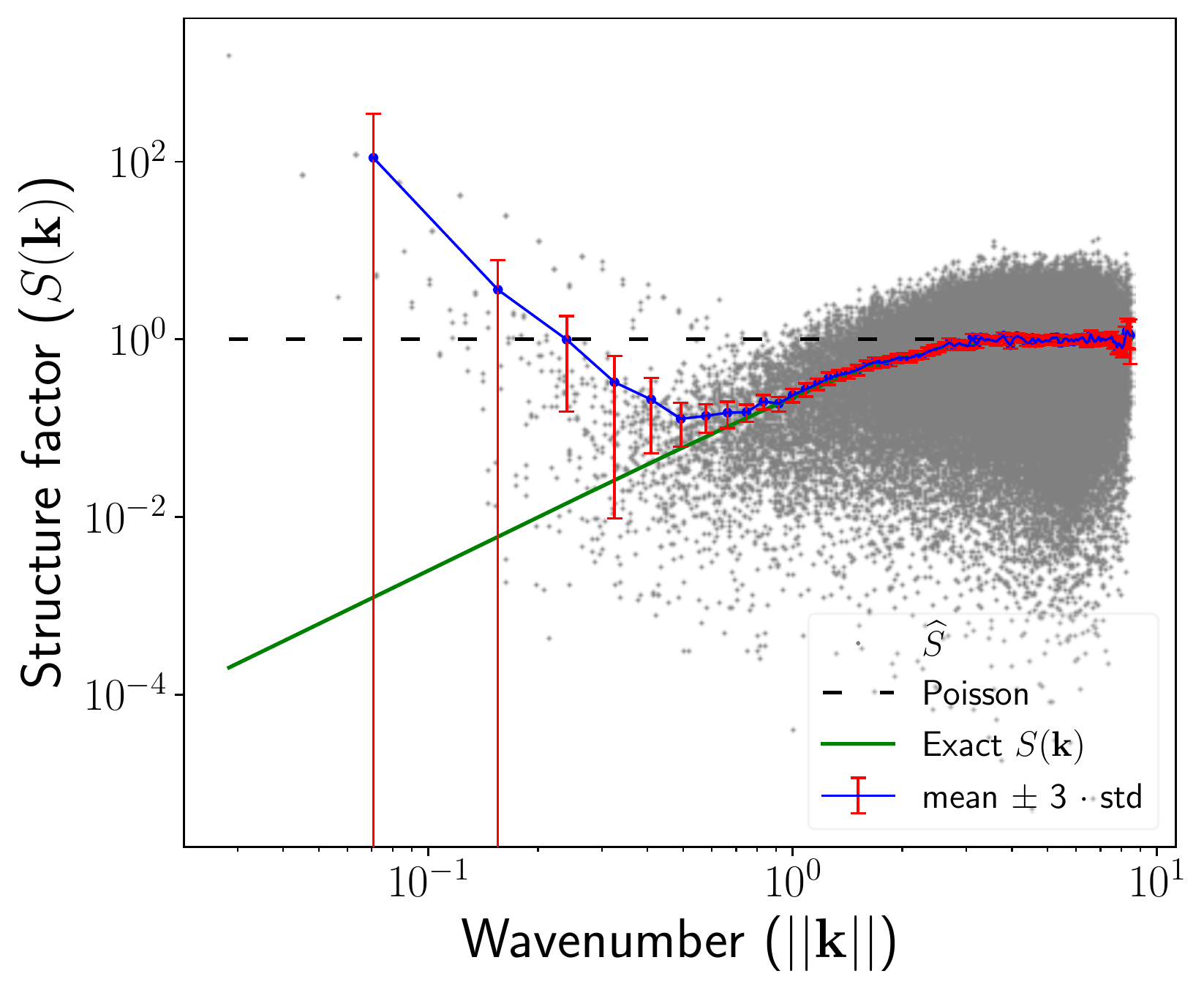}} &
    \raisebox{-\height}{\includegraphics[width=1\linewidth]{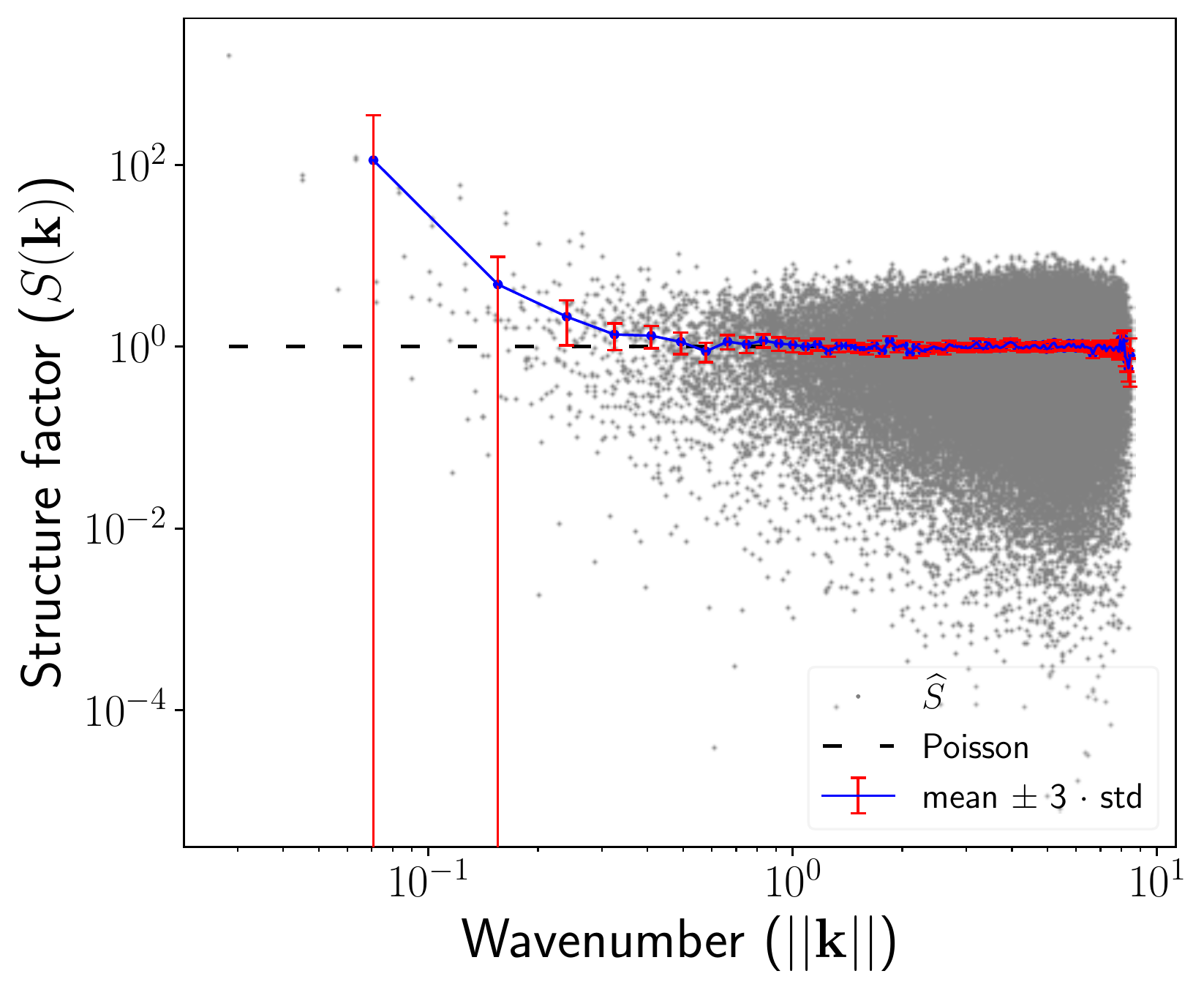}}    &
    \raisebox{-\height}{\includegraphics[width=1\linewidth]{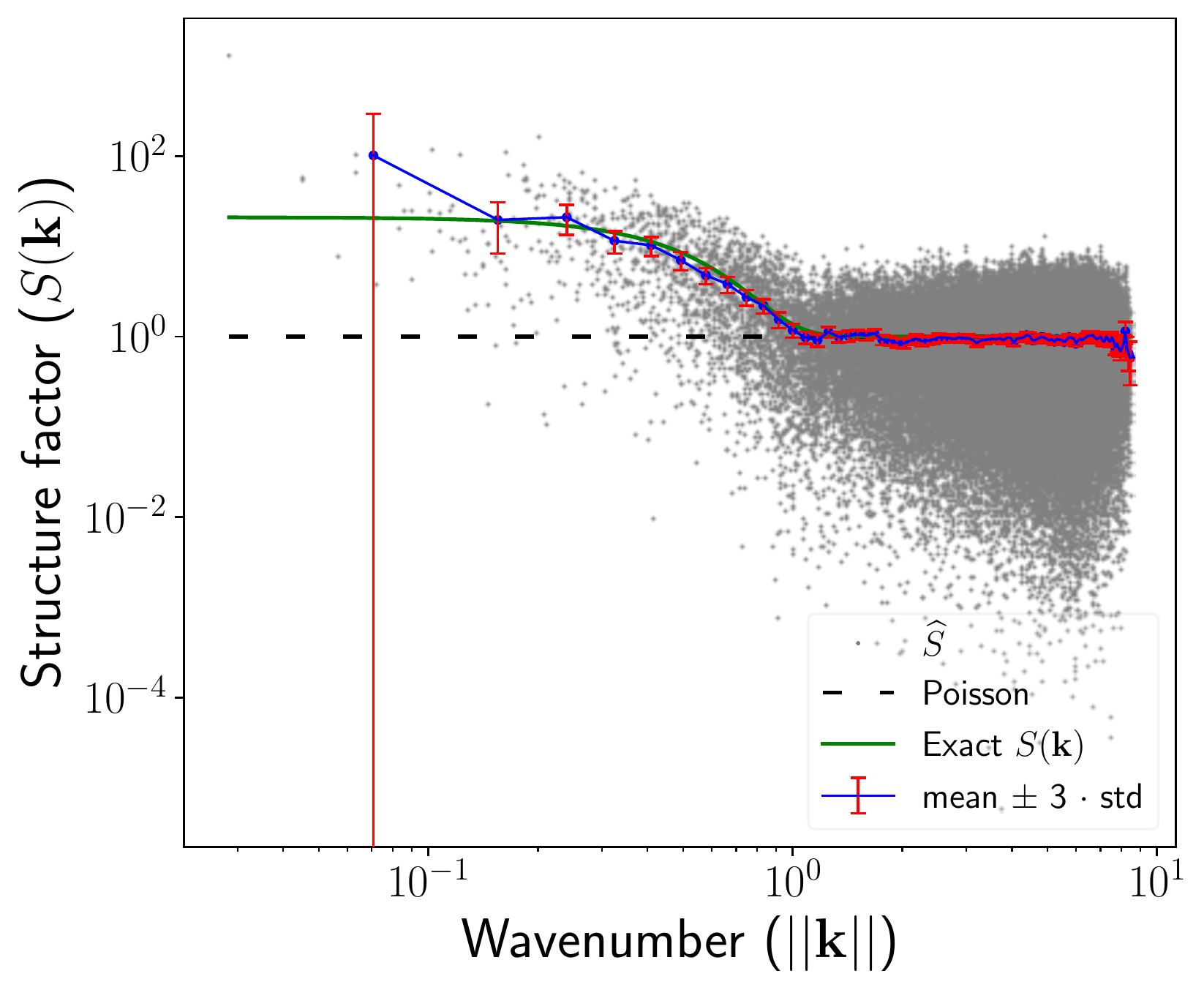}}
  \end{tabular}
  \vspace{-0.2cm}
  \begin{tabular}{p{\dimexpr 0.03\textwidth-\tabcolsep}p{\dimexpr 0.23\textwidth-\tabcolsep}p{\dimexpr 0.23\textwidth-\tabcolsep}p{\dimexpr 0.23\textwidth-\tabcolsep}p{\dimexpr 0.23\textwidth-\tabcolsep}}
    \multirow{9}{*}{\rotatebox[origin=l]{90}{$\widehat{S}_{\mathrm{SI}}(\frac{2\pi\bfn}{L} )$}}   &
    \raisebox{-\height}{\includegraphics[width=1\linewidth]{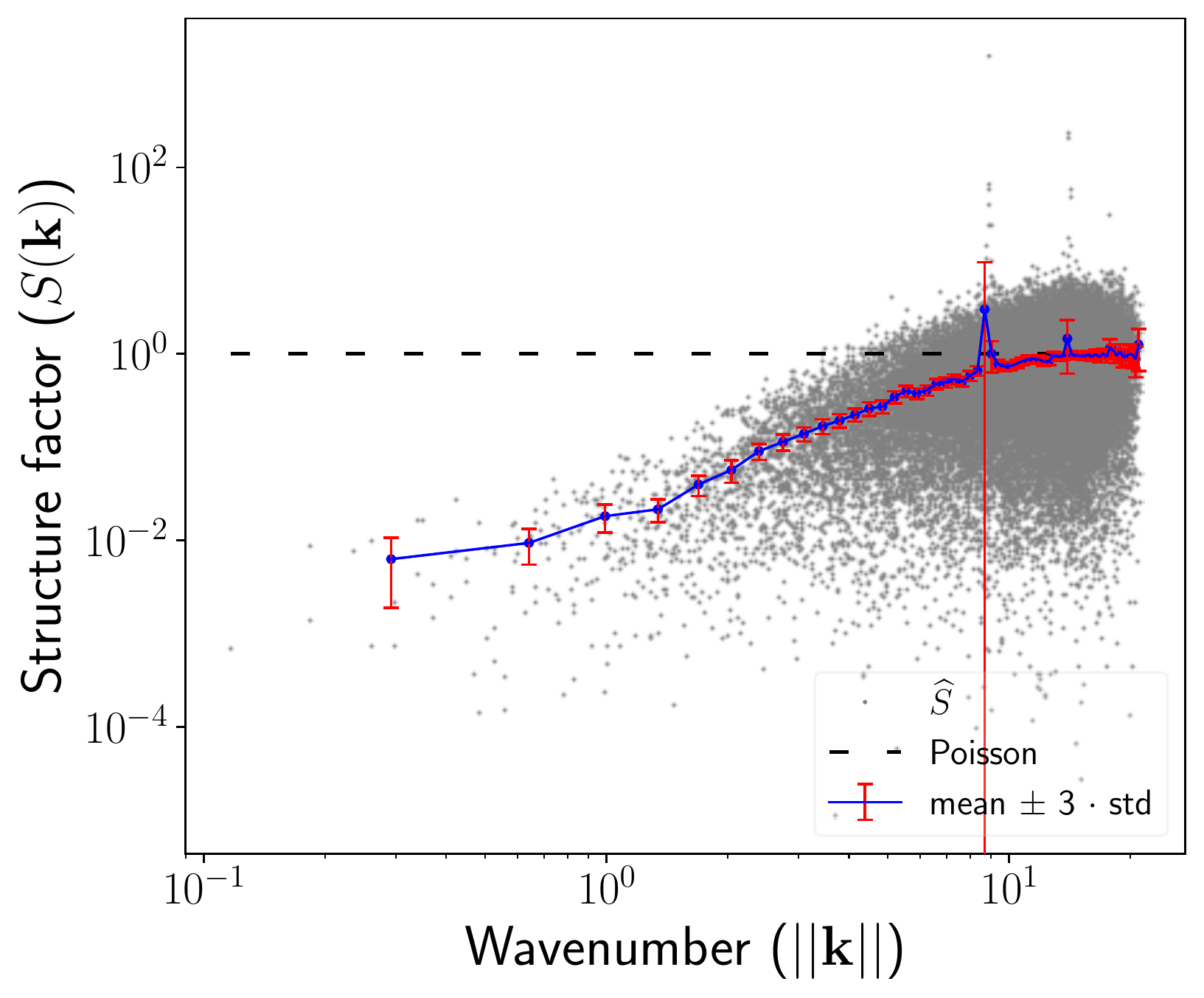}}     &
    \raisebox{-\height}{\includegraphics[width=1\linewidth]{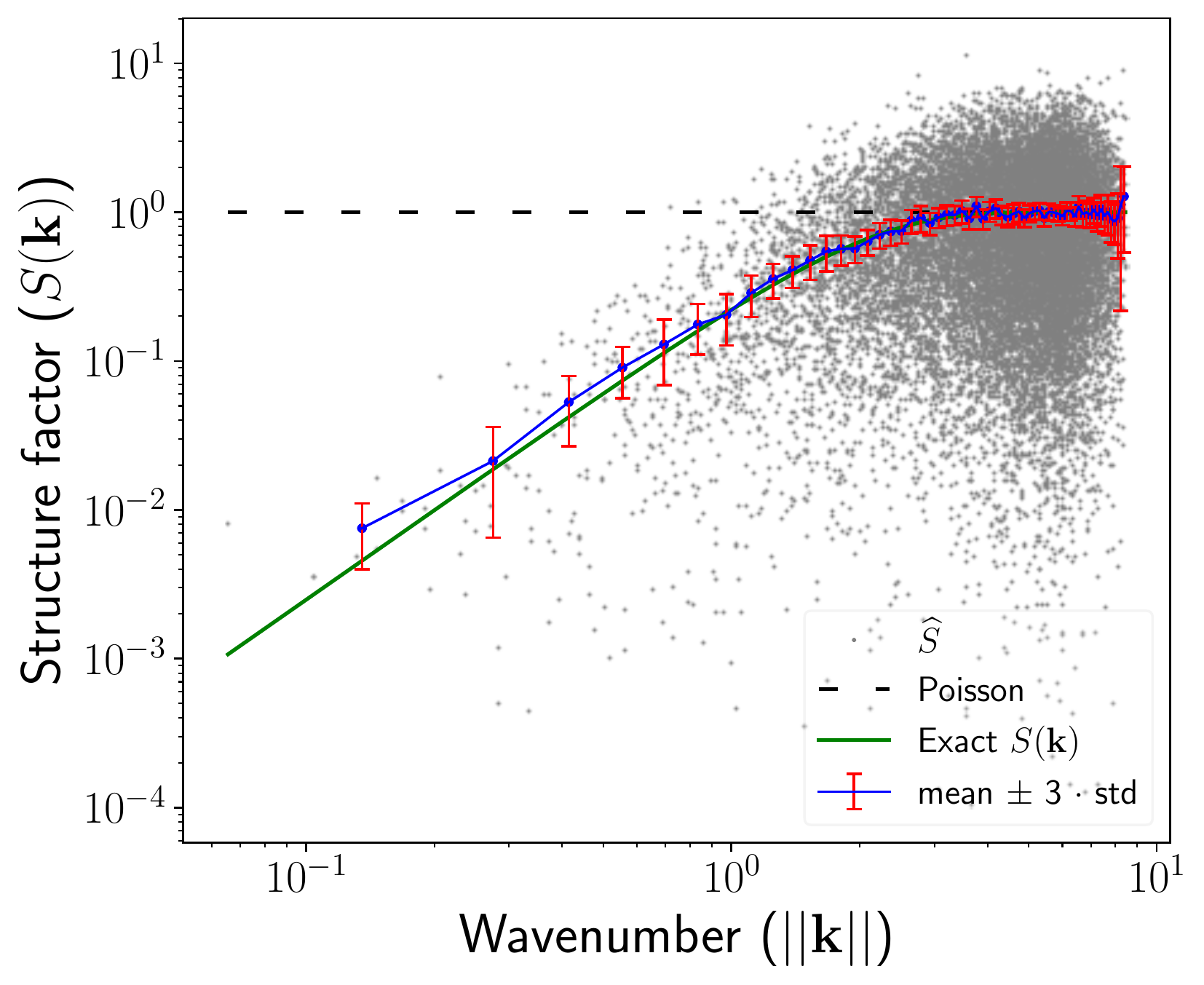}} &
    \raisebox{-\height}{\includegraphics[width=1\linewidth]{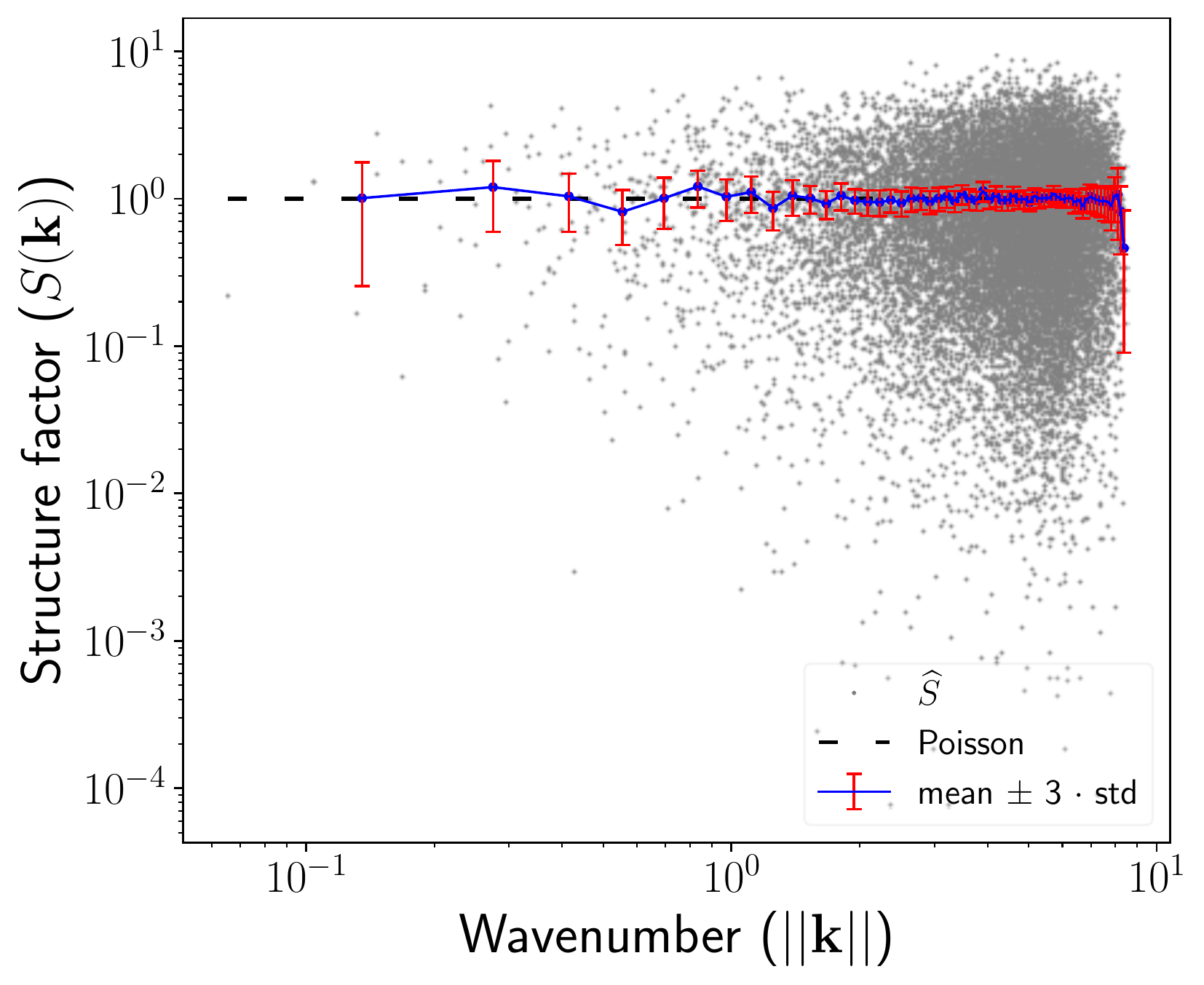}}    &
    \raisebox{-\height}{\includegraphics[width=1\linewidth]{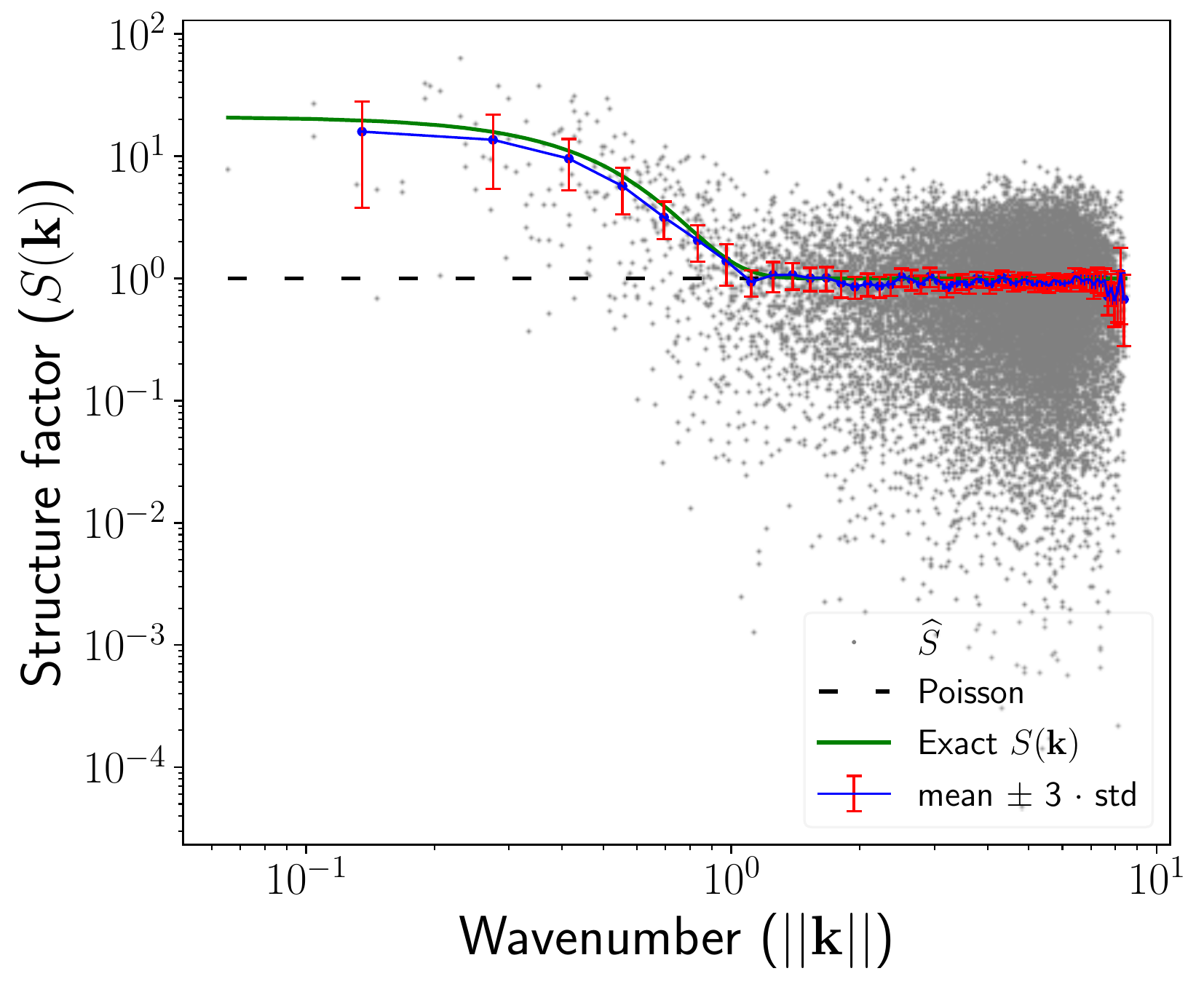}}
  \end{tabular}
  \vspace{-0.2cm}
  \begin{tabular}{p{\dimexpr 0.03\textwidth-\tabcolsep}p{\dimexpr 0.23\textwidth-\tabcolsep}p{\dimexpr 0.23\textwidth-\tabcolsep}p{\dimexpr 0.23\textwidth-\tabcolsep}p{\dimexpr 0.23\textwidth-\tabcolsep}}
    \multirow{9}{*}{\rotatebox[origin=l]{90}{$\widehat{S}_{\mathrm{DDT}}(t_0, \bfk)$}}            &
    \raisebox{-\height}{\includegraphics[width=1\linewidth]{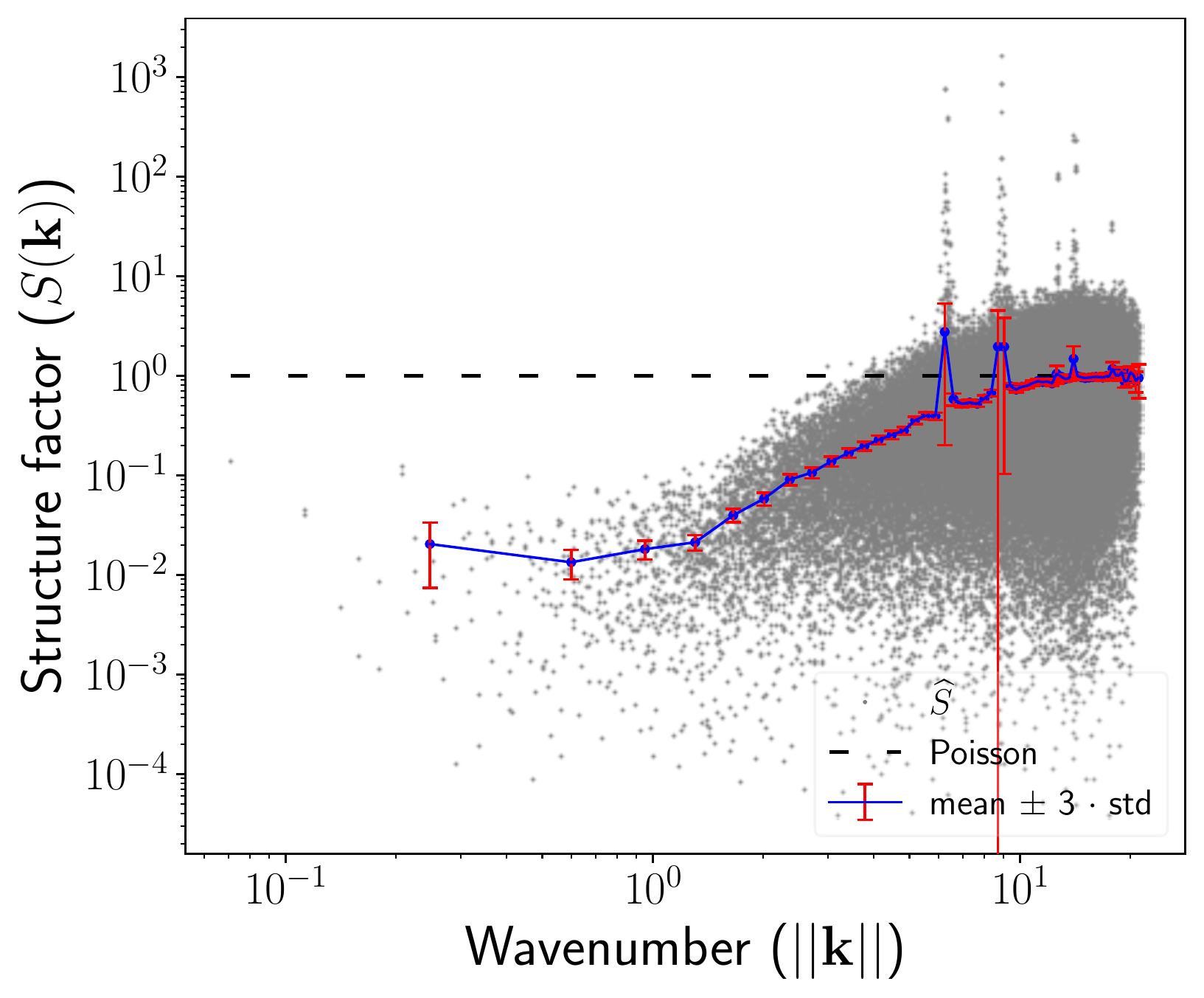}}     &
    \raisebox{-\height}{\includegraphics[width=1\linewidth]{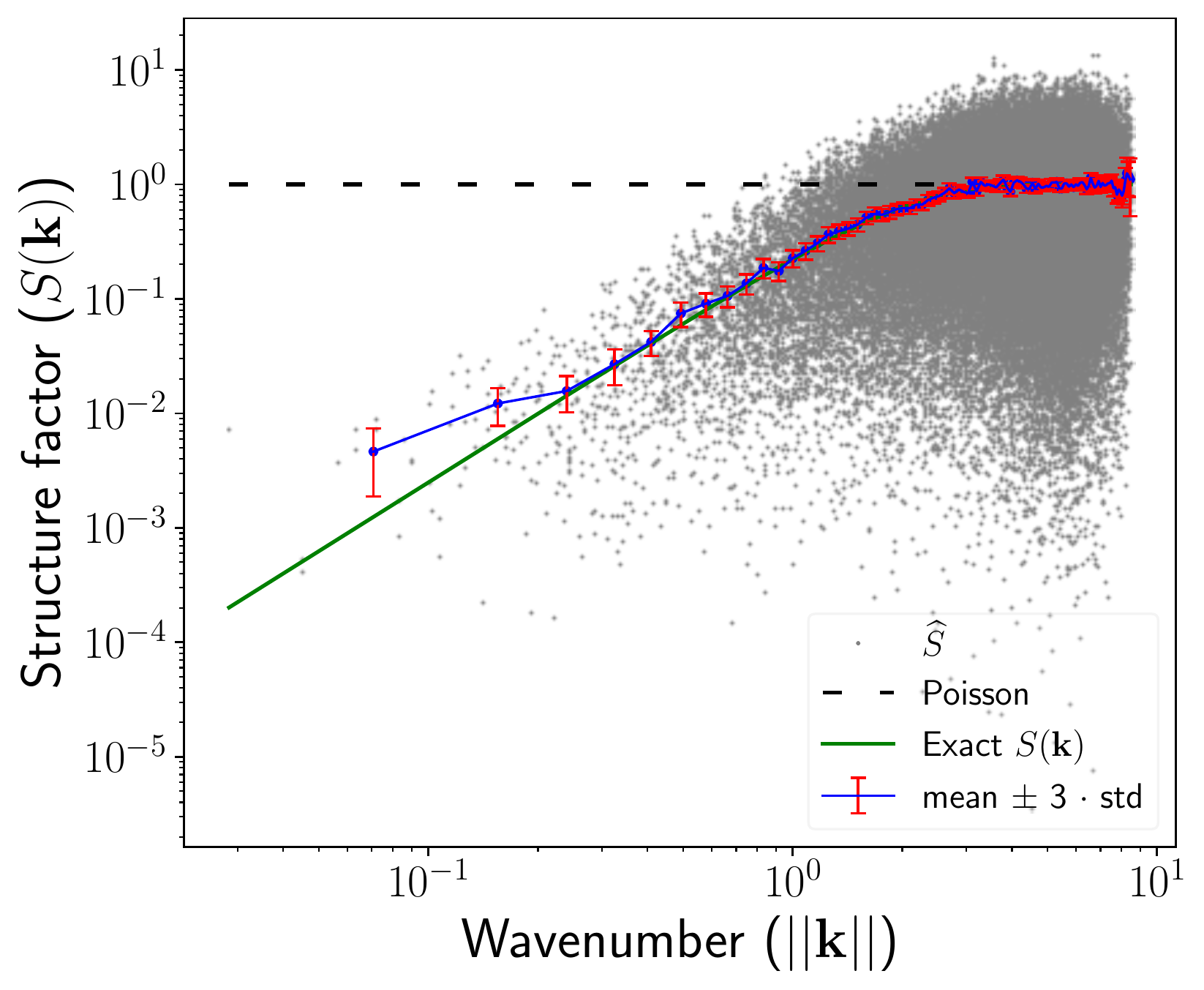}} &
    \raisebox{-\height}{\includegraphics[width=1\linewidth]{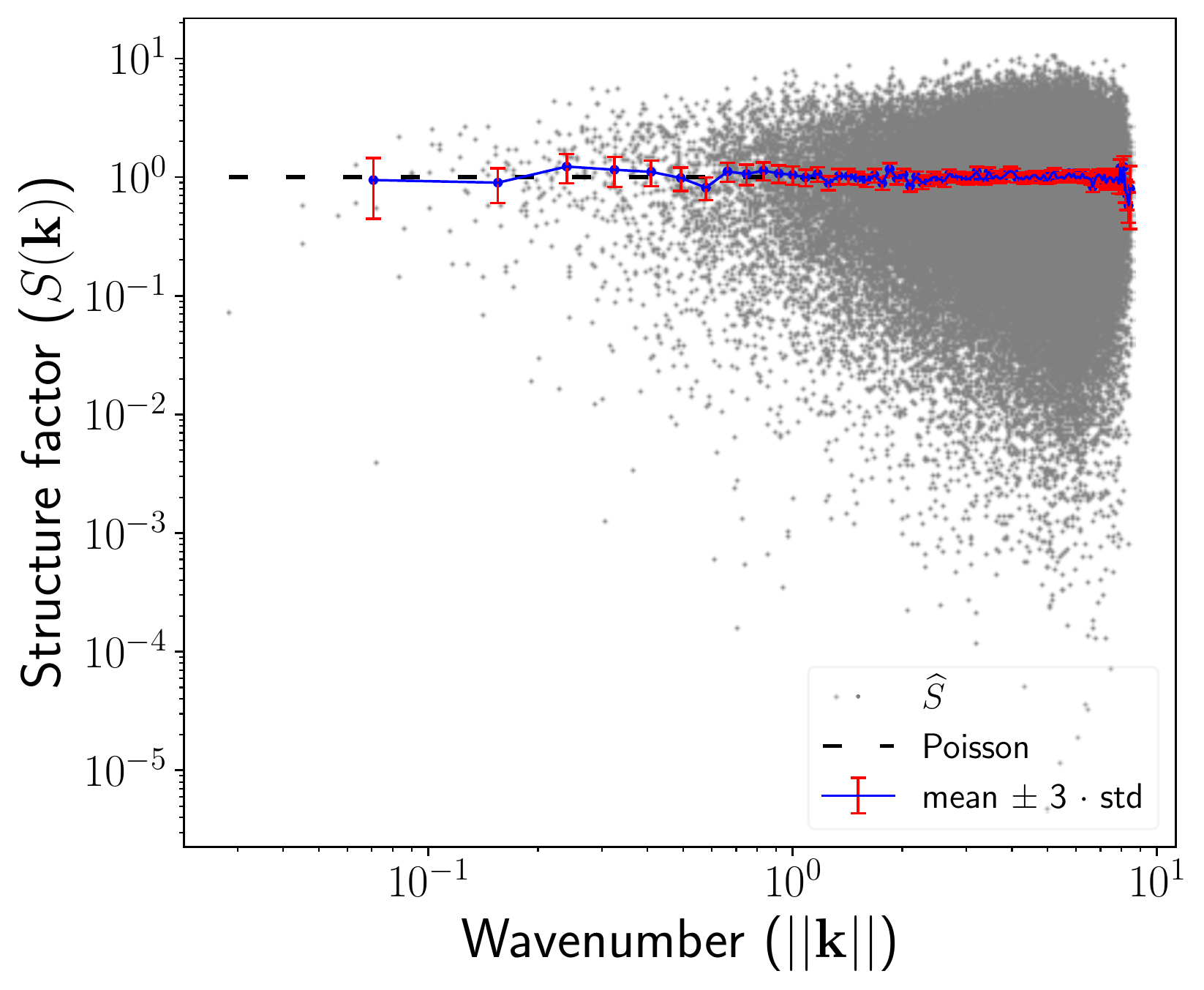}}    &
    \raisebox{-\height}{\includegraphics[width=1\linewidth]{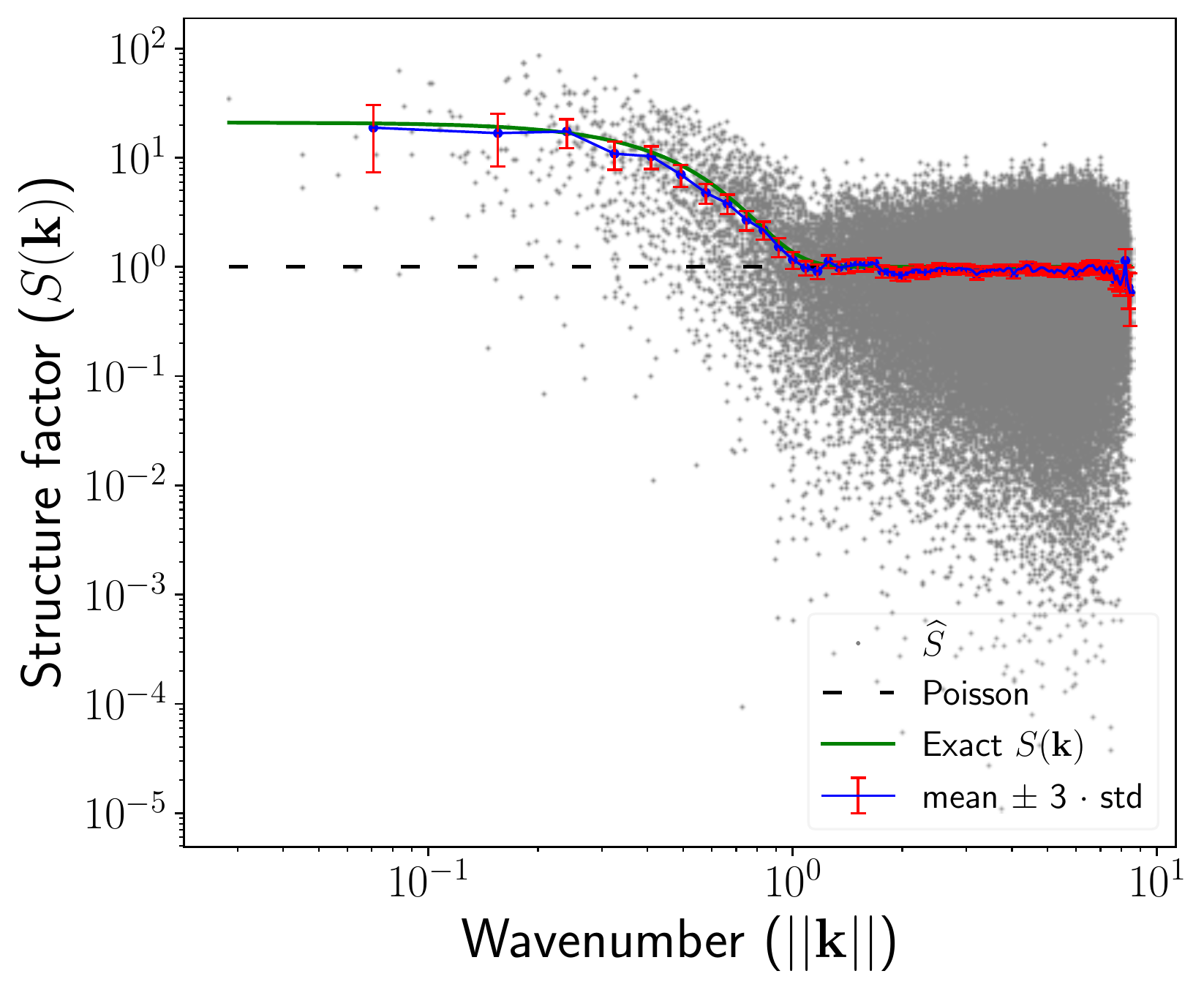}}
  \end{tabular}
  \vspace{-0.2cm}
  \begin{tabular}{p{\dimexpr 0.03\textwidth-\tabcolsep}p{\dimexpr 0.23\textwidth-\tabcolsep}p{\dimexpr 0.23\textwidth-\tabcolsep}p{\dimexpr 0.23\textwidth-\tabcolsep}p{\dimexpr 0.23\textwidth-\tabcolsep}}
    \multirow{9}{*}{\rotatebox[origin=l]{90}{$\widehat{S}_{\mathrm{UDT}}(t_0, \bfk)$}}             &
    \raisebox{-\height}{\includegraphics[width=1\linewidth]{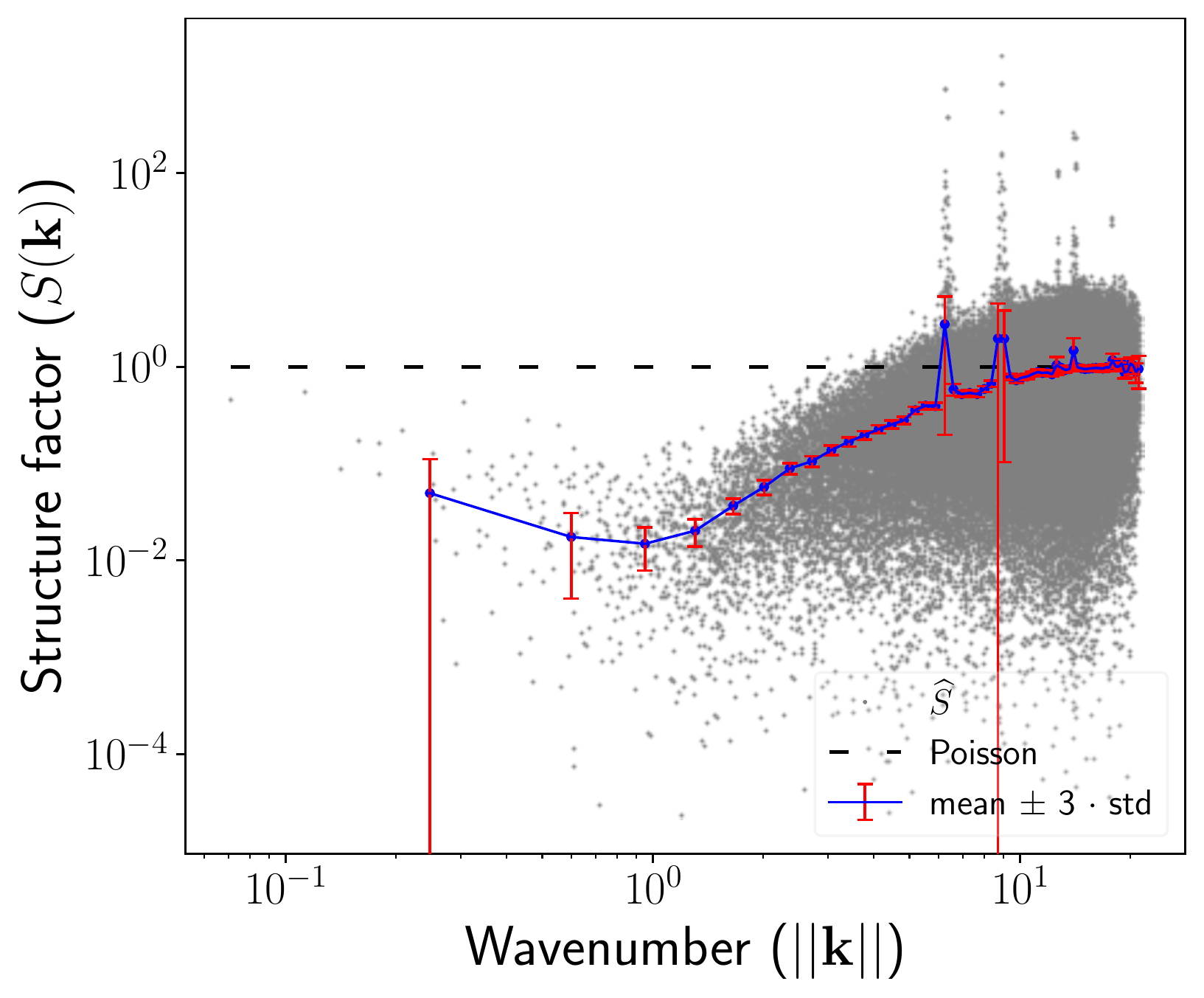}}     &
    \raisebox{-\height}{\includegraphics[width=1\linewidth]{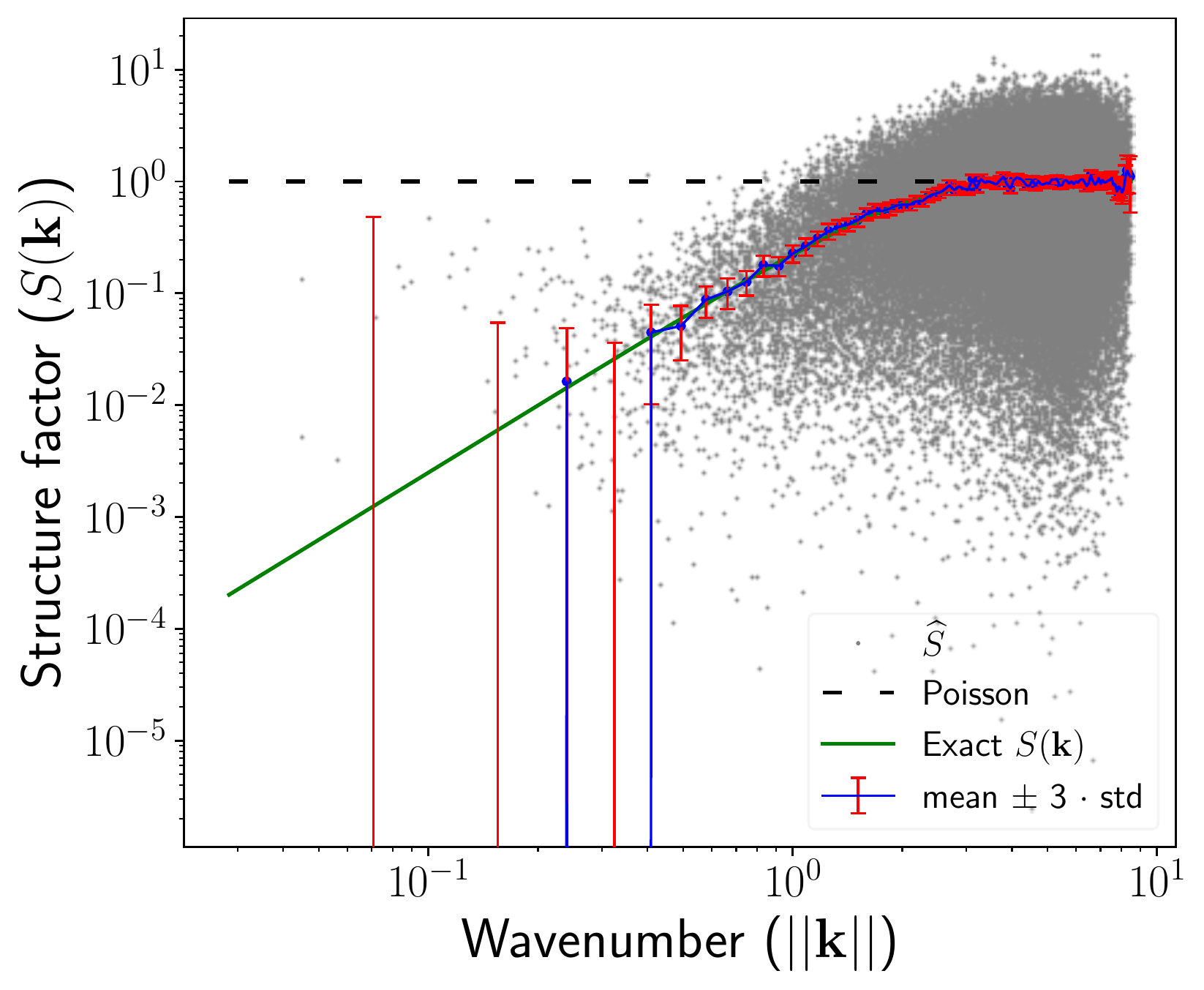}} &
    \raisebox{-\height}{\includegraphics[width=1\linewidth]{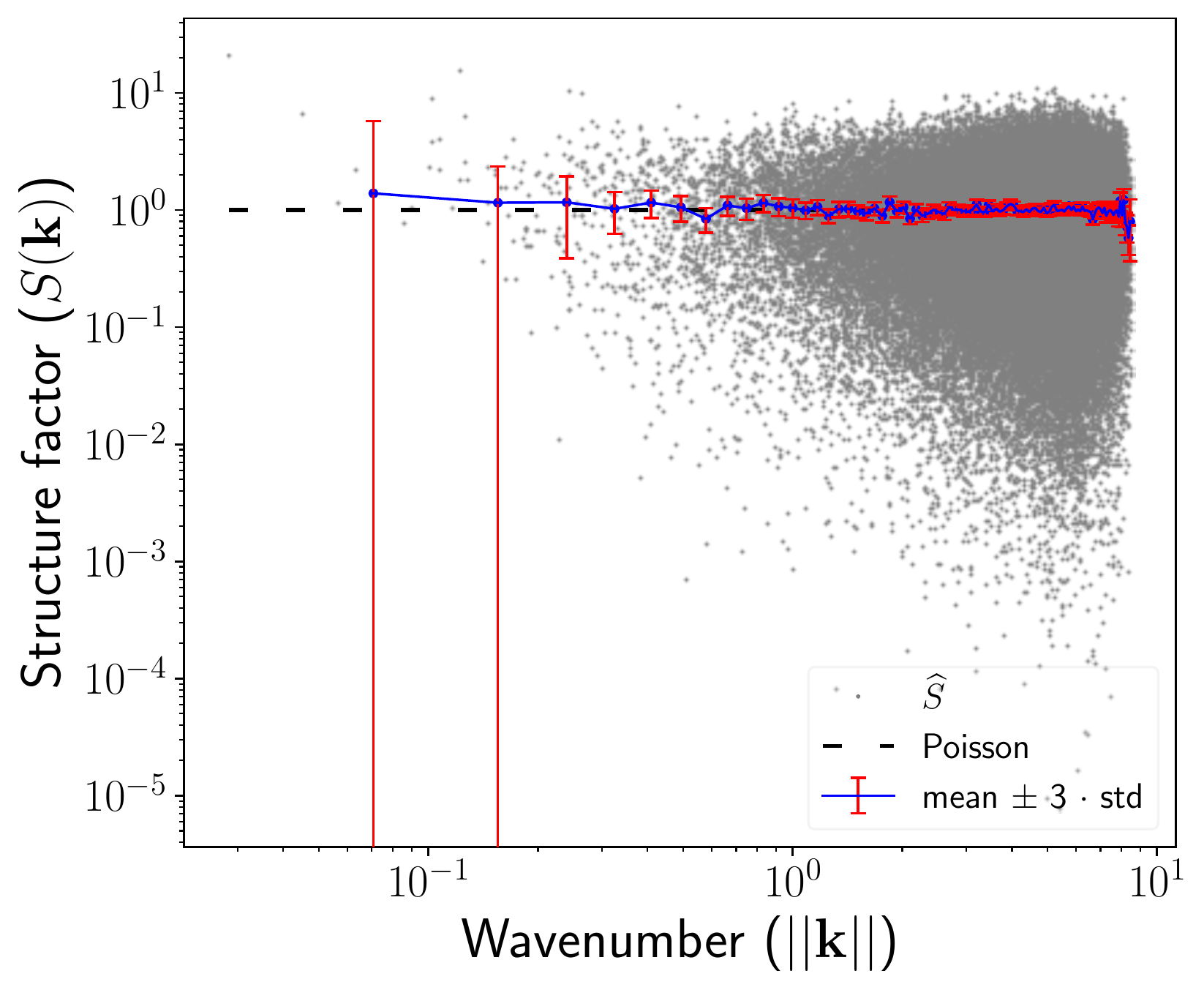}}    &
    \raisebox{-\height}{\includegraphics[width=1\linewidth]{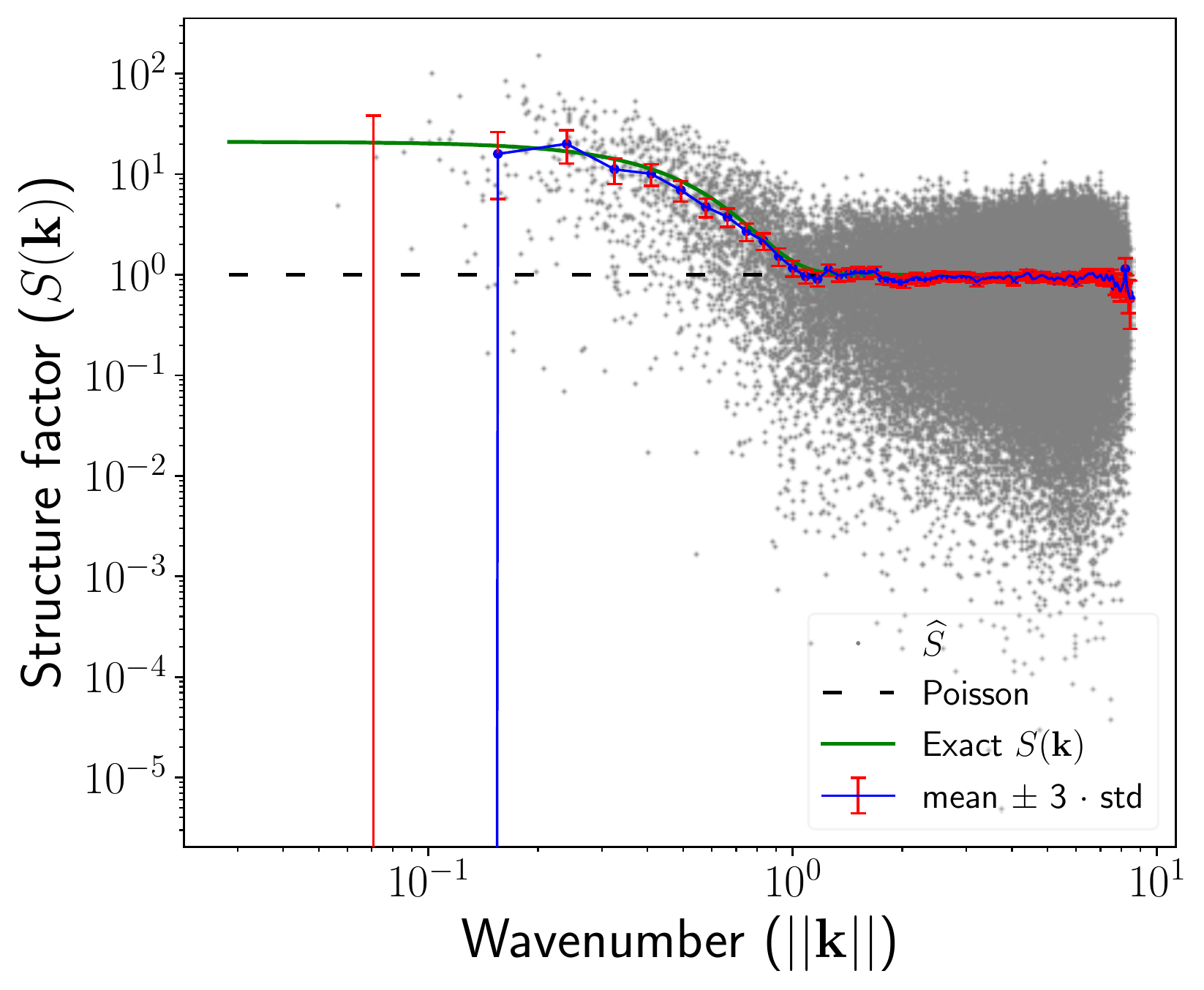}}    \\
    \caption*{}                                                                                    &
    \vspace{-0.5cm}
    \caption*{{\fontfamily{pcr}\selectfont } KLY }
                                                                                                   &
    \vspace{-0.5cm}
    \caption*{{\fontfamily{pcr}\selectfont } Ginibre}
                                                                                                   &
    \vspace{-0.5cm}
    \caption*{{\fontfamily{pcr}\selectfont } Poisson }
                                                                                                   &
    \vspace{-0.5cm}
    \caption*{{\fontfamily{pcr}\selectfont } Thomas }
  \end{tabular}
  \vspace{-0.5cm}
  \caption{Variants of the scattering intensity estimator applied to four point processes. The computation and visualization are done using \toolbox{} (see Section~\ref{ssub:test  The scattering intensity} for details)
  }
  \label{fig:scattering_intensity}
\end{figure*}
Figure~\ref{fig:scattering_intensity} illustrates the scattering intensity estimator of Section~\ref{sub:Estimators assuming stationarity}.
Columns respectively correspond to the KLY, Ginibre, Poisson, and Thomas point processes.
The first row contains a sample of each point process, observed in square windows.
The second row shows the scattering intensity $\widehat{S}_{\mathrm{SI}}$ in \eqref{eq:s_si} on arbitrary wavevectors $\bfk$, while in the third row, the estimators are only evaluated on a subset of the allowed wavevectors \eqref{eq:allowed_wave}.
The fourth and fifth rows illustrate the debiasing techniques, respectively the directly debiased scattering intensity $\widehat{S}_{\mathrm{DDT}}(t_0, \bfk)$ from \eqref{eq:s_DT}, and the undirectly debiased scattering intensity $\widehat{S}_{\mathrm{UDT}}(t_0, \bfk)$ from \eqref{eq:s_UDT}.

The clouds of grey points are the approximated structure factors of the samples observed in the first row of the figure.
For isotropic point processes, the structure factor $S$ is a radial function, so we plot $k\mapsto \hat{S}(k)$, $k\in\mathbb{R}$ and not $\mathbf{k}\mapsto \hat{S}(\mathbf{k})$, $\mathbf{k}\in\mathbb{R}^2$.
The KLY process is the only non-isotropic example: in that case, we numerically average $\hat S(\bfk)$ over vectors satisfying $\|\bfk\|_2= k$.
To regularize the obtained estimator, we bin the norm of the wavevectors regularly and provide the empirical mean (in blue) and the empirical standard deviation of the mean (red bars indicate $\pm 3$ such standard deviations).
Note that the binning can be specified by the user in our library.
On each plot, the exact structure factor is represented by a green line when it is known.
Finally, the dashed black lines are the structure factor of the homogeneous Poisson process, for reference.

While we refer to Section~\ref{sec:Comparison of the estimators} for a more detailed comparison, one can already observe from Figure~\ref{fig:scattering_intensity} that the most accurate estimators are the scattering intensity $\widehat{S}_{\mathrm{SI}}$ \eqref{eq:s_si} evaluated on the set of allowed wavevectors \eqref{eq:allowed_wave} and the debiased scattering intensity $\widehat{S}_{\mathrm{DDT}}(t_0, \bfk)$ \eqref{eq:s_DT}.
The bias at small, non-allowed wavenumbers of the scattering intensity is visible in the second row.
As for the undirectly debiased variant, it produces a few negative values, visible as large error bars on our log-log plot.

\paragraph{Using an alternate taper}
\begin{figure*}[!ht]
  \vspace*{-0.2cm}
  \begin{tabular}{p{\dimexpr 0.03\textwidth-\tabcolsep}p{\dimexpr 0.23\textwidth-\tabcolsep}p{\dimexpr 0.23\textwidth-\tabcolsep}p{\dimexpr 0.23\textwidth-\tabcolsep}p{\dimexpr 0.23\textwidth-\tabcolsep}}
    \multirow{9}{*}{\rotatebox[origin=l]{90}{Point process}}                                  &
    \raisebox{-\height}{\includegraphics[width=0.9\linewidth]{kly_pp_box.pdf}}     &
    \raisebox{-\height}{\includegraphics[width=0.9\linewidth]{ginibre_pp_box.pdf}} &
    \raisebox{-\height}{\includegraphics[width=0.9\linewidth]{poisson_pp_box.pdf}} &
    \raisebox{-\height}{\includegraphics[width=0.9\linewidth]{thomas_pp_box.pdf}}
  \end{tabular}
  \vspace{-0.2cm}
  \begin{tabular}{p{\dimexpr 0.03\textwidth-\tabcolsep}p{\dimexpr 0.23\textwidth-\tabcolsep}p{\dimexpr 0.23\textwidth-\tabcolsep}p{\dimexpr 0.23\textwidth-\tabcolsep}p{\dimexpr 0.23\textwidth-\tabcolsep}}
    \multirow{9}{*}{\rotatebox[origin=l]{90}{$\widehat{S}_{\mathrm{T}}(t,\bfk )$}}                   &
    \raisebox{-\height}{\includegraphics[width=1\linewidth]{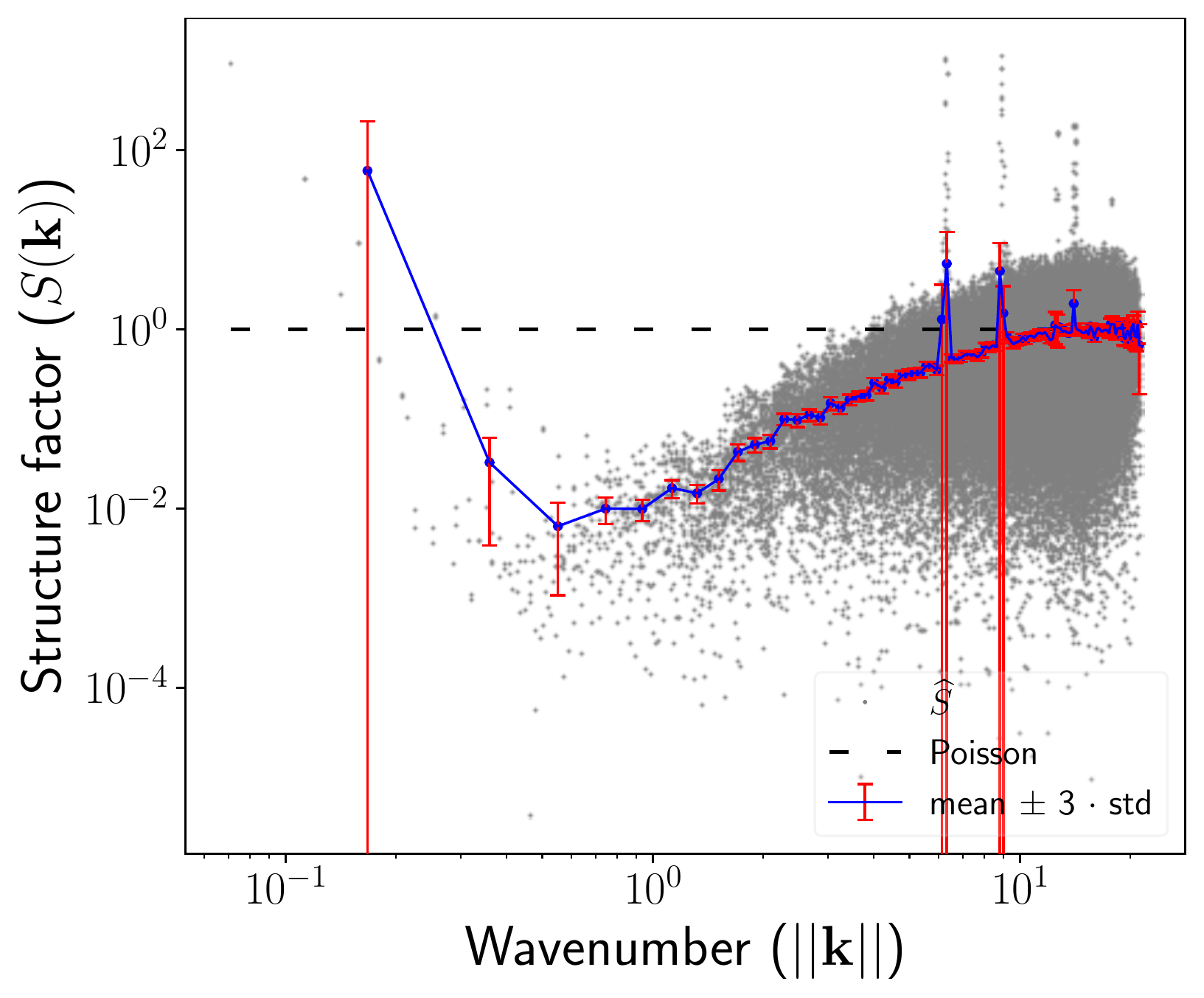}}     &
    \raisebox{-\height}{\includegraphics[width=1\linewidth]{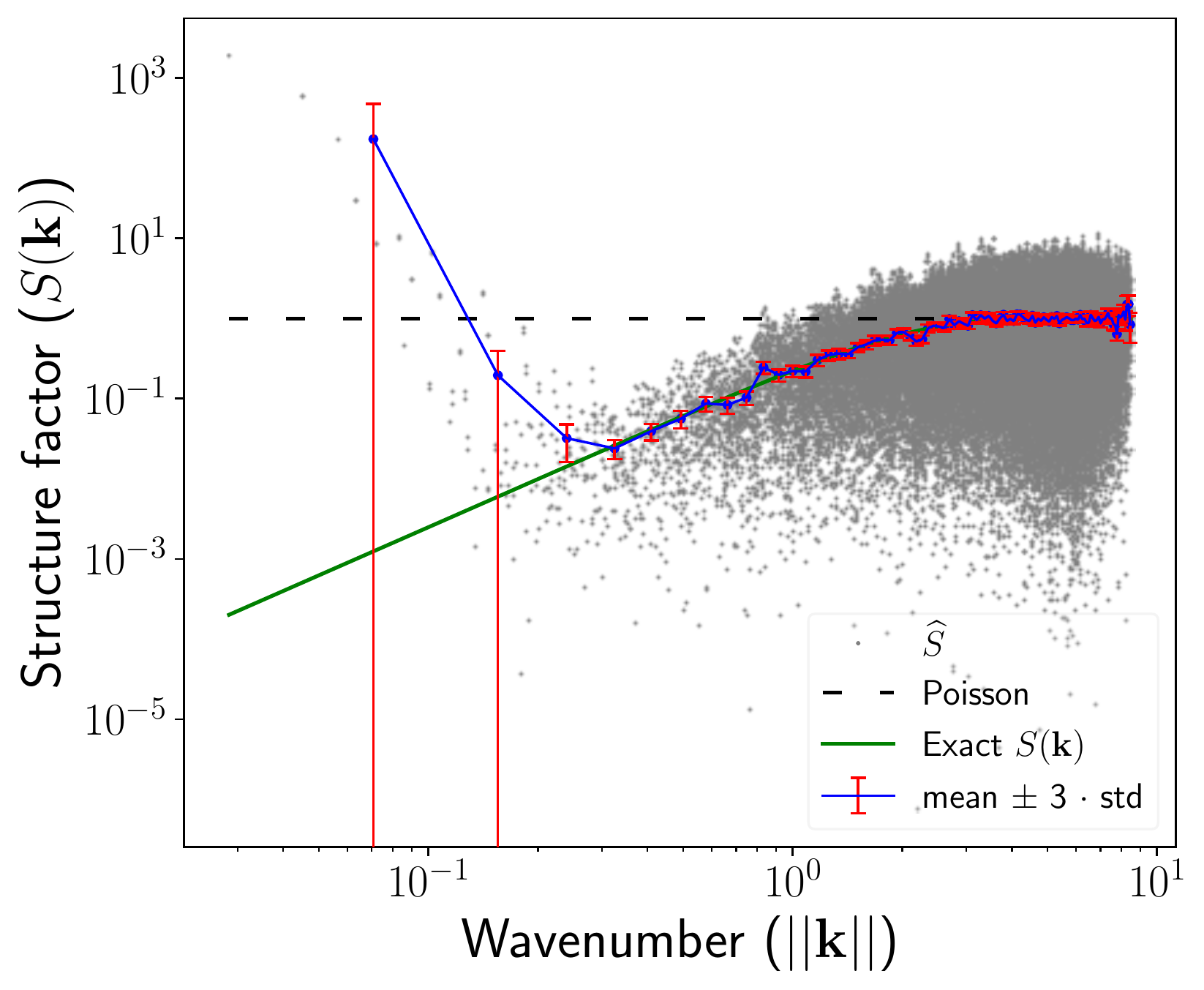}} &
    \raisebox{-\height}{\includegraphics[width=1\linewidth]{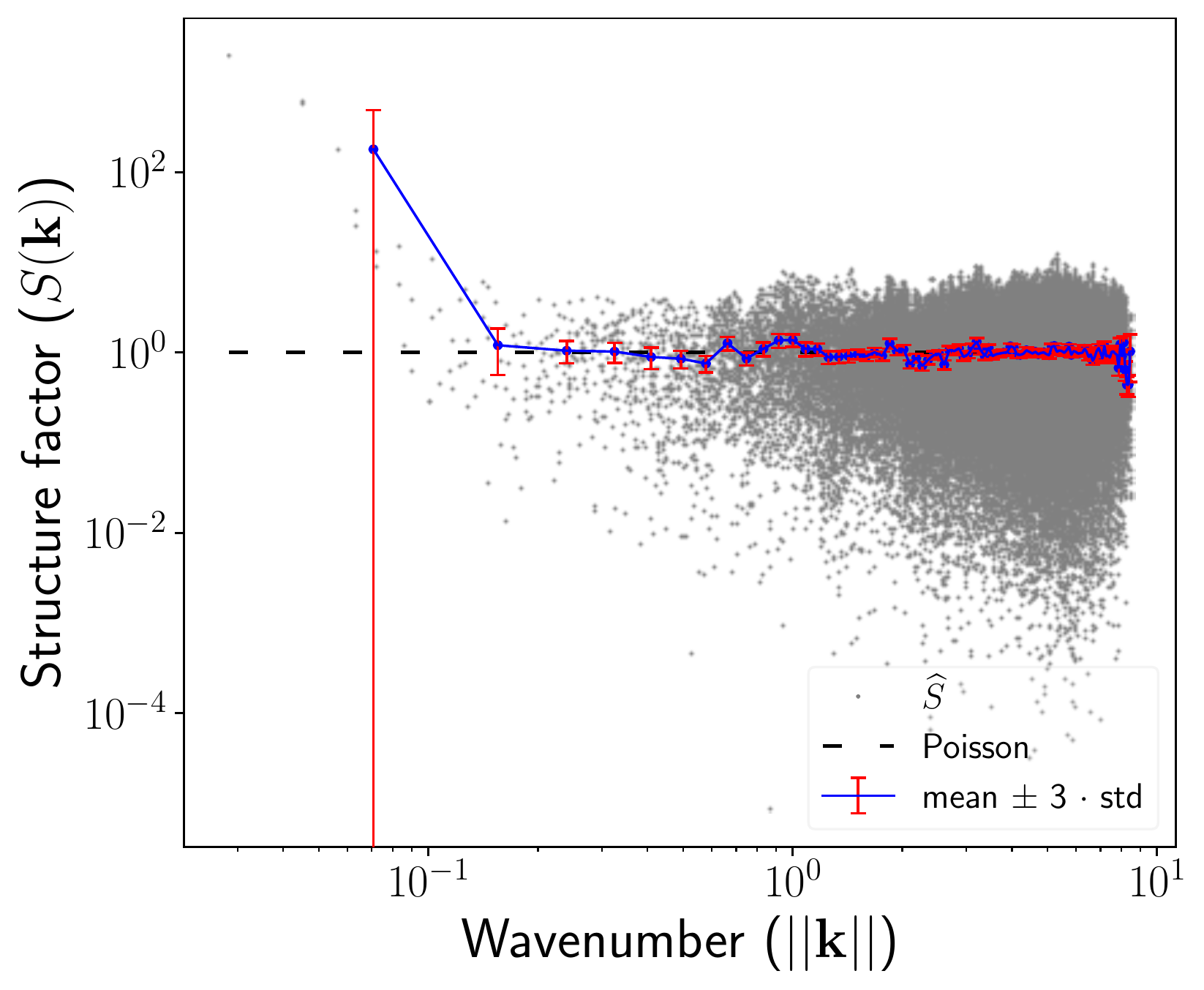}} &
    \raisebox{-\height}{\includegraphics[width=1\linewidth]{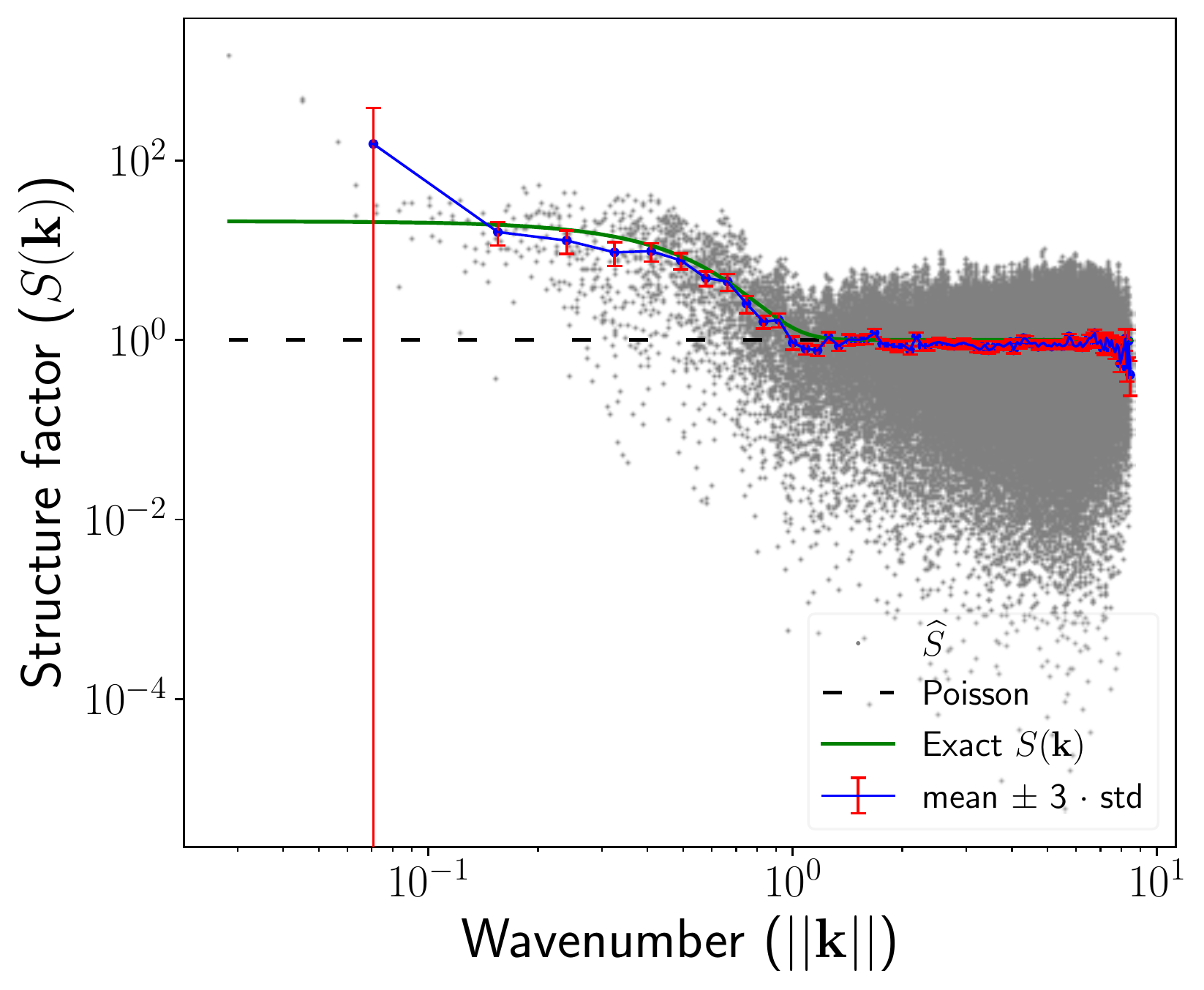}}
  \end{tabular}
  \vspace*{-0.2cm}
  \begin{tabular}{p{\dimexpr 0.03\textwidth-\tabcolsep}p{\dimexpr 0.23\textwidth-\tabcolsep}p{\dimexpr 0.23\textwidth-\tabcolsep}p{\dimexpr 0.23\textwidth-\tabcolsep}p{\dimexpr 0.23\textwidth-\tabcolsep}}
    \multirow{9}{*}{\rotatebox[origin=l]{90}{$\widehat{S}_{\mathrm{DDT}}(t, \bfk)$}}                   &
    \raisebox{-\height}{\includegraphics[width=1\linewidth]{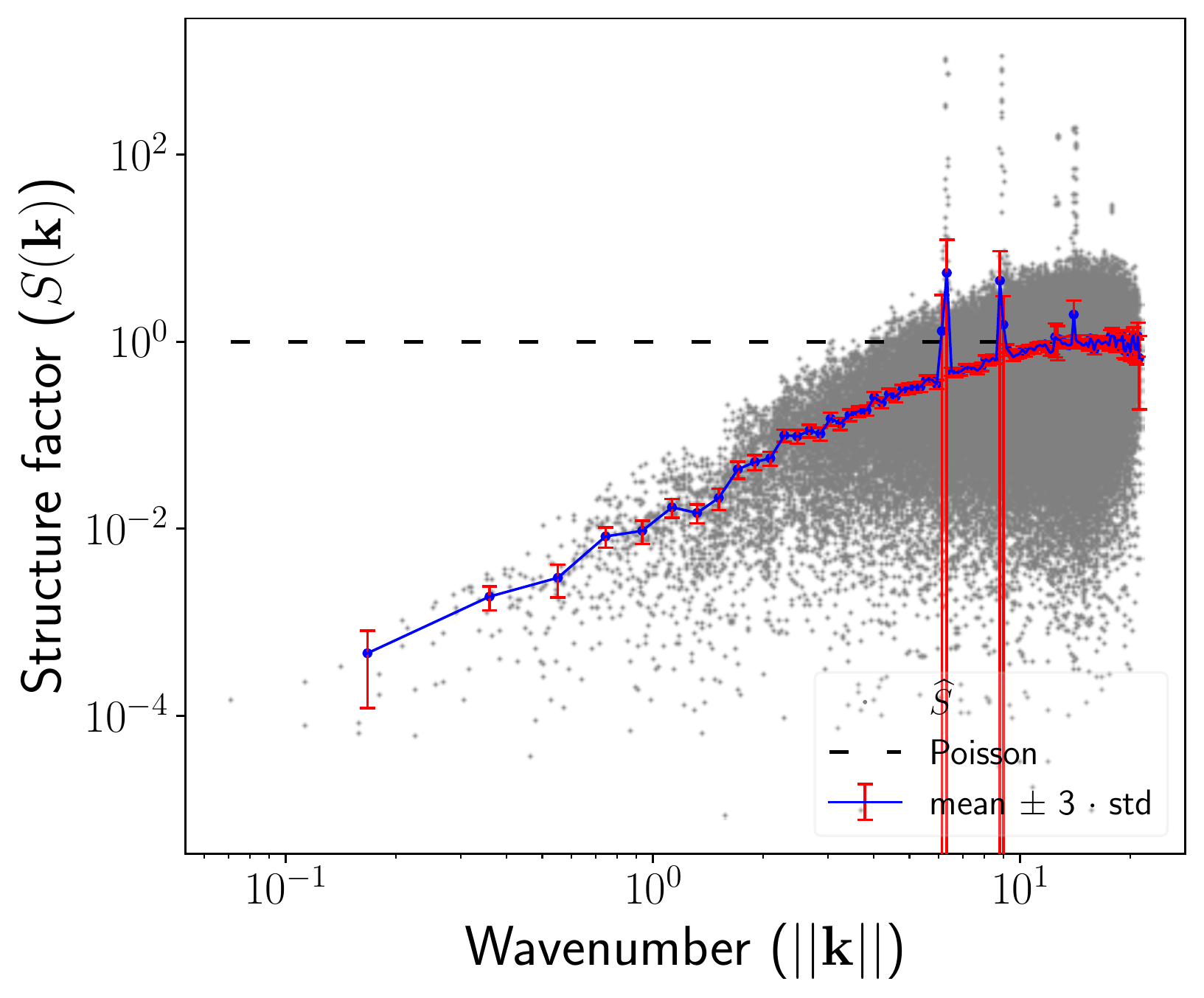}}     &
    \raisebox{-\height}{\includegraphics[width=1\linewidth]{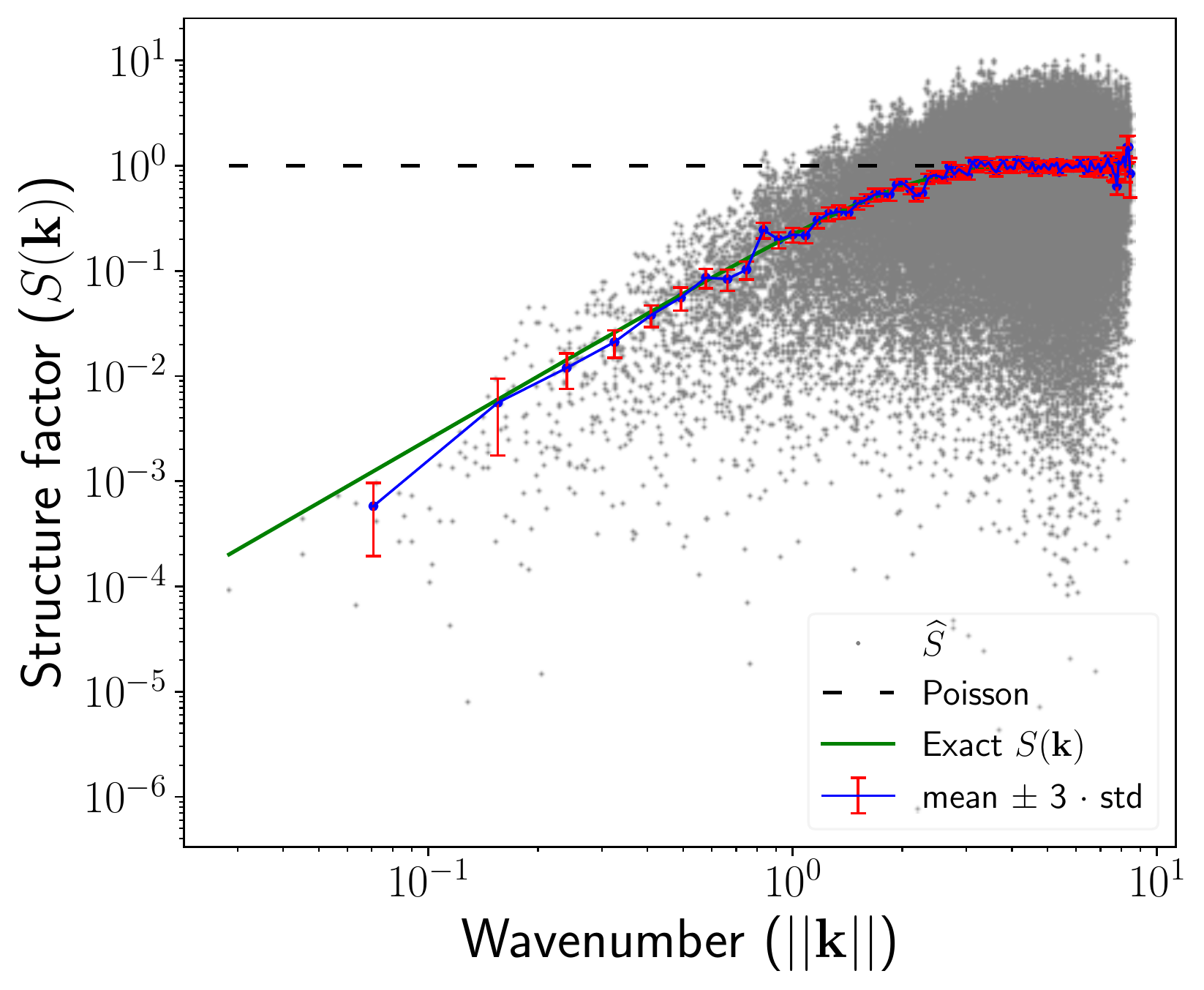}} &
    \raisebox{-\height}{\includegraphics[width=1\linewidth]{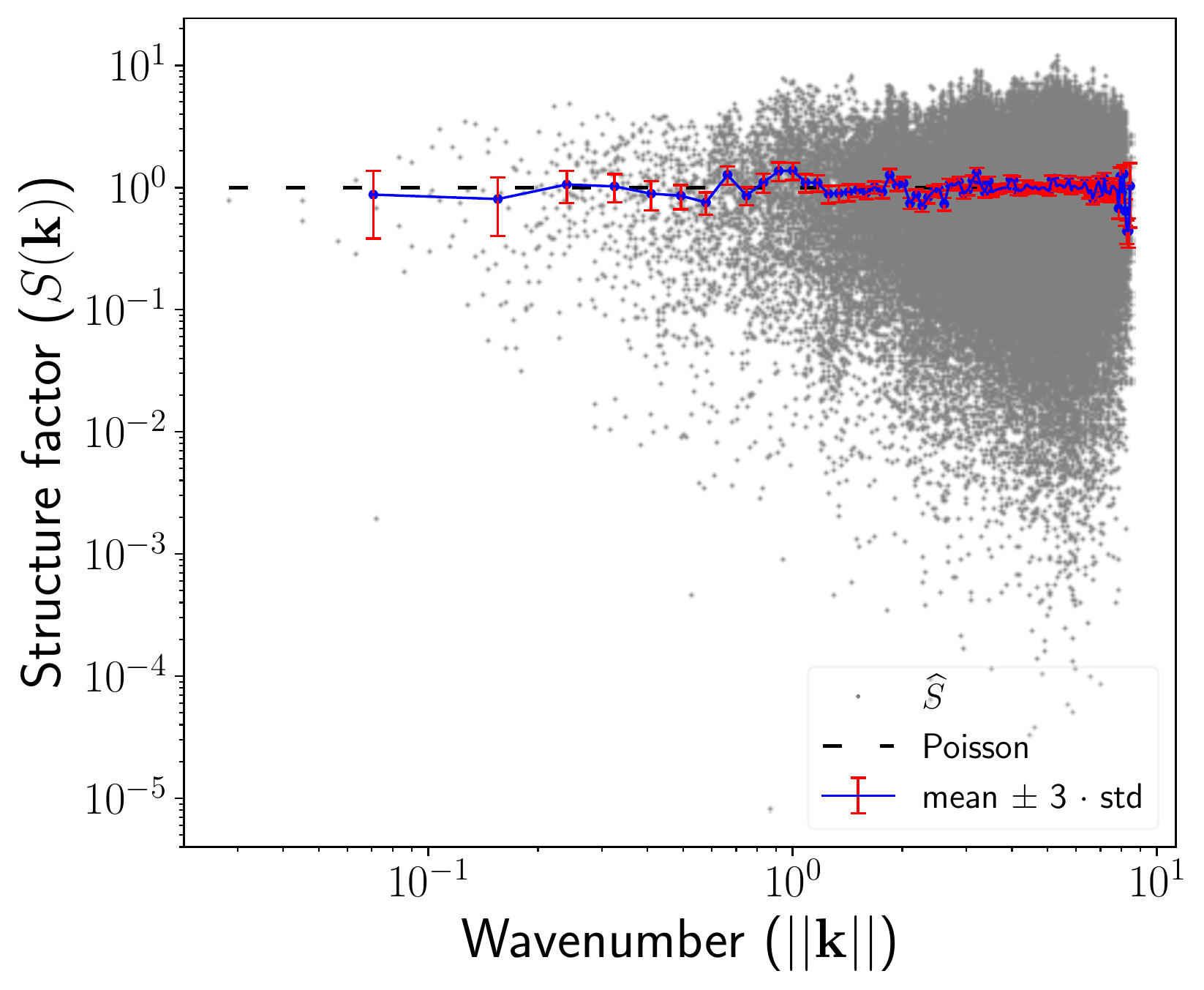}} &
    \raisebox{-\height}{\includegraphics[width=1\linewidth]{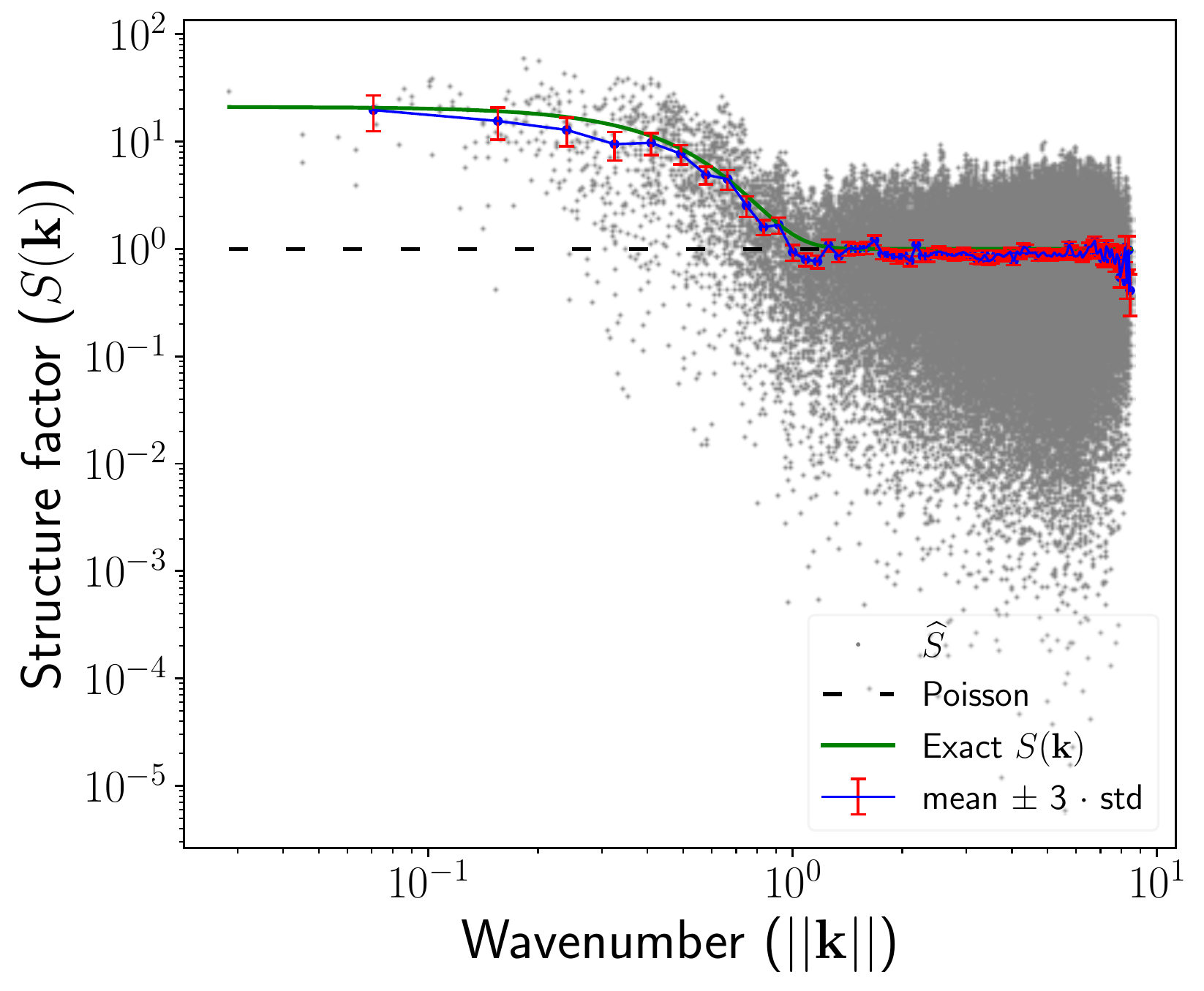}}
  \end{tabular}
  \vspace{-0.2cm}
  \begin{tabular}{p{\dimexpr 0.03\textwidth-\tabcolsep}p{\dimexpr 0.23\textwidth-\tabcolsep}p{\dimexpr 0.23\textwidth-\tabcolsep}p{\dimexpr 0.23\textwidth-\tabcolsep}p{\dimexpr 0.23\textwidth-\tabcolsep}}
    \vspace{0.2cm}
    \multirow{3}{*}{\rotatebox[origin=l]{90}{$\widehat{S}_{\mathrm{UDT}}(t, \bfk)$}}                   &
    \raisebox{-\height}{\includegraphics[width=1\linewidth]{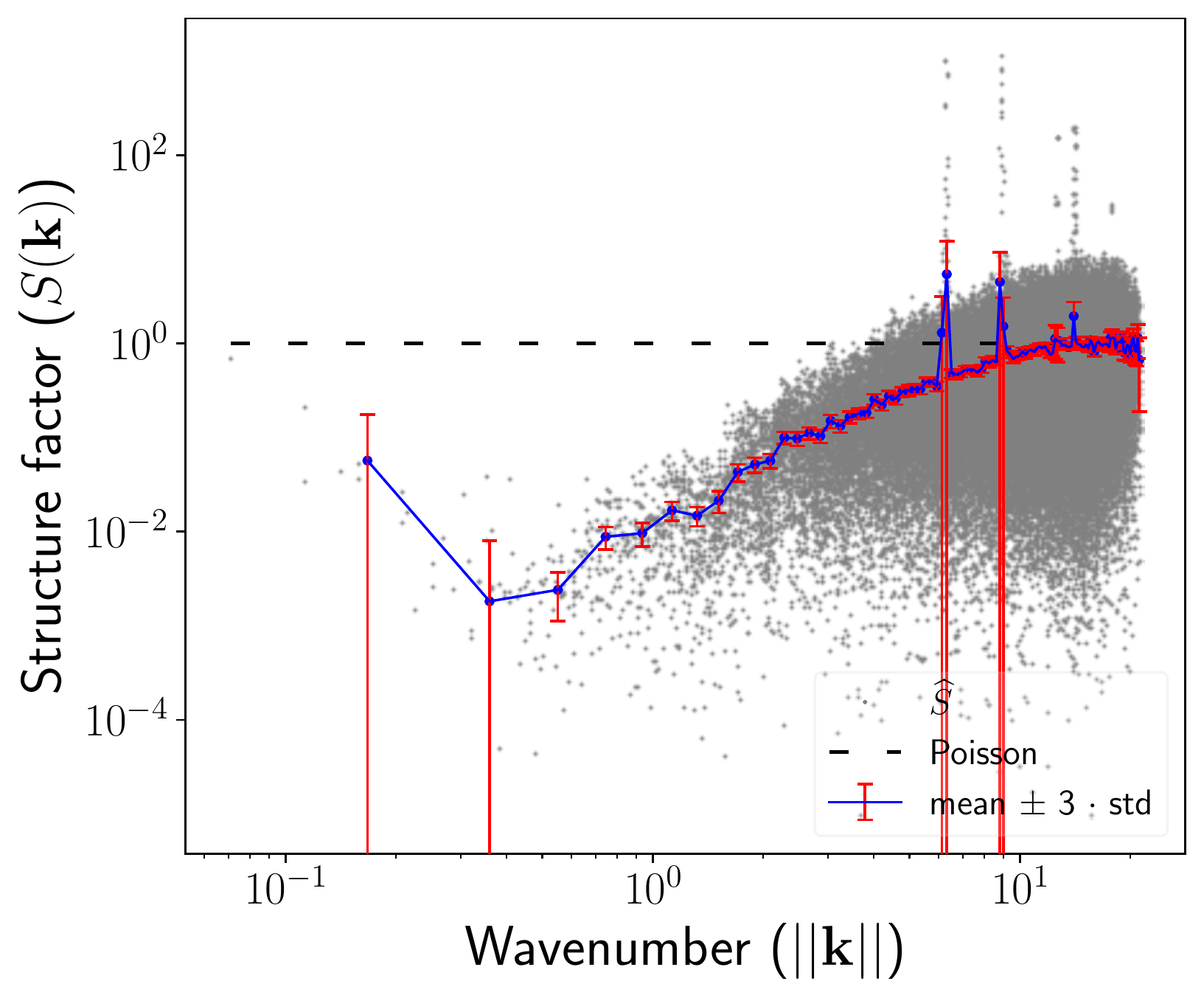}}     &
    \raisebox{-\height}{\includegraphics[width=1\linewidth]{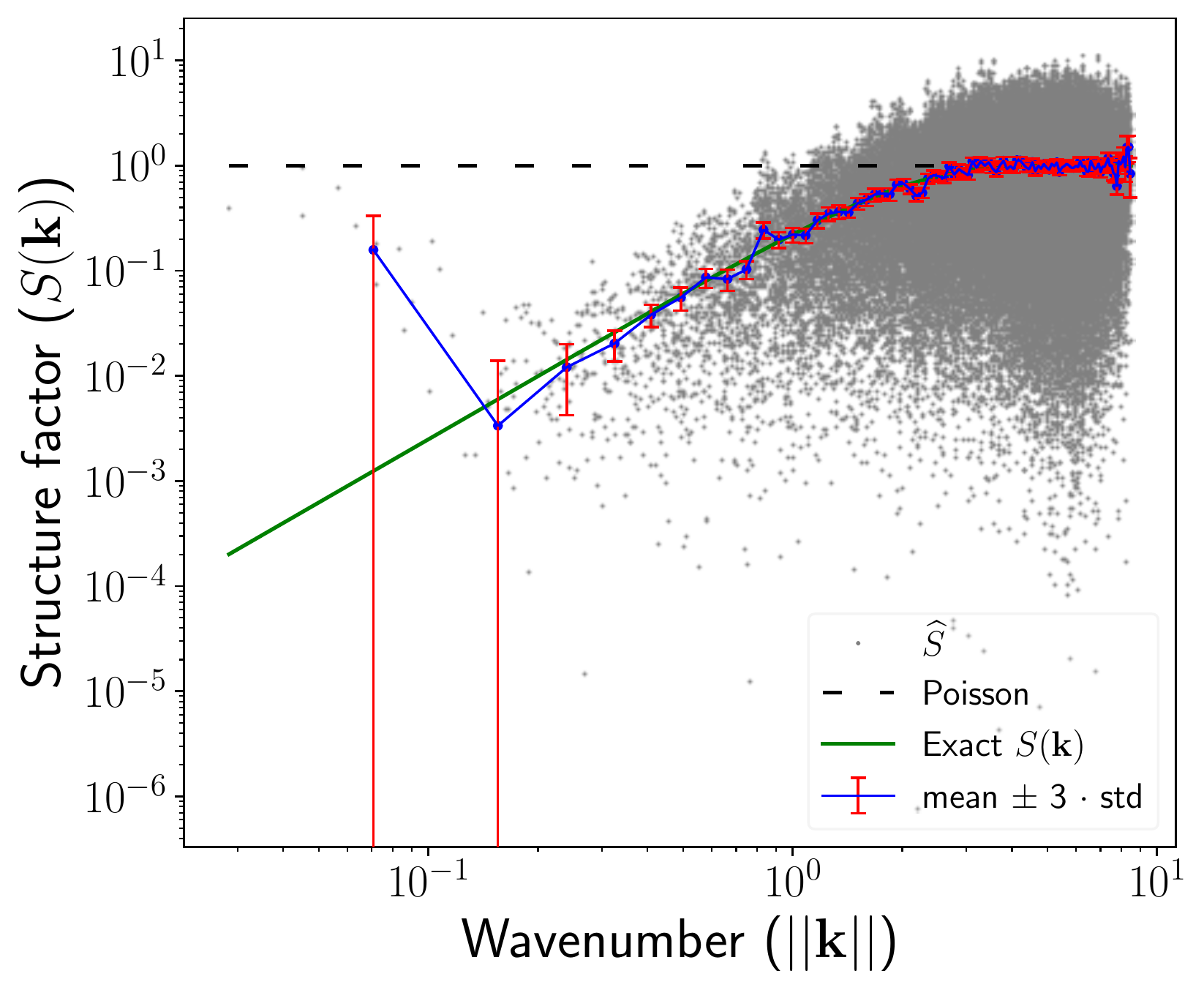}} &
    \raisebox{-\height}{\includegraphics[width=1\linewidth]{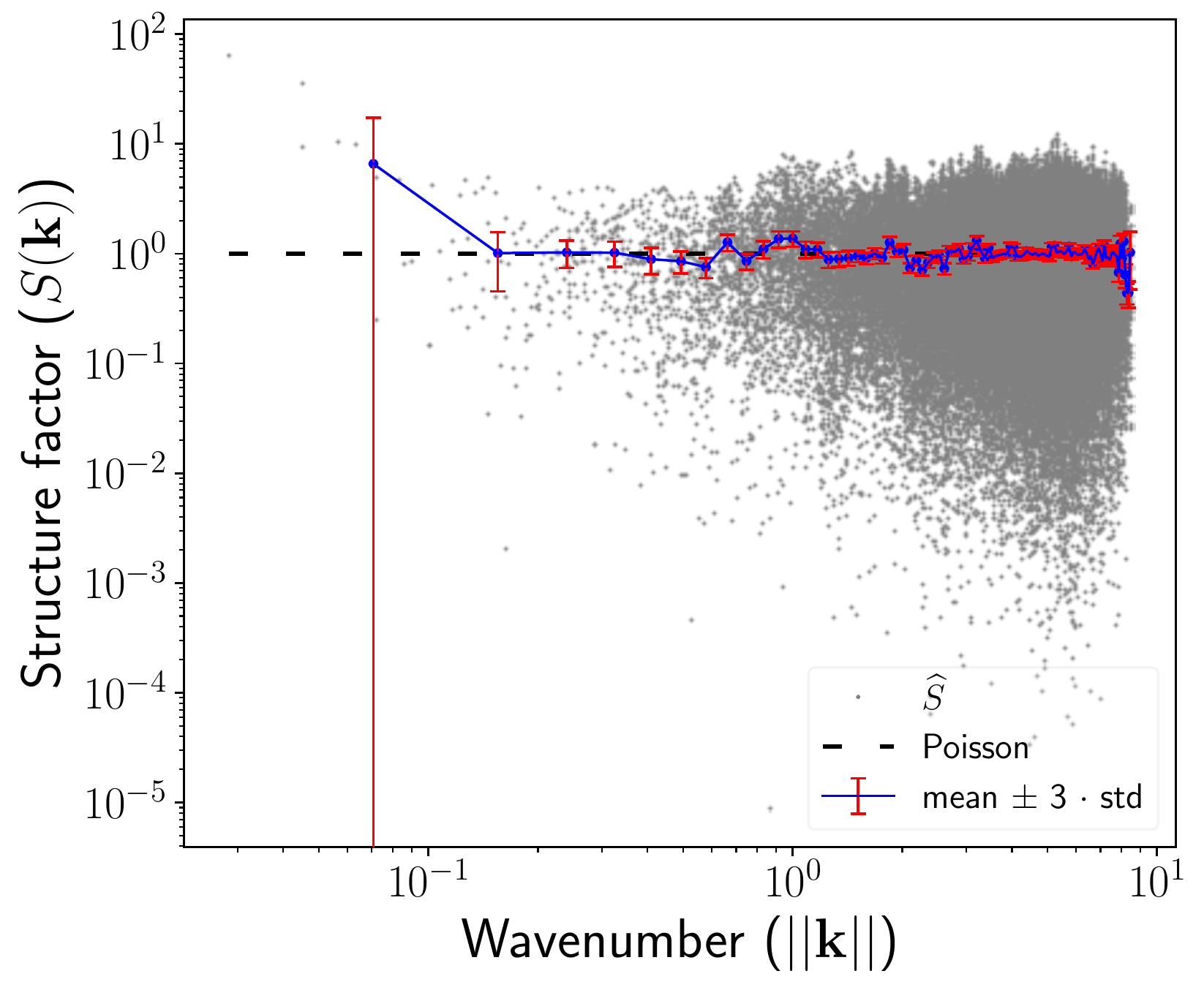}} &
    \raisebox{-\height}{\includegraphics[width=1\linewidth]{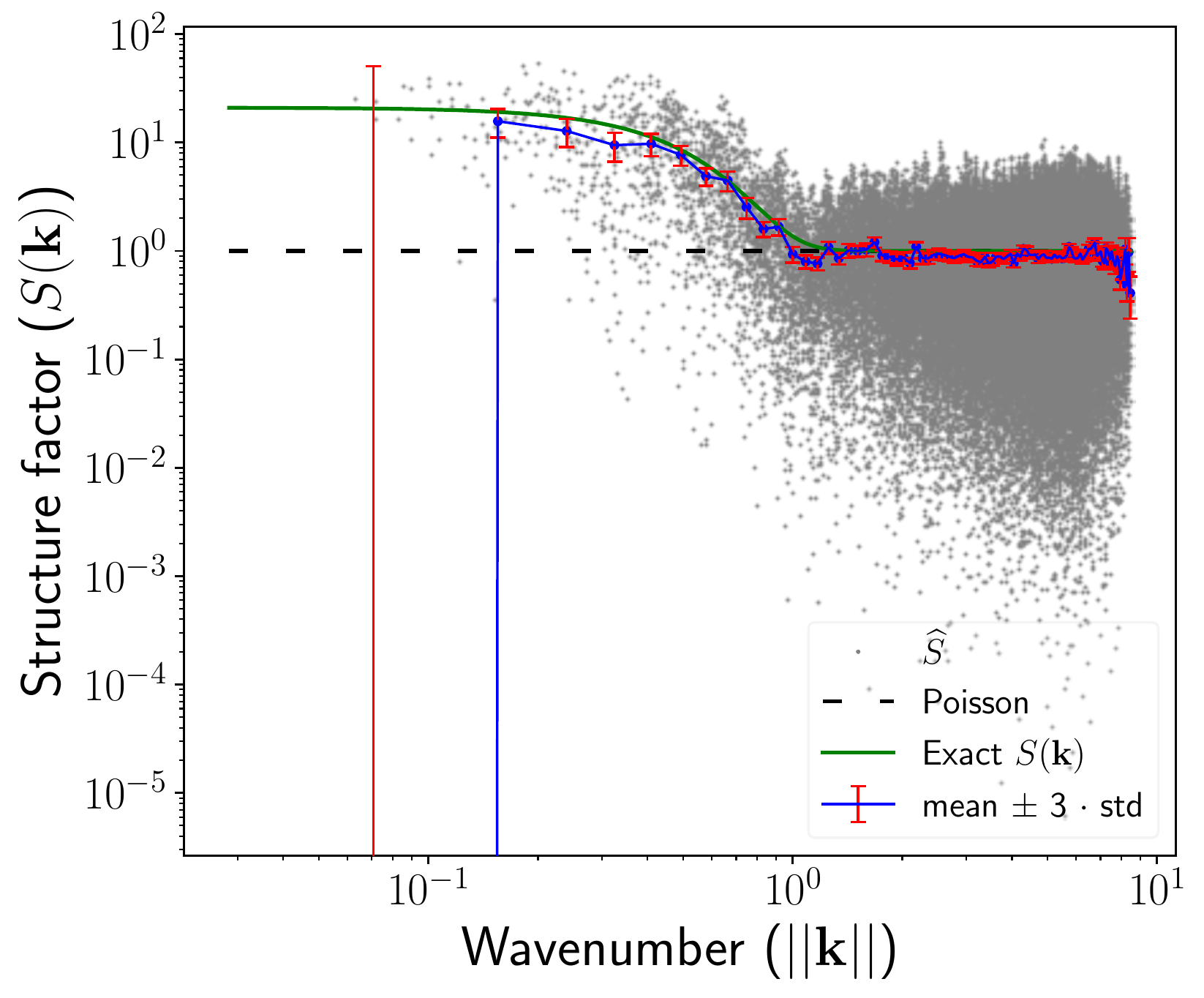}}    \\
    \caption*{}                                                                                        &
    \vspace{-0.5cm}
    \caption*{{\fontfamily{pcr}\selectfont } KLY }                                              &
    \vspace{-0.5cm}
    \caption*{{\fontfamily{pcr}\selectfont } Ginibre }                                         &
    \vspace{-0.5cm}
    \caption*{{\fontfamily{pcr}\selectfont } Poisson }                                          &
    \vspace{-0.5cm}
    \caption*{{\fontfamily{pcr}\selectfont } Thomas }
  \end{tabular}
  \vspace{-0.5cm}
  \caption{Tapered estimator and the corresponding debiased versions: KLY process (first column), Ginibre ensemble (second column), Poisson process (third column), and Thomas process (last column). The computation and visualization are done using \toolbox{}} \label{fig:s_t}
\end{figure*}
Now, as mentioned in Section~\ref{sec:Estimators of the structure factor}, the scattering intensity $\widehat{S}_{\mathrm{SI}}$ is a particular case of the tapered estimator $\widehat{S}_{\mathrm{T}}$, with the specific taper $t_0$.
We are free to use other tapers verifying \eqref{eq:limite_alpha}.

Figure~\ref{fig:s_t} shows the estimated structure factors of the same four benchmark point processes (first row), using $\widehat{S}_{\mathrm{T}}$ (second row), the corresponding directly debiased version $\widehat{S}_{\mathrm{DDT}}$ (third row), and the undirectly debiased version $\widehat{S}_{\mathrm{UDT}}$ (last row).
The taper used is the first sinusoidal taper $t_1(\bfx, W) = t(\bfx, \bfp^1, W)$ with $\bfp^1 = (1, 1)$ in \eqref{eq:sine_taper}.
The same legend applies as for Figure~\ref{fig:scattering_intensity}.

First, the asymptotic bias of $\widehat{S}_{\mathrm{T}}$ at small wavenumber $k$ is visible in the second row.
Second, for the KLY process (first column), the Ginibre ensemble (second column), and the Poisson process (third column) the estimator $\widehat{S}_{\mathrm{UDT}}$ (last row) returned a few negative values again, resulting in large inaccuracies in our log-log scale.
The directly debiased estimator $\widehat{S}_{\mathrm{DDT}}$ yields the most accurate approximation of known structure factors, consistently across point processes.

\paragraph{Averaging over multiple tapers} % (fold)
\label{ssub:test The multitapered estimator}
\begin{figure*}[!ht]
  \vspace*{-0.1cm}
  \begin{tabular}{p{\dimexpr 0.03\textwidth-\tabcolsep}p{\dimexpr 0.22\textwidth-\tabcolsep}p{\dimexpr 0.22\textwidth-\tabcolsep}p{\dimexpr 0.22\textwidth-\tabcolsep}p{\dimexpr 0.22\textwidth-\tabcolsep}}
    \multirow{9}{*}{\rotatebox[origin=l]{90}{Point process}}                                  &
    \raisebox{-\height}{\includegraphics[width=0.9\linewidth]{kly_pp_box.pdf}}     &
    \raisebox{-\height}{\includegraphics[width=0.9\linewidth]{ginibre_pp_box.pdf}} &
    \raisebox{-\height}{\includegraphics[width=0.9\linewidth]{poisson_pp_box.pdf}} &
    \raisebox{-\height}{\includegraphics[width=0.9\linewidth]{thomas_pp_box.pdf}}
  \end{tabular}
  \vspace*{-0.1cm}
  \begin{tabular}{p{\dimexpr 0.03\textwidth-\tabcolsep}p{\dimexpr 0.22\textwidth-\tabcolsep}p{\dimexpr 0.22\textwidth-\tabcolsep}p{\dimexpr 0.22\textwidth-\tabcolsep}p{\dimexpr 0.22\textwidth-\tabcolsep}}
    \multirow{9}{*}{\rotatebox[origin=l]{90}{$\widehat{S}_{\mathrm{MT}}$}}             &
    \raisebox{-\height}{\includegraphics[width=1\linewidth]{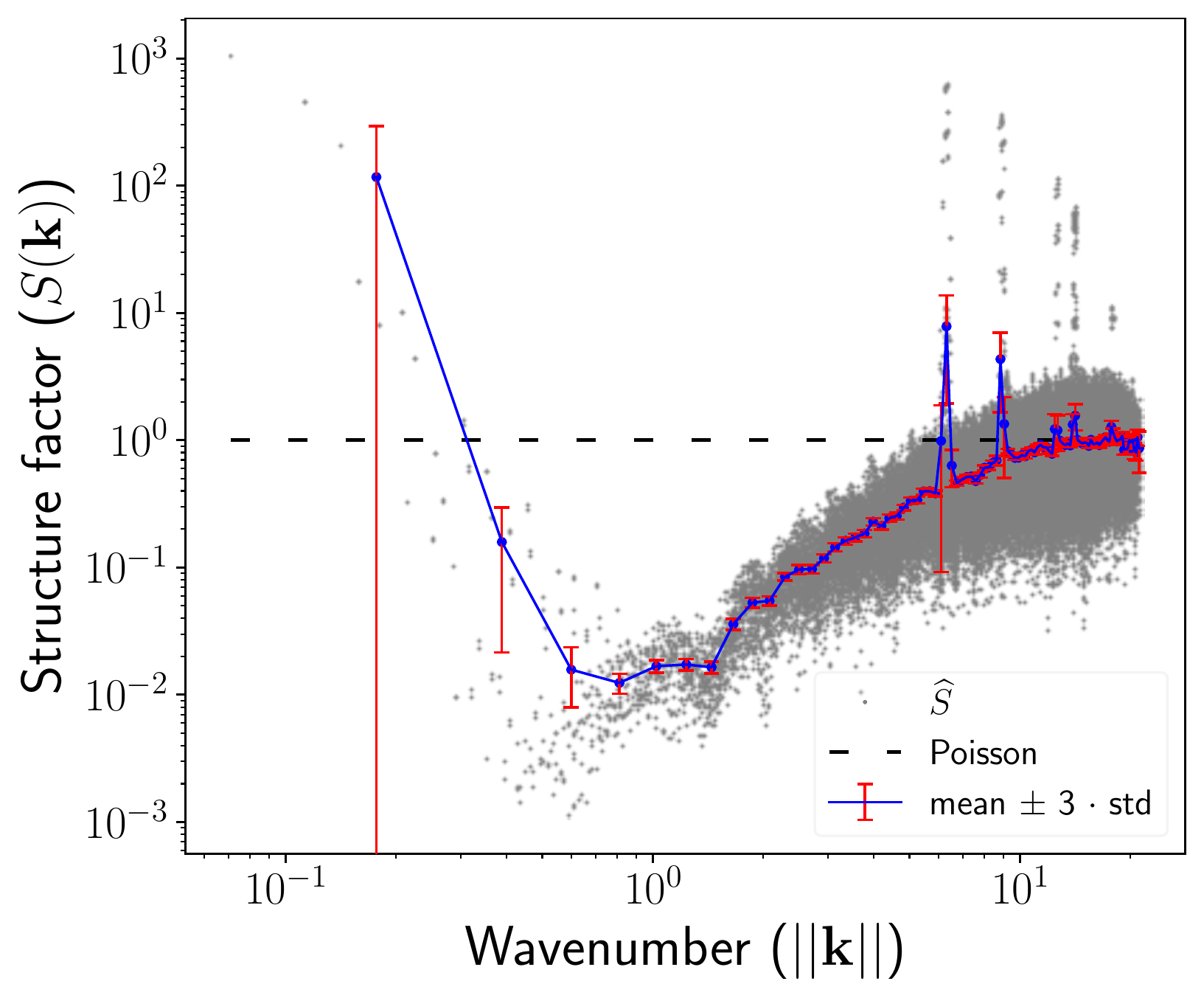}}     &
    \raisebox{-\height}{\includegraphics[width=1\linewidth]{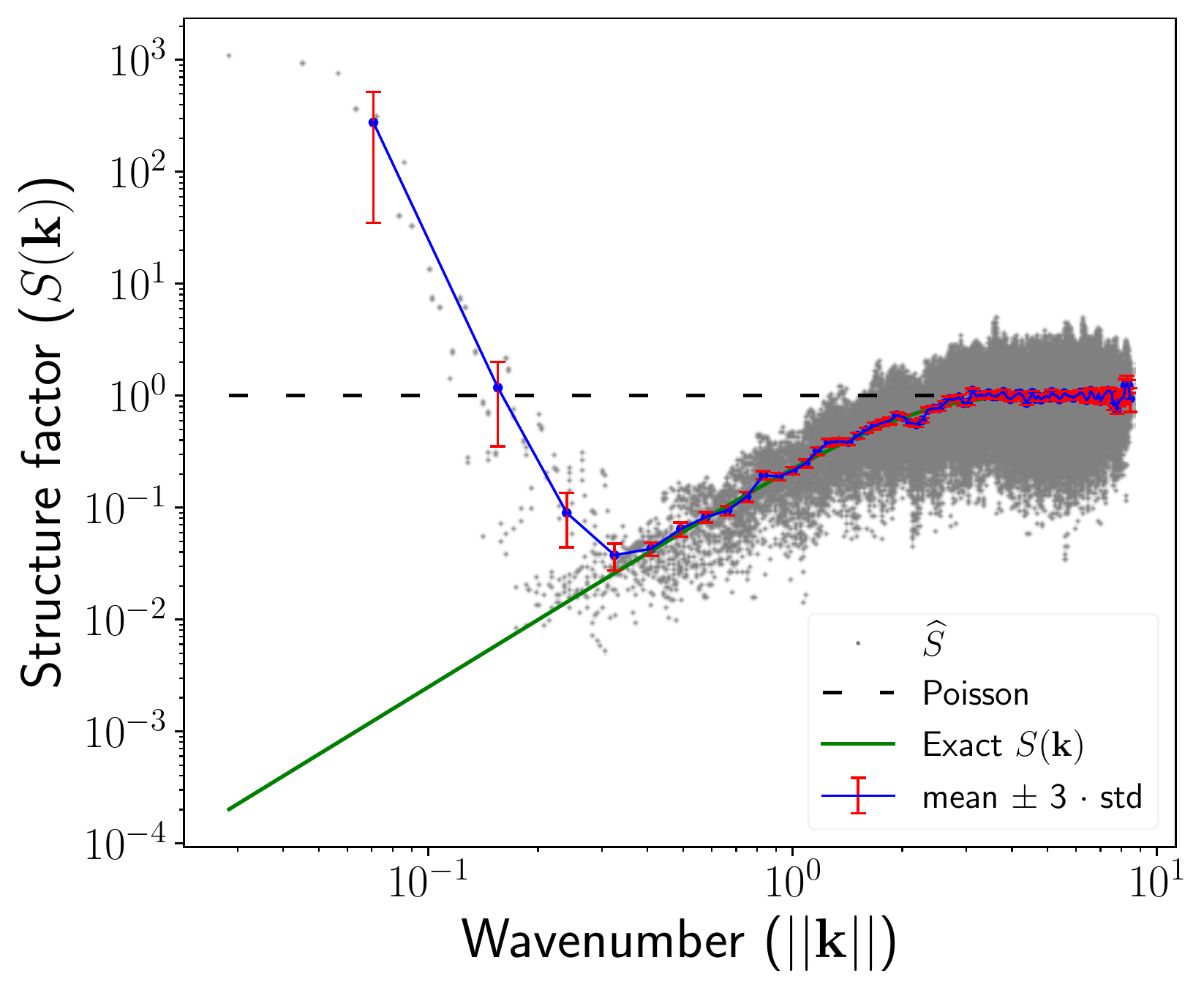}} &
    \raisebox{-\height}{\includegraphics[width=1\linewidth]{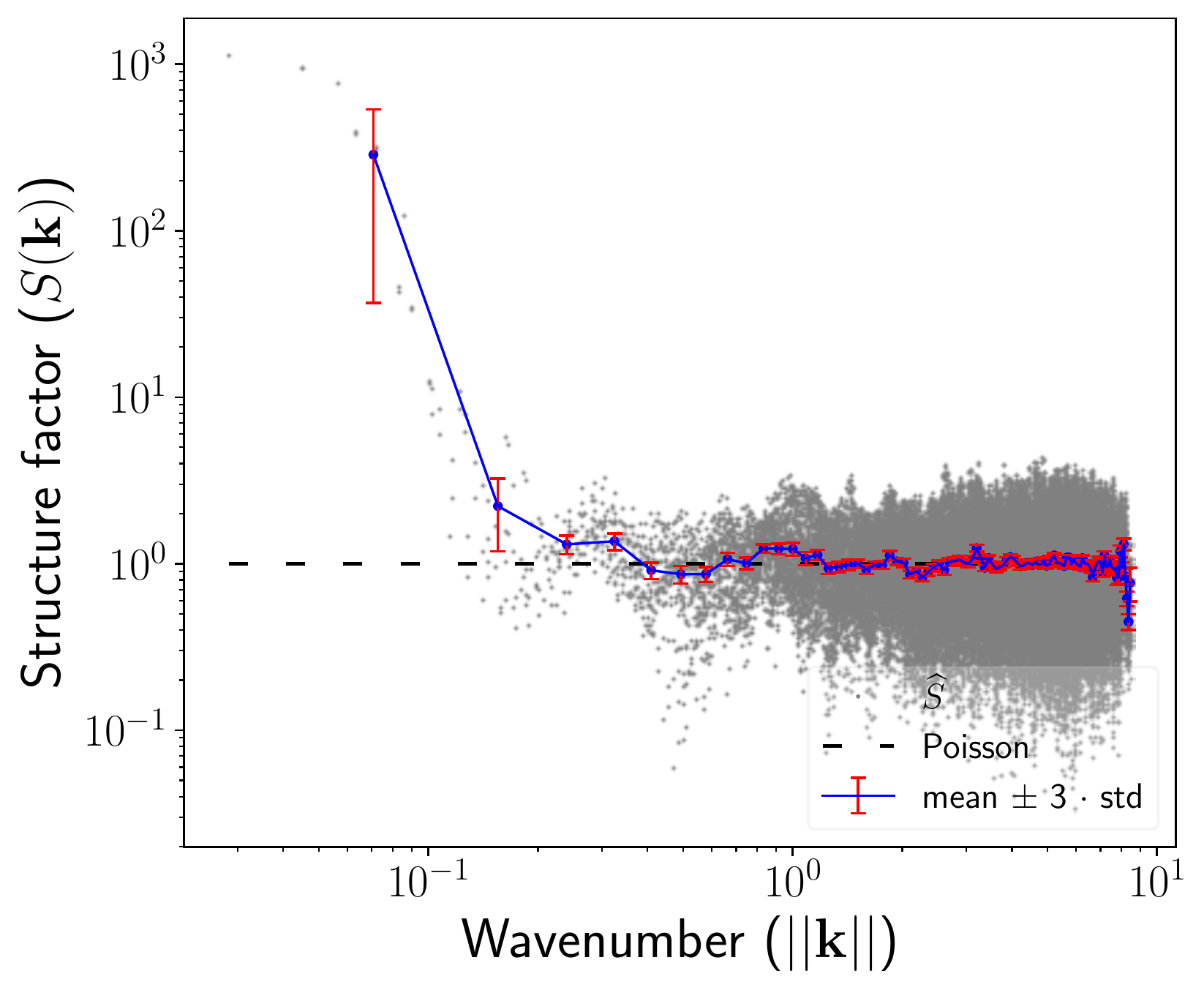}} &
    \raisebox{-\height}{\includegraphics[width=1\linewidth]{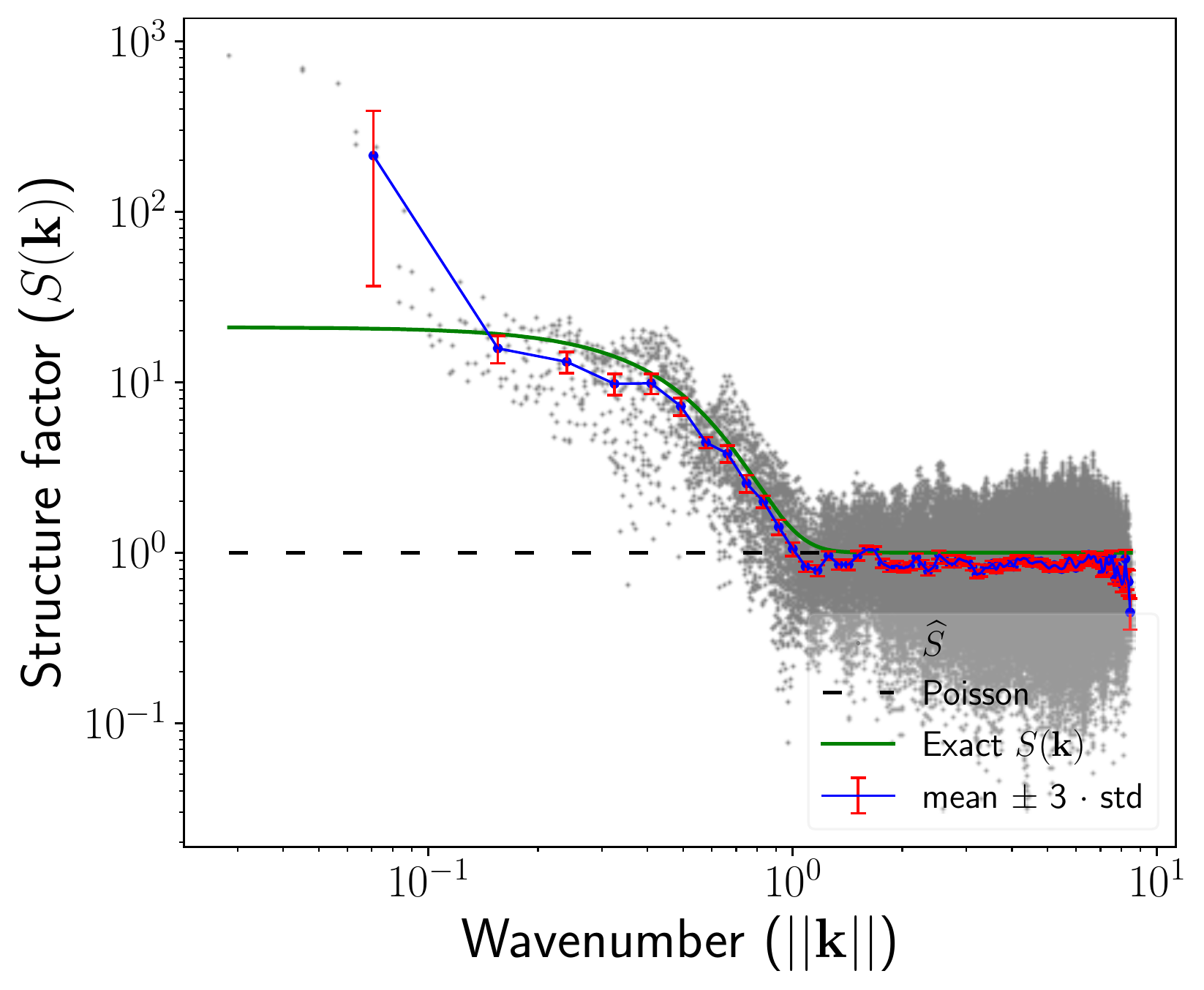}}
  \end{tabular}
  \vspace*{-0.1cm}
  \begin{tabular}{p{\dimexpr 0.03\textwidth-\tabcolsep}p{\dimexpr 0.22\textwidth-\tabcolsep}p{\dimexpr 0.22\textwidth-\tabcolsep}p{\dimexpr 0.22\textwidth-\tabcolsep}p{\dimexpr 0.22\textwidth-\tabcolsep}}
    \multirow{9}{*}{\rotatebox[origin=l]{90}{$\widehat{S}_{\mathrm{DDMT}}$}}             &
    \raisebox{-\height}{\includegraphics[width=1\linewidth]{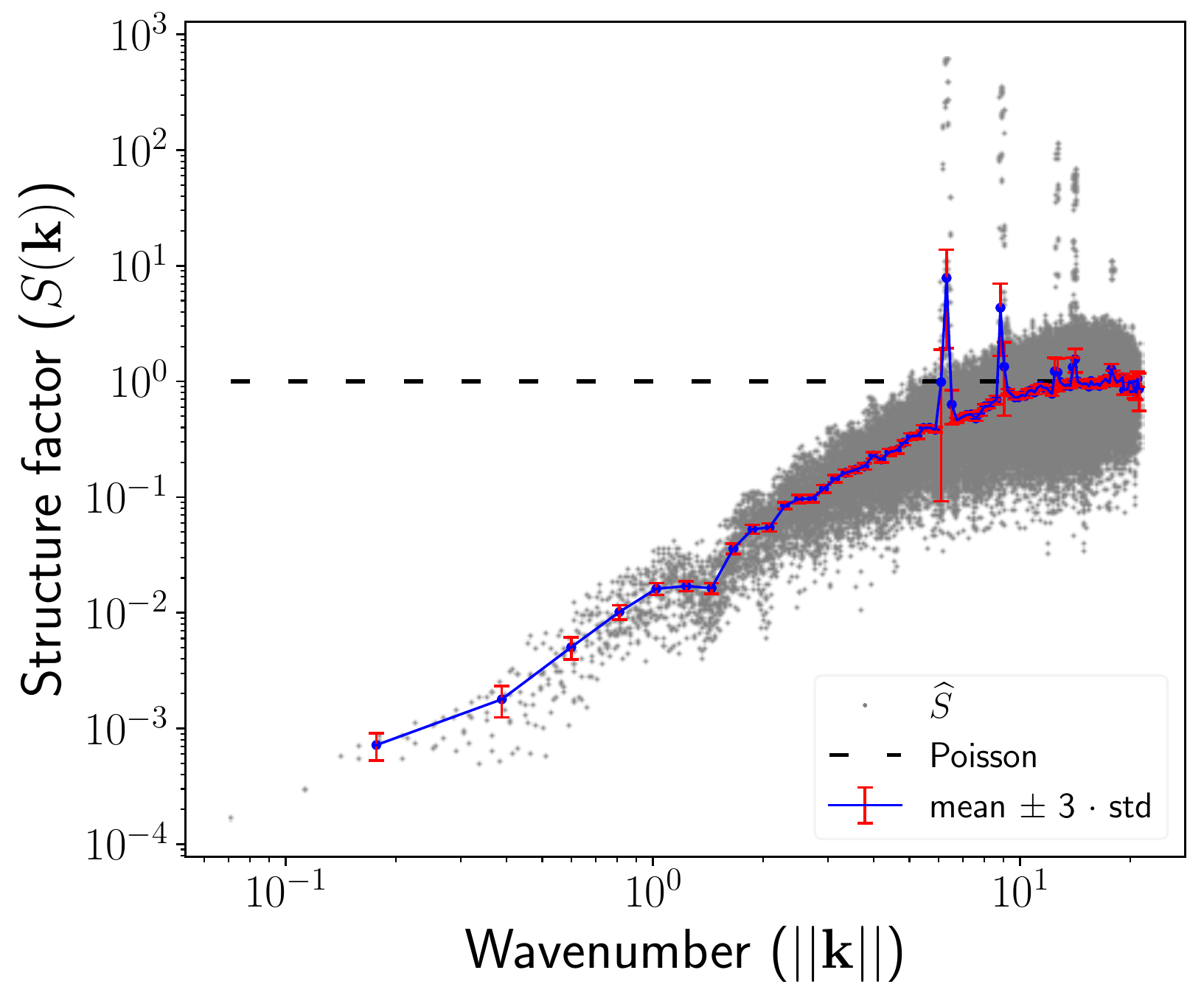}}     &
    \raisebox{-\height}{\includegraphics[width=1\linewidth]{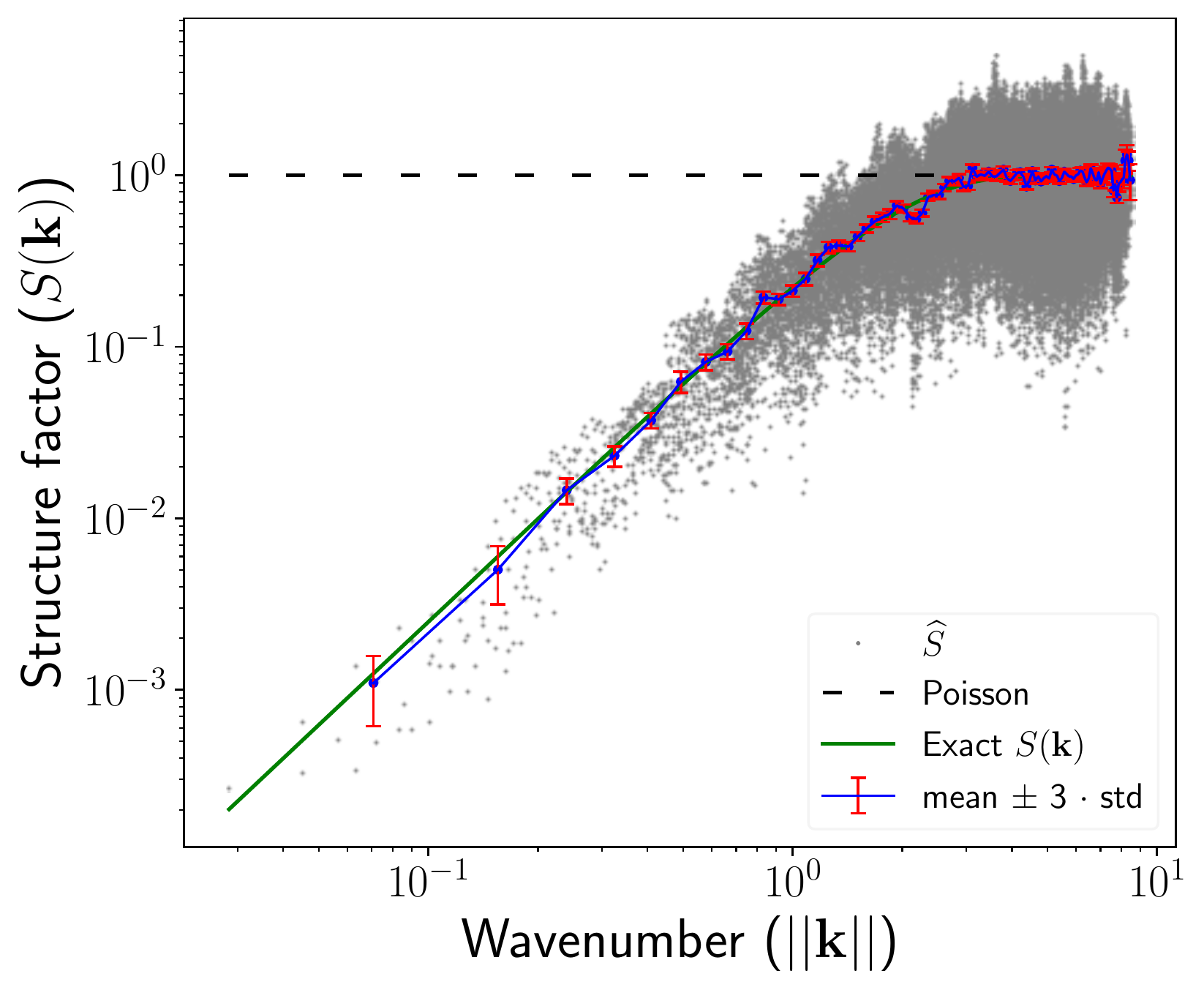}} &
    \raisebox{-\height}{\includegraphics[width=1\linewidth]{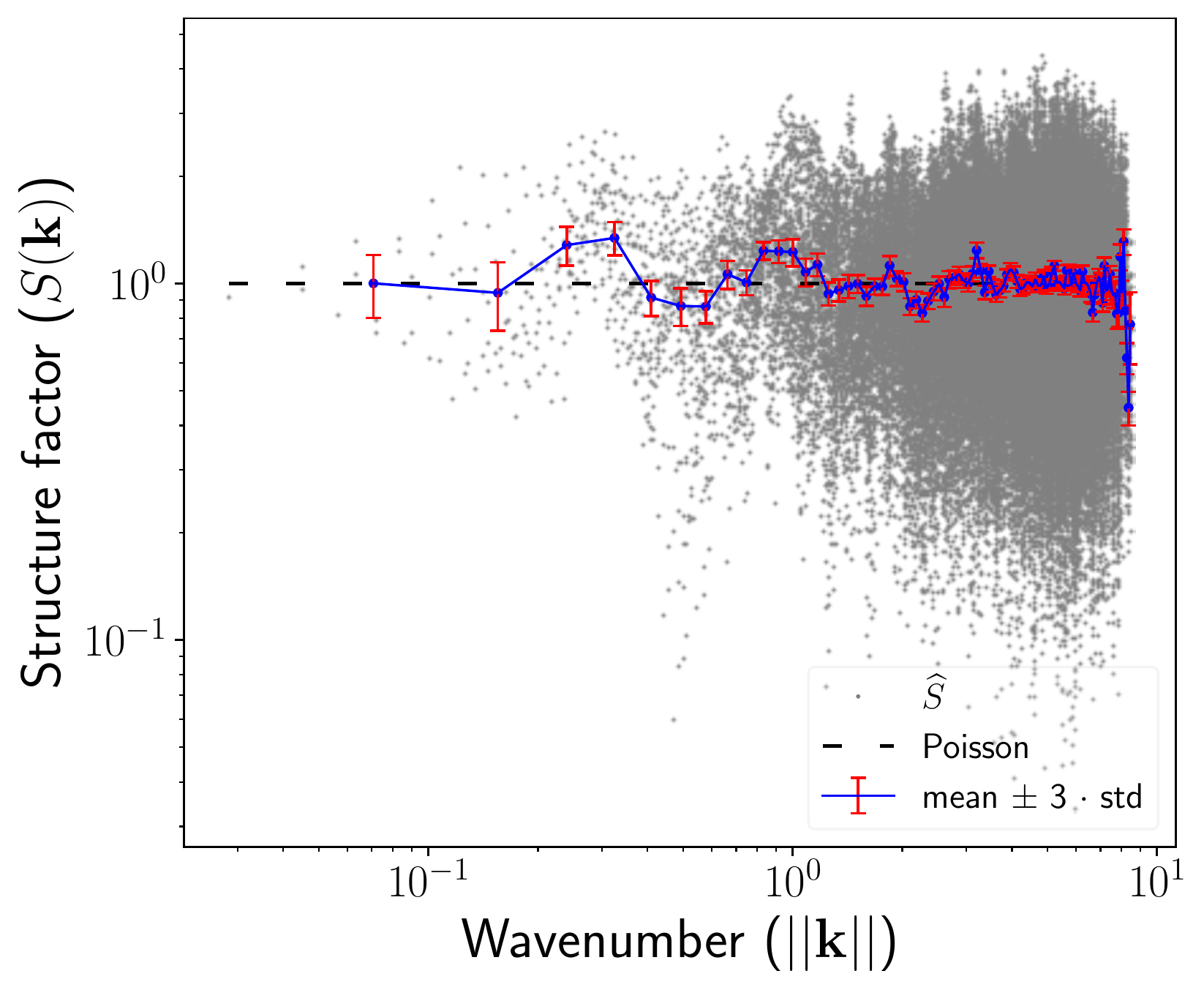}} &
    \raisebox{-\height}{\includegraphics[width=1\linewidth]{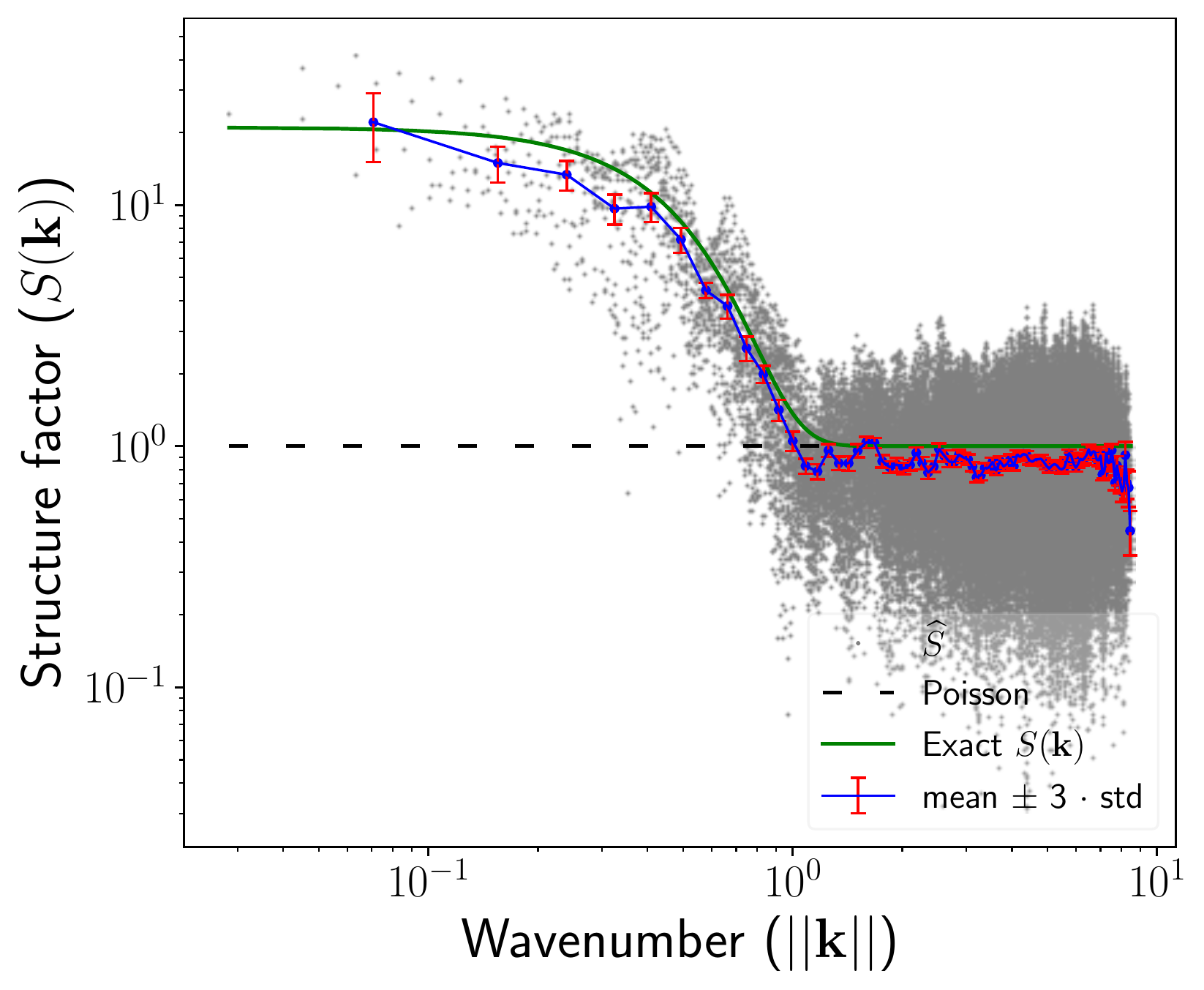}}
  \end{tabular}
  \vspace*{-0.1cm}
  \begin{tabular}{p{\dimexpr 0.03\textwidth-\tabcolsep}p{\dimexpr 0.22\textwidth-\tabcolsep}p{\dimexpr 0.22\textwidth-\tabcolsep}p{\dimexpr 0.22\textwidth-\tabcolsep}p{\dimexpr 0.22\textwidth-\tabcolsep}}
    \multirow{9}{*}{\rotatebox[origin=l]{90}{$\widehat{S}_{\mathrm{UDMT}}$}}             &
    \raisebox{-\height}{\includegraphics[width=1\linewidth]{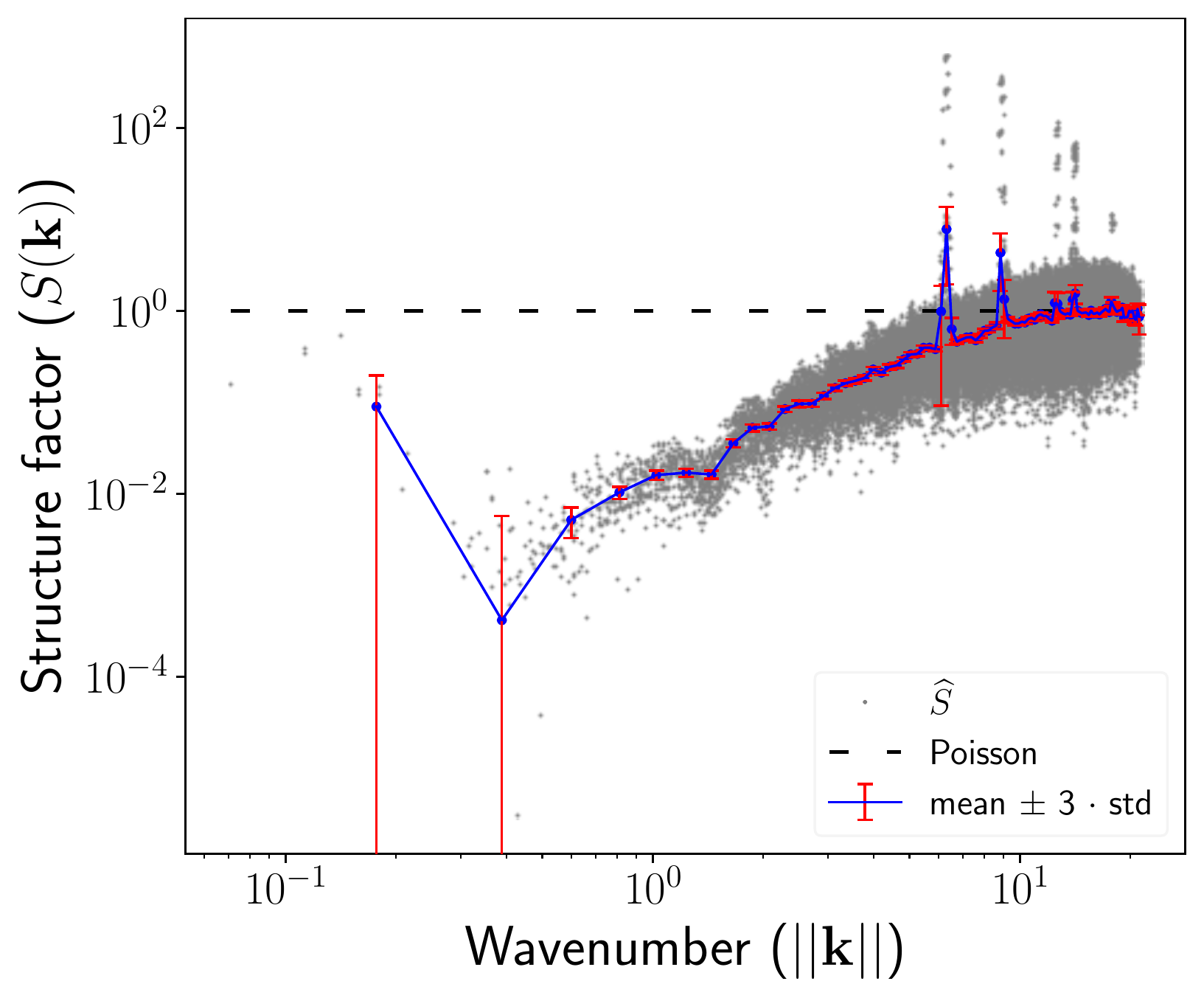}}     &
    \raisebox{-\height}{\includegraphics[width=1\linewidth]{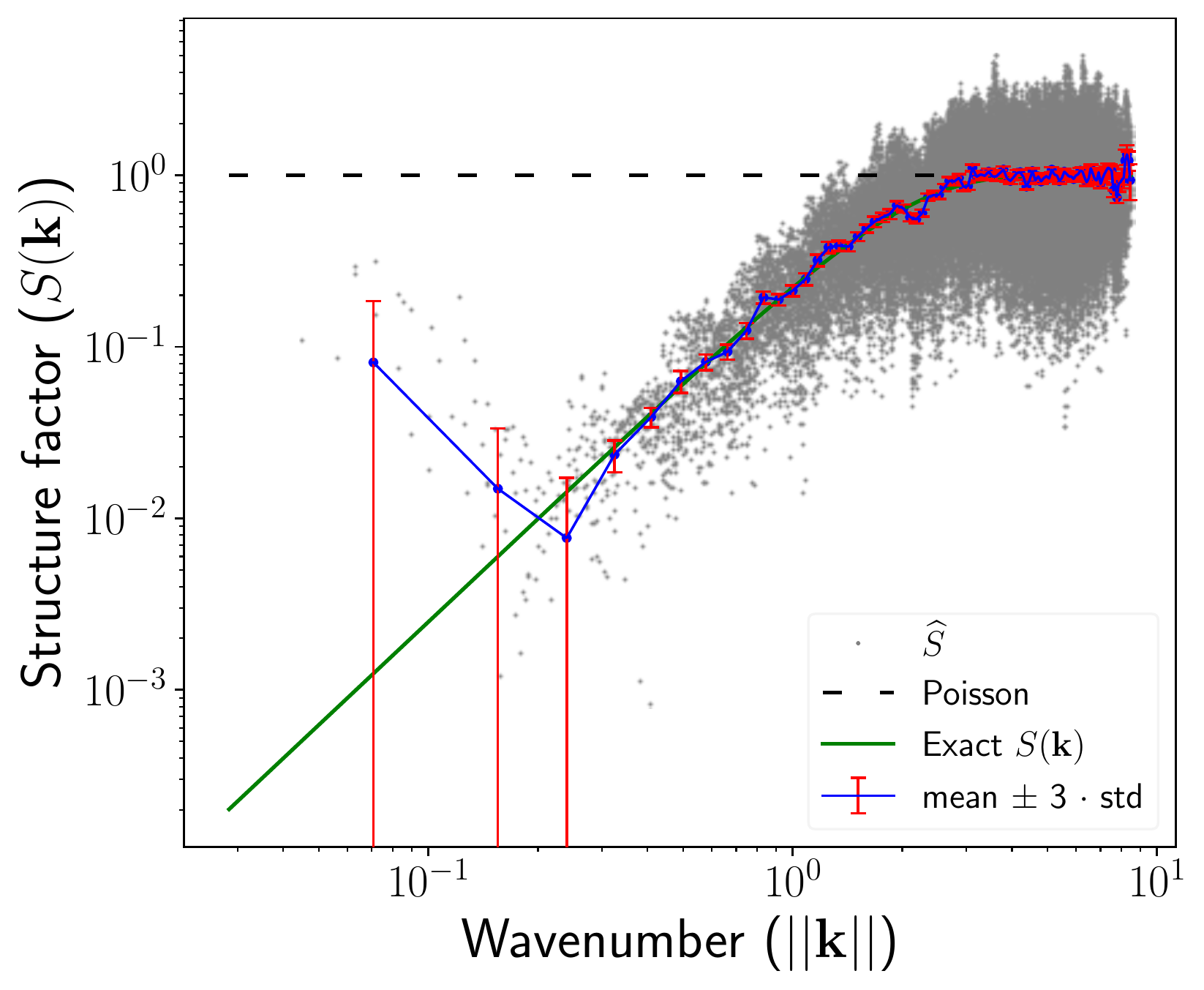}} &
    \raisebox{-\height}{\includegraphics[width=1\linewidth]{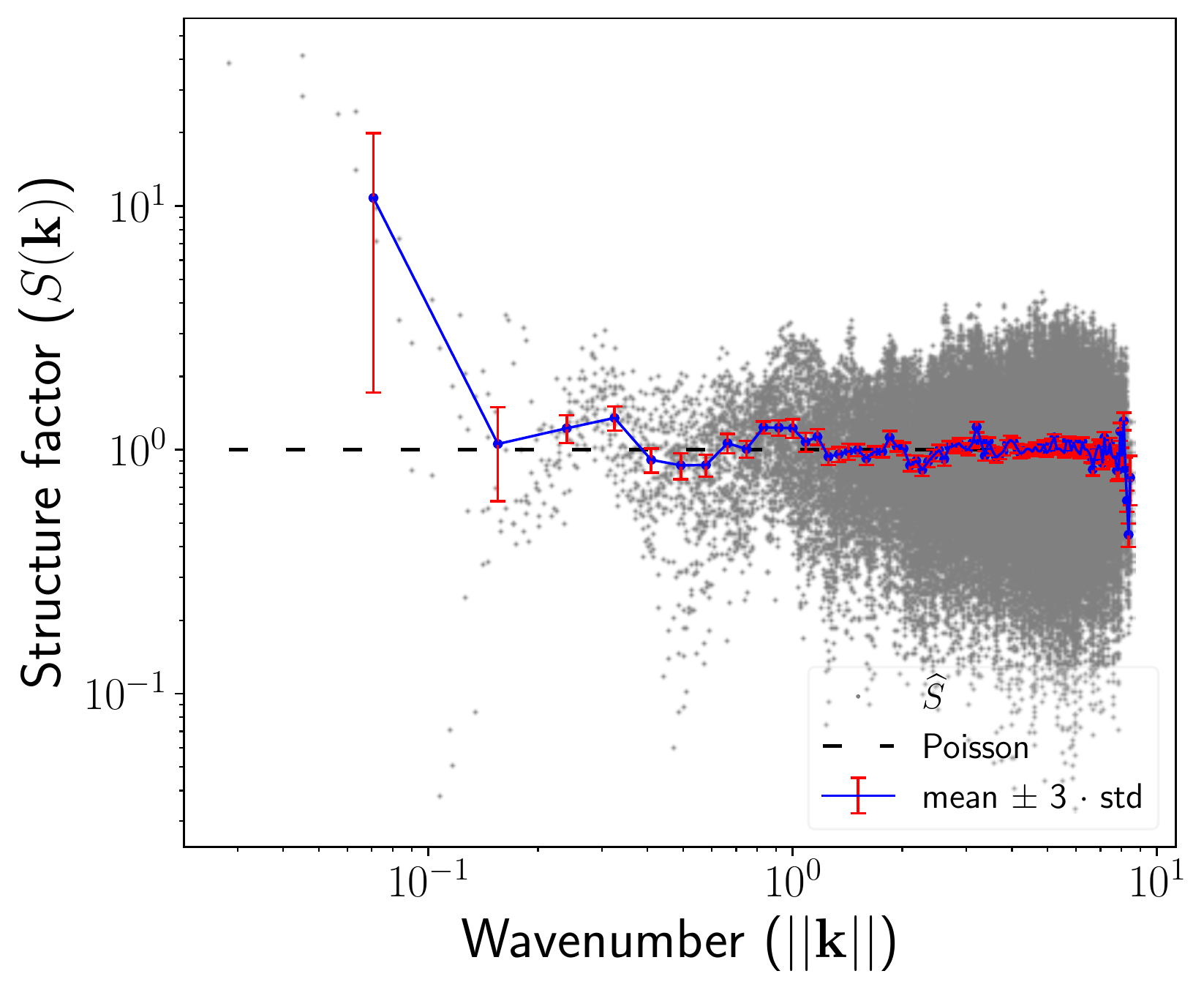}} &
    \raisebox{-\height}{\includegraphics[width=1\linewidth]{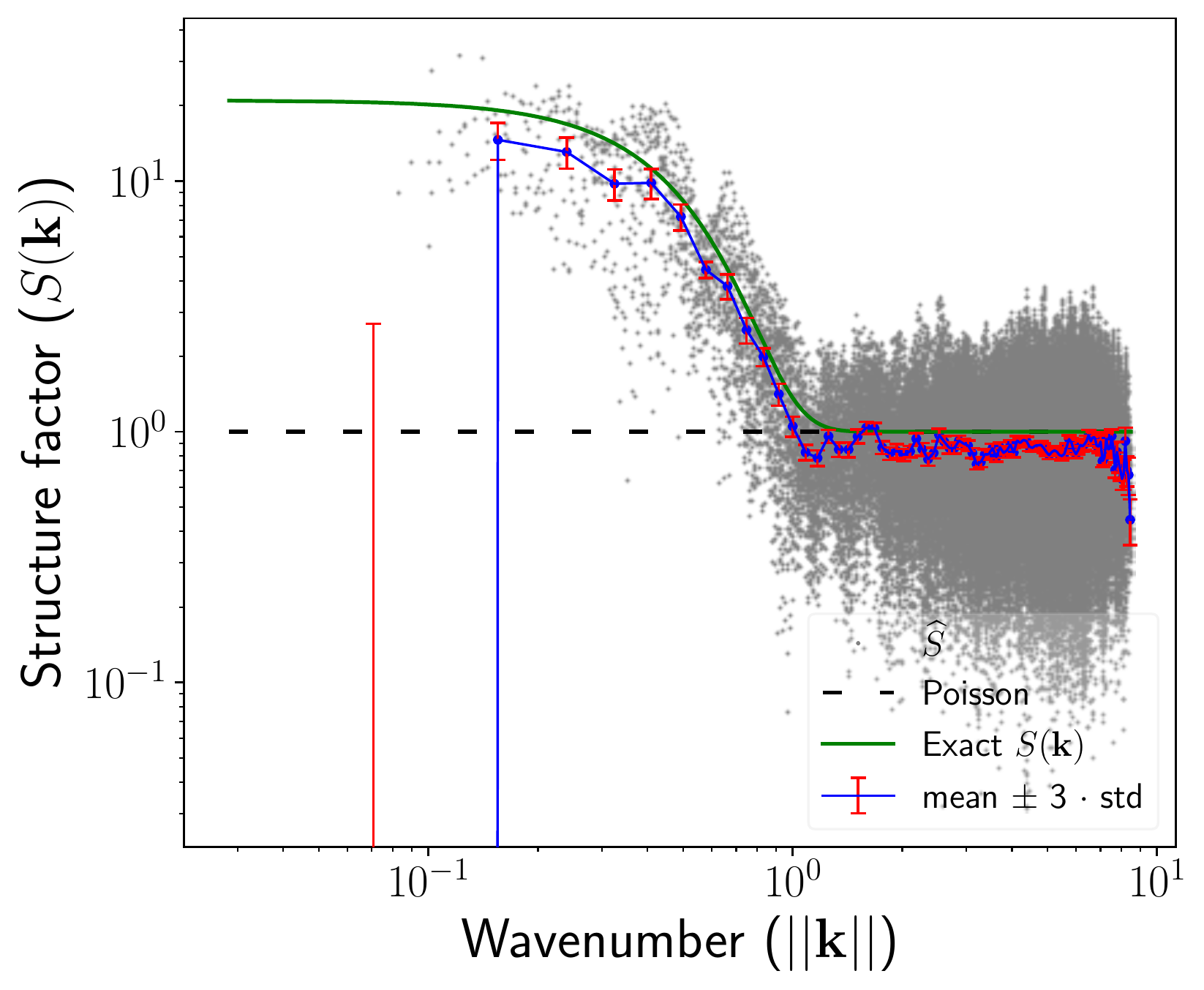}}    \\
    \caption*{}                                                                                         &
    \vspace*{-0.5cm}
    \caption*{{\fontfamily{pcr}\selectfont } KLY }                                               &
    \vspace*{-0.5cm}
    \caption*{{\fontfamily{pcr}\selectfont } Ginibre }                                          &
    \vspace*{-0.5cm}
    \caption*{{\fontfamily{pcr}\selectfont } Poisson }                                           &
    \vspace*{-0.5cm}
    \caption*{{\fontfamily{pcr}\selectfont } Thomas }
  \end{tabular}
  \vspace{-0.5cm}
  \caption{Multitapered estimator and the debiased versions: KLY process (first column), Ginibre ensemble (second column), Poisson process (third column), and Thomas process (last column). The computation and visualization are done using \toolbox{}} \label{fig:s_mtp}
\end{figure*}
The multitapered estimator $\widehat{S}_{\mathrm{MT}}$ of \eqref{eq:s_mt} is now investigated in Figure~\ref{fig:s_mtp}, using the first four sinusoidal tapers, i.e., $(t_q)_{q=1}^4$ with $t_q(\bfx, W) = t(\bfx, \bfp^q, W)$ and $\bfp^q \in \{1,2\}^2$ in \eqref{eq:sine_taper}.
We also show the results of the corresponding directly and undirectly debiased versions, $\widehat{S}_{\mathrm{DDMT}}$ and  $\widehat{S}_{\mathrm{UDMT}}$.
We again observe the bias of $\widehat{S}_{\mathrm{MT}}$ at small wavenumbers (second row), and that the negative values output by $\widehat{S}_{\mathrm{UDMT}}$ at small wavenumbers (last row) make visual assessments of hyperuniformity less straightforward.
Like with single tapers, the directly debiased estimator $\widehat{S}_{\mathrm{DDMT}}$ gives a consistently accurate approximation.
Compared to Figure~\ref{fig:s_t}, however, it is not obvious whether multitapering yields a smaller mean square error than single tapers, and a more quantitative study will investigate this in Section~\ref{sec:Comparison of the estimators}.

\subsection{Demonstrating estimators that assume isotropy} % (fold)
\label{sub:Approximating the structure factor using an ISE}

\paragraph{Bartlett's isotropic estimator} % (fold)
\begin{figure*}[!ht]
  \begin{tabular}{p{\dimexpr 0.03\textwidth-\tabcolsep}p{\dimexpr 0.22\textwidth-\tabcolsep}p{\dimexpr 0.22\textwidth-\tabcolsep}p{\dimexpr 0.22\textwidth-\tabcolsep}p{\dimexpr 0.22\textwidth-\tabcolsep}}
    \multirow{9}{*}{\rotatebox[origin=c]{90}{Point process}}                                   &
    \raisebox{-\height}{\includegraphics[width=0.9\linewidth]{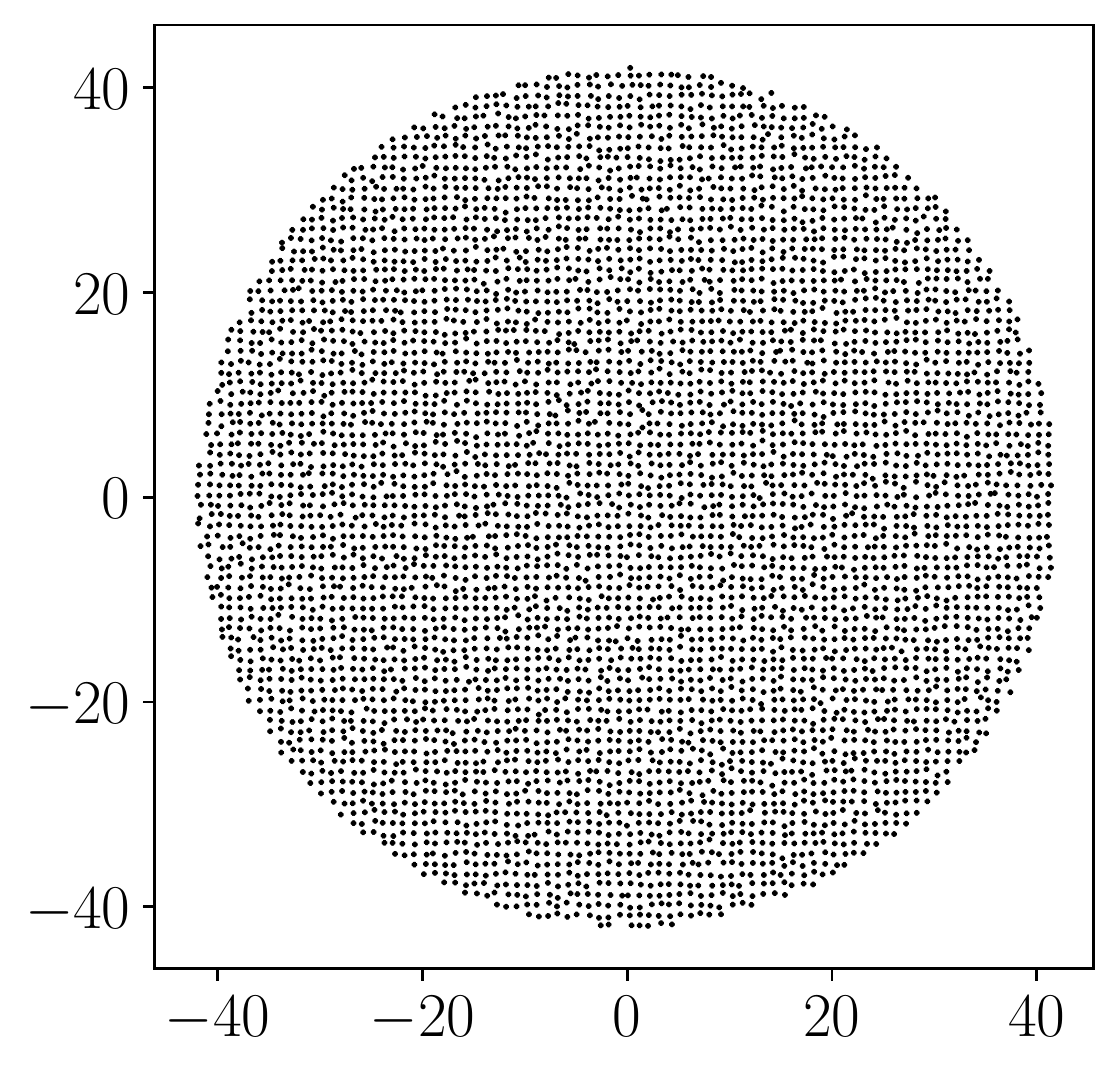}}     &
    \raisebox{-\height}{\includegraphics[width=0.9\linewidth]{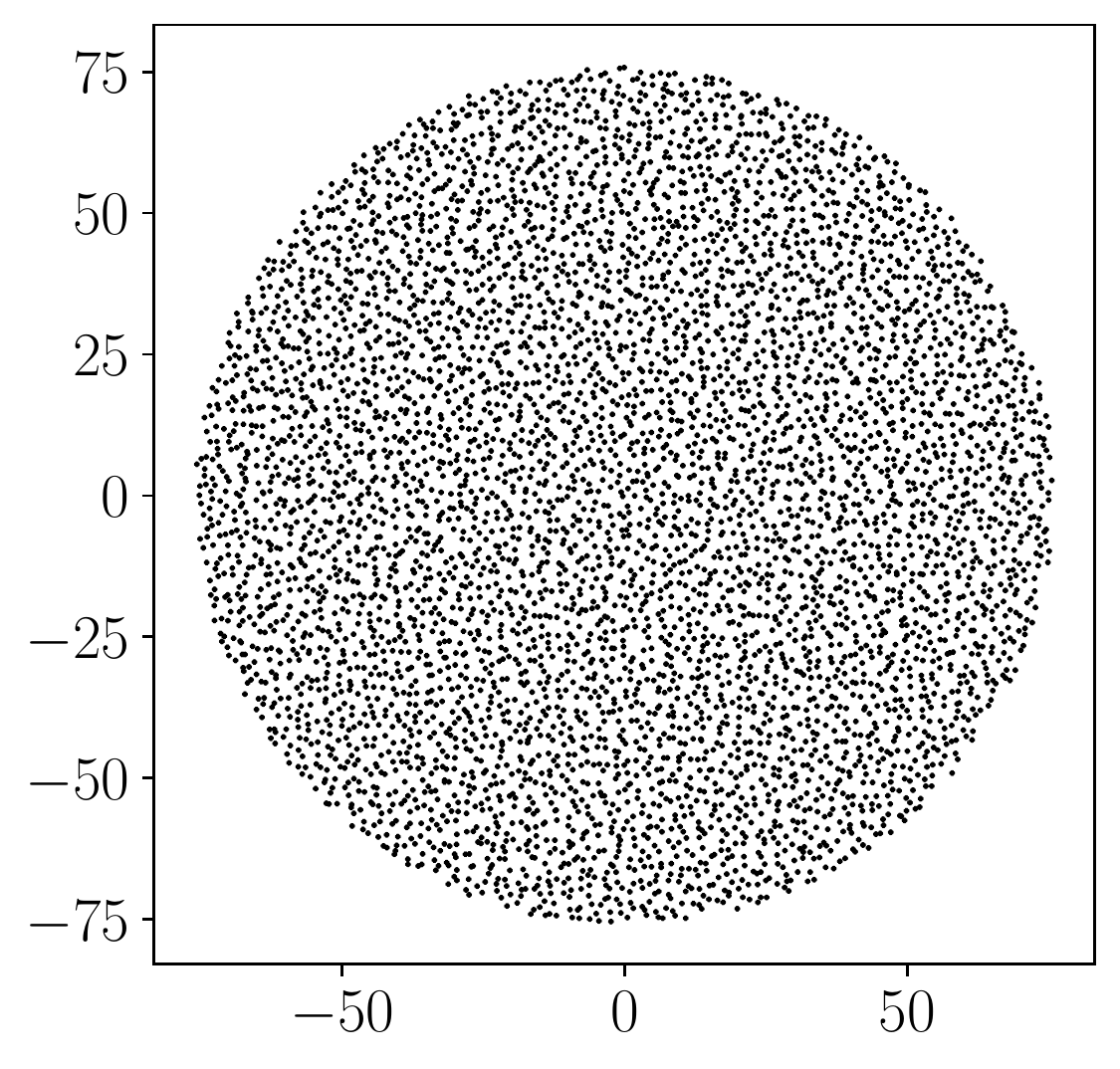}} &
    \raisebox{-\height}{\includegraphics[width=0.9\linewidth]{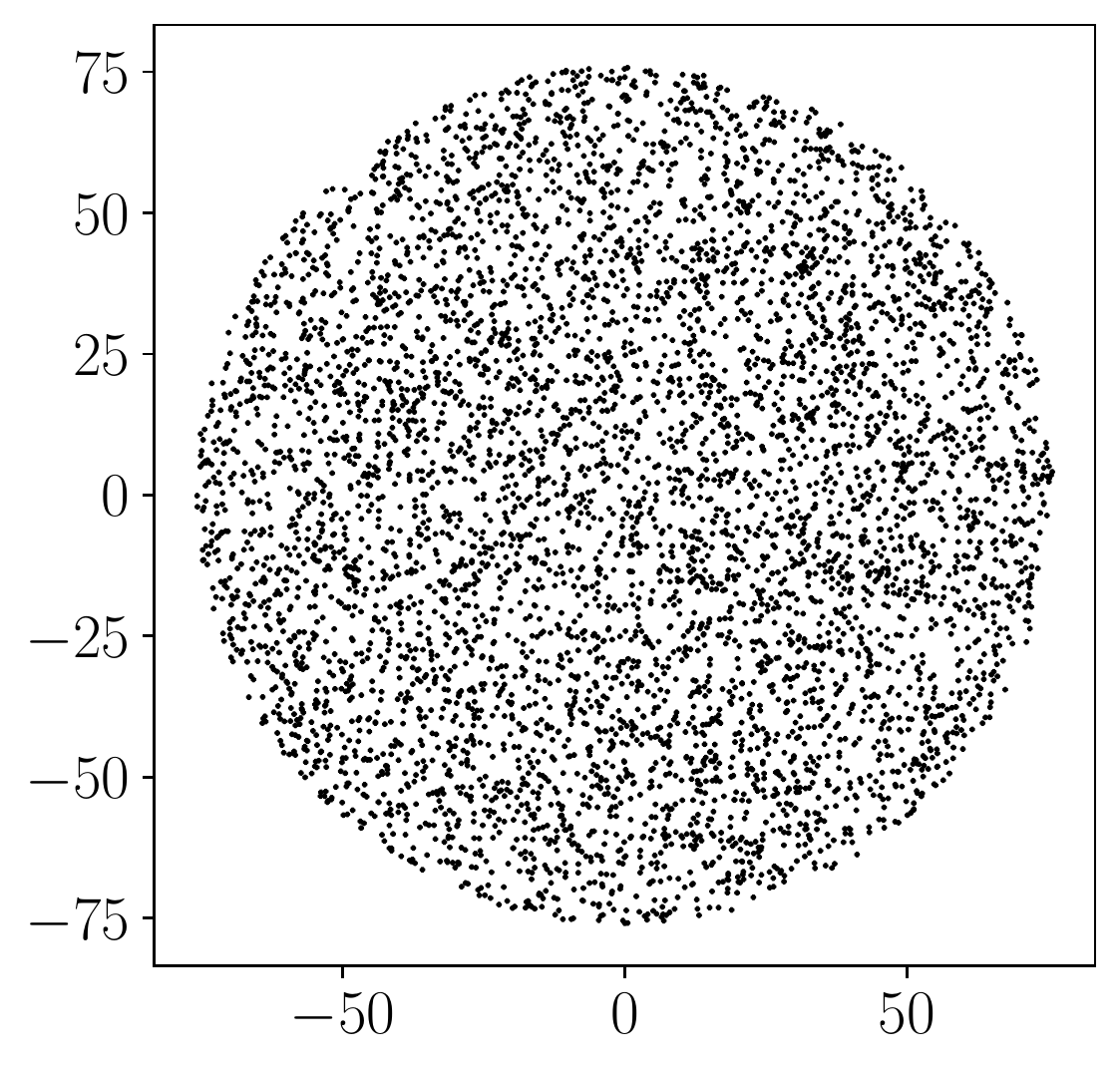}} &
    \raisebox{-\height}{\includegraphics[width=0.9\linewidth]{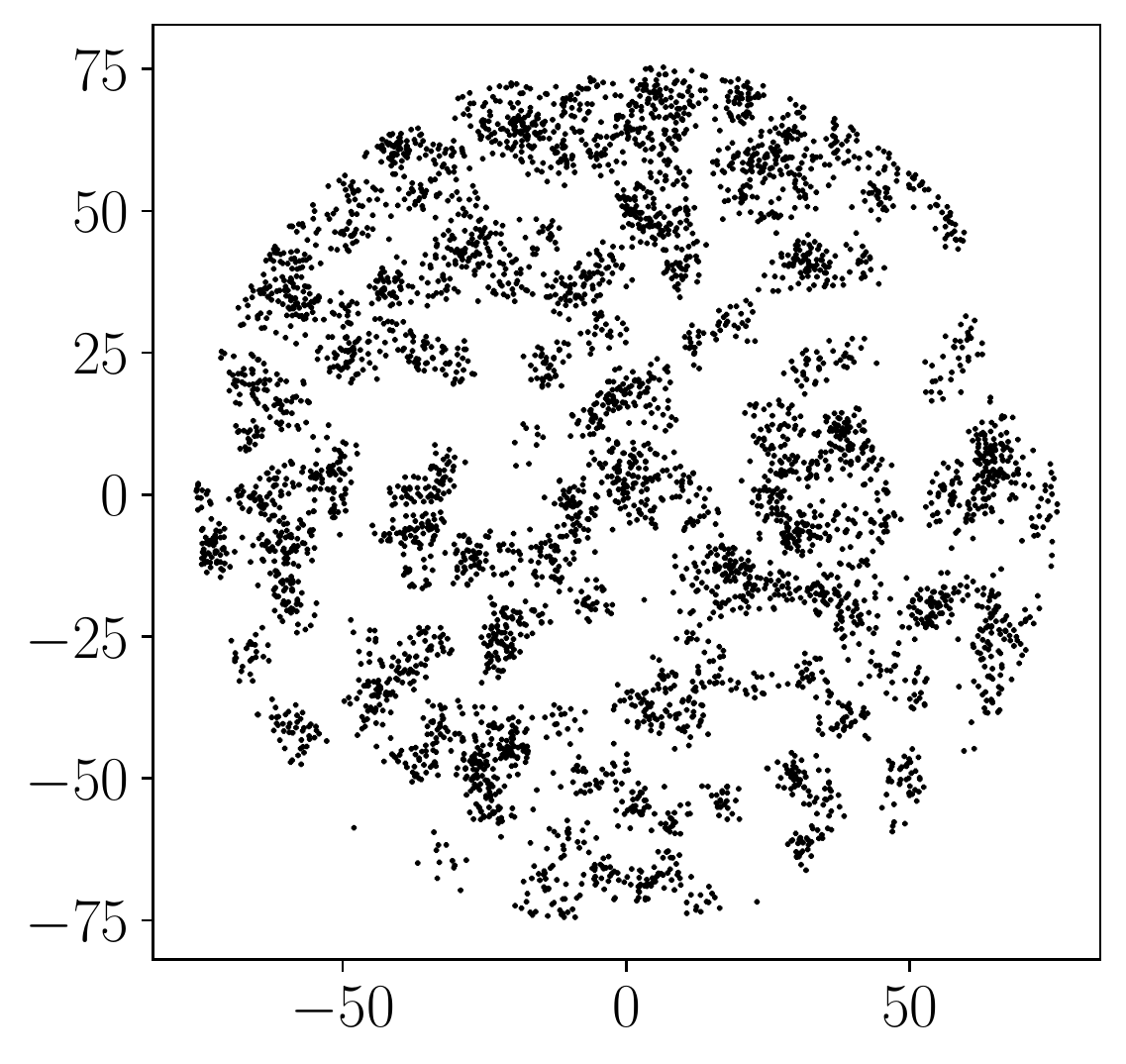}}
  \end{tabular}
  \vspace{-0.2cm}
  \begin{tabular}{p{\dimexpr 0.03\textwidth-\tabcolsep}p{\dimexpr 0.22\textwidth-\tabcolsep}p{\dimexpr 0.22\textwidth-\tabcolsep}p{\dimexpr 0.22\textwidth-\tabcolsep}p{\dimexpr 0.22\textwidth-\tabcolsep}}
    \multirow{9}{*}{\rotatebox[origin=l]{90}{$\widehat{S}_{\mathrm{BI}}(k)$}}                         &
    \raisebox{-\height}{\includegraphics[width=1\linewidth]{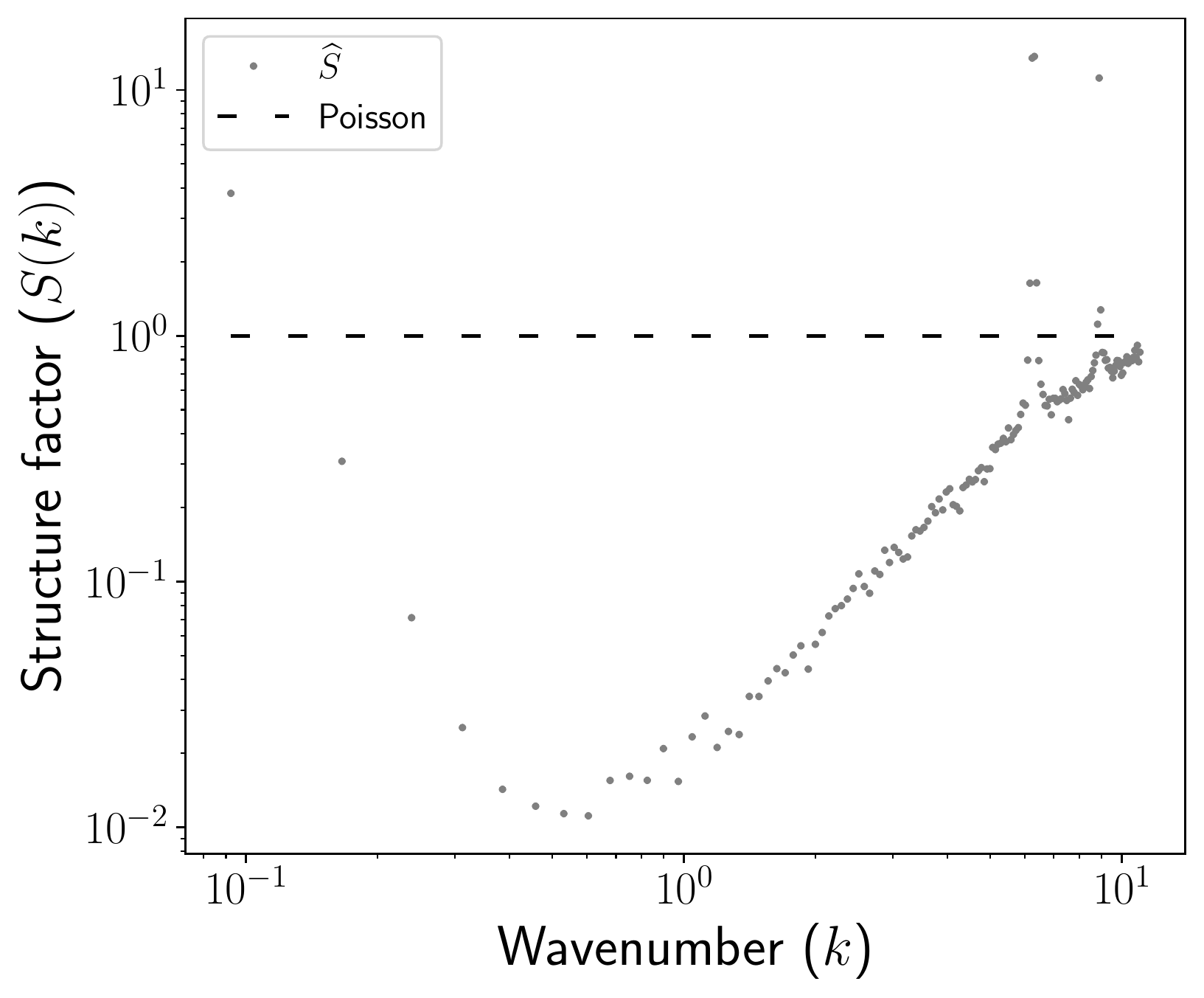}}     &
    \raisebox{-\height}{\includegraphics[width=1\linewidth]{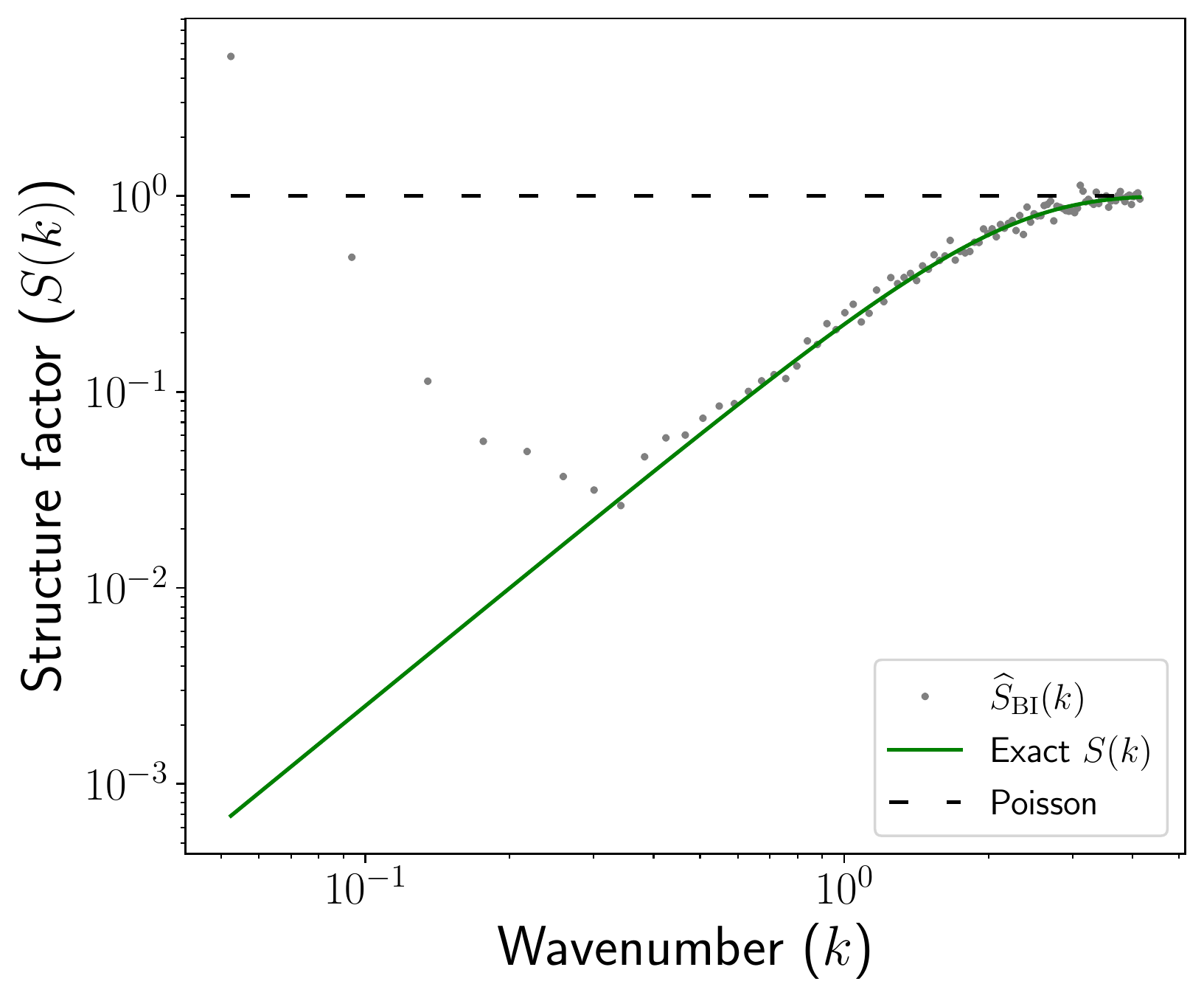}} &
    \raisebox{-\height}{\includegraphics[width=1.05\linewidth]{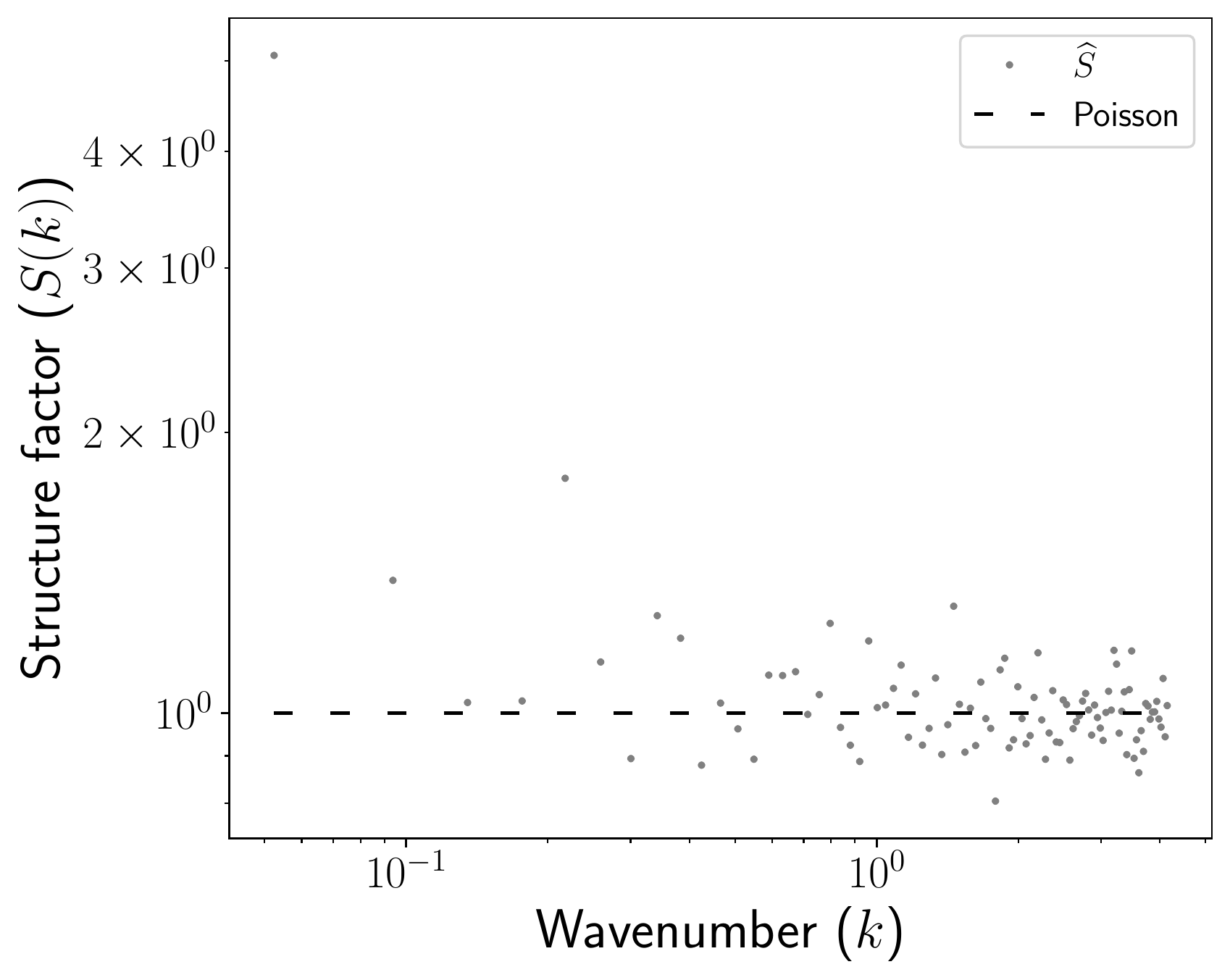}} &
    \raisebox{-\height}{\includegraphics[width=1\linewidth]{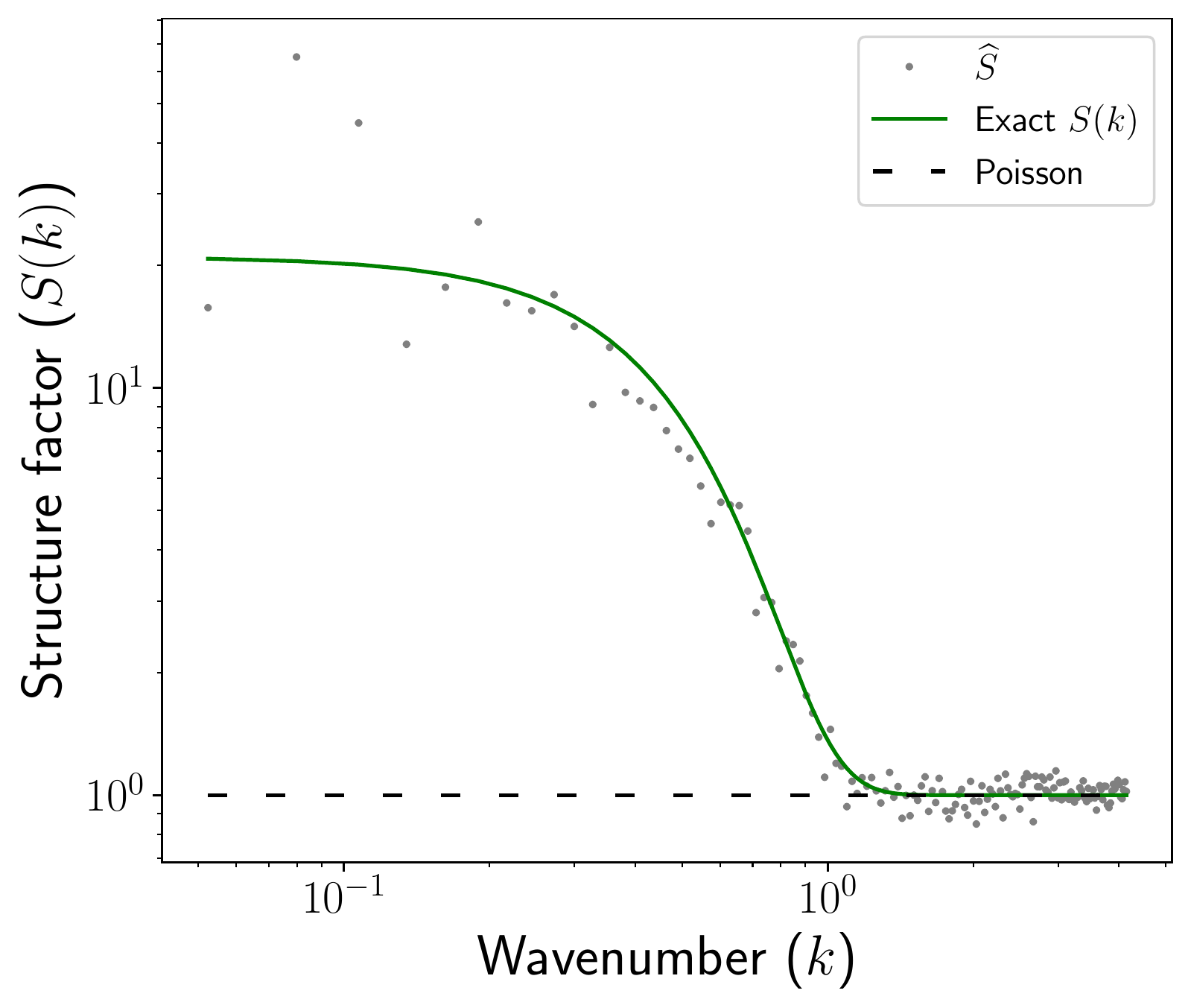}}
  \end{tabular}
  \vspace{-0.2cm}
  \begin{tabular}{p{\dimexpr 0.03\textwidth-\tabcolsep}p{\dimexpr 0.22\textwidth-\tabcolsep}p{\dimexpr 0.22\textwidth-\tabcolsep}p{\dimexpr 0.22\textwidth-\tabcolsep}p{\dimexpr 0.22\textwidth-\tabcolsep}}
    \multirow{9}{*}{\rotatebox[origin=l]{90}{$\widehat{S}_{\mathrm{BI}}(\frac{\nu}{R})$}} &
    \raisebox{-\height}{\includegraphics[width=1\linewidth]{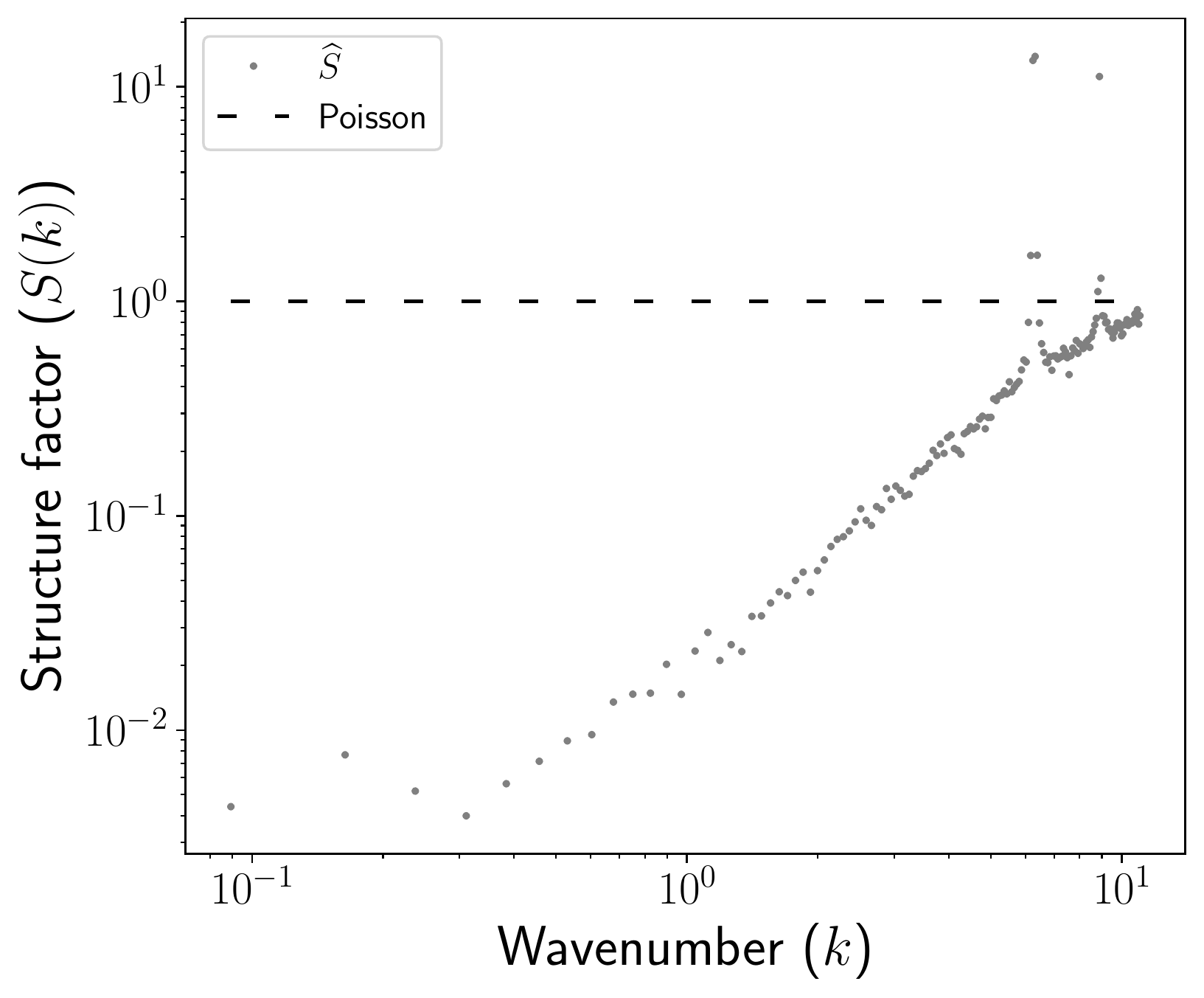}}     &
    \raisebox{-\height}{\includegraphics[width=1\linewidth]{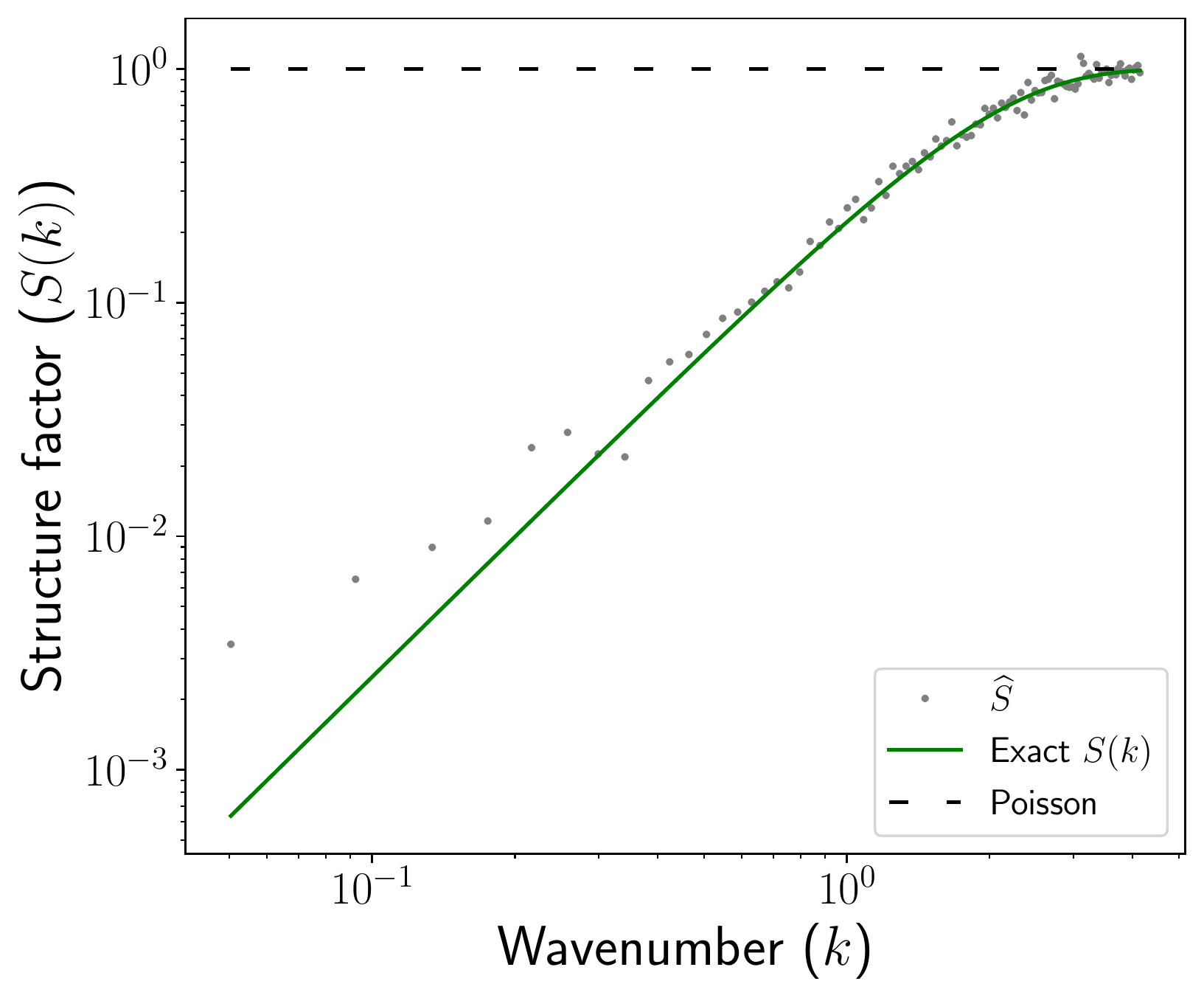}} &
    \raisebox{-\height}{\includegraphics[width=1.05\linewidth]{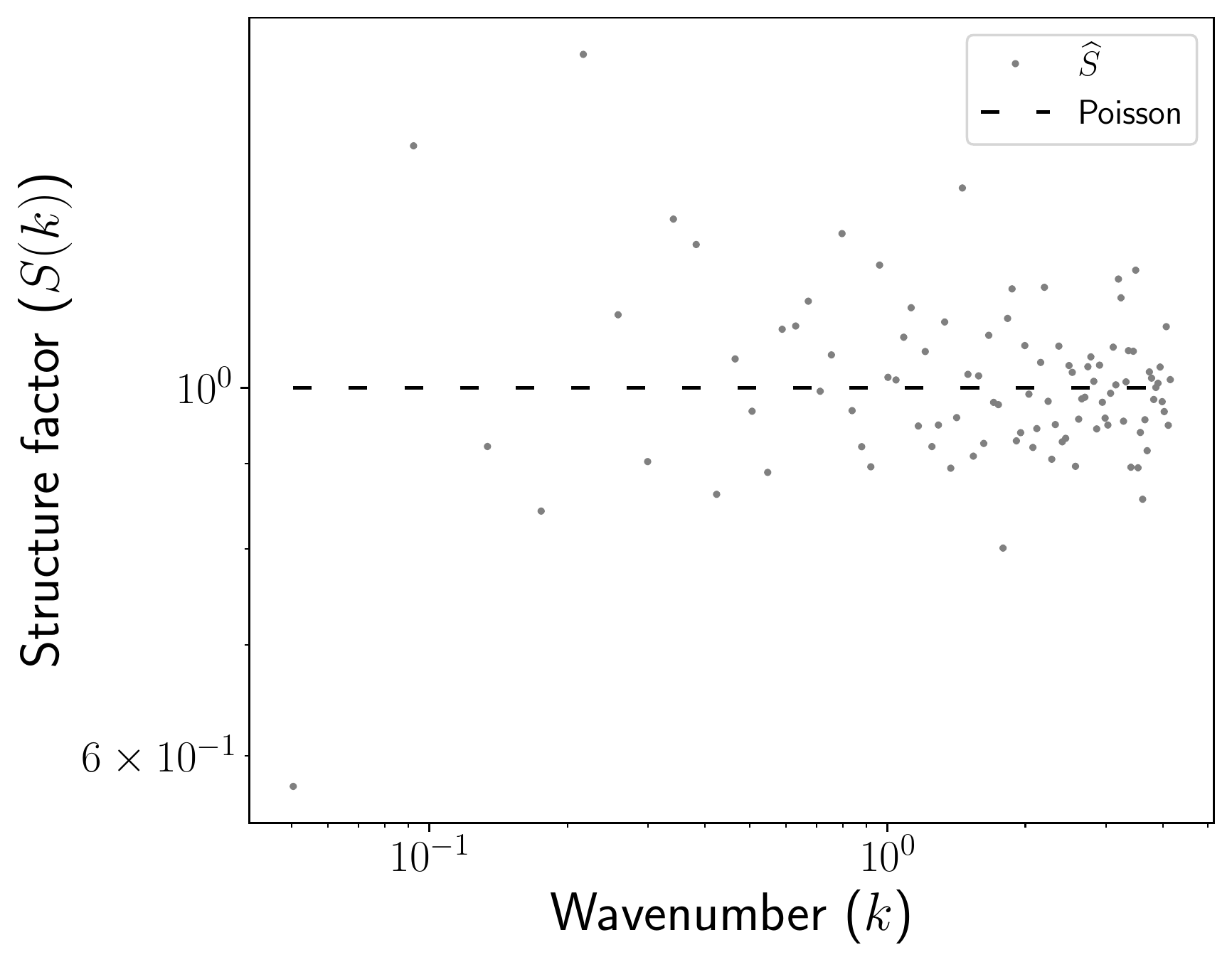}} &
    \raisebox{-\height}{\includegraphics[width=1\linewidth]{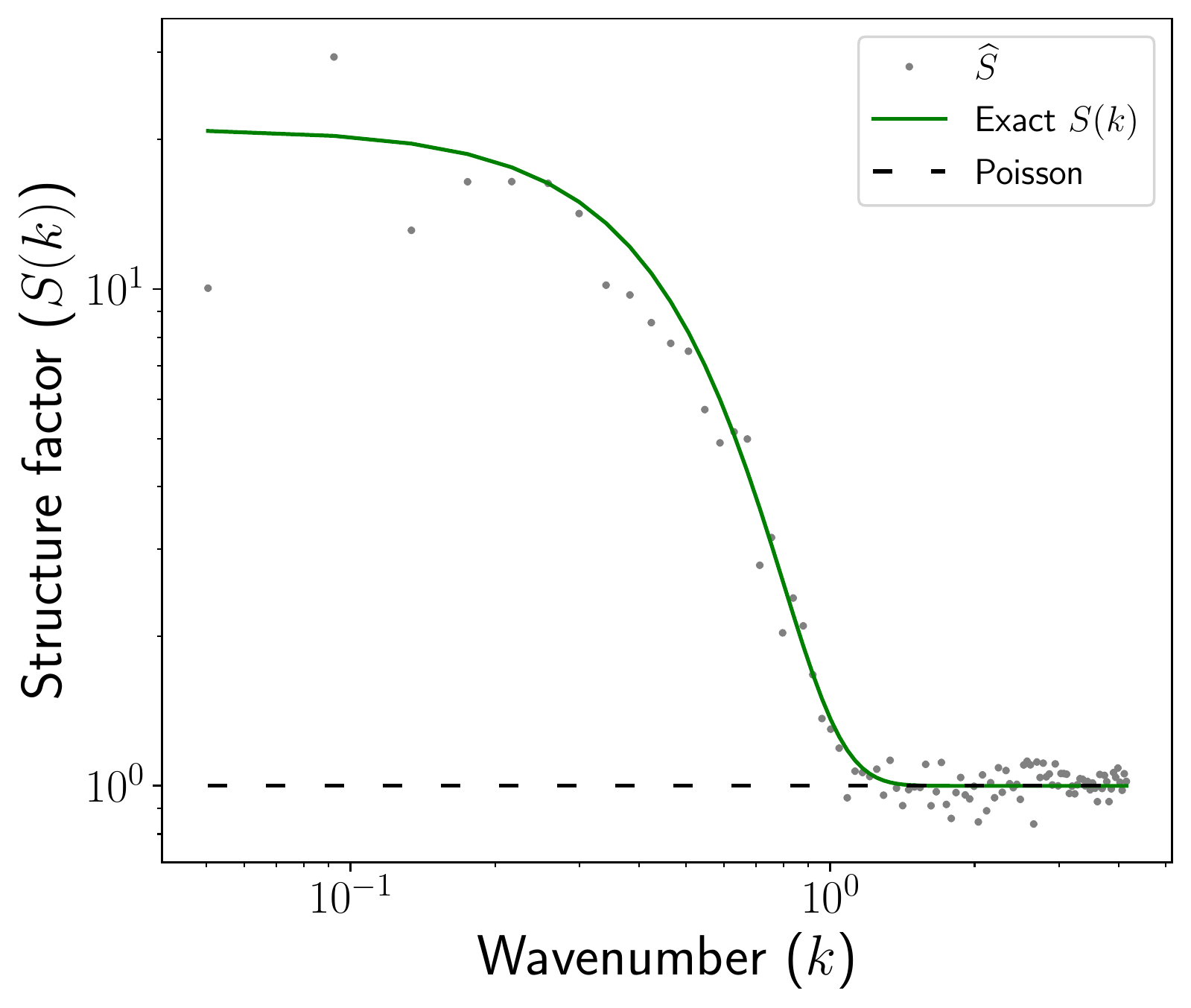}}    \\
    \caption*{}                                                                           &
    \vspace*{-0.5cm}
    \caption*{{\fontfamily{pcr}\selectfont } KLY }                                 &
    \vspace*{-0.5cm}
    \caption*{{\fontfamily{pcr}\selectfont } Ginibre }                            &
    \vspace*{-0.5cm}
    \caption*{{\fontfamily{pcr}\selectfont } Poisson }                             &
    \vspace*{-0.5cm}
    \caption*{{\fontfamily{pcr}\selectfont } Thomas }
  \end{tabular}
  \vspace{-0.5cm}
  \caption{Bartlett's isotropic estimator: KLY process (first column), Ginibre ensemble (second column), Poisson process (third column), and Thomas process (last column). The computation and visualization are done using \toolbox{}} \label{fig:s_bi}
\end{figure*}
Figure~\ref{fig:s_bi} illustrates Bartlett's isotropic estimator of Section~\ref{sub:Estimators assuming stationarity and isotropy}.
Columns respectively correspond to the KLY, Ginibre, Poisson, and Thomas
point processes.
The first row contains a sample of each point process, observed in ball windows.
The second row shows $\widehat{S}_{\mathrm{BI}}$ on arbitrary wavenumbers $k$, while in the last row, the estimator is only evaluated on a subset of the Bessel-specific allowed wavenumbers \eqref{eq:allowed_k_isotropic}.

First, we note that, unlike scattering intensity variants, plotting Bartlett's isotropic estimator $k\mapsto \widehat{S}_{\mathrm{BI}}(k)$ in \eqref{eq:s_BI} does not require binning.
On the other hand, Bartlett's estimator is significantly costlier than its scattering intensity counterpart; See Section~\ref{sub:Computational time}.
Now, we comment on the accuracy of the estimator in Figure~\ref{fig:s_bi}. Here again, small, non-allowed wavenumbers give rise to large biases for $\widehat{S}_{\mathrm{BI}}(k)$, especially for the two hyperuniform point processes (KLY and Ginibre).
When applied to allowed wavenumbers, the estimator shows accuracy across all point processes, similarly to the directly debiased tapered estimators.

\paragraph{Estimating the pair correlation function} % (fold)
\begin{figure*}[!ht]
  \begin{tabular}{p{\dimexpr 0.03\textwidth-\tabcolsep}p{\dimexpr 0.22\textwidth-\tabcolsep}p{\dimexpr 0.22\textwidth-\tabcolsep}p{\dimexpr 0.22\textwidth-\tabcolsep}p{\dimexpr 0.22\textwidth-\tabcolsep}}
    \multirow{9}{*}{\rotatebox[origin=c]{90}{Point process}}                                   &
    \raisebox{-\height}{\includegraphics[width=0.9\linewidth]{kly_pp_ball.pdf}}     &
    \raisebox{-\height}{\includegraphics[width=0.9\linewidth]{ginibre_pp_ball.pdf}} &
    \raisebox{-\height}{\includegraphics[width=0.9\linewidth]{poisson_pp_ball.pdf}} &
    \raisebox{-\height}{\includegraphics[width=0.9\linewidth]{thomas_pp_ball.pdf}}
  \end{tabular}
  \vspace{-0.2cm}
  \begin{tabular}{p{\dimexpr 0.03\textwidth-\tabcolsep}p{\dimexpr 0.22\textwidth-\tabcolsep}p{\dimexpr 0.22\textwidth-\tabcolsep}p{\dimexpr 0.22\textwidth-\tabcolsep}p{\dimexpr 0.22\textwidth-\tabcolsep}}
    \multirow{9}{*}{\rotatebox[origin=l]{90}{\texttt{pcf.ppp}}}                              &
    \raisebox{-\height}{\includegraphics[width=1\linewidth]{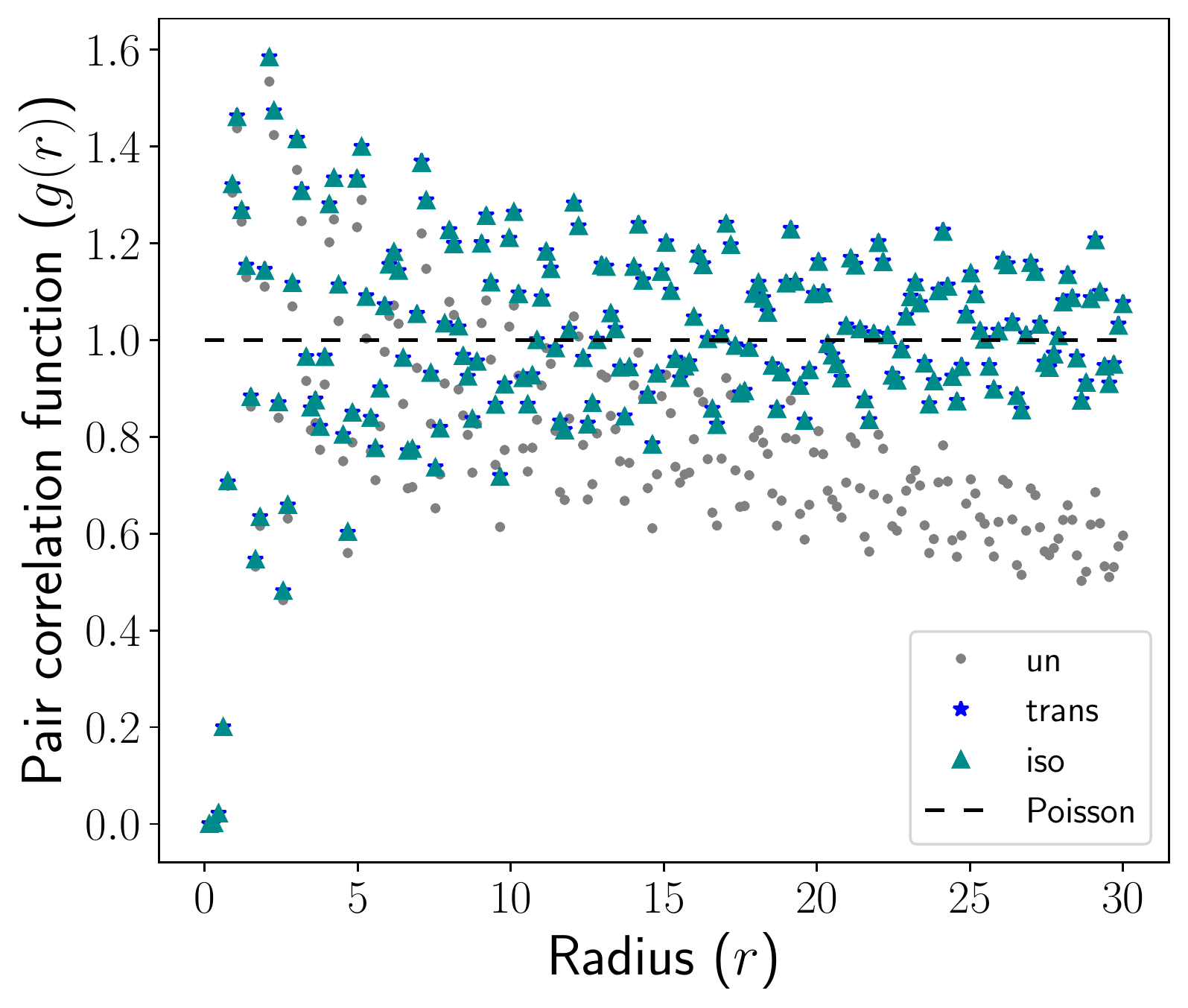}}     &
    \raisebox{-\height}{\includegraphics[width=1\linewidth]{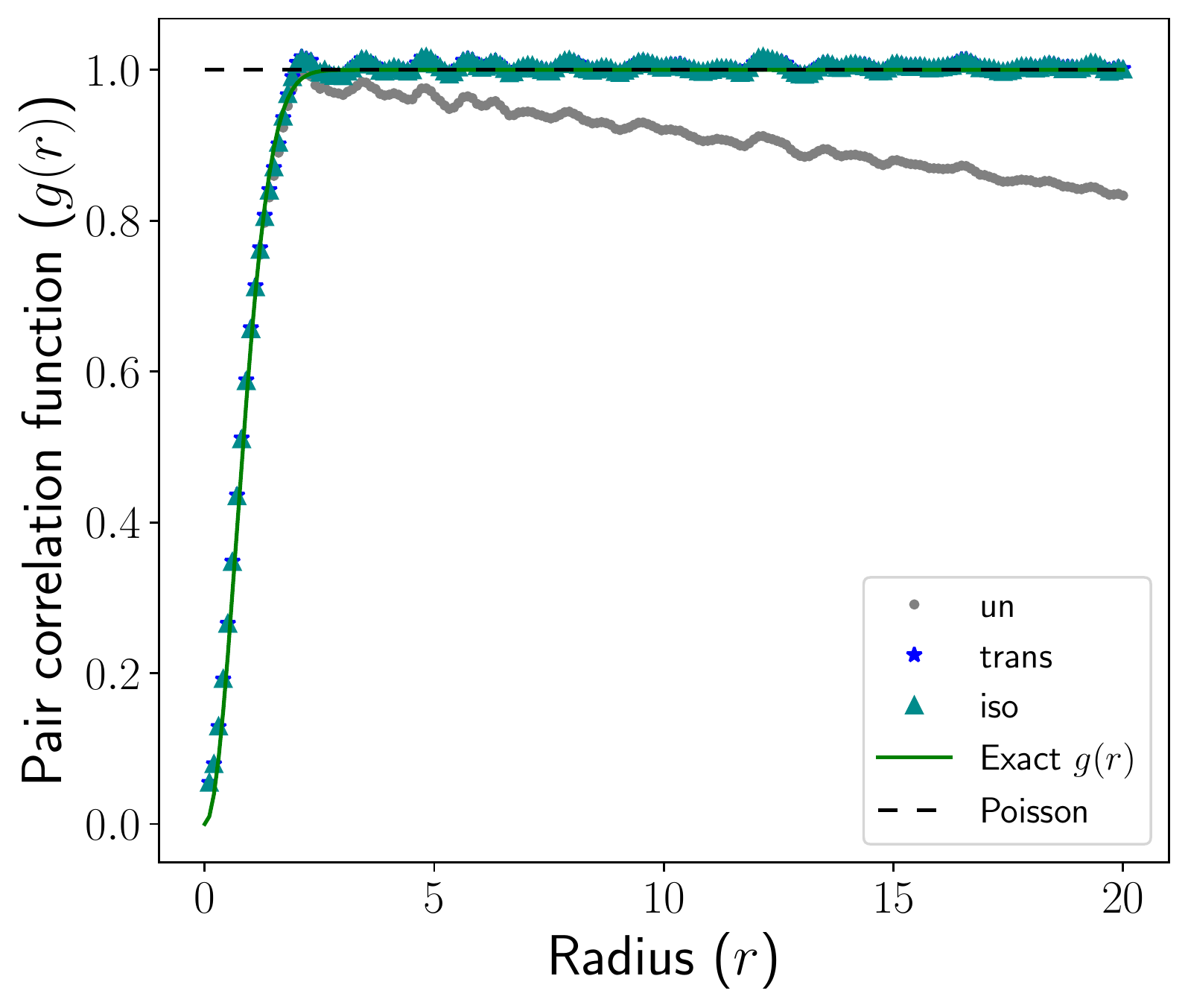}} &
    \raisebox{-\height}{\includegraphics[width=1\linewidth]{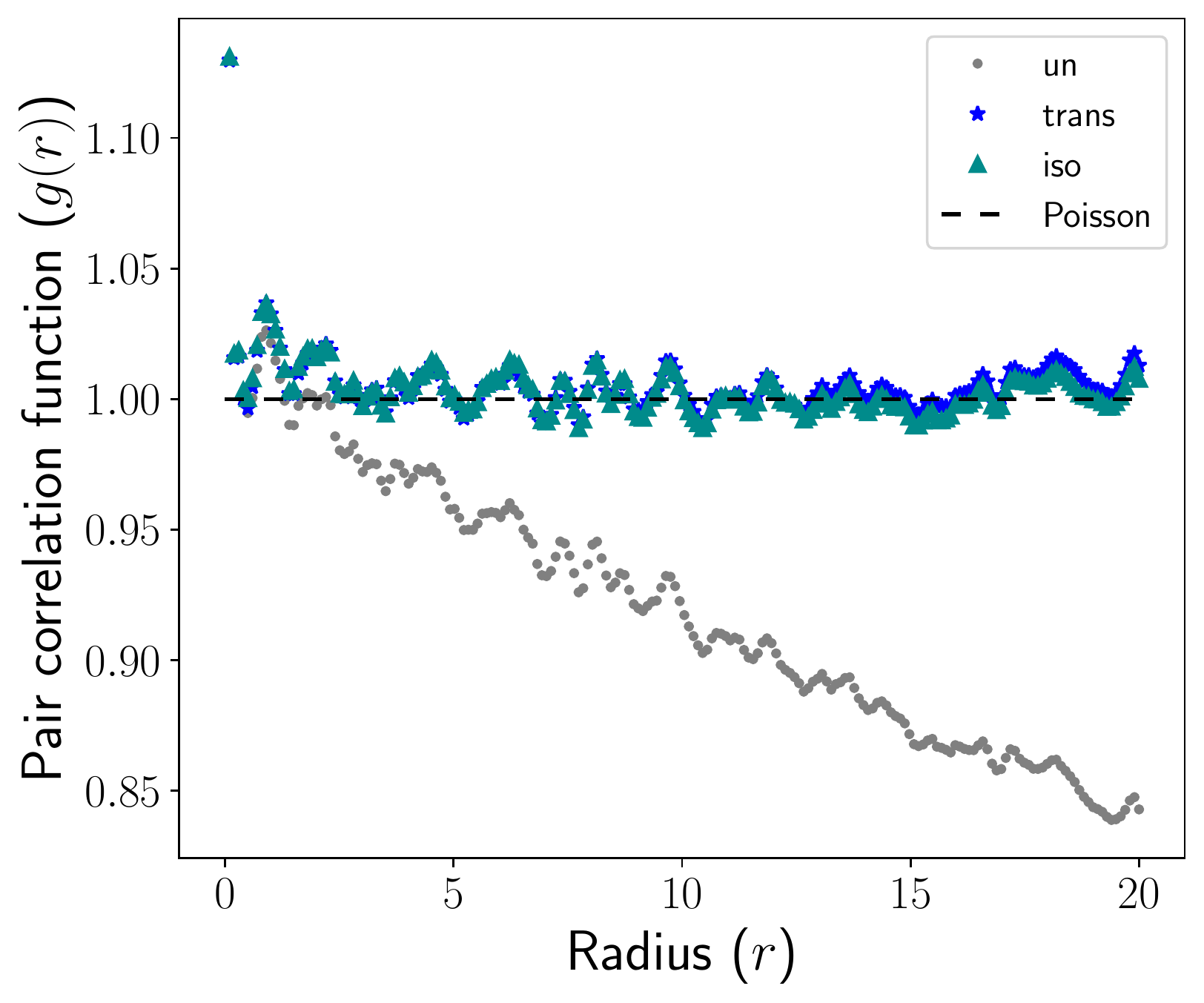}} &
    \raisebox{-\height}{\includegraphics[width=1\linewidth]{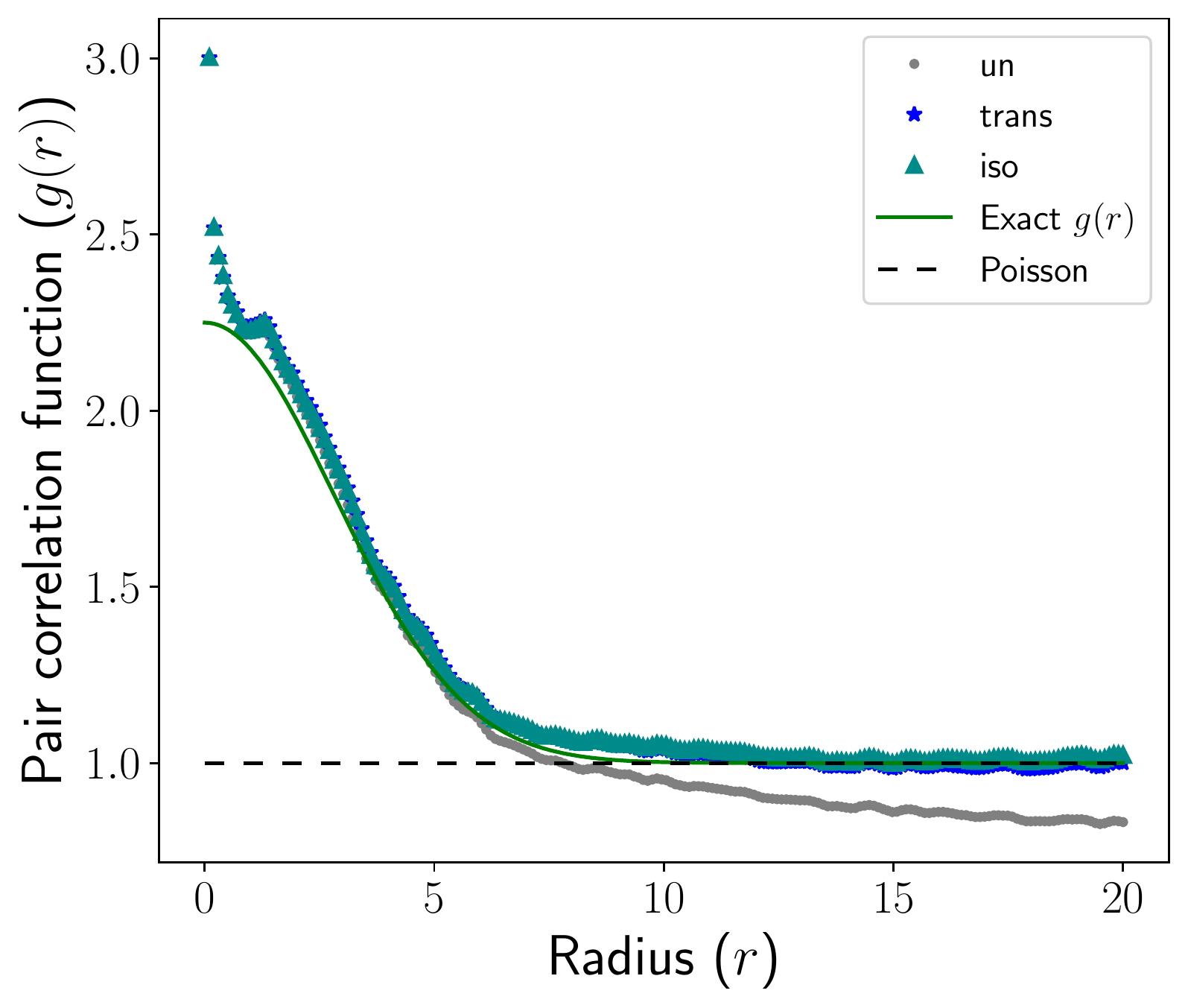}}
  \end{tabular}
  \vspace{-0.2cm}
  \begin{tabular}{p{\dimexpr 0.03\textwidth-\tabcolsep}p{\dimexpr 0.22\textwidth-\tabcolsep}p{\dimexpr 0.22\textwidth-\tabcolsep}p{\dimexpr 0.22\textwidth-\tabcolsep}p{\dimexpr 0.22\textwidth-\tabcolsep}}
    \multirow{9}{*}{\rotatebox[origin=l]{90}{\texttt{pcf.fv}}}                              &
    \raisebox{-\height}{\includegraphics[width=1\linewidth]{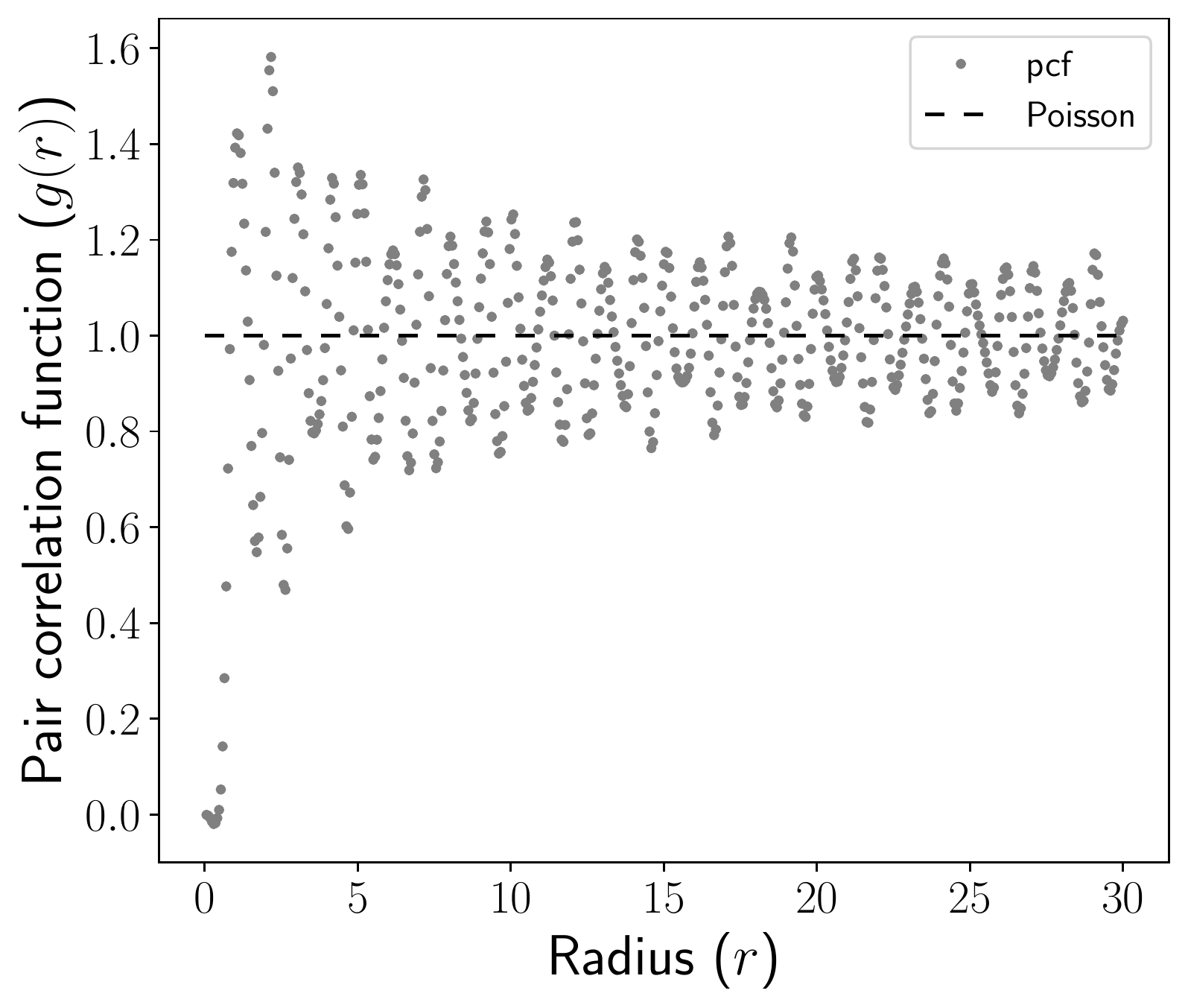}}     &
    \raisebox{-\height}{\includegraphics[width=1\linewidth]{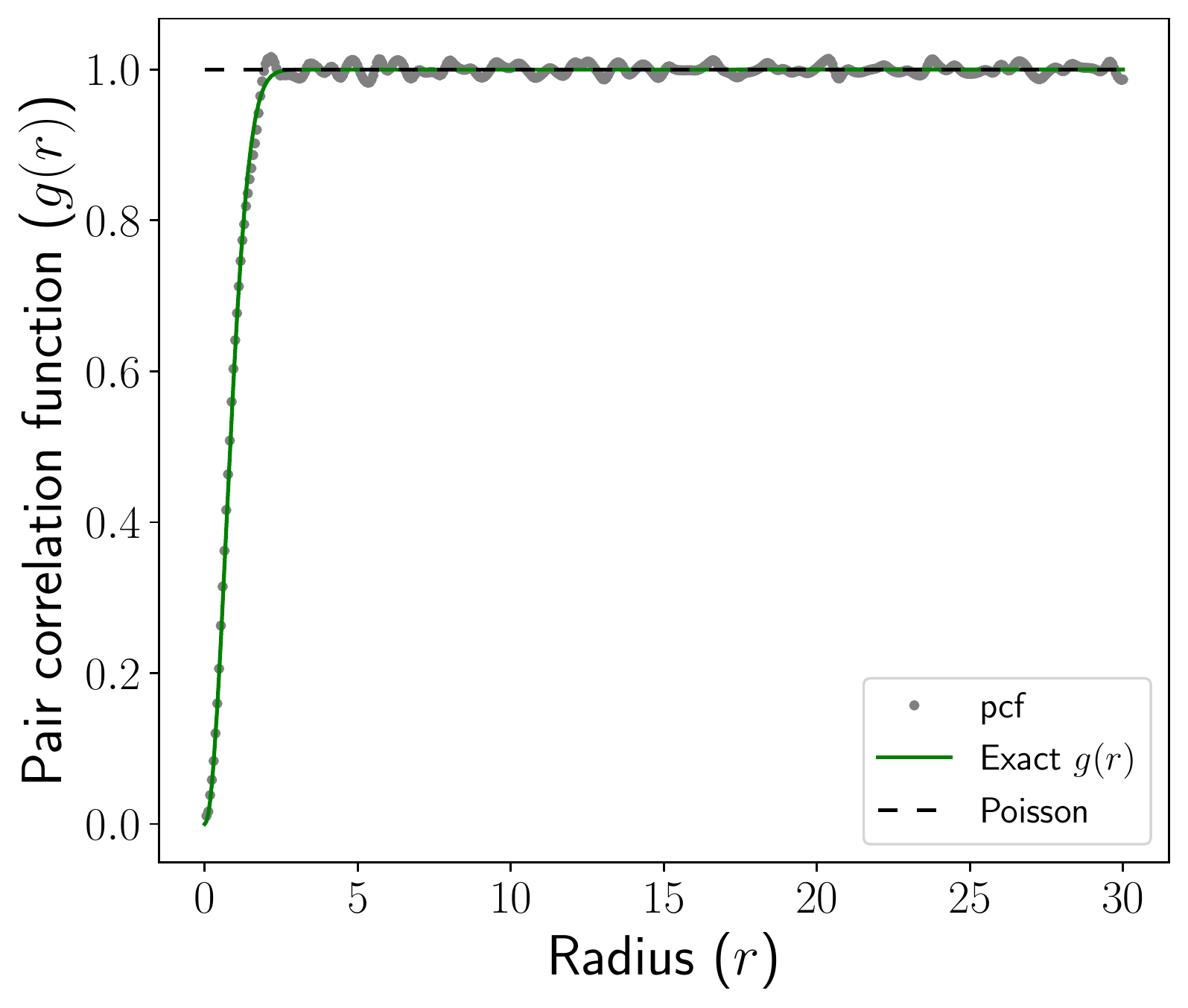}} &
    \raisebox{-\height}{\includegraphics[width=1\linewidth]{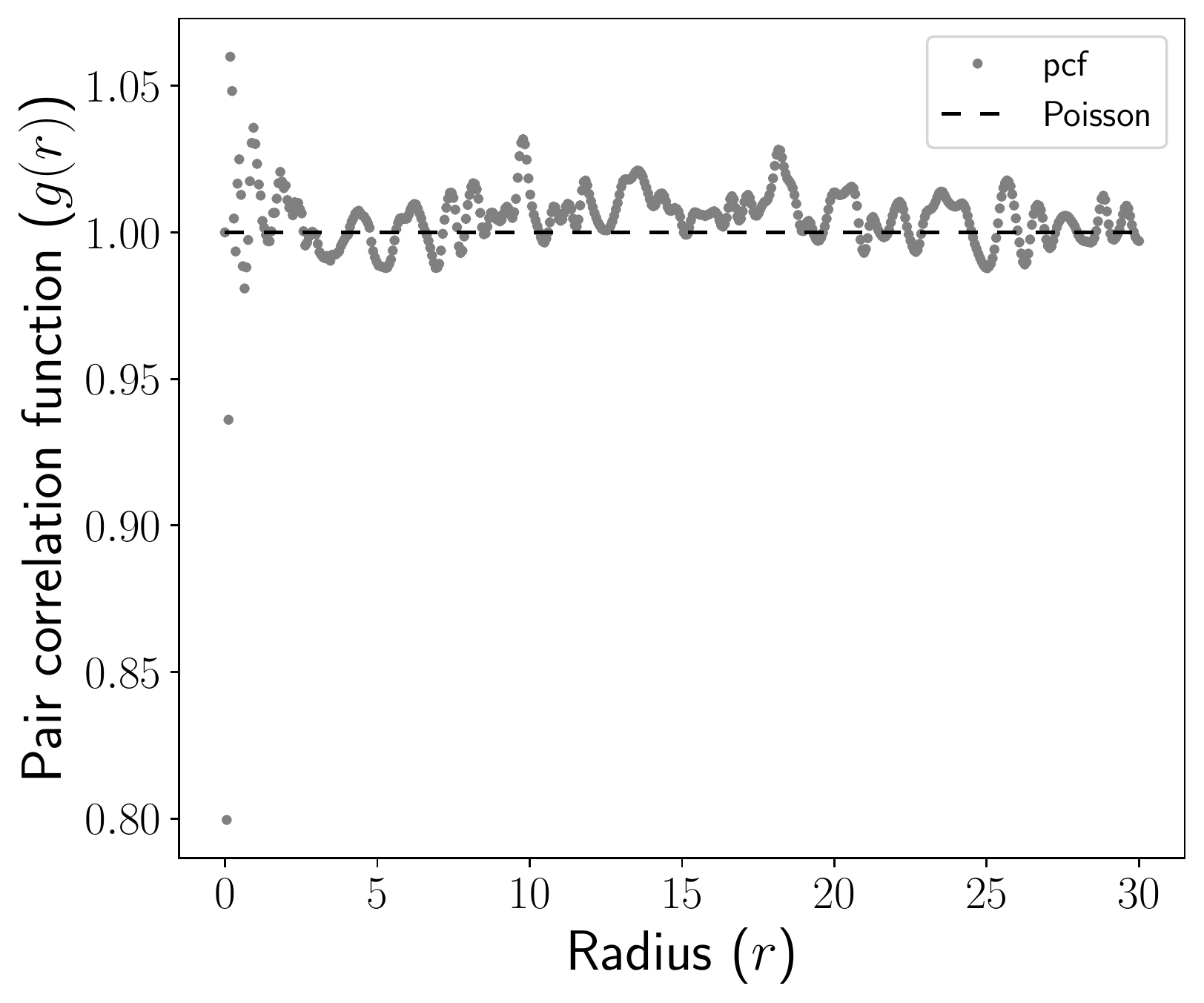}} &
    \raisebox{-\height}{\includegraphics[width=1\linewidth]{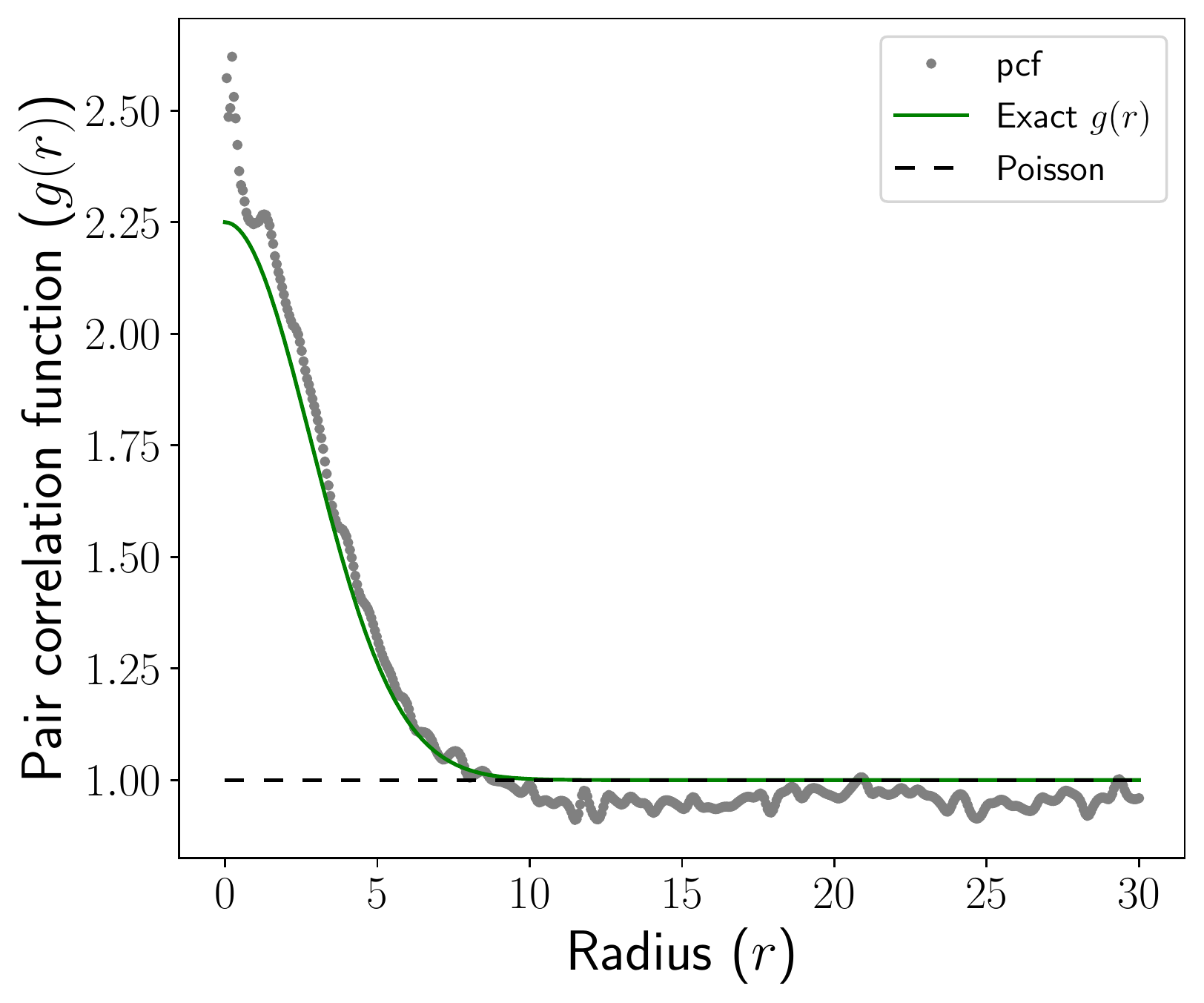}}    \\
    \caption*{}                                                                             &
    \vspace*{-0.5cm}
    \caption*{{\fontfamily{pcr}\selectfont } KLY }                                  &
    \vspace*{-0.5cm}
    \caption*{{\fontfamily{pcr}\selectfont } Ginibre }                             &
    \vspace*{-0.5cm}
    \caption*{{\fontfamily{pcr}\selectfont } Poisson }                              &
    \vspace*{-0.5cm}
    \caption*{{\fontfamily{pcr}\selectfont } Thomas }
  \end{tabular}
  \vspace{-0.5cm}
  \caption{Approximated pair correlation function: KLY process (first column), Ginibre ensemble (second column), Poisson process (third column), and Thomas process (last column). The computation and visualization are done using \toolbox{}} \label{fig:pcf}
\end{figure*}

The last two estimators of the structure factor are $\widehat{S}_{\mathrm{HO}}$ in \eqref{eq:s_ho} and $\widehat{S}_{\mathrm{HBC}}$ in \eqref{eq:s_hbc}.
These estimators require an approximation of the pair correlation function (pcf) of the point process.
We thus quickly investigate standard estimators of the pcf on our benchmark point processes.

There are two types of estimators of the pcf for stationary isotropic point processes \citep{Baddeley+Rubak+Turner:2013}: kernel density estimators applied to pairwise distances and numerical derivatives of Ripley's $K$ function.
The \texttt{R} library \href{spatstat}{https://spatstat.org/} implements both, respectively as \href{pcf.ppp}{https://www.rdocumentation.org/packages/spatstat.core/versions/2.1-2/topics/pcf.ppp}, which uses an Epanechnikov kernel and Stoyan's rule of thumb for bandwidth selection \citep[Section 7.6.2]{Baddeley+Rubak+Turner:2013}, and \href{pcf.fv}{https://www.rdocumentation.org/packages/spatstat.core/versions/2.1-2/topics/pcf.fv}, which computes the derivative of a polynomial estimator of Ripley's $K$ function.
The kernel density estimator behaves badly for small values of $r$: for many point processes, its variance becomes infinite when $r$ goes to $0$.
The derivative estimator is recommended for large datasets, where direct estimation of the pcf can be time-consuming \citep[Section 7.6.2]{Baddeley+Rubak+Turner:2013}.
Figure~\ref{fig:pcf} shows the two estimators of the pair correlation function of the benchmark point processes.
Note that we provide an independent, open-source Python interface\footnote{At \url{https://github.com/For-a-few-DPPs-more/spatstat-interface} and on PyPI.} to the $\texttt{R}$ library \texttt{spatstat}.
The second row shows the estimation of the pair correlation function using \texttt{pcf.ppp}.
This method provides a choice of boundary corrections, like \texttt{"trans"}, \texttt{"iso"}, or none (\texttt{"un"}).
For more details see \citet[Sections 7.4.4 and 7.4.5]{Baddeley+Rubak+Turner:2013}.
The last row of Figure~\ref{fig:pcf} shows the estimation of the pair correlation function using \texttt{pcf.fv}.

We observe that, for the cardinalities considered here, the choice of edge correction method is irrelevant, as long as there is one.
As expected, the uncorrected version \texttt{"un"}  underestimates the pcf as $r$ increases.
This results from counting only the pairs of points that fall inside the observation window, without correcting for border effects.
We also observe that the two methods for estimating the pcf perform similarly, and we pick \texttt{pcf.fv} for the rest of this section.
We manually remove undefined values (\texttt{NaN}, \texttt{-Inf}, or \texttt{Inf}), and we interpolate the obtained discrete approximation of $g$, in order to evaluate it at any point required by the quadratures of Section~\ref{ssub:Estimating the structure factor using Ogata quadrature}, and \ref{ssub:Estimating the structure factor using the discrete Hankel transform (DHT)}.
Finally, note that the maximum radius $r_\text{max}$ at which \texttt{spatstat} provides an approximation of $g$ is limited by the size of the observation window.
Typically, it should be less than half the window diameter for a ball window,
and less than $1/4$ of the smaller side length of the window for a rectangular window; see the \href{documentation}{https://www.rdocumentation.org/packages/spatstat.core/versions/2.3-1/topics/Kest} of \texttt{spatstat}.
For larger values than the $r_{\max}$ provided by \texttt{spatstat}, we manually set $g$ to be identically 1, which has the effect of automatically truncating quadratures that evaluate $g-1$, like Ogata's quadrature \eqref{eq:s_ho}.

\paragraph{Hankel transform quadratures} % (fold)
\label{ssub:Hankel transform quadratures}
\begin{figure*}[!ht]
  \begin{tabular}{p{\dimexpr 0.03\textwidth-\tabcolsep}p{\dimexpr 0.22\textwidth-\tabcolsep}p{\dimexpr 0.22\textwidth-\tabcolsep}p{\dimexpr 0.22\textwidth-\tabcolsep}p{\dimexpr 0.22\textwidth-\tabcolsep}}
    \multirow{9}{*}{\rotatebox[origin=c]{90}{Point process}}                                   &
    \raisebox{-\height}{\includegraphics[width=0.9\linewidth]{kly_pp_ball.pdf}}     &
    \raisebox{-\height}{\includegraphics[width=0.9\linewidth]{ginibre_pp_ball.pdf}} &
    \raisebox{-\height}{\includegraphics[width=0.9\linewidth]{poisson_pp_ball.pdf}} &
    \raisebox{-\height}{\includegraphics[width=0.9\linewidth]{thomas_pp_ball.pdf}}
  \end{tabular}
  \vspace{-0.2cm}
  \begin{tabular}{p{\dimexpr 0.03\textwidth-\tabcolsep}p{\dimexpr 0.22\textwidth-\tabcolsep}p{\dimexpr 0.22\textwidth-\tabcolsep}p{\dimexpr 0.22\textwidth-\tabcolsep}p{\dimexpr 0.22\textwidth-\tabcolsep}}
    \multirow{8}{*}{\rotatebox[origin=l]{90}{$\widehat{S}_{\mathrm{HO}}(k)$}}             &
    \vspace{-0.5cm}
    \raisebox{-\height}{\includegraphics[width=1\linewidth]{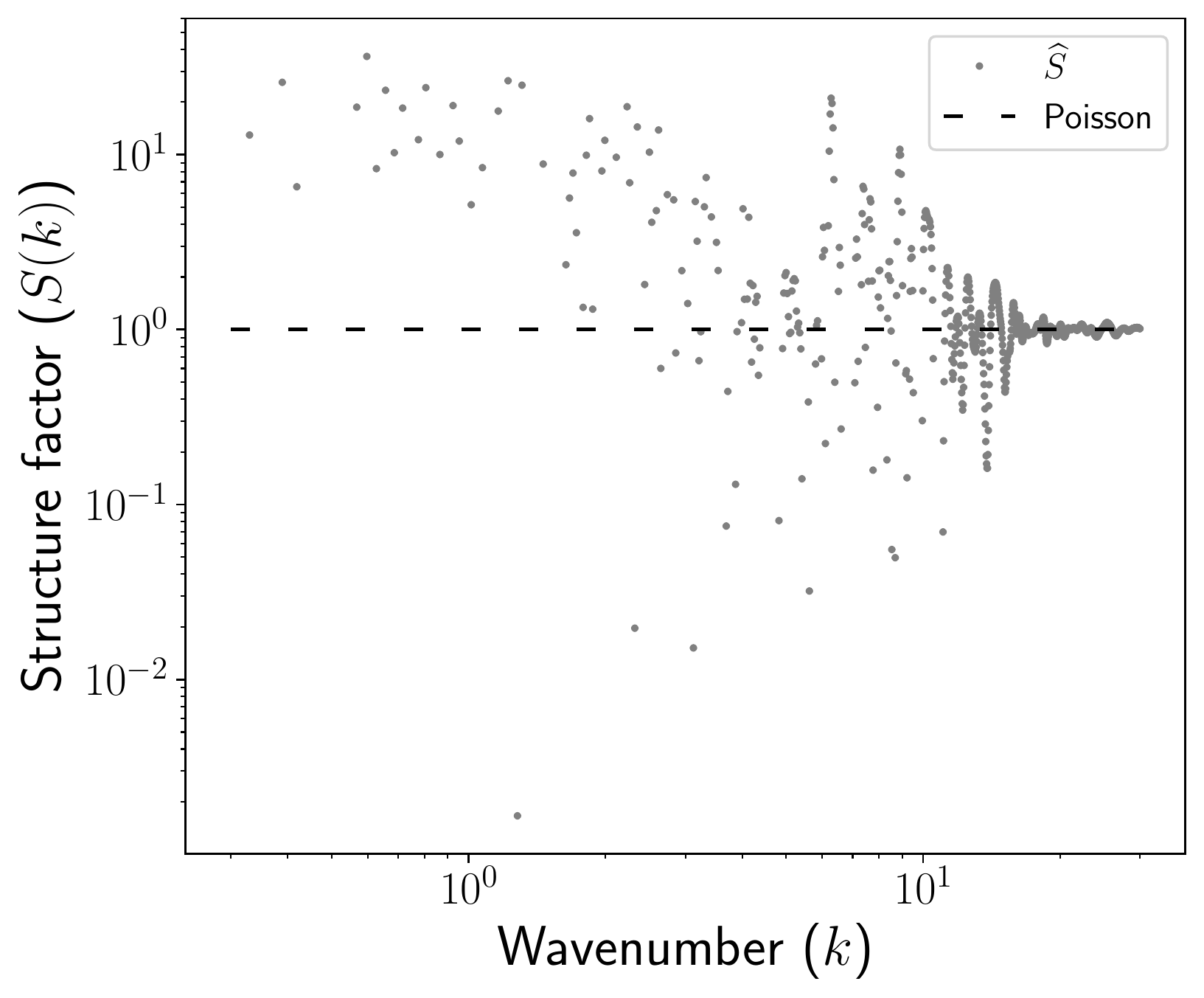}}     &
    \vspace{-0.5cm}
    \raisebox{-\height}{\includegraphics[width=1\linewidth]{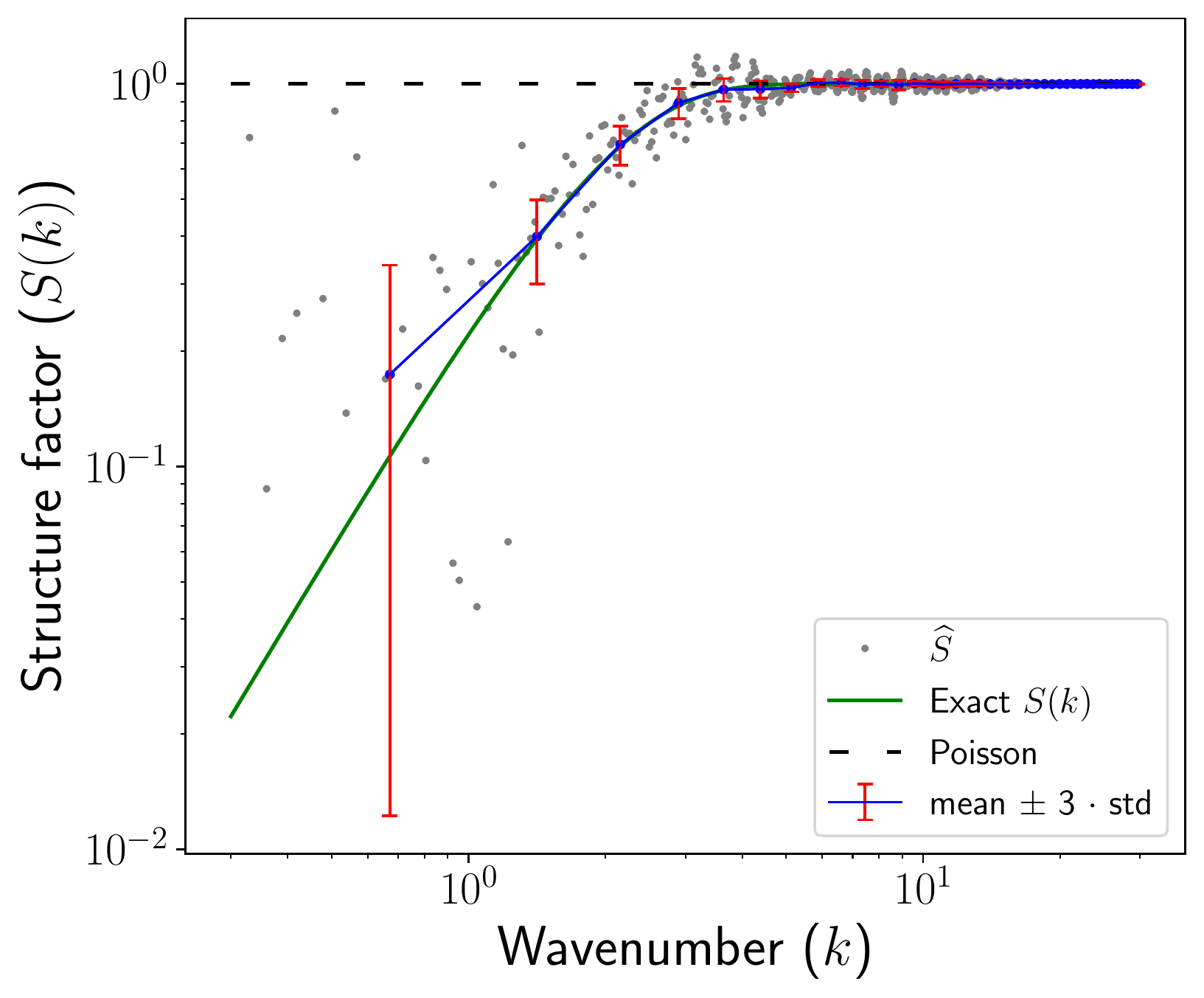}} &
    \vspace{-0.5cm}
    \raisebox{-\height}{\includegraphics[width=1\linewidth]{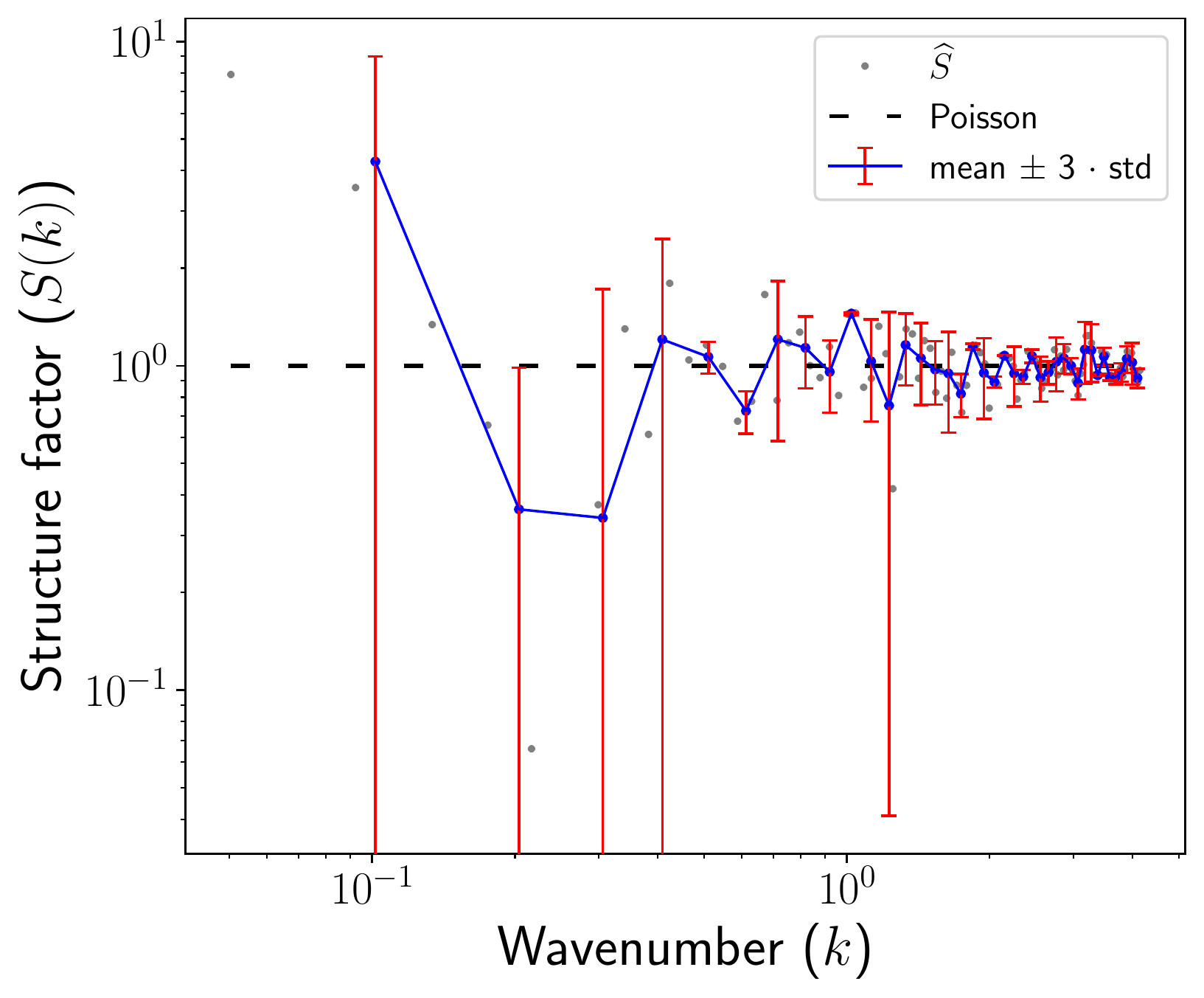}} &
    \vspace{-0.5cm}
    \raisebox{-\height}{\includegraphics[width=1\linewidth]{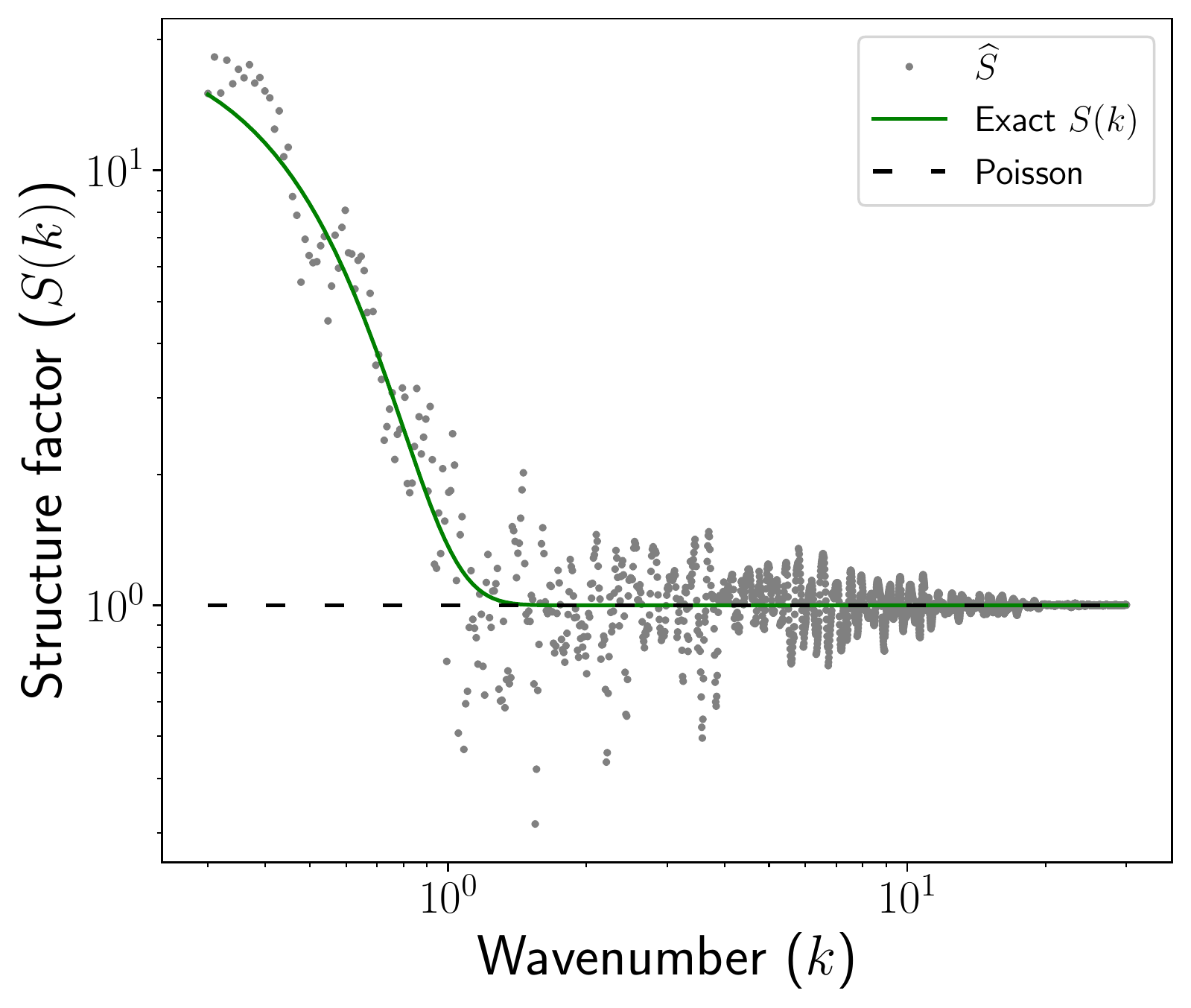}}
  \end{tabular}
  \vspace{-0.2cm}
  \begin{tabular}{p{\dimexpr 0.03\textwidth-\tabcolsep}p{\dimexpr 0.22\textwidth-\tabcolsep}p{\dimexpr 0.22\textwidth-\tabcolsep}p{\dimexpr 0.22\textwidth-\tabcolsep}p{\dimexpr 0.22\textwidth-\tabcolsep}}
    \multirow{9}{*}{\rotatebox[origin=l]{90}{$\widehat{S}_{\mathrm{HBC}}(k)$}}             &
    \raisebox{-\height}{\includegraphics[width=1\linewidth]{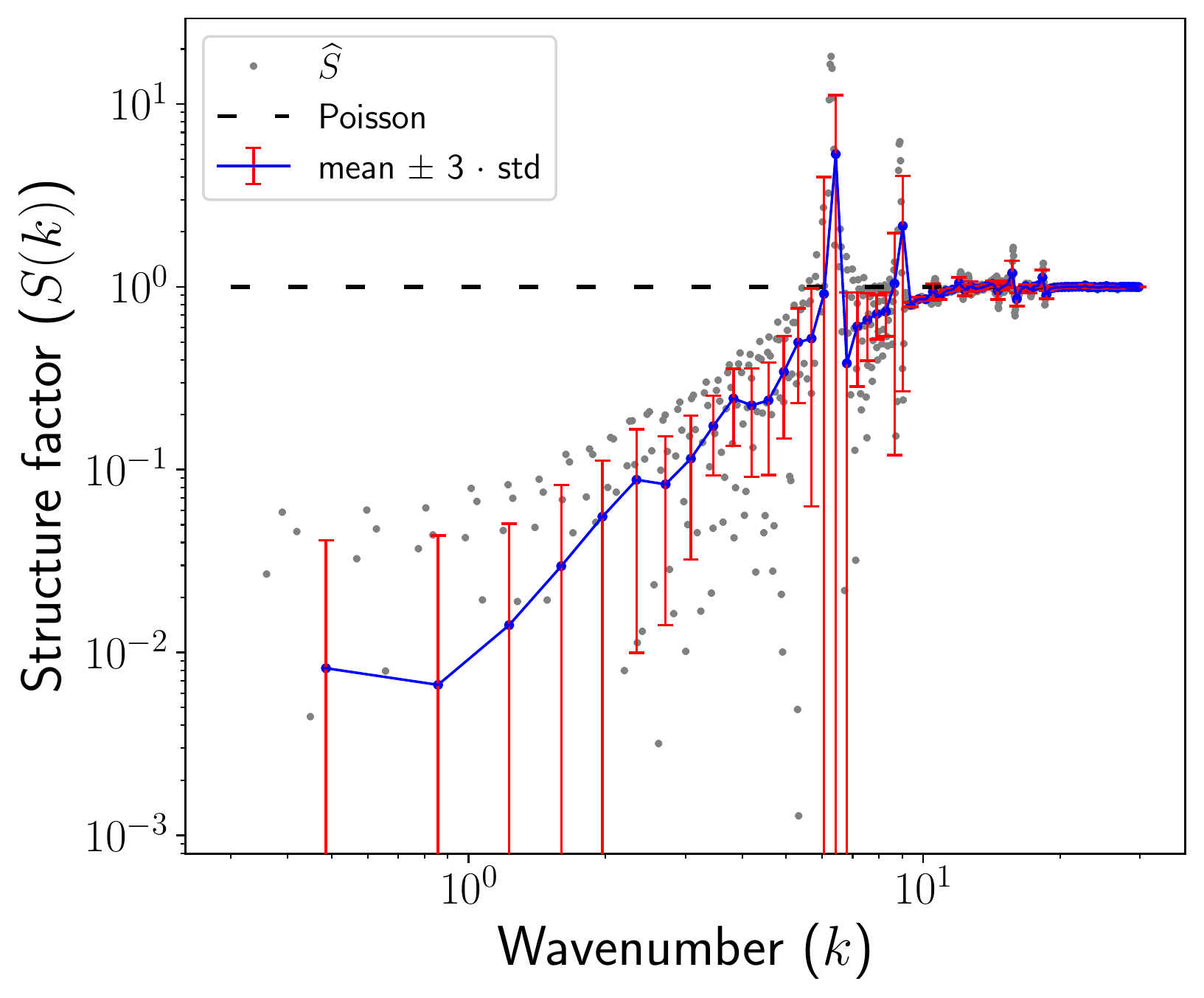}}     &
    \raisebox{-\height}{\includegraphics[width=1\linewidth]{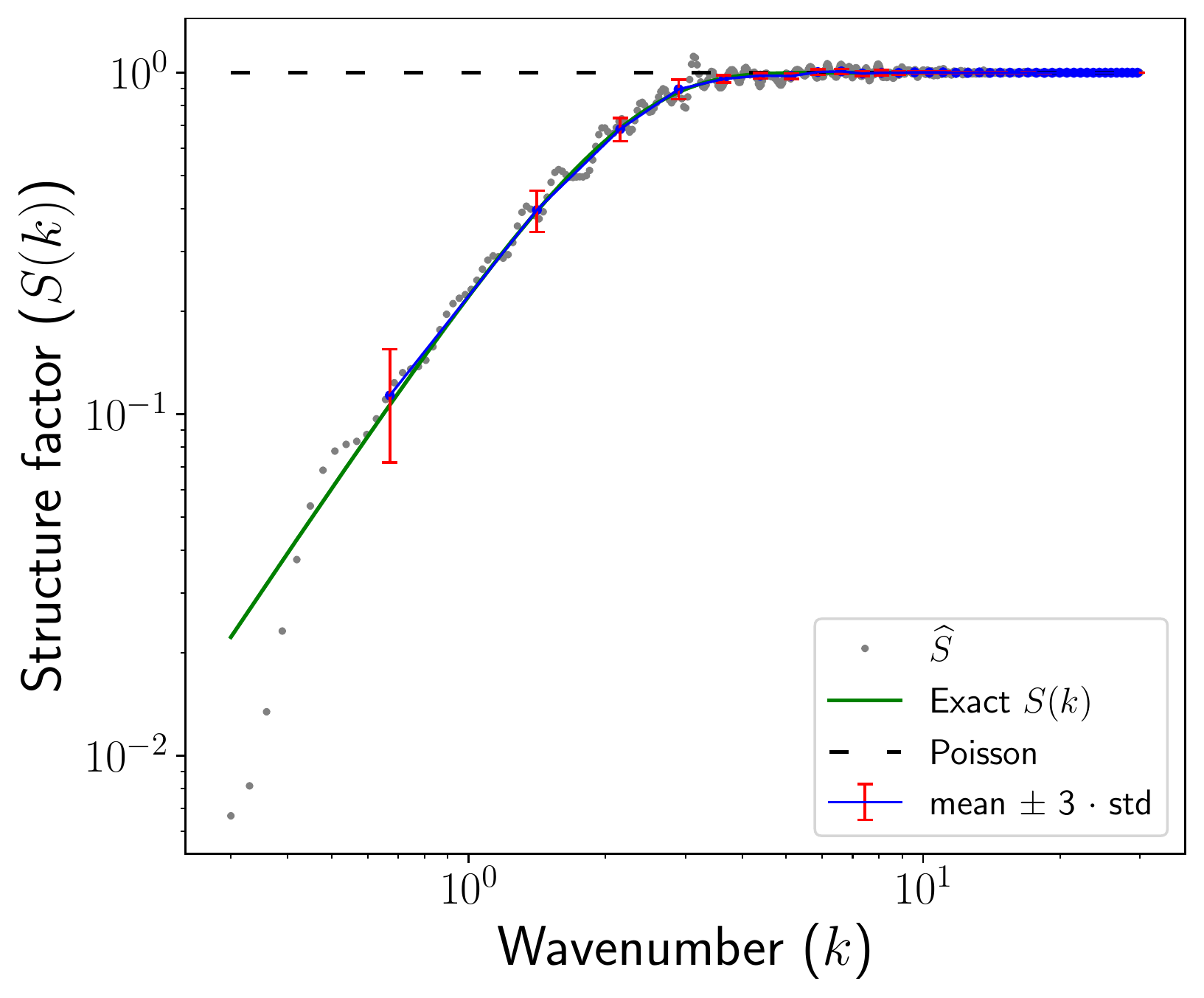}} &
    \raisebox{-\height}{\includegraphics[width=1\linewidth]{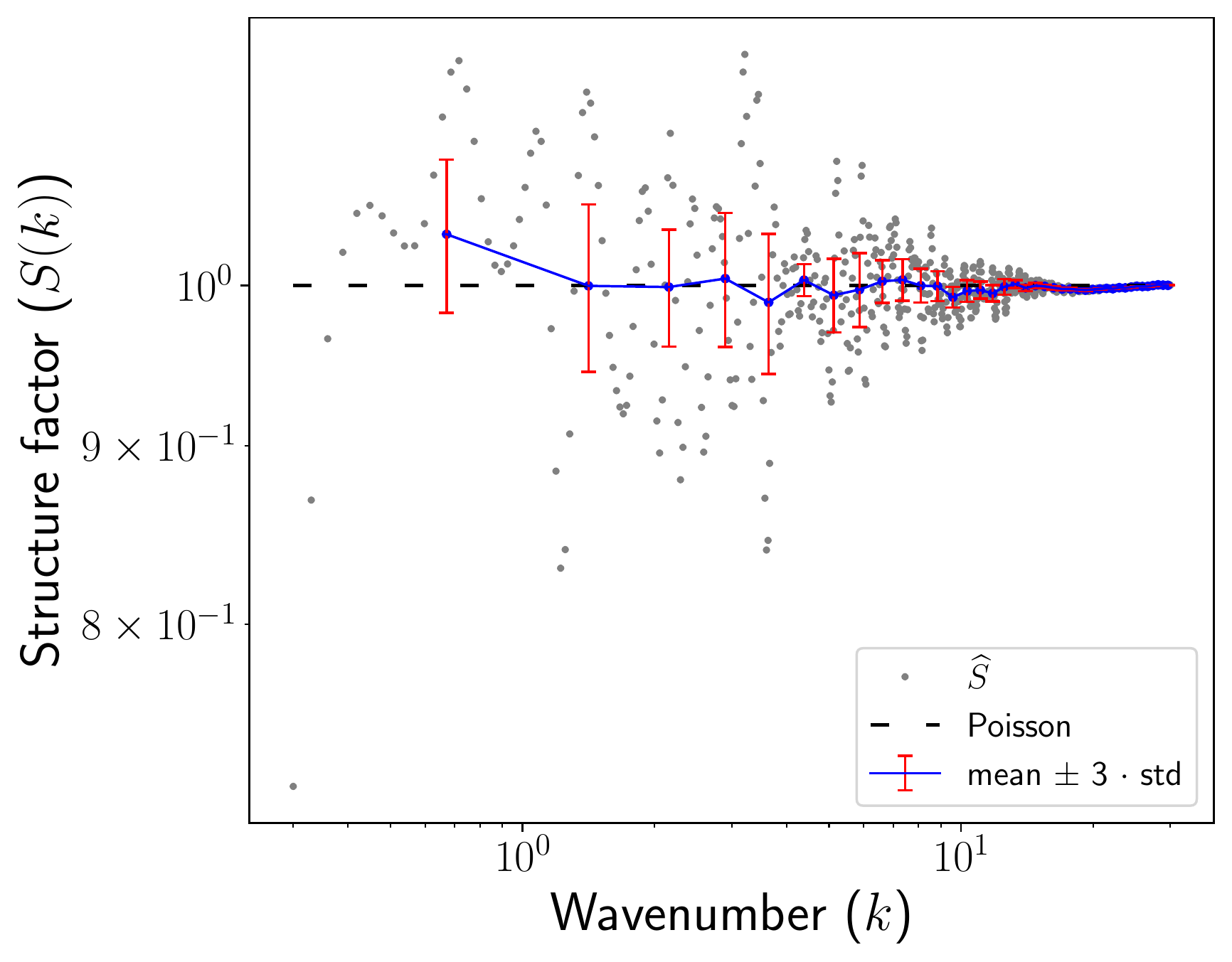}} &
    \raisebox{-\height}{\includegraphics[width=1\linewidth]{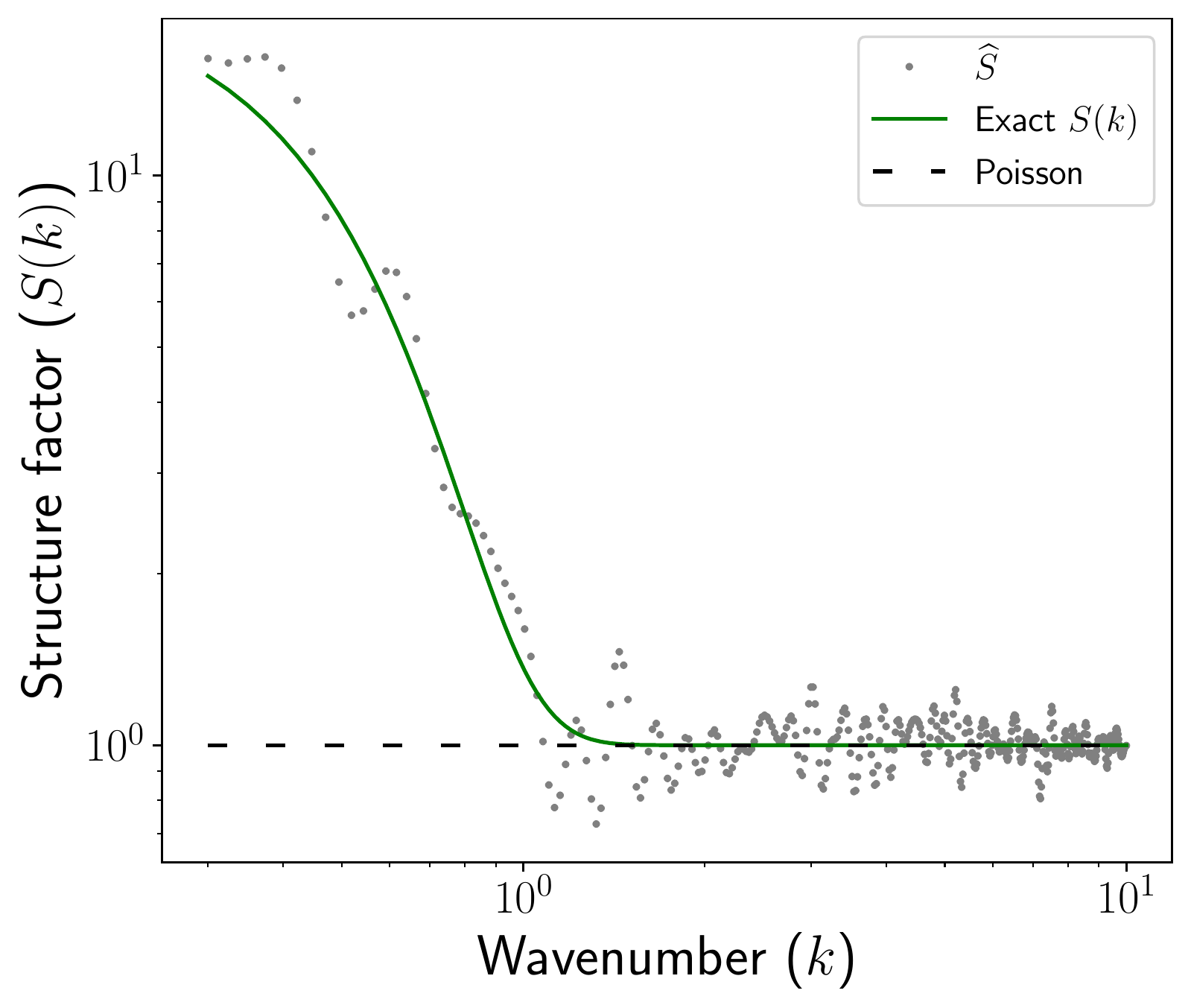}}    \\
    \caption*{}                                                                            &
    \vspace{-0.5cm}
    \caption*{{\fontfamily{pcr}\selectfont } KLY }                                 &
    \vspace{-0.5cm}
    \caption*{{\fontfamily{pcr}\selectfont } Ginibre}                            &
    \vspace{-0.5cm}
    \caption*{{\fontfamily{pcr}\selectfont } Poisson }                             &
    \vspace{-0.5cm}
    \caption*{{\fontfamily{pcr}\selectfont } Thomas }
  \end{tabular}
  \vspace{-0.5cm}
  \caption{Estimation using Hankel transform quadratures: KLY process (first column), Ginibre ensemble (second column), Poisson process (third column), and Thomas process (last column). The computation and visualization are done using \toolbox{}} \label{fig:s_h}
\end{figure*}

Figure~\ref{fig:s_h} shows the results of Ogata's $\widehat{S}_{\mathrm{HO}}$ (second row) and Baddour-Chouinard's $\widehat{S}_{\mathrm{HBC}}$ (last row) on our four benchmark point processes from Section~\ref{sub:point_processes}.
The legend is the same as for Figure~\ref{fig:scattering_intensity}; see Section~\ref{ssub:test  The scattering intensity}.
For the accuracy of the estimators, we can see that $\widehat{S}_{\mathrm{HO}}$ failed to approximate the structure factor of the KLY process.
Even the results of $\widehat{S}_{\mathrm{HBC}}$ seem to be unreliable.
The non-isotropy of the KLY process may be the reason for the fluctuations of its approximated pair correlation function in Figure~\ref{fig:pcf}, leading to the inaccuracies of the quadratures.
For the remaining point processes, $\widehat{S}_{\mathrm{HBC}}$ seems to give more accurate results than $\widehat{S}_{\mathrm{HO}}$.

\subsection{Hyperuniformity diagnostics} % (fold)
\label{sub:Hyperuniformity tests}
We now demonstrate the estimation of the hyperuniformity index $H$ from Section~\ref{sub:Effective hyperuniformity}, the multiscale hyperuniformity test from Section~\ref{sec:The coupled sum estimator and a test of hyperuniformity}, and the estimation of the decay rate $\alpha$ from Section~\ref{sub:hyperuniformity} using $A=50$ samples from the point processes of Section \ref{sub:point_processes}.

\paragraph{Effective hyperuniformity}

\begin{figure}[!ht]
  \centering
  \begin{subfigure}[b]{0.2\textwidth}
    \centering
    \includegraphics[width=0.8\textwidth]{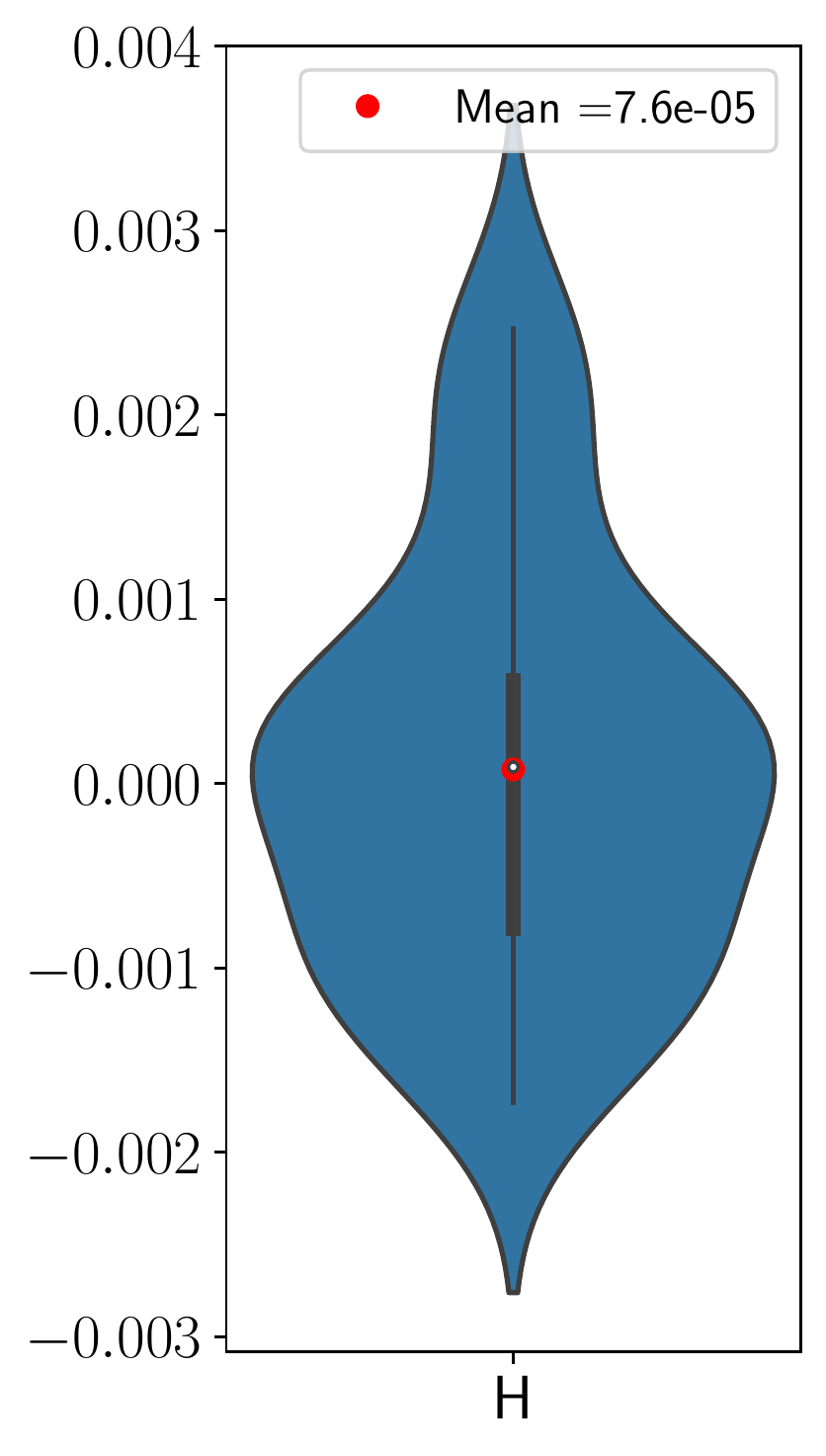}
    \caption{KLY }
    \label{fig:H_index_kly}
  \end{subfigure}
  \hfill
  \begin{subfigure}[b]{0.2\textwidth}
    \centering
    \includegraphics[width=0.8\textwidth]{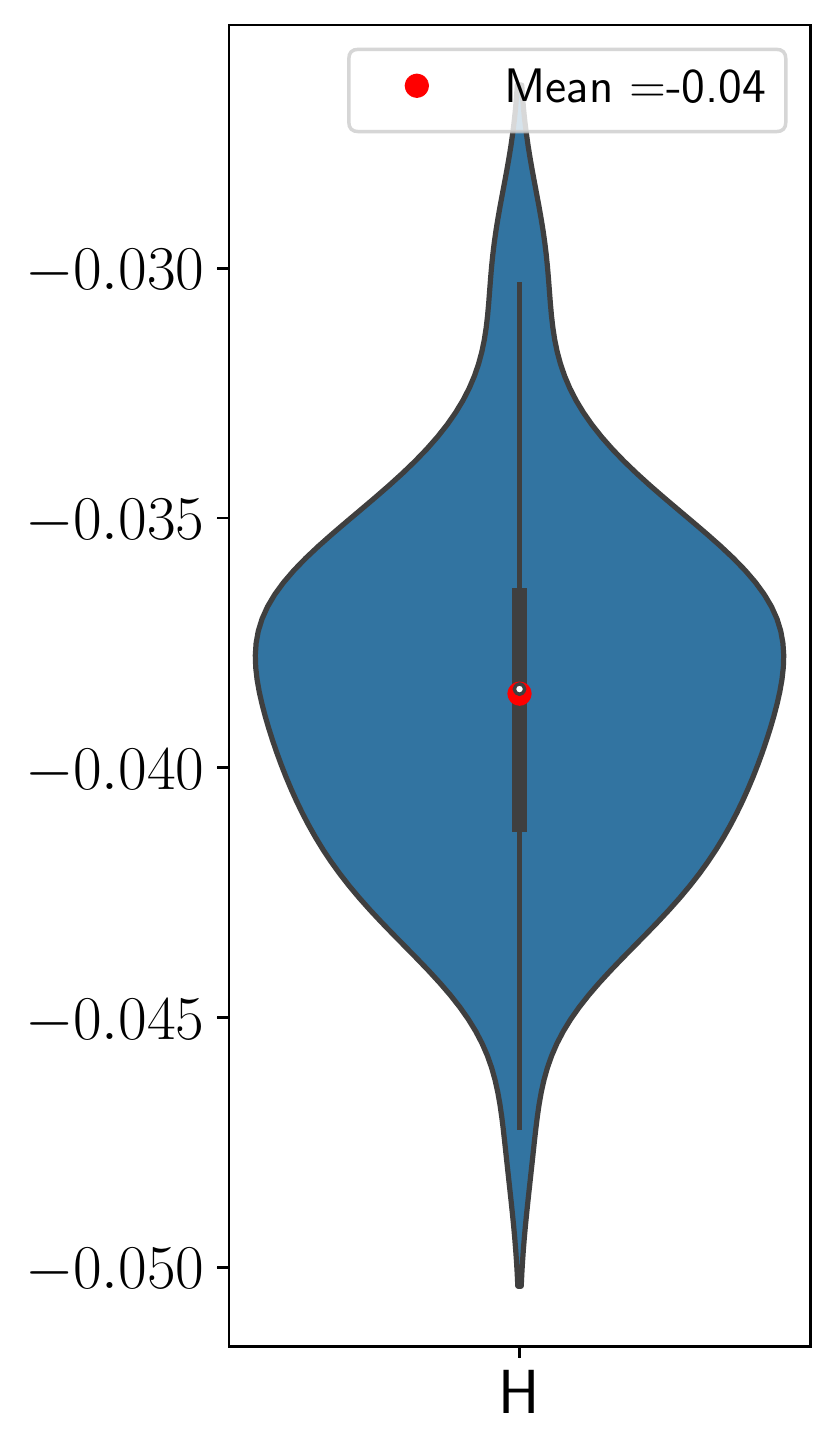}
    \caption{Ginibre }
    \label{fig:H_index_ginibre}
  \end{subfigure}
  \hfill
  \begin{subfigure}[b]{0.2\textwidth}
    \centering
    \includegraphics[width=0.68\textwidth]{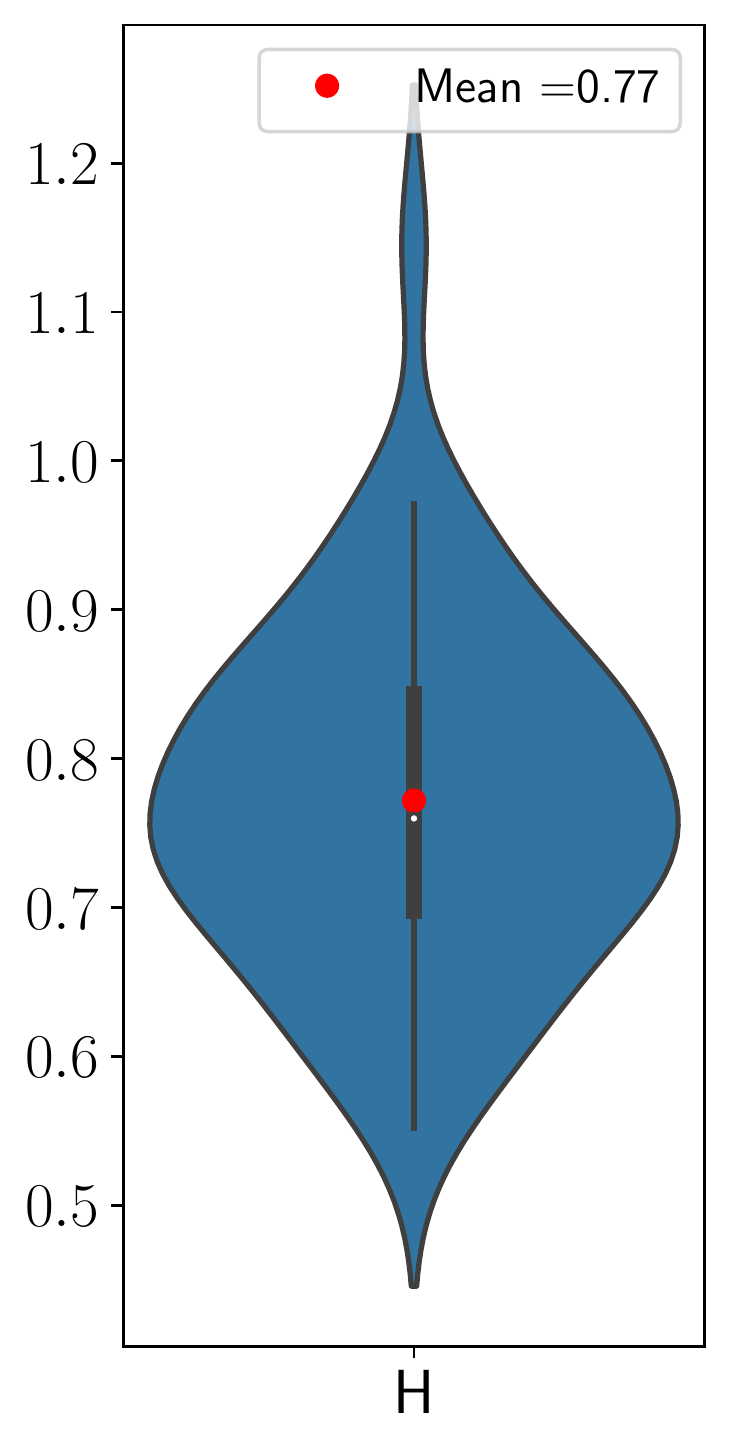}
    \caption{Poisson }
    \label{fig:H_index_poisson}
  \end{subfigure}
  \hfill
  \begin{subfigure}[b]{0.2\textwidth}
    \centering
    \includegraphics[width=0.7 \textwidth]{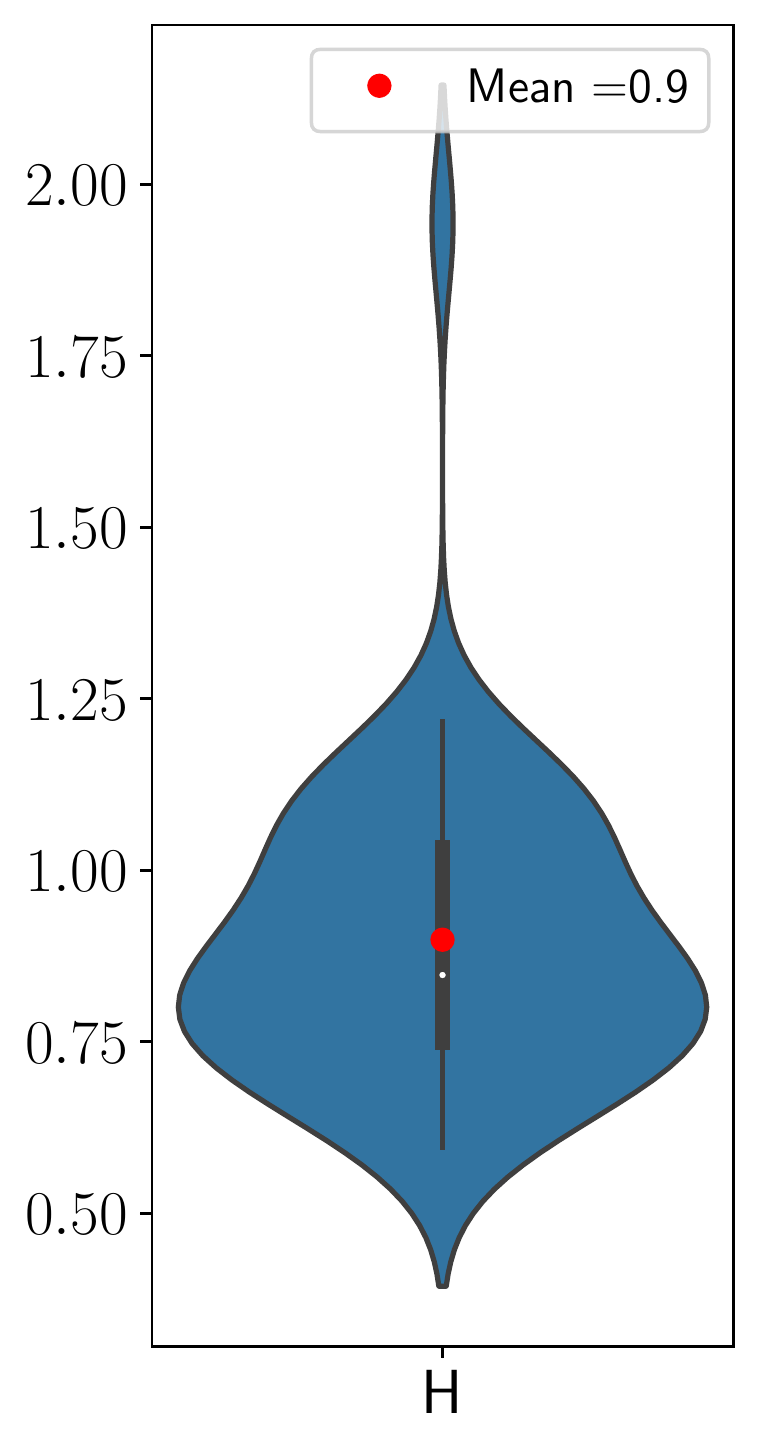}
    \caption{Thomas }
    \label{fig:H_index_thomas}
  \end{subfigure}
  \caption{Violin plots of $H$ across $A=50$ samples from the KLY, Ginibre, Poisson, and Thomas point process. Note the different $y$-scales. The
    computation is done using \toolbox{}.}
  \label{fig:H index}
\end{figure}
Figure~\ref{fig:H index} illustrates the violin plots\footnote{
A violin plot gathers a box plot and a kernel density estimator of the assumed underlying density.
  The former shows the median (white point), the interquartile range (thick black bar in the center), and the rest of the distribution except for points determined as outliers (thin black line in the center).
  We also add the mean (red point).
}
of $H$ across $A=50$ samples from the benchmark point processes.
We used the results of $\widehat{S}_{\mathrm{BI}}$, across $A$ samples of roughly $10^4$ points each (see Section \ref{sub:Approximating the structure factor using an ISE}).
To fit the line required to compute $H$, we considered the wavenumbers up to $0.6$ for the Thomas process and $1$ for the remaining point processes.
These values were chosen manually: the trade-off is to remain close to zero while including enough data points to fit a line.
The violin plots of Figure~\ref{fig:H index} indicate that, consistently across realizations of the Poisson and Thomas point processes, $H$ is larger than, say, $0.5$.
This is a strong hint that these point processes are not hyperuniform.
On the contrary, for Ginibre, $H$ is even slightly negative, hinting at hyperuniformity.
For the KLY process, although $H$ is close to zero, we note that a threshold of $10^{-3}$ would not lead to the same answer across all 50 realizations.
\paragraph{Multiscale hyperuniformity test}
\begin{table}[!ht]
  \centering
   \caption{Multiscale hyperuniformity test}
  \label{tab:Multiscale hyperuniformity test}
  \small
  \begin{tabular}{|l|c|c|c|c|}
    \hline
    \rule{0pt}{15pt}
    {}
     & $\bar{Z}_{A}$
     & $CI[\bbE [Z]]$
     & $\bar{Z}_{A}$
     & $CI[\bbE [Z]]$
    \\
    \hline
    KLY
    & $0.003$
    & $[ -0.003, 0.009]$
    & $ 0.003$
    & $[-0.0003 , 0.007]$
    \\
    \hline
     Ginibre
     & $0.015$
     & $[-0.021,  0.051]$
     & $0.007$
     & $[-0.003, 0.011]$
    \\
    \hline
    Poisson
    & $0.832$
    & $[0.444, 1.220]$
    & $0.781$
    & $[0.560, 1.001]$
    \\
    \hline
    Thomas
    & $0.928$
    & $[ 0.788 , 1.068]$
    & $1$
    & $[0.999 , 1 ]$
     \\
    \hline
    $\widehat{S}$
     & \multicolumn{2}{|c|}{$\widehat{S}_{\mathrm{SI}}$}
     & \multicolumn{2}{|c|}{$\widehat{S}_{\mathrm{BI}}$}
    \\
    \hline
  \end{tabular}%
\end{table}
\begin{figure}[!ht]
  \begin{tabular}{p{0.02\textwidth-\tabcolsep}p{0.2\textwidth}p{0.2\textwidth}p{0.2\textwidth}p{0.2\textwidth}}
    \multirow{8}{*}{\rotatebox[origin=c]{90}{  $\widehat{S}_{\mathrm{SI}}$}}                                  &
    \raisebox{-\height}{\includegraphics[width=0.75\linewidth]{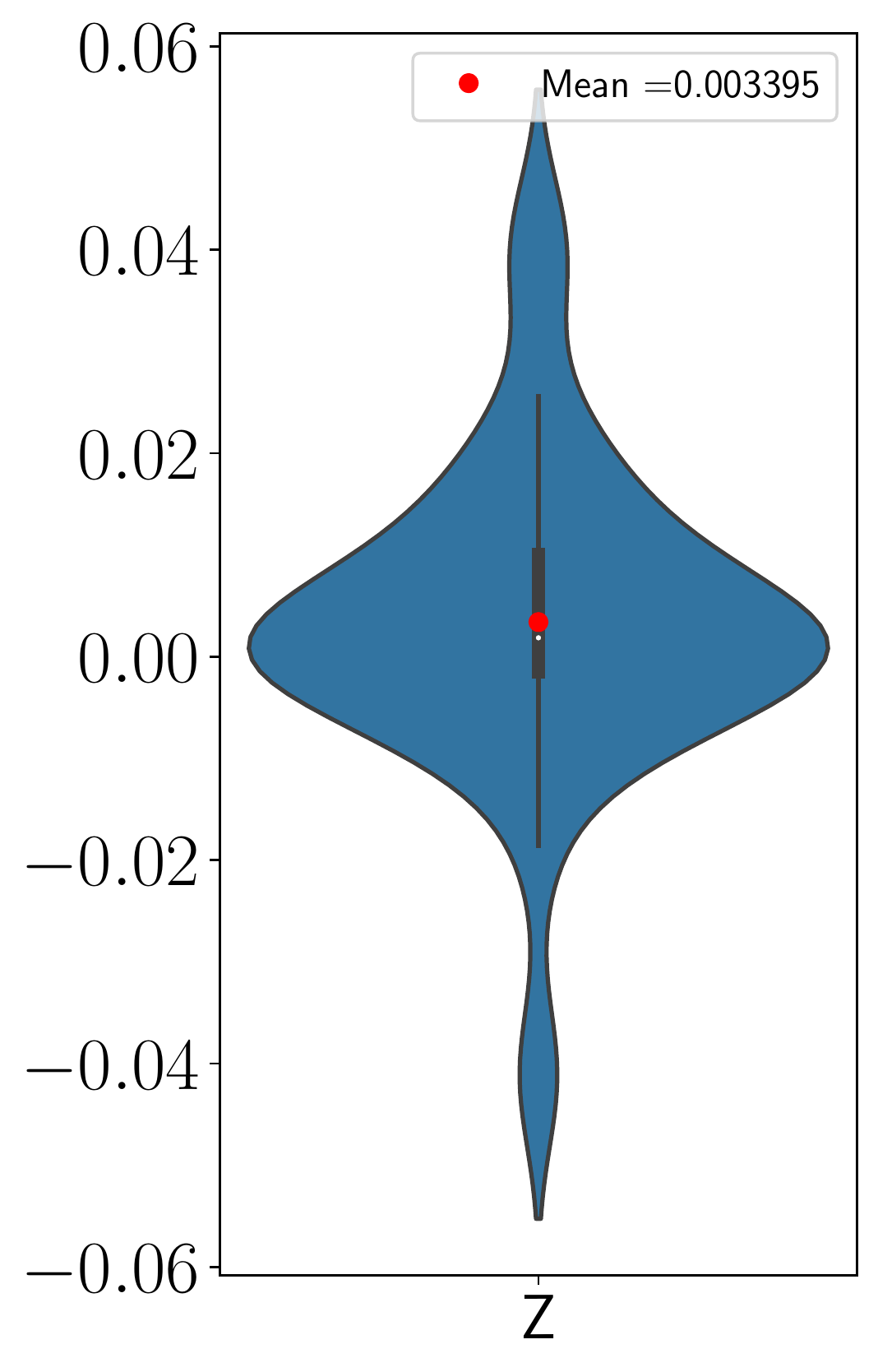}} &
    \raisebox{-\height}{\includegraphics[width=0.72\linewidth]{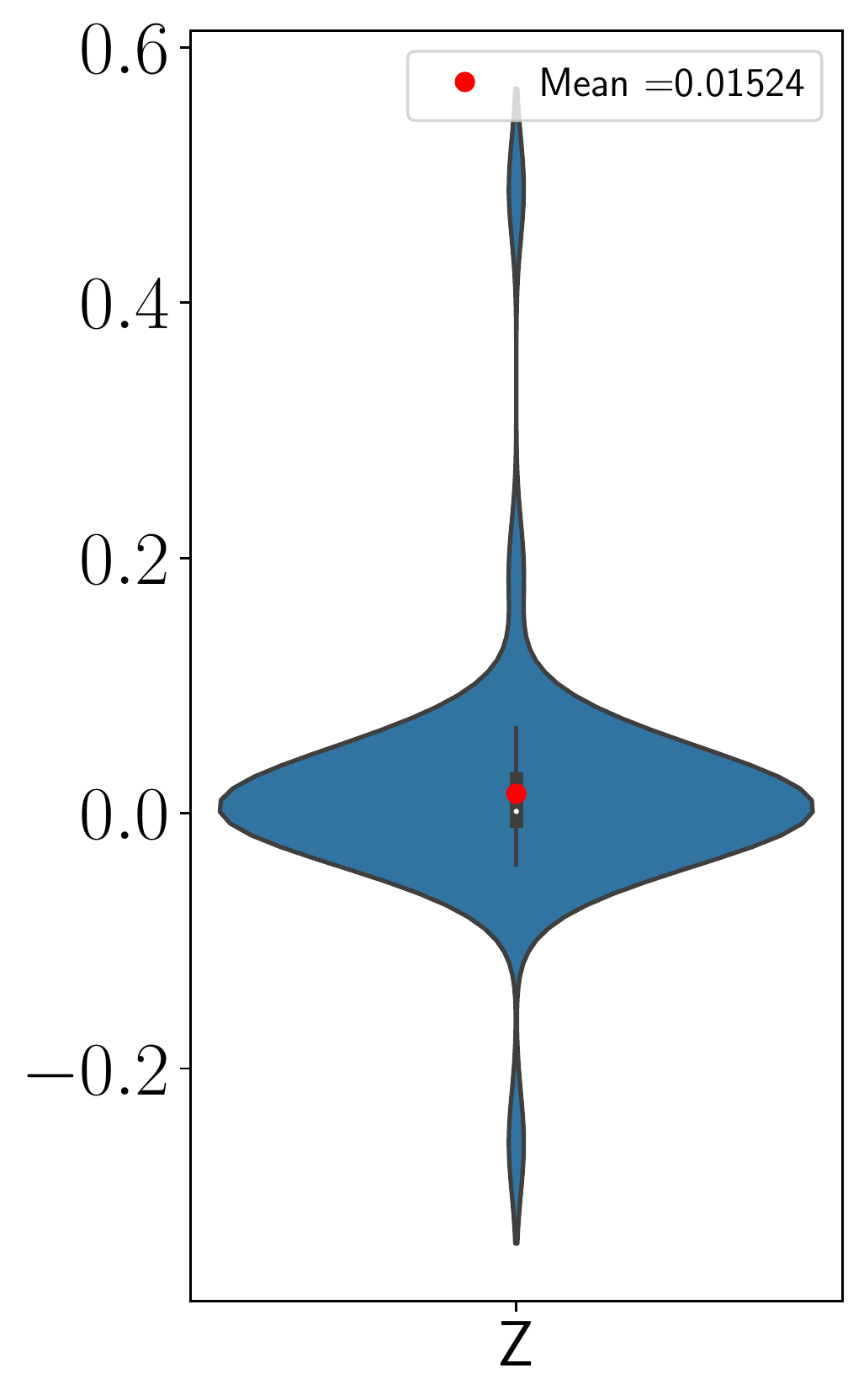}} &
    \raisebox{-\height}{\includegraphics[width=0.67\linewidth]{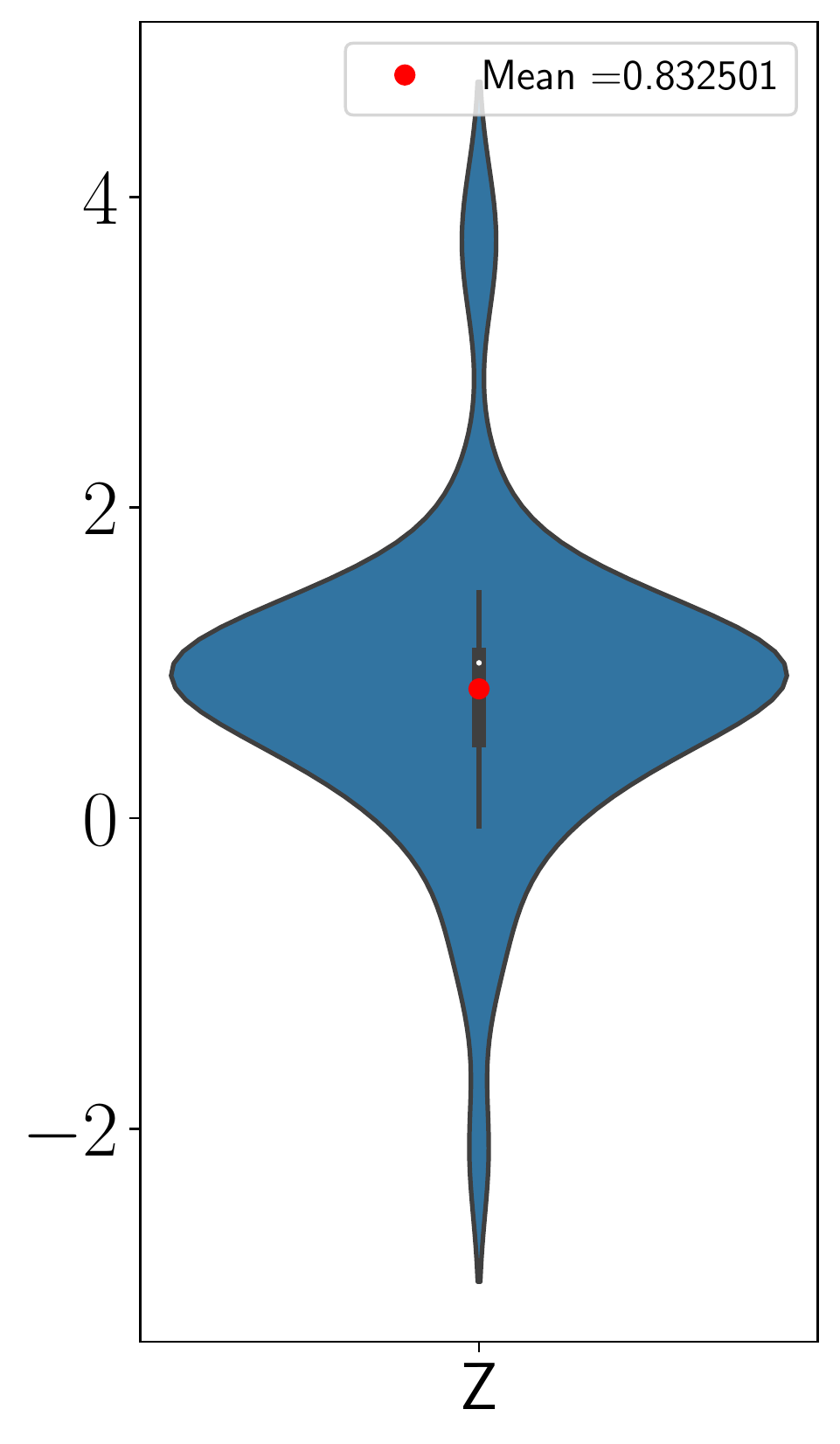}}
    &
    \raisebox{-\height}{\includegraphics[width=0.71\linewidth]{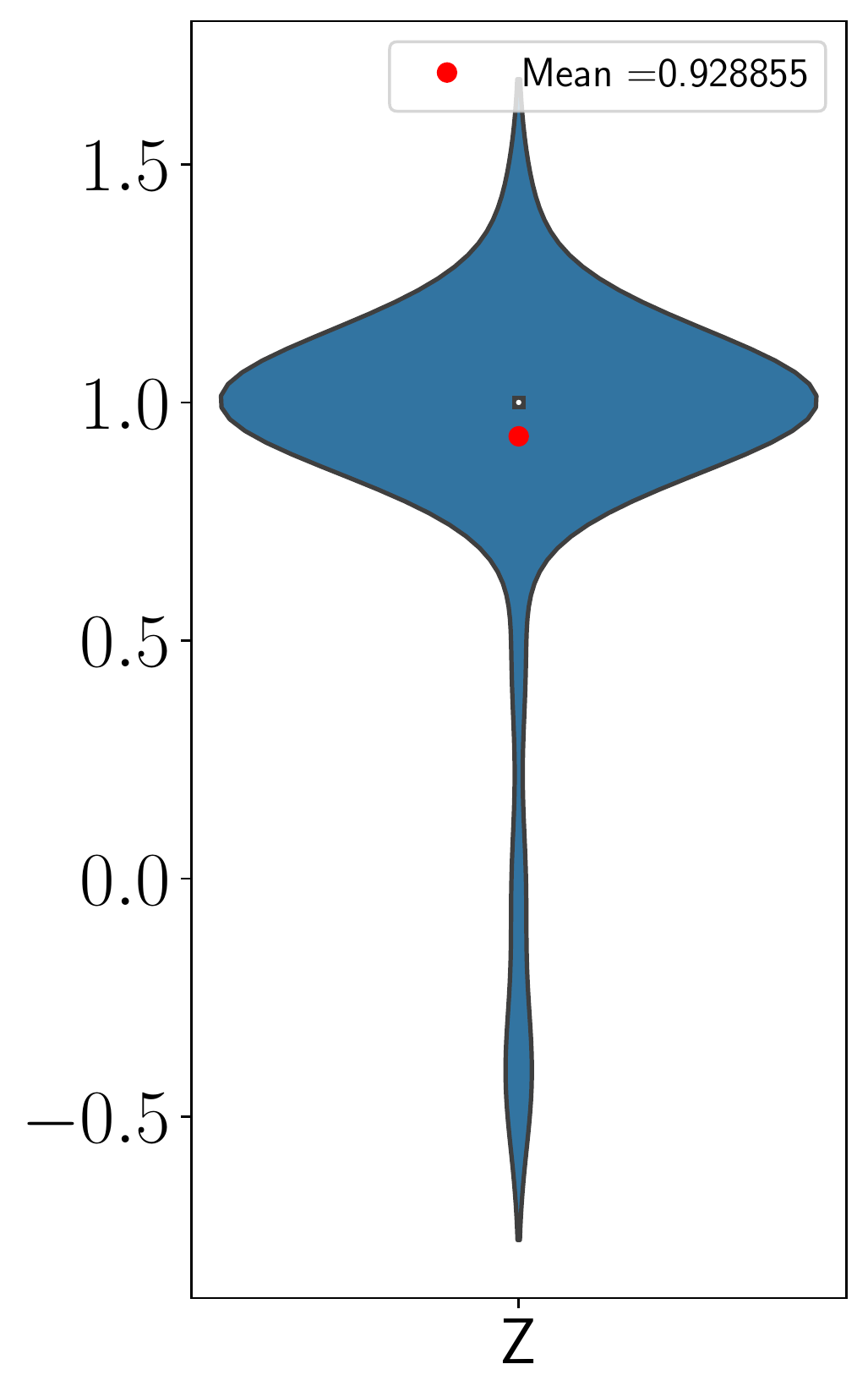}}
  \end{tabular}
  \begin{tabular}{p{0.02\textwidth-\tabcolsep}p{ 0.2\textwidth}p{0.2\textwidth}p{0.2\textwidth}p{0.2\textwidth}}
    \multirow{8}{*}{\rotatebox[origin=c]{90}{$\widehat{S}_{\mathrm{BI}}$}}                                           &
    \raisebox{-\height}{\includegraphics[width=0.7\linewidth]{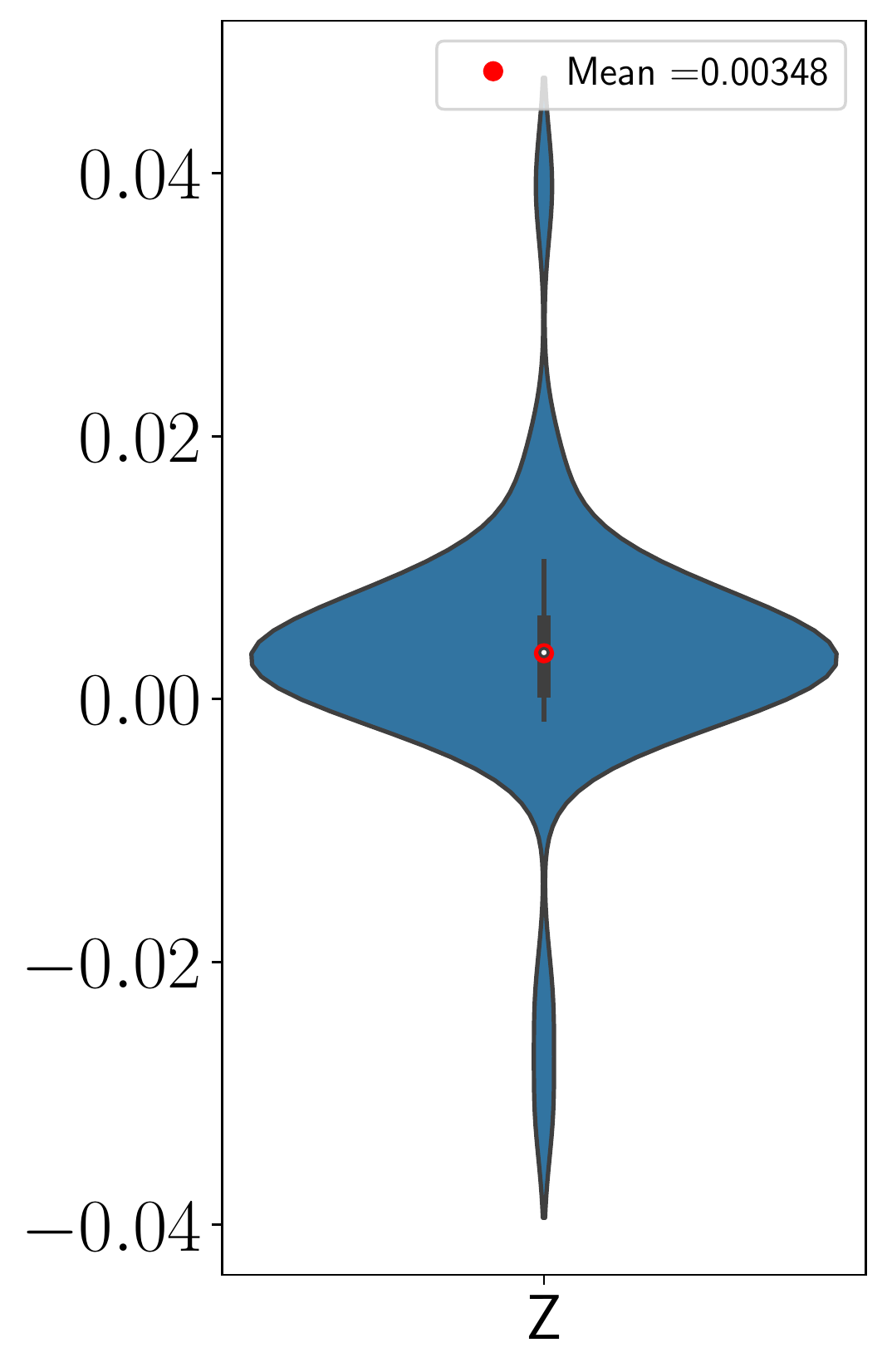}}  &
    \raisebox{-\height}{\includegraphics[width=0.74\linewidth]{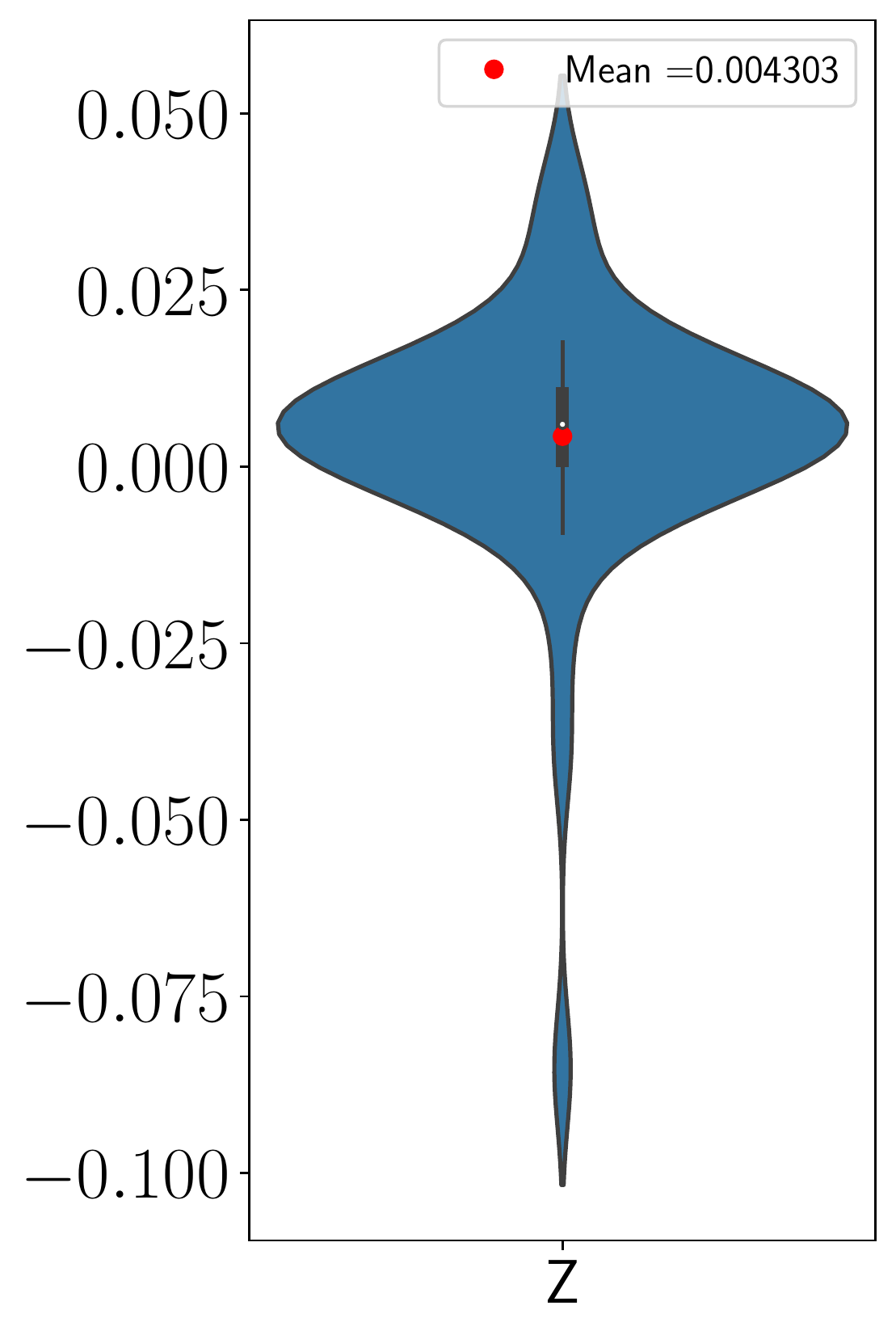}} &
    \raisebox{-\height}{\includegraphics[width=0.65\linewidth]{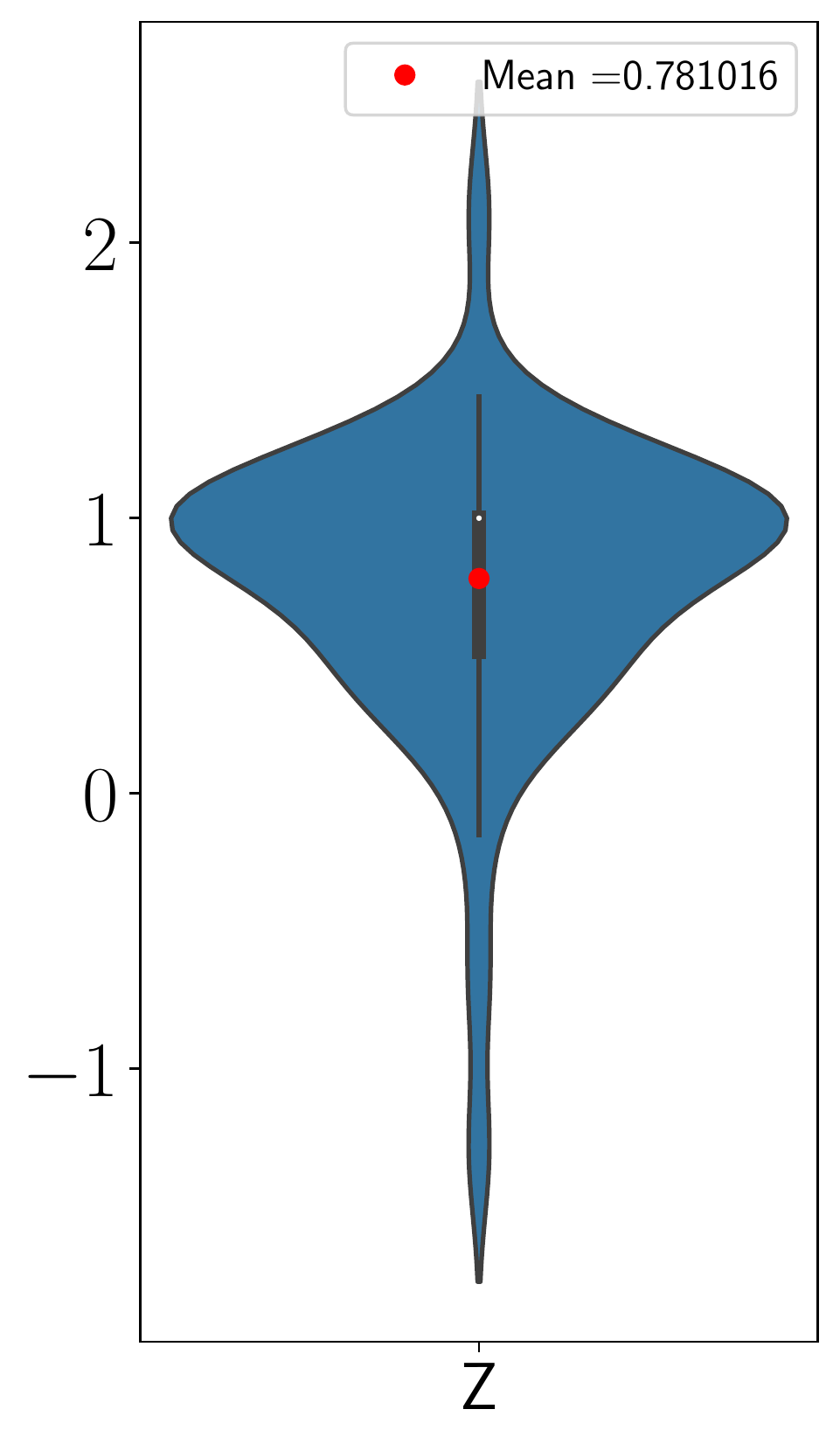}}
    &
    \raisebox{-\height}{\includegraphics[width=0.7\linewidth]{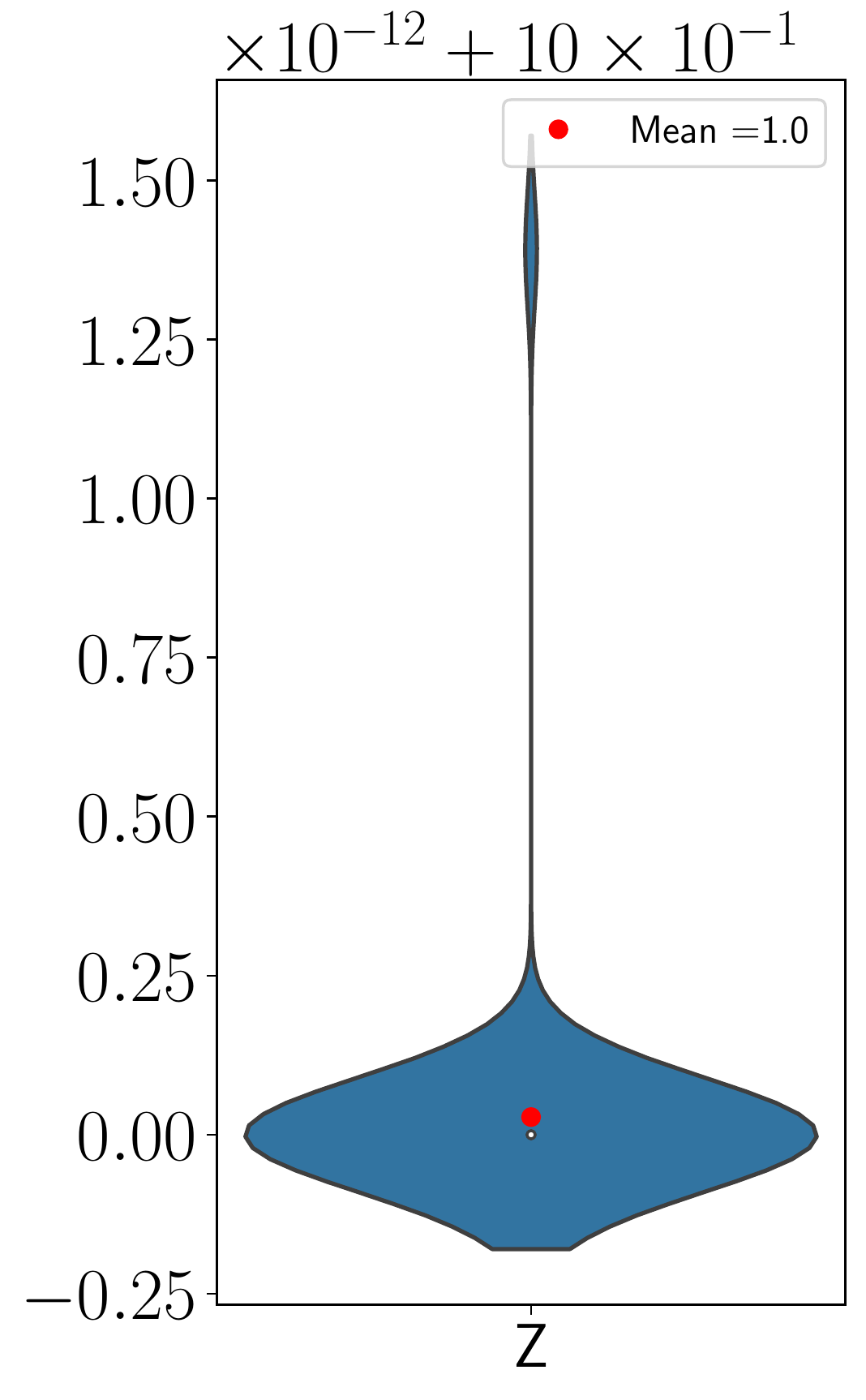}}   \\
    \caption*{}                                                                                              &
    \vspace{-0.5cm}
    \caption*{{\fontfamily{pcr}\selectfont } KLY }&
    \vspace{-0.5cm}
    \caption*{{\fontfamily{pcr}\selectfont } Ginibre }                                              &
    \vspace{-0.5cm}
    \caption*{{\fontfamily{pcr}\selectfont } Poisson }                                               &
    \vspace{-0.5cm}
    \caption*{{\fontfamily{pcr}\selectfont } Thomas }
  \end{tabular}
  \vspace{-0.95cm}
  \caption{Violin plots of $Z$ across $50$ samples from the KLY, Ginibre, Poisson, and Thomas point process using $\widehat{S}_{\mathrm{SI}}$ (first line) and $\widehat{S}_{\mathrm{BI}}$ (last line). The
  computation is done using \toolbox{}} \label{fig:violin_z}
\end{figure}

Table~\ref{tab:Multiscale hyperuniformity test} summarizes the results of the multiscale hyperuniformity test of Section~\ref{sub:Multiscale hyperuniformity test}, for the scattering intensity and Bartlett's isotropic estimators, on rectangular windows for the former and ball windows for the latter.
To compute $\bar{Z}_A$, we used $A$ draws of $(\calX_a, M_a)$.
In practice, the Ginibre ensemble cannot be sampled on arbitrarily large windows on a personal computer.
We thus proceed as follows.
Let $L_{\text{max}}$ be the maximum sidelength, respectively maximum radius $R_{\text{max}}$, of the box- (respectively ball-) shaped window on which the point process can be sampled in practice.
Since $M_a$ is Poisson with parameter $\lambda$, we choose $\lambda$ such that the probability that $W_{M_a}$ is larger than $W_{L_{\text{max}}}$ (respectively $W_{R_{\text{max}}}$) is smaller than $10^{-4}$.
Precisely, for the scattering intensity, $\lambda=85$, and the sidelength of the box window ranges from $L_{\text{min}}=20$ to $L_{\text{max}}= 140$, with a unit stepsize for the Ginibre, the Poisson and the Thomas process.
The wavevectors used $\{\bfk_m^{\text{min}}\}_{m \geq 1}$ are the minimum wavevectors of \eqref{eq:res_allowed_wave} corresponding to $\{W_m\}_{m\geq 1}$.
Finally, the asymptotic confidence interval \eqref{e:confidence_interval}, denoted by $CI[\bbE [Z]]$ in Table~\ref{tab:Multiscale hyperuniformity test}, has a $99.7\%$ asymptotic level since we use three standard deviations.
For Bartlett's estimator, we use ball windows with a minimum radius $R_{\text{min}}= 30$, a maximum radius $R_{\text{max}}=100$, a unit step size, $\lambda=50$, and the minimum wavenumbers of \eqref{eq:allowed_k_isotropic} corresponding to the subwindows for the Ginibre, the Poisson and the Thomas process.
For the KLY process having bigger intensity than the other benchmarks point processes, we use smaller parameters $L_{\text{max}}= 80$, $R_{\text{max}}=56$ and $L_0=R_0=20$.

We can see from Table~\ref{tab:Multiscale hyperuniformity test} that the test successfully rejects hyperuniformity for the Poisson and Thomas point processes, and does not reject for the Ginibre and the KLY process, as expected.
Moreover, we note that $ \widehat{S}_{\mathrm{BI}}$ provides tighter confidence intervals.
Yet, one should bear in mind that, because $\widehat{S}_{\mathrm{SI}}$ uses a rectangular window and Ginibre is naturally sampled on disk-like windows, $\widehat{S}_{\mathrm{BI}}$ has access here to a sample of larger cardinality than $\widehat{S}_{\mathrm{SI}}$. Moreover, $\widehat{S}_{\mathrm{BI}}$ is also computationally more expensive to evaluate.

\begin{figure}[!ht]
  \centering
    \includegraphics[width=0.45 \textwidth]{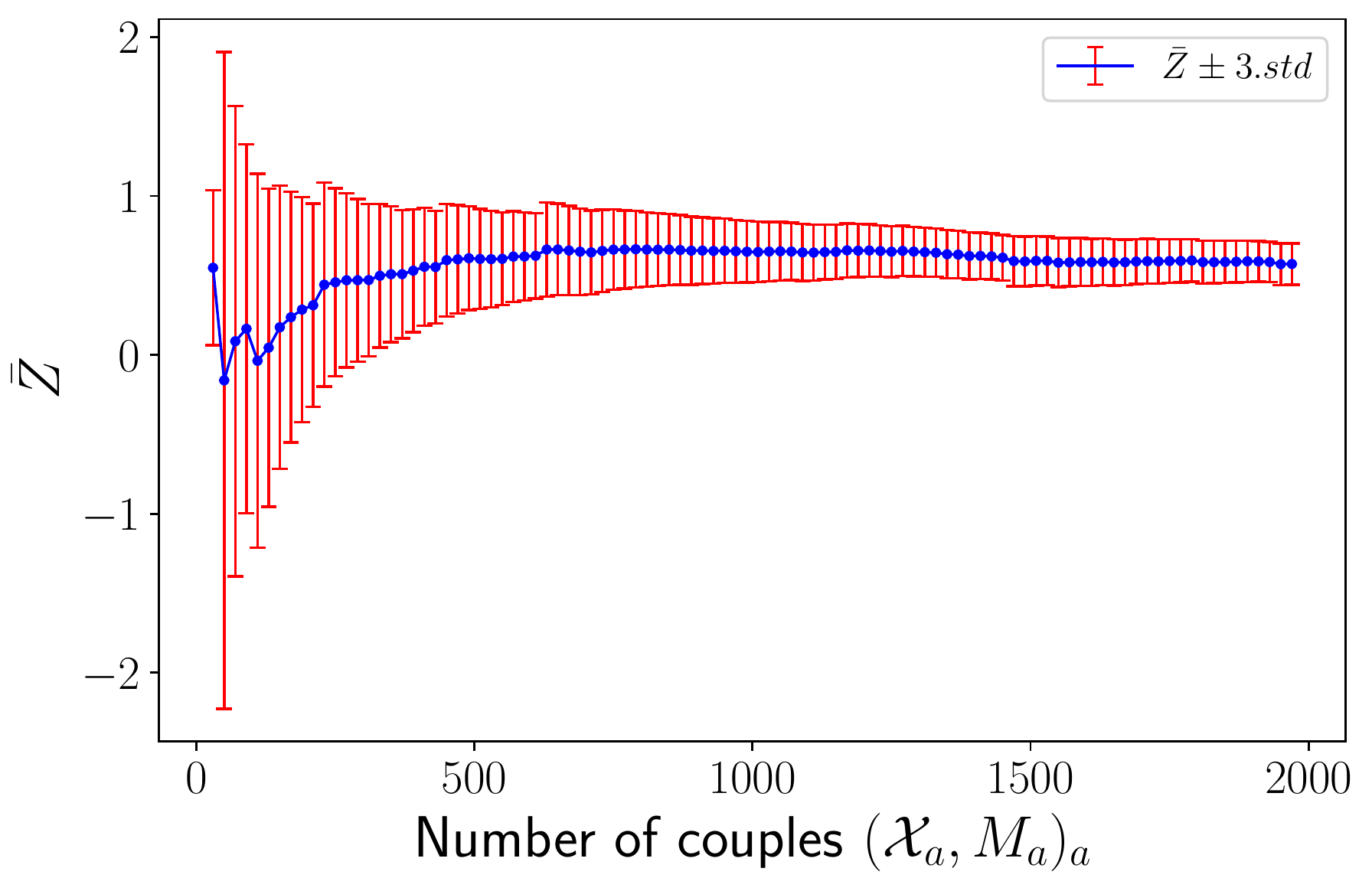}
    \caption{$CI[\bbE [\bar{Z}]]$ for a Poisson point process with the scattering intensity, as a function of the number of realizations of $Z$.}
    \label{fig:ci_poisson_decrease}
\end{figure}

Note that the coverage of the confidence interval on which our test is based is only controlled when $A$ goes to infinity.
For a fixed $\lambda$, we thus recommend choosing $A$ as large as possible.
In particular, the overly wide confidence interval for the Poisson process in Table~\ref{tab:Multiscale hyperuniformity test} calls for increasing $A$.
Indeed, the violin plot of the $50$ realizations of $Z$ (Figure~\ref{fig:violin_z}) has large support.
Increasing $A$ naturally reduces the size of that support; see Figure~\ref{fig:ci_poisson_decrease}.
Note in passing how $\bar{Z}_{A}$ does not converge to $S(\mathbf{0})=1$, which is an effect of our capping the estimated structure factor in \eqref{e:capped_estimates}.
Overall, there is no free lunch: our test might fail in diagnosing hyperuniformity if $A$ or $\lambda$ is too small.

Finally, we demonstrate the test on three thinned versions of the Ginibre samples with retaining probabilities $p = \{0.9, 0.5, 0.1\}$, as described in Section~\ref{ssub:Thinning the Ginibre point process}.
The corresponding point processes have respectively $S(0)= \{0.1, 0.5, 0.9\}$.
We have noticed that the estimator $ \widehat{S}_{\mathrm{BI}}$ provides tighter confidence intervals, and thus focused here on $ \widehat{S}_{\mathrm{BI}}$ to compute $\bar{Z}_{50}$.
We keep the same parameters as before.
Note that we use the same $50$ realizations of the Ginibre ensemble across the four different values of $p$ (including $p=1$), so that, strictly speaking, the asymptotic confidence intervals are to be understood with a Bonferroni correction.

\begin{figure}[!ht]
  \centering
  \begin{subfigure}[b]{0.2\textwidth}
    \centering
    \includegraphics[width=0.72\textwidth]{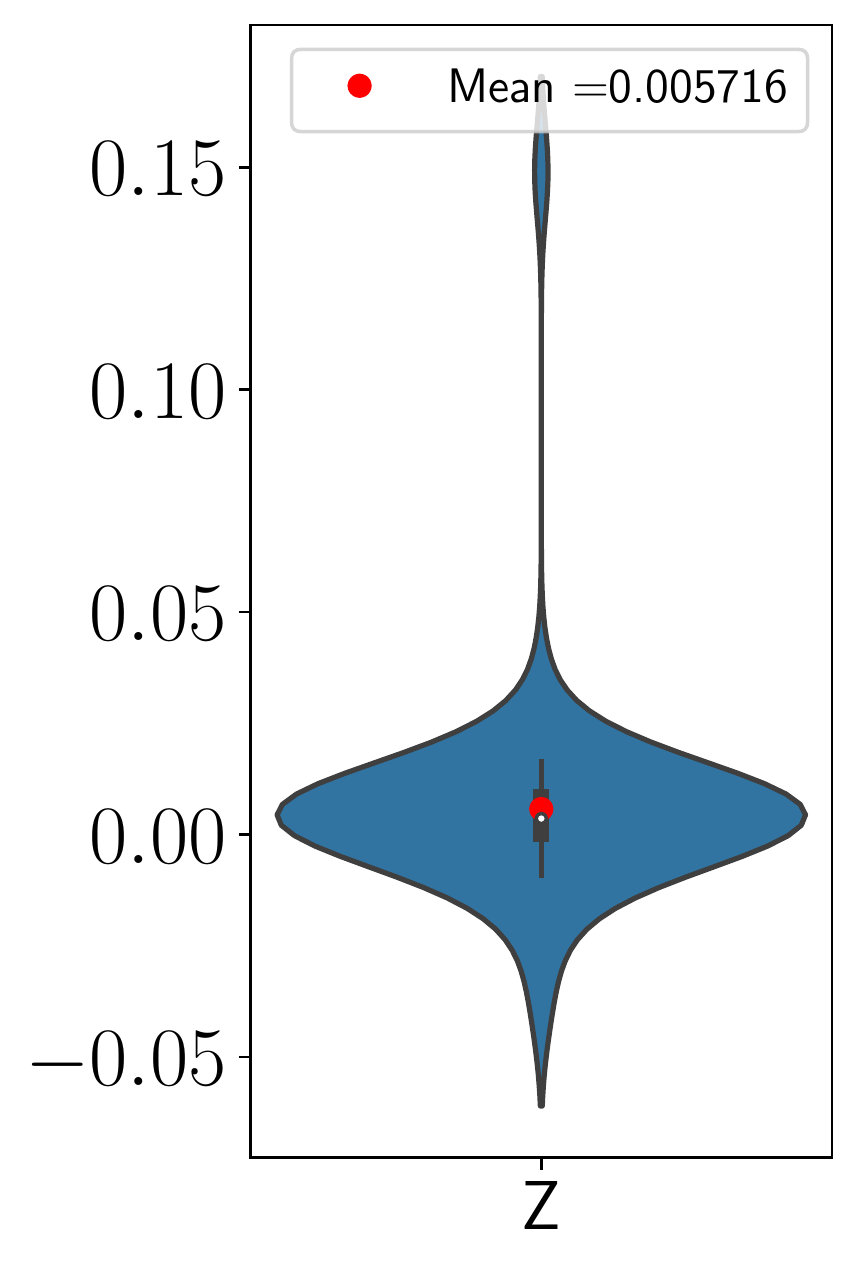}
    \caption{Ginibre }
    %\label{fig:H_index_kly}
  \end{subfigure}
  \hfill
  \begin{subfigure}[b]{0.2\textwidth}
    \centering
    \includegraphics[width=0.68\textwidth]{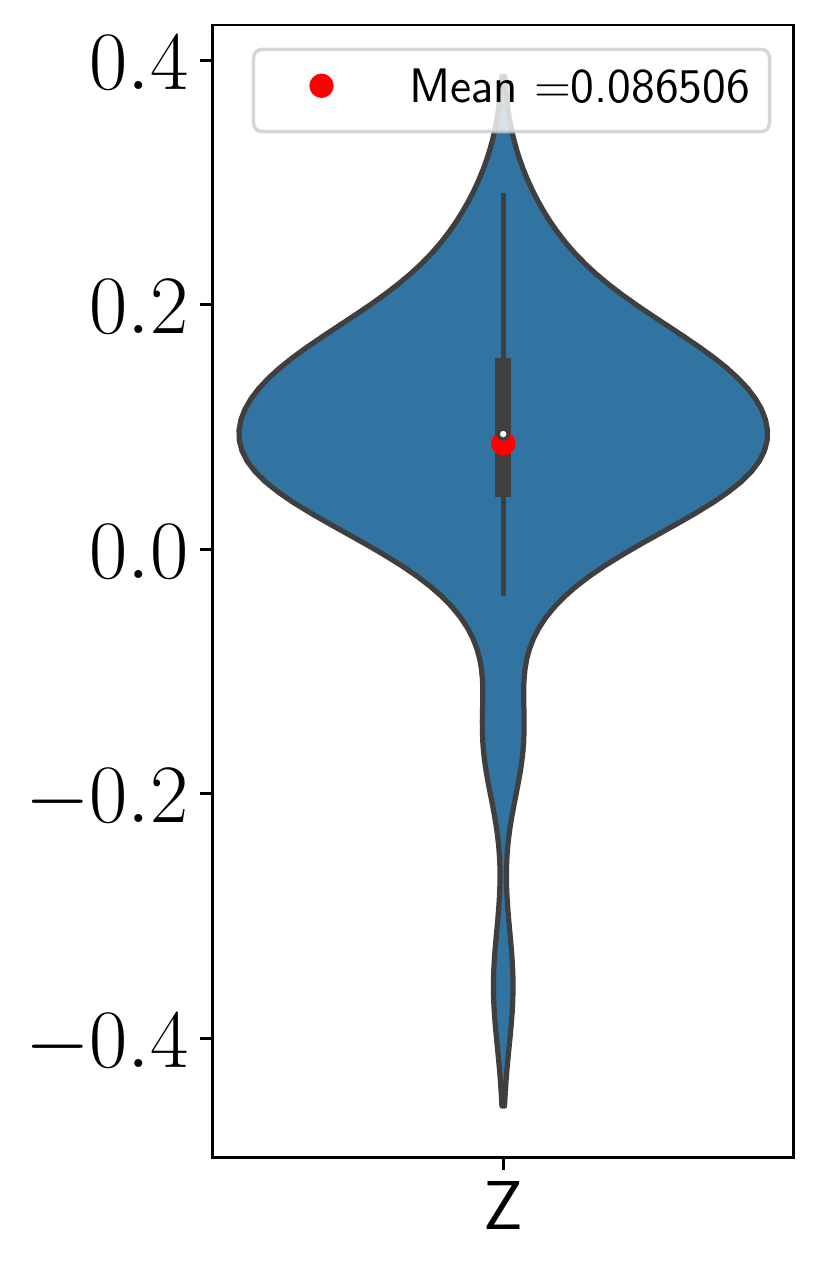}
    \caption{$p=0.9$ }
    %\label{fig:H_index_ginibre}
  \end{subfigure}
  \hfill
  \begin{subfigure}[b]{0.2\textwidth}
    \centering
    \includegraphics[width=0.61\textwidth]{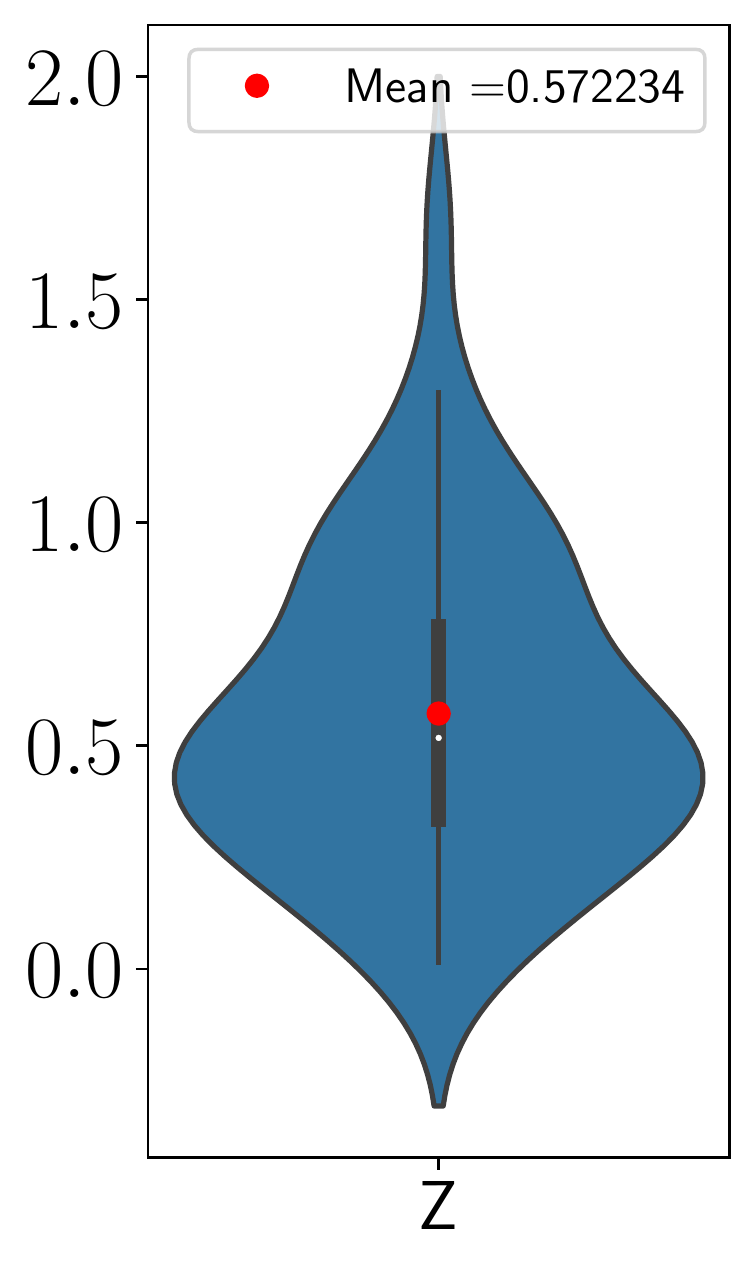}
    \caption{$p=0.5$ }
    %\label{fig:H_index_poisson}
  \end{subfigure}
  \hfill
  \begin{subfigure}[b]{0.2\textwidth}
    \centering
    \includegraphics[width=0.62 \textwidth]{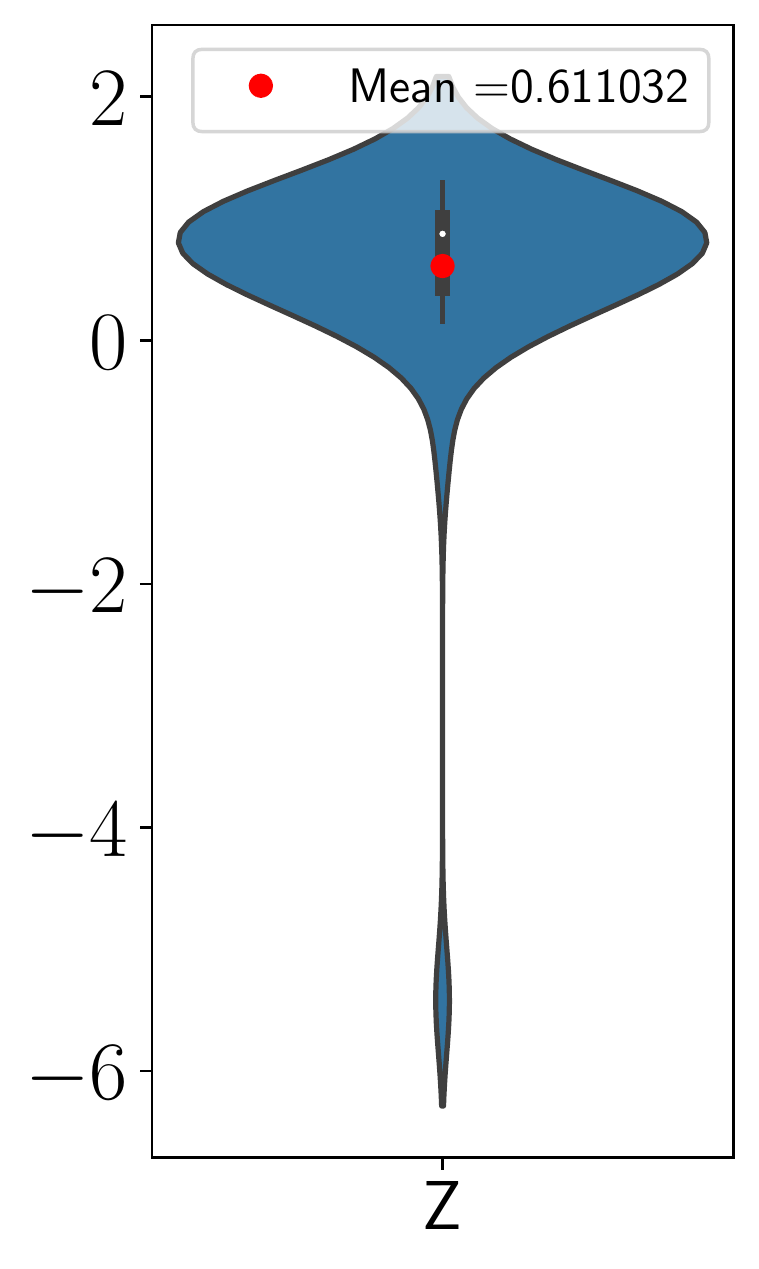}
    \caption{$p=0.1$ }
    %\label{fig:multiscale_test_ginibre_thinning}
  \end{subfigure}
  \caption{Violin plots of $Z$ obtained using $\widehat{S}_{\mathrm{BI}}$ across $A=50$ samples from the Ginibre and the corresponding independent thinning with retaining probability $p$. The
  computation is done using \toolbox{}.}
  \label{fig:multiscale_test_ginibre_thinning}
\end{figure}

Table \ref{tab:Multiscale hyperuniformity test thinning} summarizes the obtained results, and Figure \ref{fig:multiscale_test_ginibre_thinning} shows the corresponding violin plots.
The test successfully rejects the hyperuniformity for the three thinned versions.
Note that the test is more sensitive when using $ \widehat{S}_{\mathrm{SI}}$ within different trials.
Nevertheless, the failure of the test can be noticed from the wide confidence intervals obtained in general.
As mentioned before, getting a wide confidence interval call for increasing $A$.
We recommend using $\widehat{S}_{\mathrm{BI}}$, if possible.
Finally,  as expected for the window sizes/intensity that we consider, the case $0.9<p<1$ remains difficult: in preliminary experiments, we did not reject hyperuniformity without hand-tuning the test's parameters to reach the desired conclusion.
We leave this critical case to future work.

\begin{table}[!ht]
  \centering
   \caption{Multiscale hyperuniformity test obtained using $\widehat{S}_{\mathrm{BI}}$ on the thinned Ginibre process.}
  \label{tab:Multiscale hyperuniformity test thinning}
  \small
  \begin{tabular}{|l|c|c|c|c|}
    \hline
    \rule{0pt}{15pt}
    {}
     & $\bar{Z}_{A}$
     & $CI[\bbE [Z]]$
    \\
    \hline
    Ginibre
    & $0.0057$
    & $[ -0.0042, 0.0156]$
    \\
    \hline
     Thinning $p=0.9$
     & $0.0865$
     & $[0.0411,  0.1318]$
    \\
    \hline
    Thinning $p=0.5$
    & $0.5722$
    & $[0.4227, 0.7217]$
    \\
    \hline
    Thinning $p=0.1$
    & $0.611$
    & $[  0.2082 , 1.0137]$
     \\
    \hline
  \end{tabular}%
\end{table}

%-------------------
\paragraph{Hyperunifomity class}

\begin{figure}[!ht]
  \centering
  \begin{subfigure}[b]{0.42\textwidth}
    \centering
    \includegraphics[width=0.31 \textwidth]{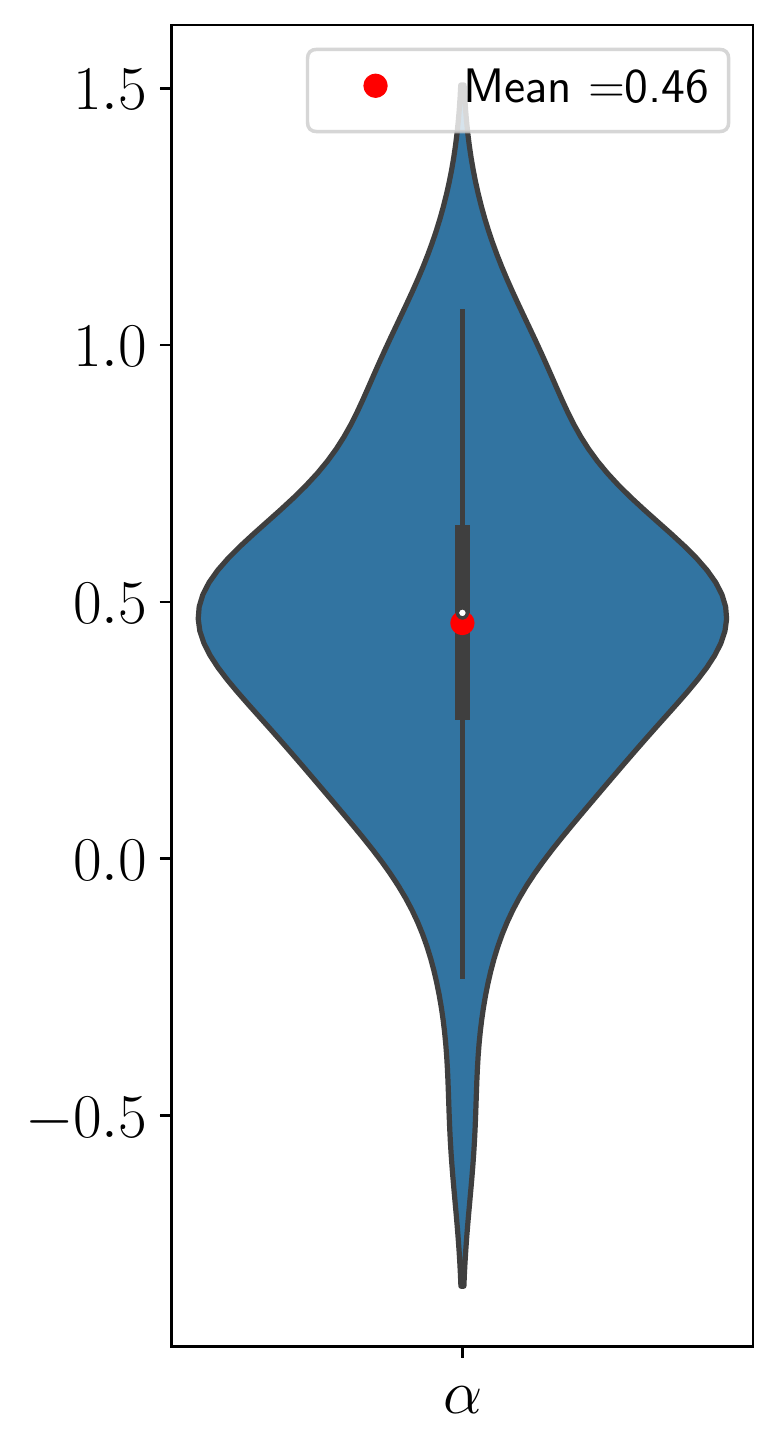}
    \caption{KLY}
    \label{fig:power_decay_kly}
  \end{subfigure}
  \begin{subfigure}[b]{0.42\textwidth}
    \centering
    \includegraphics[width= 0.3 \textwidth]{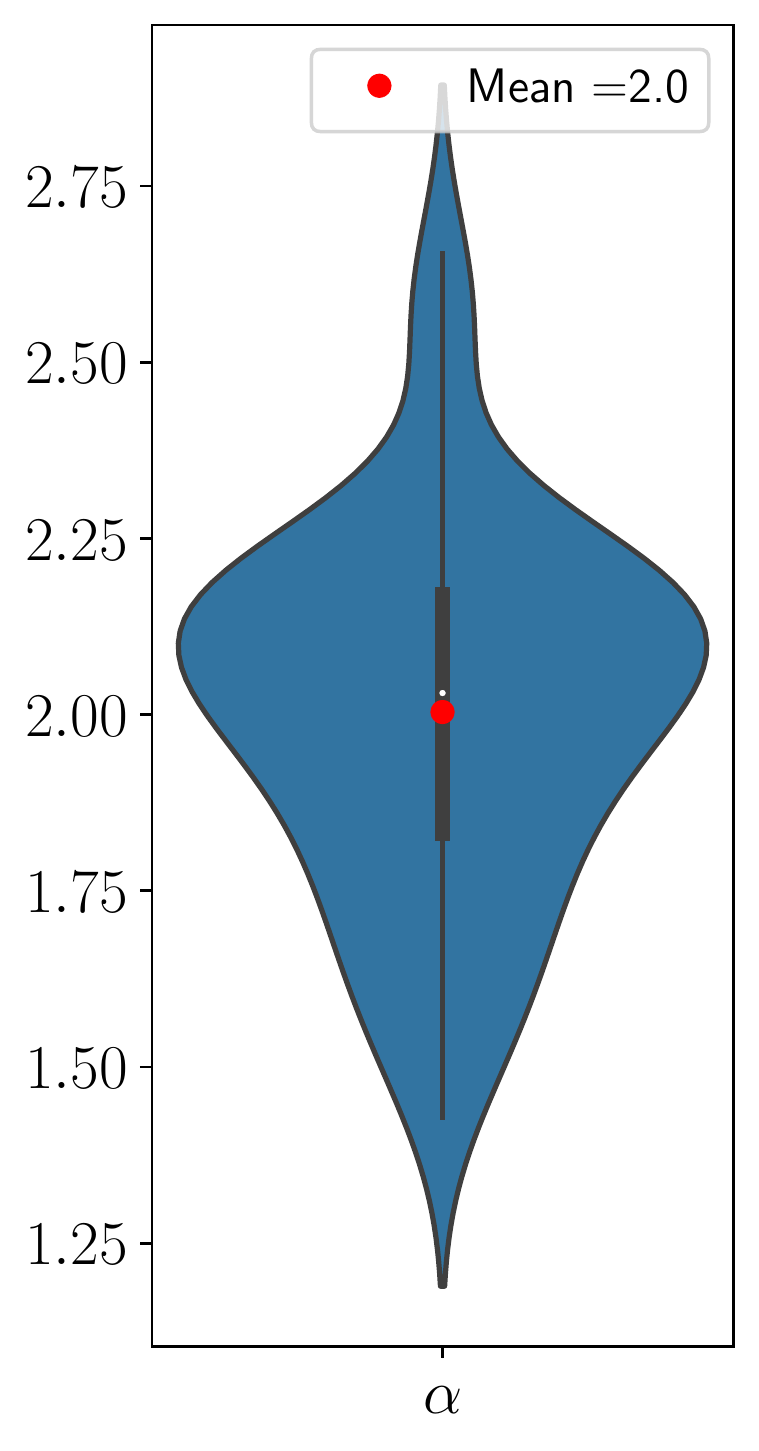}
    \caption{Ginibre}
    \label{fig:power_decal_ginibre}
  \end{subfigure}

  \caption{Violin plots of $\alpha$ across $A$ samples of the KLY and the Ginibre point process.
    The computation is done using \toolbox{}}
  \label{fig:power decay}
\end{figure}
Figure~\ref{fig:power decay} shows the violin plots of the estimated power decay $\alpha$ across $A$ samples, of roughly $10^4$ points each, of the KLY and the Ginibre point processes.
We used the results of $\widehat{S}_{\mathrm{BI}}$.
To approximate the decay rate $\alpha$ of the structure factor, the maximum wavenumber used to fit the polynomial is equal to $0.45$ for the Ginibre ensemble, and to $0.6$ for the KLY process.
Again, these thresholds are fit manually and represent a trade-off between having enough points to fit our monomial and being close to zero.

Results are shown in Figure~\ref{fig:power decay}.
There is limited evidence that the KLY process is indeed hyperuniform: while the support of the distribution is large and includes 0, most estimated values of $\alpha$ are positive, and concentrate around $0.5$.
This could be further taken as a hint that KLY belongs to Class \rom{3} in Table~\ref{tab:class_of_hyperuniformity}.
Nevertheless, KLY is known to be Class \rom{1} when matching a Poisson process with a grid \citep{Klatt+Last+Yogeshwaran:2020}, so we'd expect $\alpha$ to be bigger than $1$.
The misleading concentration around $0.5$ may result from the non-isotropy of the KLY process and calls for caution when making claims on the hyperuniformity class from regression diagnostics.
For the Ginibre ensemble, the concentration of values of $\alpha$ around $2$ successfully reflects the known power decay (Section~\ref{sub:point_processes}).

%-----------------------------------------
\section{A quantitative comparison of the estimators}
% sectionIllustrating the toolbox (endsection nameComputational techniques} % (fold)
\label{sec:Comparison of the estimators}

We now compare the cost and accuracy of all estimators more quantitatively.

\subsection{A note on computational costs} % (fold)
\label{sub:Computational time}

For a given wavevector, evaluating any single-tapered estimator requires a sum of $N$ terms.
Multitapering naturally multiplies the cost by the number of tapers, but it can be trivially parallelized, especially since the number of tapers remains low in practice \citep{Rajala+Olhede+John:2020}.

On the other hand, for a sample of $N$ points, Bartlett's estimator $\widehat{S}_{\mathrm{BI}}$ is a sum of $\calO ((N^2- N)/2)$ evaluations of a Bessel function.
This makes Bartlett’s estimator significantly costlier than its scattering intensity counterpart.
As an example, for a realization of $N=10^4$ points from a point process in $\bbR^2$, using a modern laptop, the evaluation of $\widehat{S}_{\mathrm{SI}}$ at a single wavevector took about one millisecond, compared to 2 seconds for $\widehat{S}_{\mathrm{BI}}$.
Note however that for a similar accuracy on an isotropic point process, $\widehat{S}_{\mathrm{BI}}$ needs to be evaluated at fewer wavevectors than, say, the scattering intensity.
Indeed, a single value of the former should be compared to a binwise average of the latter.

Now, for Hankel quadrature estimators, the main bottleneck is the approximation of the pcf.
A kernel density estimator based on $N$ points is again a quadratic computation.
In our experience, combining the estimator \texttt{pcf.fv}, based on numerically differentiating an estimated Ripley's K function, and a Hankel quadrature to build $\widehat{S}_{\mathrm{HO}}$ and $\widehat{S}_{\mathrm{HBC}}$ is the least expensive pipeline.

\subsection{Measuring the accuracy of the estimators near zero}
\label{statistical tests}
To confirm the intuitions gained from Section~\ref{sec:Illustrating the toolbox}, in particular that the multitapered variant of the scattering intensity and Bartlett's estimator dominate the others when it comes to visual diagnostics of hyperuniformity, we now compare the mean square error of the different estimators of the structure factor, integrated near 0.

For a subdivision $[k_1,k_2) \cup \dots \cup [k_{J-1},k_J)$ of some interval $[k_1,k_J)$, the integrated MSE reads
\begin{align*}
  \mathrm{iMSE}(\widehat{S}) & = \int_{k_1}^{k_J} \mathrm{MSE}(\widehat S)(k) \mathrm d k                                                                                                \\
                             & = \int_{k_1}^{k_J} \mathbb{E}[\widehat{S}(k) - S(k)]^2 \mathrm{d}k                                                                                             \\
                             & = \mathbb{E} \left[ \int_{k_1}^{k_J} [\widehat{S}(k) - S(k)]^2 \mathrm{d}k \right]  .
\end{align*}
A crude numerical integration using the trapezoidal rule gives,
 \begin{equation}
\mathrm{iMSE}(\widehat{S}) \approx \mathbb{E} \bigg[ \sum_{j=1}^{J-1} \frac12 (k_{j+1} - k_{j})
 \left\{[\widehat{S}(k_{j+1}) - S(k_{j+1})]^2 + [\widehat{S}(k_{j}) - S(k_{j})]^2 \right \} \bigg],
  \label{e:trapeze_rule}
  \end{equation}
The expectation in \eqref{e:trapeze_rule} is under the investigated point process.
Assuming it is easy to sample from the point process, we can estimate the iMSE in \eqref{e:trapeze_rule} by an empirical average $\widehat{\mathrm{iMSE}}$ over point process samples.
Similarly, the difference of the ${\mathrm{iMSE}}$s of two different estimators of $S$ can be approximated by an empirical average of differences of $\widehat{\mathrm{iMSE}}$s.
This yields a natural (paired) Student test to assess whether the difference of the ${\mathrm{iMSE}}$s of two different estimators of $S$ is 0.

\subsection{Comparing variants of the scattering intensity} % (fold)
\label{sub:Comparaison of the results}

In Section~\ref{sec:Illustrating the toolbox}, we derived the intuition that $\widehat{S}_{\mathrm{SI}}$ on its allowed values, $\widehat{S}_{\mathrm{DDT}}(t_0, \bfk)$ , $\widehat{S}_{\mathrm{DDT}}(t_1, \bfk)$, and $\widehat{S}_{\mathrm{DDMT}}((t_q)_{q=1}^4, \bfk)$ gave the most accurate approximations among the estimators of Section~\ref{sec:Estimators of the structure factor}.
We further expect $\widehat{S}_{\mathrm{DDMT}}((t_q)_{q=1}^4, \bfk)$ to have the smallest integrated MSE among them.
To test this, we run three paired, one-sided Student tests, comparing $\widehat{S}_{\mathrm{DDMT}}((t_q)_{q=1}^4, \bfk)$ to each of the three former estimators.

Specifically, we use $50$ independent realizations of approximately $5800$ points each, from the Ginibre, Poisson, and Thomas point processes, for which we know the exact structure factor.
The intensity of all processes is $\rho=1/\pi$.
The additional parameters of the Thomas process are, as used throughout the paper, $\rho_{\text{parent}}= 1/(20\pi)$, and $\sigma=2$.
We evaluated $\widehat{S}_{\mathrm{SI}}$ \eqref{eq:s_si} on its allowed wavevectors \eqref{eq:allowed_wave}, between $k_1=0.1$ and $k_J=2.8$.
For the other estimators, $\widehat{S}_{\mathrm{DDT}}(t_0, \bfk)$, $\widehat{S}_{\mathrm{DDT}}(t_1, \bfk)$ \eqref{eq:s_DT}, and $\widehat{S}_{\mathrm{DDMT}}((t_q)_{q=1}^4, \bfk)$ \eqref{eq:s_mt}, we used arbitrary wavevectors keeping the same range of the wavenumbers used for $\widehat{S}_{\mathrm{SI}}$.
The tapers used are $t_0$ \eqref{eq:t_0}, and the first 4 tapers $(t_q)_{q=1}^4$ of the family of sinusoidal tapers \eqref{eq:sine_taper}.
As we got different approximated values of $S$ for different wavevectors of the same wavenumber, for each wavenumber we set the estimated structure factor to the sample mean of these values.

\begin{table}[!ht]
  \centering
   \caption{Paired $t$-tests
  }
  \label{tab:t_score_DSE}
  \small
  \resizebox{\columnwidth}{!}{%
  \begin{tabular}{|l|c|c|c|c|c|c|c|}
    \hline
    \rule{0pt}{15pt}
    Estimators
     & $T$-score
     & $p$-value
     & $T$-score
     & $p$-value
     & $T$-score
     & $p$-value
    \\
    \hline
    $\widehat{S}_{\mathrm{DDMT}}$, $\widehat{S}_{\mathrm{SI}}$
     & $-29.53$
     & $3 \times 10^{-33}$
     & $-41.59$
     & $3\times 10^{-40}$
     & $-9.24$
     & $10^{-12}$
    \\
    \hline
    $\widehat{S}_{\mathrm{DDMT}}$,  $\widehat{S}_{\mathrm{DDT}}$
     & $-22.40$
     & $ 10^{-27}$
     & $-30.42$
     & $8 \times 10^{-34}$
     & $-6.38$
     & $2 \times 10^{-8}$
    \\
    \hline
    $\widehat{S}_{\mathrm{DDMT}}$, $\widehat{S}_{\mathrm{DDT}}$
     & $-12.18$
     & $9 \times 10^{-17}$
     & $-25.39$
     & $3\times 10^{-30}$
     & $-7.16$
     & $10^{-9}$
    \\
    \hline
     & \multicolumn{2}{|c|}{Ginibre }
     & \multicolumn{2}{|c|}{Poisson }
     & \multicolumn{2}{|c|}{Thomas }
    \\
    \hline
  \end{tabular}%
  }
  \footnotetext{$T$-scores of the $\widehat{\mathrm{iMSE}}$ and the associated $p$-values}
\end{table}

Table~\ref{tab:t_score_DSE} summarizes the results of the paired one-sided Student tests.
For each point process, applying a Bonferroni correction, we can simultaneously reject at the level $0.01$ the three hypotheses that there is no difference in mean between the estimated iMSEs.
This confirms our intuition that $\widehat{S}_{\mathrm{DDMT}}((t_q)_1^4)$ yields the smallest integrated MSE among the considered variants of the scattering intensity.
In particular, multitapering helps.

For information, the estimated iMSEs are given in Table~\ref{tab:summary_statistics_DSE}, in the form of a confidence interval (CI) of the $\widehat{\mathrm{iMSE}}$s plus or minus 3 empirical standard deviations of the mean.
There is roughly a factor 4 between the $\widehat{\mathrm{iMSE}}$ of the directly debiased multitapered estimator and each of the other three, which confirms its domination.
For reference, we have also indicated the contribution of the sample integrated variance ($\widehat{\mathrm{iVar}}$) to each $\widehat{\mathrm{iMSE}}$.
Clearly, the variance is the biggest contributor, and the squared bias is at least one order of magnitude smaller, for all point processes.

\begin{figure*}[!ht]
  \begin{tabular}{p{\dimexpr 0.04\textwidth-\tabcolsep}p{ 0.3\textwidth}p{0.3\textwidth}p{ 0.3\textwidth}}
    \multirow{9}{*}{\rotatebox[origin=c]{90}{Point process}}                                  &
    \raisebox{-\height}{\includegraphics[width=0.7\linewidth]{ginibre_pp_box.pdf}} &
    \raisebox{-\height}{\includegraphics[width=0.7\linewidth]{poisson_pp_box.pdf}} &
    \raisebox{-\height}{\includegraphics[width=0.7\linewidth]{thomas_pp_box.pdf}}
  \end{tabular}
  \begin{tabular}{p{\dimexpr 0.03\textwidth-\tabcolsep}p{\dimexpr 0.3\textwidth-\tabcolsep}p{\dimexpr 0.3\textwidth-\tabcolsep}p{\dimexpr 0.3\textwidth-\tabcolsep}}
    \multirow{9}{*}{\rotatebox[origin=c]{90}{$\bar{\widehat{S}}$}}                                           &
    \raisebox{-\height}{\includegraphics[width=0.8\linewidth]{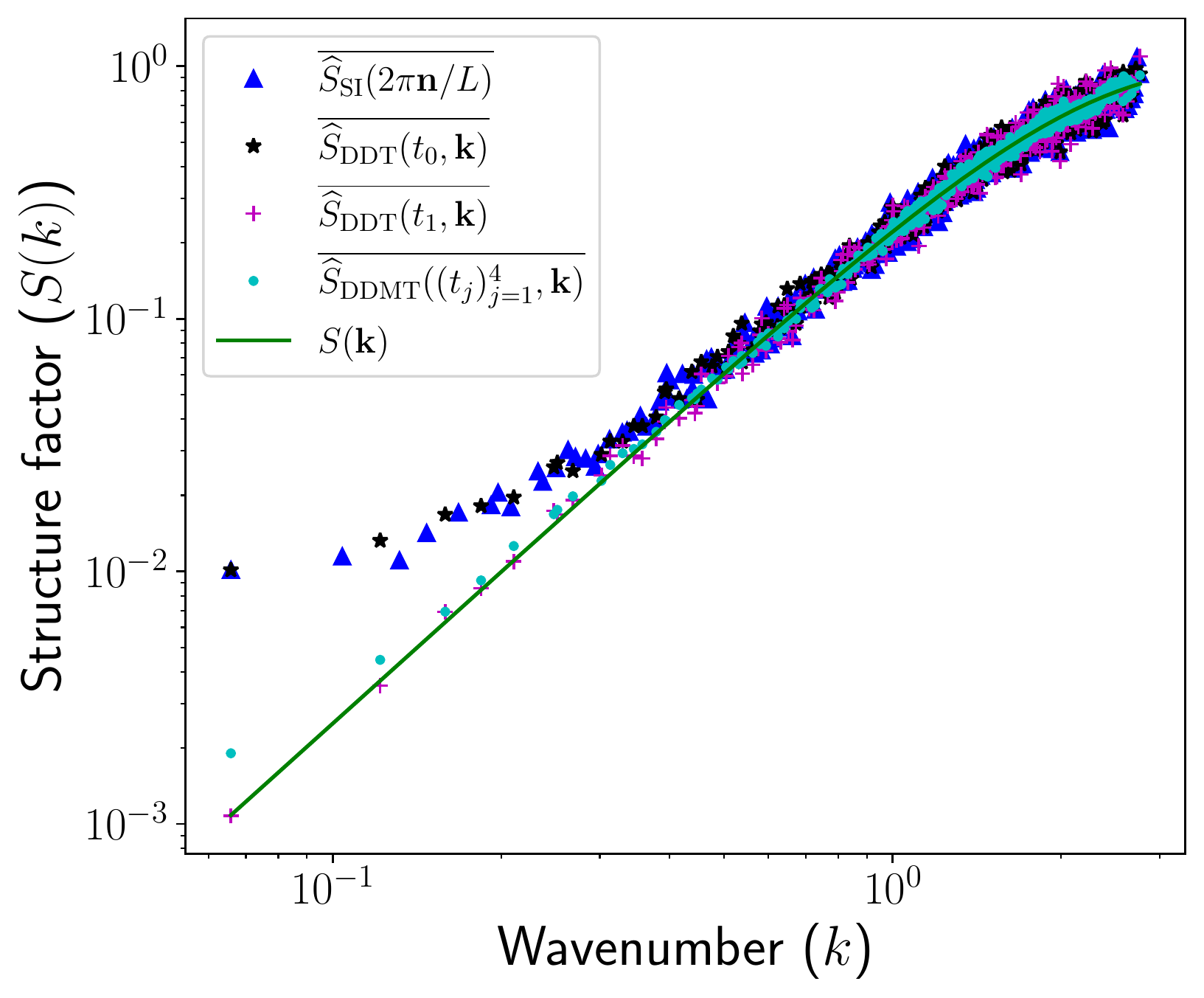}}  &
    \raisebox{-\height}{\includegraphics[width=0.86\linewidth]{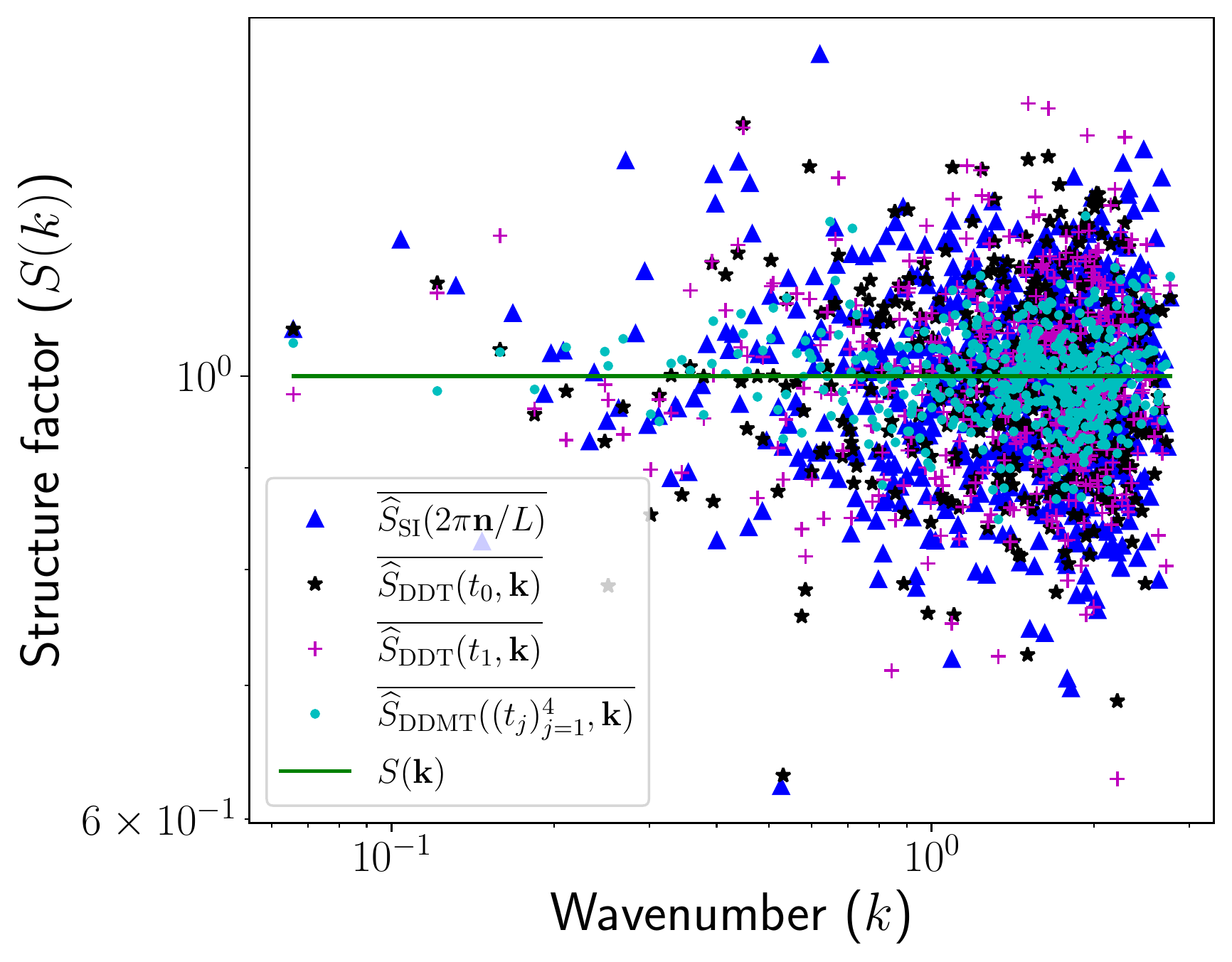}} &
    \raisebox{-\height}{\includegraphics[width=0.8\linewidth]{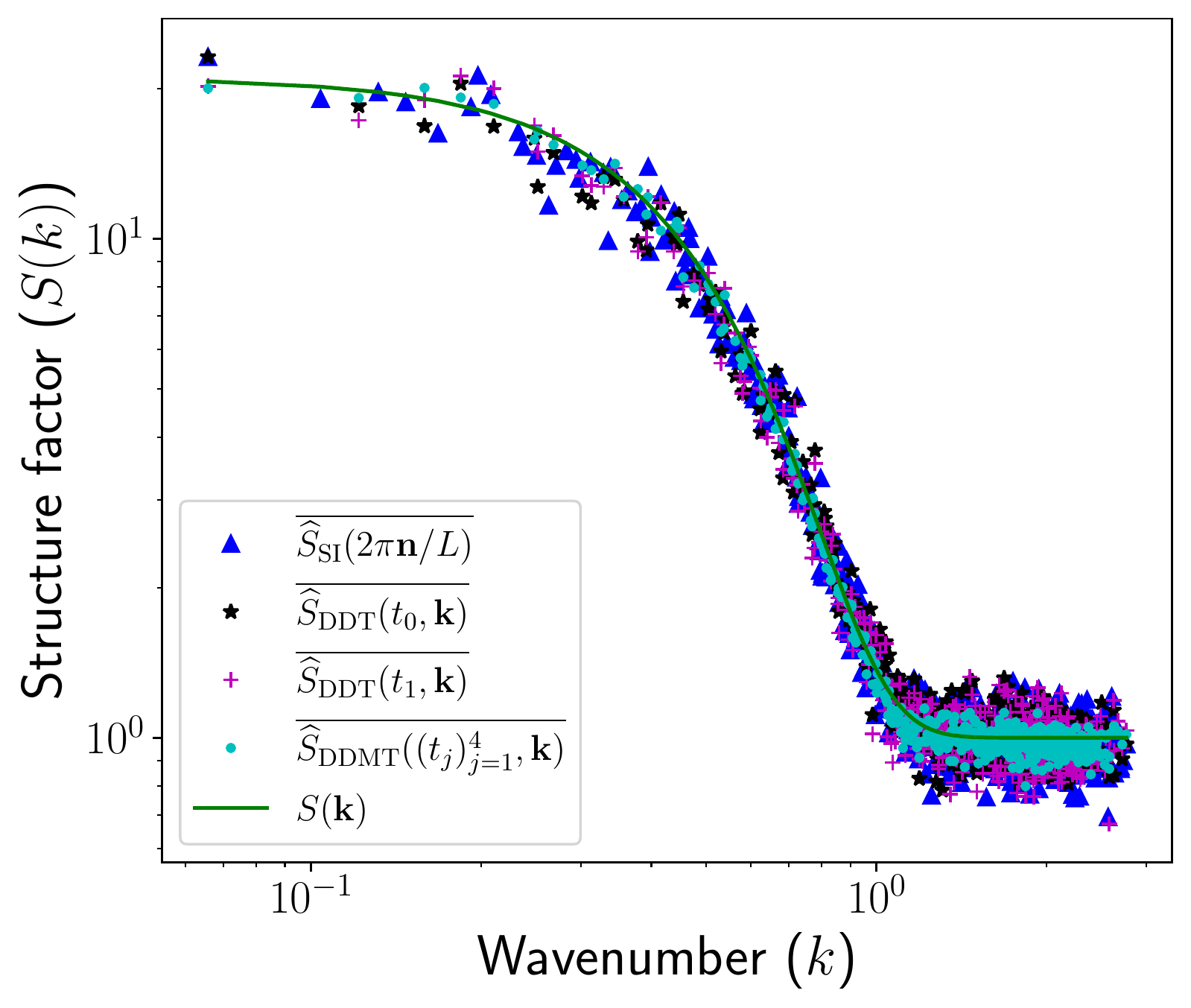}}     \\
    \caption*{}                                                                                              &
    \vspace{-0.5cm}
    \caption*{{\fontfamily{pcr}\selectfont } Ginibre}                                              &
    \vspace{-0.5cm}
    \caption*{{\fontfamily{pcr}\selectfont } Poisson }                                               &
    \vspace{-0.5cm}
    \caption*{{\fontfamily{pcr}\selectfont } Thomas}
  \end{tabular}
  \vspace{-0.8cm}
  \caption{Pointwise averages of the variants of the scattering intensity, numerically rotation-averaged, across 50 independent realizations of the Ginibre, Poisson, and Thomas point processes. The computation and visualization are done using \toolbox{}} \label{fig:mean_estimators}
\end{figure*}

Finally, Figure~\ref{fig:mean_estimators} shows the pointwise average of each estimator, across the $50$ point process realizations.
A reduction of bias at small $k$ for $\widehat{S}_{\mathrm{DDT}}(t_1, \bfk)$ and $\widehat{S}_{\mathrm{DDMT}}((t_q)_{q=1}^4, \bfk)$ is visible for Ginibre.

\begin{table*}[!ht]
  \centering
   \caption{Sample integrated variance and MSE
  }
  \label{tab:summary_statistics_DSE}
  \small
  \begin{tabular}{|l|c|c|c|c|c|c|c|}
    \hline
    \rule{0pt}{15pt}
    Estimators
     & $\widehat{\mathrm{iVar}}$
     & $\mathrm{CI}[\widehat{\mathrm{iMSE}}]$
     & $\widehat{\mathrm{iVar}}$
     & $\mathrm{CI}[\widehat{\mathrm{iMSE}}]$
     & $\widehat{\mathrm{iVar}}$
     & $\mathrm{CI}[\widehat{\mathrm{iMSE}}]$
    \\
    \hline
    $\widehat{S}_{\mathrm{SI}}(2 \pi \bfn / L)$
     & 0.32
     & $0.32 \pm 0.02$
     & 1.31
     & $1.34 \pm 0.06$
     & 69.51
     & $70.71 \pm 17.95$
    \\
    \hline
    $\widehat{S}_{\mathrm{DDT}}(t_0)$
     & 0.32
     & $0.33 \pm 0.03$
     & 1.44
     & $1.47 \pm 0.1$
     & 72.15
     & $73.63 \pm 26.12$
    \\
    \hline
    $\widehat{S}_{\mathrm{DDT}}(t_1)$
     & 0.34
     & $0.35 \pm 0.06$
     & 1.47
     & $1.50 \pm 0.14$
     & 79.29
     & $80.51 \pm 27.20$
    \\
    \hline
    $\widehat{S}_{\mathrm{DDMT}}((t_q)_1^4)$
     & 0.08
     & $ \mathbf{0.08} \pm 0.007$
     & 0.37
     & $\mathbf{0.38} \pm 0.02$
     & 17.90
     & $\mathbf{18.19} \pm 4.19$
    \\
    \hline
     & \multicolumn{2}{|c|}{Ginibre }
     & \multicolumn{2}{|c|}{Poisson }
     & \multicolumn{2}{|c|}{Thomas }
    \\
    \hline
  \end{tabular}
\end{table*}

%end subsection

\subsection{Comparing estimators that assume isotropy} % (fold)
\label{sub:Comparaison between the ISE}
We now run the same comparison on $\widehat{S}_{\mathrm{BI}}$ \eqref{eq:s_BI}, $\widehat{S}_{\mathrm{HO}}$ \eqref{eq:s_ho}, and $\widehat{S}_{\mathrm{HBC}}$ \eqref{eq:s_hbc}.
For this study, we have sampled $50$ independent realizations of approximately $5800$ points each, from the Ginibre, Poisson, and Thomas processes, with the same parameters as in Section~\ref{sub:Comparaison of the results}, except that the observation window for all point processes is now a 2-dimensional ball window centered at the origin.
For $\widehat{S}_{\mathrm{HO}}$, and $\widehat{S}_{\mathrm{HBC}}$ we used the method \texttt{pcf.fv} to approximate the pair correlation function with maximal approximation radius $r_{\max}= 30$.

We have noted in Section \ref{sec:Illustrating the toolbox} that $\widehat{S}_{\mathrm{HBC}}$ is more robust to non-isotropy than $\widehat{S}_{\mathrm{HO}}$, and that $\widehat{S}_{\mathrm{BI}}$ gave the tightest approximations in this family of estimators.

\begin{table}[!ht]
  \centering
   \caption{Paired $t$-tests }
  \label{tab:t_score_ISE}
  \small
  \begin{tabular}{|l|c|c|c|c|c|c|c|}
    \hline
    \rule{0pt}{15pt}
    Estimators
     & $T$-score
     & $p$-value
     & $T$-score
     & $p$-value
     & $T$-score
     & $p$-value
    \\
    \hline
    $\widehat{S}_{\mathrm{BI}}$ vs. $\widehat{S}_{\mathrm{HO}}$
     & $-12.24$
     & $7 \times 10^{-17}$
     & $-6.60$
     & $  10^{-8}$
     & $-5.32$
     & $10^{-6}$
    \\
    \hline
    $\widehat{S}_{\mathrm{BI}}$ vs.  $\widehat{S}_{\mathrm{HBC}}$
     & $-25.51$
     & $2 \times 10^{-30}$
     & $-5.32$
     & $10^{-6}$
     & $-5.16$
     & $2 \times 10^{-6}$
    \\
    \hline
     & \multicolumn{2}{|c|}{Ginibre }
     & \multicolumn{2}{|c|}{Poisson}
     & \multicolumn{2}{|c|}{Thomas}
    \\
    \hline
  \end{tabular}%
  \footnotetext{$T$-scores of the $\widehat{\mathrm{iMSE}}$ and the associated $p$-values}
\end{table}

Table~\ref{tab:t_score_ISE}, summarizes the results of two one-sided paired $t$-tests per point process.
For each point process, we can again simultaneously reject at level $0.01$ the hypotheses that there is no difference between the estimates iMSEs.
This confirms the claim that Bartlett's estimator is the most accurate near 0, among estimators that assume isotropy.
For reference, Table~\ref{tab:summary_statistics_ISE} gives the same summary statistics as Table~\ref{tab:summary_statistics_DSE} did for scattering intensity variants.
Bartlett's estimator yields $\widehat{\mathrm{iMSE}}$s one order of magnitude smaller than both Hankel transform quadratures, for both Ginibre and Poisson, and a factor 3 for Thomas.
Again, the integrated MSE is mostly variance.

\begin{table*}[!ht]
  \centering
   \caption{Sample integrated variance and MSE }
  \label{tab:summary_statistics_ISE}
  \small
  \begin{tabular}{|l|c|c|c|c|c|c|c|}
    \hline
    \rule{0pt}{15pt}
    $\widehat{S}$
     & $\widehat{\mathrm{iVar}}$
     & $\mathrm{CI}[\widehat{\mathrm{iMSE}}]$
     & $\widehat{\mathrm{iVar}}$
     & $\mathrm{CI}[\widehat{\mathrm{iMSE}}]$
     & $\widehat{\mathrm{iVar}}$
     & $\mathrm{CI}[\widehat{\mathrm{iMSE}}]$
    \\
    \hline
    $\widehat{S}_{\mathrm{BI}}$
     & $3.9 \times 10^{-3}$
     & $\mathbf{4.0 \times 10^{-3}} \pm 3 \times 10^{-4} $
     & 0.057
     & $\mathbf{0.058} \pm 9 \times 10^{-3}$
     & 11.25
     & $\mathbf{11.65} \pm 4.71$
    \\
    \hline
    $\widehat{S}_{\mathrm{HO}}$
     & $ 0.37$
     & $ 0.38 \pm 0.09$
     & $ 2.12$
     & $ 2.14 \pm 0.93$
     & $43.63$
     & $ 46.70 \pm 18.40$
    \\
    \hline
    $\widehat{S}_{\mathrm{HBC}}$
     & 0.03
     & $0.03 \pm 0.01$
     & 2.44
     & $2.45 \pm 1.33$
     & 57.62
     & $63.02 \pm 28.62$
    \\
    \hline
     & \multicolumn{2}{|c|}{Ginibre}
     & \multicolumn{2}{|c|}{Poisson}
     & \multicolumn{2}{|c|}{Thomas}
    \\
    \hline
  \end{tabular}
\end{table*}

\begin{figure*}[!ht]
  \vspace{-0.7cm}
  \begin{tabular}{p{0.04\textwidth-\tabcolsep}p{ 0.3\textwidth}p{ 0.3\textwidth}p{ 0.3\textwidth}}
    \multirow{9}{*}{\rotatebox[origin=c]{90}{Point process}}                                   &
    \raisebox{-\height}{\includegraphics[width=0.7\linewidth]{ginibre_pp_ball.pdf}} &
    \raisebox{-\height}{\includegraphics[width=0.7\linewidth]{poisson_pp_ball.pdf}} &
    \raisebox{-\height}{\includegraphics[width=0.7\linewidth]{thomas_pp_ball.pdf}}
  \end{tabular}
  \begin{tabular}{p{ 0.03\textwidth-\tabcolsep}p{ 0.3\textwidth}p{ 0.3\textwidth}p{ 0.3\textwidth}}
    \multirow{9}{*}{\rotatebox[origin=c]{90}{$\bar{\widehat{S}}$}}                                           &
    \raisebox{-\height}{\includegraphics[width=0.8\linewidth]{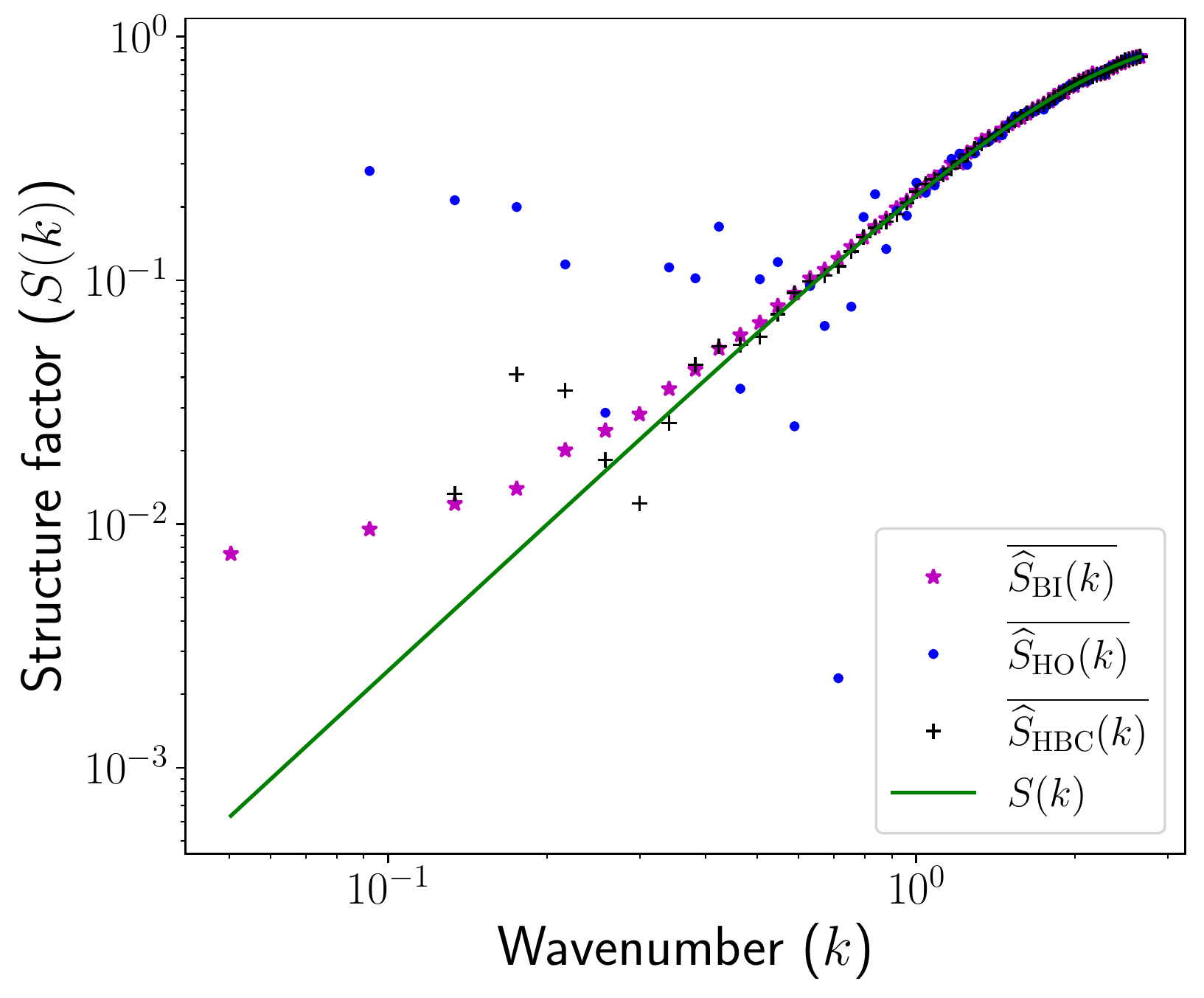}}  &
    \raisebox{-\height}{\includegraphics[width=0.86\linewidth]{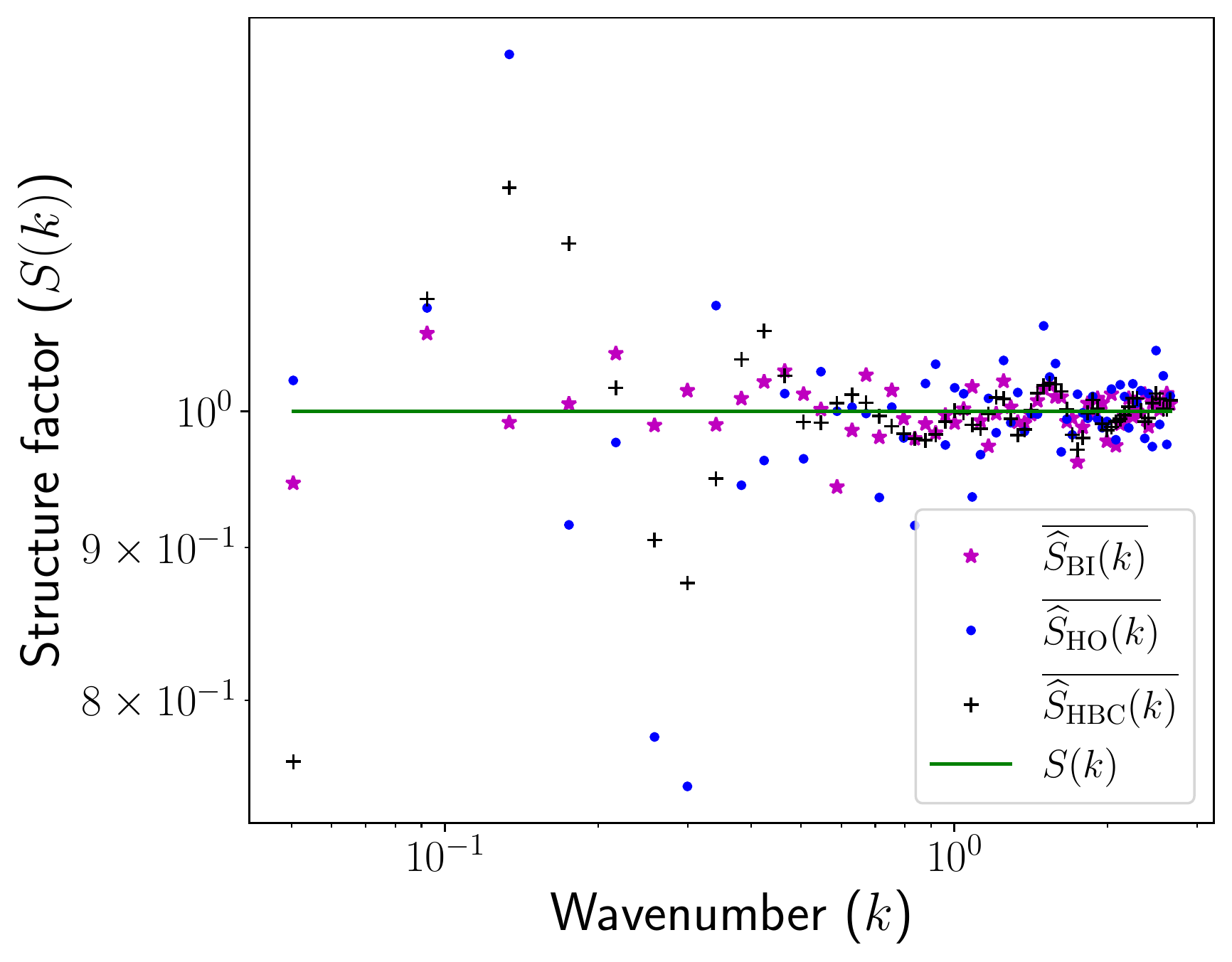}} &
    \raisebox{-\height}{\includegraphics[width=0.8\linewidth]{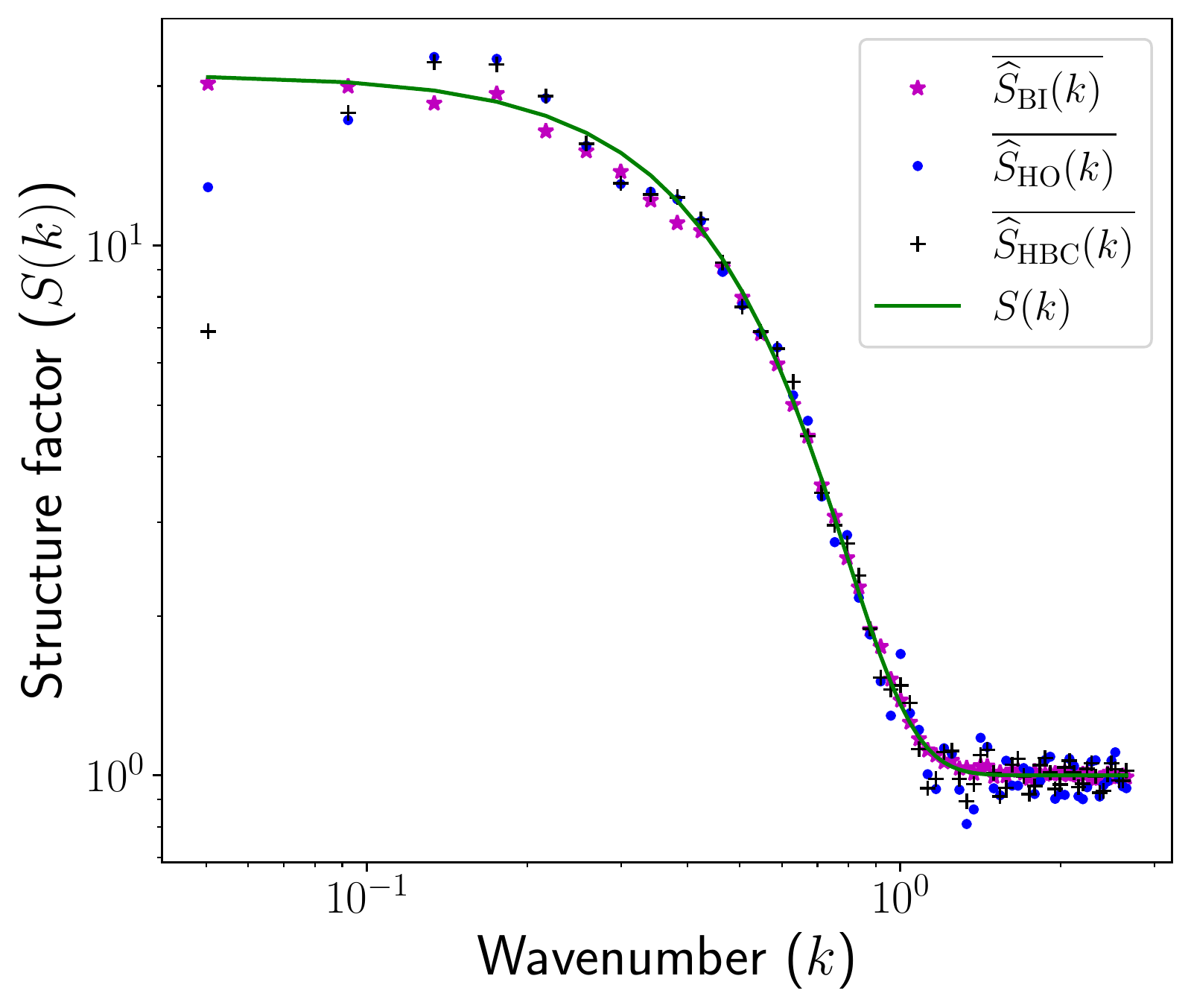}}     \\
    \caption*{}                                                                                              &
    \vspace{-0.5cm}
    \caption*{{\fontfamily{pcr}\selectfont } Ginibre }                                              &
    \vspace{-0.5cm}
    \caption*{{\fontfamily{pcr}\selectfont } Poisson }                                               &
    \vspace{-0.5cm}
    \caption*{{\fontfamily{pcr}\selectfont } Thomas}
  \end{tabular}
  \vspace{-0.8cm}
  \caption{Pointwise averages of the three estimators that assume isotropy, across 50 independent samples of the Ginibre, Poisson, and Thomas point processes. The computation and visualization are done using \toolbox{}}
  \label{fig:mean_estimators_ISE}
\end{figure*}

Finally, Figure~\ref{fig:mean_estimators_ISE} shows the pointwise average of each estimator, across the $50$ point process realizations.
The accuracy of Ogata's quadrature at small wavenumbers is poor. The bias of the two Hankel quadrature estimators is well visible at small $k$ for all point processes.
This is likely due to poor estimation of the pair correlation function at large scales.

\section{Summary and discussion}
\label{Summary and discussion}
We surveyed estimators of the structure factor of a stationary point process, along with numerical diagnostics of hyperuniformity.
We also proposed the first (asymptotic) test of hyperuniformity, based on the debiasing techniques of \citep{Rhee+Glynn:2015}.
Interestingly, while we were working on the revision of the current paper, \cite*{Klatt+al:2022} proposed a different test for Class \rom{1} hyperuniformity, which only requires one (large) sample.
A rigorous comparison of both tests will require a careful examination of trade-offs in computing time and test power and may be explored in future work.

We provided an open-source Python toolbox \toolbox{} containing all estimators and diagnostics and used it to benchmark known estimators.
The two estimators that fared best in our benchmark are the multitapered variant of the scattering intensity and, in the case of isotropic point processes, Bartlett's isotropic estimator.
Further comparing these two estimators is a natural avenue for further work, but needs careful thinking.
For starters, the two estimators are not defined on similar windows, and they do not require the same number of evaluations for a similar accuracy.
Then, it is not clear how to best choose the tapers in the multitaper estimator, not even mentioning that a multitapered version of Bartlett's estimator can also in principle be derived.
On the other hand, the computational cost of Bartlett's estimator should be lowered, e.g. by subsampling pairs of points.

Estimators based on Hankel transform quadratures, which rely on first estimating the pair correlation function $g$, comparatively showed poor performance.
But it is still possible that estimating $g$ first can help estimating $S$ if, say, edge correction plays an important role.
If the user only has data collected on a non-rectangular, non-ball window, we would then recommend trying the Hankel-Baddour-Chouinard estimator.
For our benchmarks, though, the large cardinalities and the regular windows involved do not build upon this strong point of quadrature-based estimators.
It would also be interesting to investigate edge correction,
 or regularization schemes for the pair correlation function, that are tailored to structure factor estimation.

On another note, since the structure factor is a Fourier transform, one might be tempted to periodize one's sample before computing estimators.
However, we do not see a clear argument for (or against) periodization yet.
First, if one is using the scattering intensity applied on any of the restricted allowed wavectors in $\bbA_L^{res}$, then periodizing the point process has no effect on the estimator.
The case of wavevectors in $\bbA_L$, and of other estimators, is less straightforward.
Second, adding periodic boundaries forces an arbitrary regularity at long distances, which intuitively would impact any empirical diagnostic of hyperuniformity.

On the point process side, we hope that our survey and software can contribute to the standardization and reproducibility of empirical investigations involving structure factors, including the study of hyperuniform point processes.

\paragraph*{Acknowledgements}
We thank Jean-Fran\c cois Coeurjolly, Michael Andreas Klatt, Günter Last, David Dereudre, and Simon Coste for insightful discussions along this project.
The motivation for our multiscale test was sparked at the Karlsruhe workshop on New trends in point process theory in March 2022, by the lively discussions with the participants about a preliminary draft of our paper. We know from personal communication that, by then, Günter Last, Andreas Klatt, and Norbert Henze were independently working on their own test, whose preprint \cite*{Klatt+al:2022} came out as we were answering the referees on our own manuscript.
Finally, we thank a referee for suggesting thinning a hyperuniform point process to assess the detection performance of our test. It has been independently brought to our attention that this procedure can also be found in \citep{Klatt+al:2022,Torquato+al:2018}.

\paragraph*{Funding}
This work is supported by ERC-2019-STG-851866 and ANR-20-CHIA-0002.
%Declaration
\section*{Declarations}
\paragraph*{Conflicts of interest/Competing interests}
The authors have no conflicts
of interest to declare that relate to the content of this article.
\paragraph*{Code availability}

The code is published as an open-source \texttt{Python} toolbox under the project name \texttt{structure-factor}.
The package is licensed under the \href{MIT license}{https://github.com/For-a-few-DPPs-more/structure-factor/blob/main/LICENSE} and is available on \href{Github}{https://github.com/For-a-few-DPPs-more/structure-factor/tree/tapered_periodogram}\footnote{\url{https://github.com/For-a-few-DPPs-more/structure-factor}} and \href{PyPI}{https://pypi.org/project/structure-factor/}\footnote{\url{https://pypi.org/project/structure-factor/}}.

\bibliography{sn-bibliography}

\appendix
\section{Proof of Proposition~\ref{prop:hu_test}}
\label{proof:prop3}

\begin{proof}
  Let $M \in L^p$ with $p>0$. We first prove that $Z_m\rightarrow Z$ in $L^p$.
  As we have $Z_m \rightarrow Z$ a.s., it is enough to show that $Z_m$ is uniformly bounded in $L^p$.
  For a realization $M'$ of $M$ we have,
  $$\lvert Z_m \rvert \leq \sum_{j=1}^{m \wedge M'} \frac{\lvert Y_{j} - Y_{j-1} \rvert }{\bbP (M \geq j)} \leq \frac{M'}{\bbP(M \geq M')}.$$
  By assumption $M \in L^p$ so $Z_m$ is uniformly bounded in $L^p$.
  This proves the first part of the proposition.

  Before proving the additional two points, note that, since $S$ is continuous,
  \begin{equation}
  \label{eq:conv_S_m}
  \bbE [\widehat{S}_m(\bfk_m^{\text{min}})] \xrightarrow[m\rightarrow \infty]{} S(\bf0).
  \end{equation}
  Now, let us prove the first point of the proposition.
  Assume that $M \in L^1$ and $\calX$ is hyperuniform, so that $S(\mathbf{0} )= 0$.
  Since $\hat S_m$ is nonnegative, Equation \eqref{eq:conv_S_m} yields
  $$\widehat{S}_m(\bfk_m^{\text{min}}) \xrightarrow[m\rightarrow \infty]{ L^1} 0.$$
  Moreover, letting $f:x\mapsto 1\wedge x$, $\vert f(x)\vert\leq x$ on $\mathbb{R}^+$, so that
  $$\mathbb{E} [\vert f(\widehat{S}_m(\bfk_m^{\text{min}}))\vert] \leq \mathbb{E} [ \widehat{S}_m(\bfk_m^{\text{min}}) ]\rightarrow 0,$$
%  Moreover, $f:x\mapsto 1\wedge x$ is uniformly continuous and $(Y_m)_m$ are uniformly integrable in $L^p$, so
  and \begin{equation}
  \label{eq:conv_Y_m}
  Y_m = f(\widehat{S}_m(\bfk_m^{\text{min}})) \xrightarrow[m\rightarrow \infty]{ L^1} 0.
  \end{equation}
  Since $\bbE [Y_m]= \bbE[Z_m]$ and $Z_m$ converges in $L^1$ to $Z$, by unicity of the limit, we have $\bbE [Z]=0$.
  %Using Equations~\eqref{eq:conv_Z_m}, and \eqref{eq:conv_Y_m}, with $p=1$ we get $\bbE [Z]=0$.

  It remains to show the last point of the proposition.
  Assume again that $M \in L^1$, but that $\calX$ is not hyperuniform, so that $S( \mathbf{0}) >0$.
  Reasoning by contradiction, assume that $\bbE [Z]=0$.
  As $\bbE [Y_m]= \bbE[Z_m]$ and $Z_m$ converges in $L^1$ to $Z$, we get
  \begin{equation}
    \label{eq:conv_exp_1}
      \bbE [\widehat{S}_m(k_m^{\text{min}}) \mathds{1}_{\{ \widehat{S}_m(k_m^{\text{min}})<1 \}}] \xrightarrow[m \rightarrow \infty]{} 0,
  \end{equation}
  and
  \begin{equation}
  \label{eq:conv_exp_2}
  \bbE [ \mathds{1}_{\{ \widehat{S}_m(k_m^{\text{min}}) \geq 1 \}}] \xrightarrow[m \rightarrow \infty]{} 0.
  \end{equation}
  Meanwhile,
  $$\bbE [\widehat{S}_m(k_m^{\text{min}})] \xrightarrow[m \rightarrow \infty]{} S(\mathbf{0} ) >0.$$
  Using Equation~\eqref{eq:conv_exp_1}, we get
  \begin{equation}
  \label{eq:to_contradict}
  \bbE [\widehat{S}_m(k_m^{\text{min}}) \mathds{1}_{\{ \widehat{S}_m(k_m^{\text{min}}) \geq 1 \}}] \xrightarrow[m \rightarrow \infty]{} S(\mathbf{0}) >0.
  \end{equation}
  Finally, Cauchy-Schwarz, together with Condition \eqref{e:boundedness_condition} and Equation~\eqref{eq:conv_exp_2} yield
  \begin{align*}
   \bbE [\widehat{S}_m(k_m^{\text{min}}) \mathds{1}_{\{ \widehat{S}_m(k_m^{\text{min}}) \geq 1 \}}]
   &\leq \bbE ^{1/2}[\widehat{S}^2_m(k_m^{\text{min}})] \times \bbE^{1/2} [\mathds{1}_{\{ \widehat{S}_m(k_m^{\text{min}}) \geq 1 \}}]
   \\
   & \rightarrow 0,
  \end{align*}
  which contradicts Equation~\eqref{eq:to_contradict} and ends the proof.
\end{proof}
 \section{Validity of Assumption~\eqref{e:boundedness_condition}} \label{appB: condition_of_prop3}
 In what follows, we show that Assumption~\eqref{e:boundedness_condition} is satisfied for a homogeneous Poisson point process $\cal X$ of intensity $\rho$, $\widehat{S}= \widehat{S}_{\mathrm{SI}}$, and $W_m$ are increasing rectangular windows.

Let $N_m = \leb{\calX \cap W_m}$.
Then
\begin{align*}
  \rho^2  \leb{W_m}^2 \bbE  [\widehat{S}^2_m(\bfk) ]
  & = \bbE  \left (\left \lvert \sum_{\bfx \in  \calX \cap W_m } e^{- \i \langle \bfk, \bfx \rangle } \right \rvert^2 \right )^2  \\
  &= \bbE   \Big[ N_m^2 + \big (\sum_{\bfx, \bfy \in  \calX \cap W_m }^{\bfx \neq \bfy} e^{- \i \langle \bfk, \bfx - \bfy \rangle } \big ) ^2 + 2 N_m \sum_{\bfx, \bfy \in  \calX \cap W_m }^{\bfx \neq \bfy} e^{- \i \langle \bfk, \bfx - \bfy \rangle } \Big]
  \\
  &= \bbE   \Big[ 2 N_m^2 - N_m
  \\
  & + \sum_{\bfx, \bfy \in  \calX \cap W_m }^{\bfx \neq \bfy} 2 N_me^{- \i \langle \bfk, \bfx - \bfy \rangle }
  + e^{- \i 2 \langle \bfk, \bfx - \bfy \rangle }
  \\
  &+
  \sum_{\bfx, \bfy, \bfz \in  \calX \cap W_m }^{\bfx \neq \bfy \neq \bfz} 2 e^{- \i \langle \bfk, \bfx - \bfy \rangle }
  + e^{- \i \langle \bfk, 2 \bfx - \bfy - \bfz \rangle } + e^{ \i \langle \bfk, 2 \bfx - \bfy - \bfz \rangle }
  \\
  &+
  \sum_{\bfx, \bfy, \bfz, \bft \in  \calX \cap W_m }^{\bfx \neq \bfy \neq \bfz \neq \bft} e^{- \i \langle \bfk, \bfx - \bfy + \bfz - \bft\rangle }
  \Big ]
  \\
  &=
  \bbE   \Big[ 2 N_m^2 - N_m +  \sum_{\bfx, \bfy \in  \calX \cap W_m }^{\bfx \neq \bfy} 4 e^{- \i \langle \bfk, \bfx - \bfy \rangle }
  + e^{- \i 2 \langle \bfk, \bfx - \bfy \rangle }
  \\
  &+
  \sum_{\bfx, \bfy, \bfz \in  \calX \cap W_m }^{\bfx \neq \bfy \neq \bfz} 4 e^{- \i \langle \bfk, \bfx - \bfy \rangle }
  + e^{- \i \langle \bfk, 2 \bfx - \bfy - \bfz \rangle } + e^{ \i \langle \bfk, 2 \bfx - \bfy - \bfz \rangle }
  \\
  &+
  \sum_{\bfx, \bfy, \bfz, \bft \in  \calX \cap W_m }^{\bfx \neq \bfy \neq \bfz \neq \bft} e^{- \i \langle \bfk, \bfx - \bfy + \bfz - \bft\rangle }
   \Big ]
   \\
   &= \bbE [2 N_m^2 - N_m]
   \\
   &+ \int_{W_m \times W_m} \Big (4 e^{- \i \langle \bfk, \bfx - \bfy \rangle }
   + e^{- \i 2 \langle \bfk, \bfx - \bfy \rangle } \Big) \rho^2 \mathrm{d}\bfx \mathrm{d}\bfy
   \\
   & +
   \int_{W_m \times W_m \times W_m} \Big (4 e^{- \i \langle \bfk, \bfx - \bfy \rangle }
   + e^{- \i \langle \bfk, 2 \bfx - \bfy - \bfz \rangle }
   \\
   & \hspace{3cm}+ e^{ \i \langle \bfk, 2 \bfx - \bfy - \bfz \rangle } \Big) \rho^3 \mathrm{d}\bfx \mathrm{d}\bfy \mathrm{d}\bfz\\
   &+ \int_{W_m \times W_m \times W_m \times W_m} e^{- \i \langle \bfk, \bfx - \bfy + \bfz - \bft\rangle } \rho^4  \mathrm{d}\bfx \mathrm{d}\bfy \mathrm{d}\bfz \mathrm{d}\bft.
\end{align*}
The last line was obtained using the definition of the $n$-th product density $\rho^{(n)}$ and that for any $n \geq 1$, $\rho^{(n)}$ of $\calX$ simplifies to $\rho^n$.
It is a well-known property of homogeneous Poisson point processes \cite[Section 2.3.3]{Chiu+Stoyan+al:2013}.
Now, using the parity of $\mathds{1}_{W_m}$ and that $N_m$ is a Poisson r.v., we get
\begin{align}
  \label{eq:var_s_m}
  \nonumber
  \bbE  [\widehat{S}^2_m(\bfk_m^{\text{min}}) ]=
  & \frac{1}{(\rho \leb{W_m})^2 }
  \Big [
  \rho \leb{W_m} + 2 (\rho \leb{W_m})^2  \\
  & + \rho^4 \calF^4  (\mathds{1}_{W_m})(\bfk_m^{\text{min}})
  \nonumber
  \\
  & + \rho^2 \Big (4 \calF^2(\mathds{1}_{W_m})(\bfk_m^{\text{min}}) + \calF^2(\mathds{1}_{W_m})(2 \bfk_m^{\text{min}})\Big )
  \nonumber
  \\
  &+ \rho^3 \Big ( 4 \leb{W_m} \calF^2(\mathds{1}_{W_m})(\bfk_m^{\text{min}}) +
  \nonumber
  \\
  &\hspace{0.7cm} 2 \calF(\mathds{1}_{W_m})(2 \bfk_m^{\text{min}}) \calF^2(\mathds{1}_{W_m})(\bfk_m^{\text{min}}) \Big )
  \Big].
\end{align}
Upon noting that $\bfk_m^{\text{min}} = (\frac{2 \pi}{L_1}, \cdots, \frac{2 \pi}{L_d})$ and
$$\calF(\mathds{1}_{W_m})(\bfk)= \prod_{j=1}^d\frac{\sin(k_j L_j/2)}{k_j/2}.$$
Equation~\eqref{eq:var_s_m} simplifies to
$$\bbE  [\widehat{S}^2_m(\bfk_m^{\text{min}}) ] = \frac{1}{\rho \leb{W_m}} + 2. $$
Thus Assumption~\eqref{e:boundedness_condition} holds.
\end{document}